\documentclass[aps,physrev,twocolumn,unsortedaddress,eqsecnum]{revtex4-2}
\usepackage{dcolumn}
\usepackage{bm}
\usepackage{epsfig}
\usepackage{physics,verbatim}
\usepackage{epstopdf}
\usepackage{color}
\usepackage{float}

\newcommand{\eq}[1]{(\ref{eq:#1})}

\def\a{{\alpha}}
\begin{document}

\title{\textbf{Quantum field theory approach for multistage chemical kinetics in liquids}}

\author{Roman~V.~Li}
\affiliation{Voevodsky Institute of Chemical Kinetics and Combustion, Novosibirsk, 630090, Russian Federation}

\author{Oleg~A.~Igoshin}
\affiliation{Department of Bioengineering}
\affiliation{Department of BioSciences}
\affiliation{Department of Chemistry}
\affiliation{Center for Theoretical Biological Physics, Rice University, Houston, TX 77005, USA}
\author{Evgeny~B.~{Krissinel}}
\affiliation{Research Complex at Harwell, Scientific Computing, Science and Technology Facilities Council, Didcot, OX11 0FA, United Kingdom}

\author{Pavel~A.~Frantsuzov}
\affiliation{Voevodsky Institute of Chemical Kinetics and Combustion, Novosibirsk, 630090, Russian Federation}
\email{frantsuzov@kinetics.nsc.ru}

\date{\today }

\vspace{12pt}

\begin{abstract}
    Reaction-diffusion processes play an important role in a variety of physical, chemical, and biological systems. Conventionally, the kinetics of these processes are described by the law of mass action. However, there are various cases where these equations are insufficient. A fundamental challenge lies in accurately accounting for the microscopic correlations that inevitably arise in bimolecular reactions. While approaches to describe microscopic correlations in many specific cases exist, no general theory for multistage reactions has been established. In this article, we apply the quantum field theory approach to derive kinetic equations for general multistage reactive systems termed CMET (complete modified encounter theory).  CMET can be formulated as a set of coupled partial differential equations that can be easily integrated numerically, thereby serving as a versatile tool for investigating reaction-diffusion processes. Across multiple case studies, we demonstrated that CMET reproduces the kinetics predicted by many other theories within their respective scopes of applicability.
\end{abstract}
\maketitle

\section{Introduction}
\label{sec:Introduction}
A broad spectrum of physical, chemical, and biological phenomena can be well described in terms of particles that undergo diffusion in a medium while participating in elementary reactions, such as first-order (unimolecular) and second-order (bimolecular) transformations. Representative examples span multiple disciplines, including photodissociation in liquids \cite{Harris_1988}, electron-hole recombination in semiconductors \cite{Seeger}, radical pair reactions in spin chemistry \cite{Salikhov}, and reaction–diffusion processes observed in soft-matter systems, living cells, and biological tissues \cite{GrzybowskiiACIEE2010}. The kinetics of such systems have been studied for more than a century, dating back to the pioneering works of Smoluchowski \cite{Sm}, Fisher \cite{FisherAE1937}, Kolmogorov, Petrovskii, and Piskunov \cite{KolmogorovBMU1937},  Zeldovich and Frank-Kamenetsky \cite{ZeldovichAP1938},  Debye \cite{DebyeTES1942}, Turing \cite{TuringPTRSL1952} and their contemporaries; however, our understanding remains incomplete. In particular, the interplay between spatial transport, microscopic correlations, and reaction dynamics continues to reveal unexpected deviations from the classical laws of mass action (LMA). Phenomenological LMA  approaches typically employ a set of coupled  equations  describing the reaction-diffusion process\cite{Zhabotinsky}:
\begin{widetext}
\begin{equation}
\frac{\partial}{\partial t} C_i(\mathbf r,t) =
D_i\nabla^2 C_i(\mathbf r,t)+\sum_{j} \big[ q_{i,j}C_j(\mathbf r,t) - q_{j,i}C_i(\mathbf r,t) \big]+
\sum_{jkl} \big[ k_{ij,kl}C_k(\mathbf r,t)C_l(\mathbf r,t)-
k_{kl,ij}C_i(\mathbf r,t)C_j(\mathbf r,t) \big]
,
\label{eq:RDE}
\end{equation}
\end{widetext}
where $C_i(\mathbf r,t)$ is the concentration of species $A_i$ at position $\mathbf r$ and time $t$, $D_i$ is its diffusion coefficient, $q_{i,j}$ is the unimolecular rate constant for $A_j \to A_i$, and $k_{kl,ij}$ is the bimolecular rate constant for $A_k + A_l \to A_i + A_j$. The first term on the right-hand side accounts for spatial diffusion, the second for gain and loss due to unimolecular transformations, and the third for analogous changes from bimolecular reactions. While Eq. (\ref{eq:RDE}) offers a compact and widely used macroscopic description, deriving it from microscopic dynamics and identifying its limitations remain central challenges in nonequilibrium statistical mechanics.

While the mass-action description of bimolecular reaction kinetics for reaction-diffusion systems is widely used, it is well known to be limited to cases where reactants' diffusion is so fast that spatial correlations in their positions can be ignored. Indeed, to describe the reaction rate in the last term of Eq.(\ref{eq:RDE}) as the product of local concentration of reactants, one has to assume that their positions before the reaction are uncorrelated. This assumption is analogous to that made in the formulation of the Boltzmann equation in kinetic theory, which postulates that particle positions/velocities are uncorrelated before collisions \cite{ehrenfest2014conceptual}. However, both physical and chemical interactions between the particles render this assumption invalid \cite{Murphy_2024}, particularly in the diffusion-limited regime, where spatial correlations persist over extended timescales. For example, consider the case of geminate recombination when the reactants are generated as products of a prior reaction and are consequently formed in close proximity \cite{BurshteinACP2004}. As a result, these reactants have an enhanced probability of recombining, which is not accounted for in simple mass-action kinetics.

Several studies have demonstrated significant deviations from mass-action kinetics in a wide range of fields and applications. Examples include absorption of molecules into metal-organic frameworks \cite{Ma2021,Zeng2018}, metabolic channeling, regulation of signal transduction,  pharmacokinetics \cite{Klann2011}, and catalysis \cite{Lee2016}. Geminate recombination plays an essential role in electron-hole recombination kinetics in lead halide perovskites \cite{GoldsmithJPCC2017,GulbinasAEM2017,HempelArXiv2024} and in organic collar cells \cite{KosterNatureComm2017}.

To address the limitations of mass-action kinetics arising from spatial correlations, one can begin with a rigorous microscopic description based on the Bogoliubov–Born–Green–Kirkwood–Yvon (BBGKY) hierarchy, which governs the time evolution of many-particle distribution functions \cite{Balescu}. The BBGKY framework naturally captures the interplay between diffusion and reactions at all correlation orders but leads to an infinite hierarchy of coupled equations. To render the problem tractable, various closure approximations have been proposed \cite{Forster1974, lu2025}. The resulting formalism often yields integral-differential equations for concentrations that feature memory kernels, accounting for the delayed effects of correlations and reaction history \cite{Forster1974}. The alternative form extends the classical reaction–diffusion description to a coupled system of equations involving both concentrations and pair correlation functions, explicitly accounting for spatial correlations between reactants \cite{lu2025}. However, despite these efforts, a comprehensive first-principles derivation of kinetic equations for arbitrary reaction networks—including unimolecular and bimolecular reactions with physical interactions between particles remains an open problem.

Quantum field theory (QFT) approaches are promising for formulating a first-principles derivation. Doi's pioneering work showed that the formalism of quantum field theory can be applied to classical many-particle systems \cite{DoiJPA1976} and especially to diffusion-influenced chemical reactions \cite{DoiJPA1976s}. A similar approach was independently suggested by Zeldovich and Ovchinnikov  \cite{ZeldovichJETP1978}.
Subsequently, quantum field theory methods were applied to describe specific diffusion-influenced reactions in multiple publications
\cite{YashinJSP1985,CardyJSP1995,LeeCardyJSP1995-er,RudavetsJPA1993,TauberJPA2005,LucivjanskyTMP2011,LucivjanskyTMP2011s,HanaiJSP2023}
 (see also the review Ref. \onlinecite{GlasserRMP1998}).
Despite the effectiveness of this approach, quantum field theory methods have not yet been used to describe the general case of multistage reactions.

In this paper, we apply the second-quantization approach \cite{DoiJPA1976} to derive rigorously the kinetic equations for dilute solutions of an arbitrary multistage reactive system. To this end, we construct a diagrammatic technique to describe reaction kinetics and apply a regular expansion of the collision integral in terms of the small parameter. As a result, we obtained universal kinetic equations for concentrations. We demonstrate that the resulting equations contain additional terms compared to the phenomenologically formulated equations in Ref. \onlinecite{FrantsuzovCPL2000}, and thus refer to them as the complete modified encounter theory (CMET). The simulation results demonstrate that the CMET kinetic equations accurately describe all known kinetic regimes of multistage diffusion-influenced reactions, including non-stationary and stationary bulk recombination, geminate recombination, and the fluctuation asymptotic regime.

\section{Theoretical Background}
To orient readers, we begin with a brief overview of theoretical approaches for formulating kinetic equations for reactive systems beyond LMA \eq{RDE}, including the resulting equations.

In his pioneering work, Marian  Smoluchowski \cite{Sm} introduced the concept of time-dependent rate  $k(t)$ in diffusion-influenced reaction.
In the Smoluchowski approach (SM), the reaction rate is determined from the diffusion equation for a pair distribution function with an absorbing boundary condition at the contact distance.  The approach was then considerably improved by taking into account  the partially absorbing boundary condition \cite{CK,Waite,WF} and
remote reactions such as electron or energy transfer  \cite{KMR,TunBag} (see also Ref. \onlinecite{Rice} and references therein).

In the case of the irreversible reaction $A+B  \rightarrow C+B$, Smoluchowski theory results in the differential form of the kinetic equation for the $A$'s concentration ($C_A$):
\begin{equation}
\frac d{dt} C_A(t) = -k(t) C_A(t) C_B
\label{eq:DET}
\end{equation}
The time-dependent rate constant is determined as
\begin{equation}
    k(t) = \int_b^{\infty} W(r) n(r, t)\dd^3 r,
   \label{eq:Kt}
\end{equation}
where $b$ is the contact distance, $W(r)$ is the distance-dependent reaction rate between particles $A$ and $B$ when the distance between them is $r$ and  $n(r, t)$ is defined as
$$P_{AB}(r,t)=n(r,t) C_A(t) C_B$$
where $P_{AB}(r,t)$ is a pair distribution function of $A$ and $B$. The function $n(r,t)$ obeys the following equation
\begin{equation}
    \pdv{t} n(r, t) =  D \nabla \left(\nabla n(r, t)+ \left[\nabla U(r)\right]n(r, t) \right) - W(r)n(r, t).
     \label{eq:n}
\end{equation}
Here, $D=D_A+D_B$ is the diffusion coefficient of the relative motion of the pair
and $U(r)$ is the interaction potential in units of $kT$.
The following initial condition corresponds to the absence of the correlation between particles at $t=0$:
$$ n(r, 0) = 1.$$
Boundary conditions corresponding to the absence of the correlation between particles at $r\to\infty$ and no flux at $r=b$ are typically imposed:
\begin{eqnarray*}
    \begin{gathered}
        \lim_{r\rightarrow\infty}n(r, t) = 1; \\ \left(\nabla n(r, t)+ \left[\nabla U(r)\right]n(r, t)\right)\bigg|_{r=b}=0,
    \end{gathered}
\end{eqnarray*}
 In the long-time limit, Eq.(\ref{eq:Kt}) becomes  equivalent to the LMA  kinetic equation, with the steady-state rate constant
 \begin{equation}
 k=\lim_{t\rightarrow \infty} k(t)=4\pi R_D D
 \label{RD}
 \end{equation}
Here, the last equality defines an effective reaction radius $R_D$. In the case of the absorbing condition \cite{Sm}, the effective reaction radius is equal to the contact distance $b$.
Eqs.(\ref{eq:DET}-\ref{eq:Kt}) are only valid when the probability of a three-particle encounter is small. It is equivalent to the following applicability condition:
\begin{equation}
 \xi  \ll 1
 \label{Cond}
 \end{equation}
where parameter $\xi$ is defined as
 \begin{equation}
    \xi = \frac 4 3 \pi R_D^3 C_B=\frac {k^3} {48 \pi^2 D^3}  C_B
    \label{xiAB}
 \end{equation}

It was shown \cite{AllBlum,Szabo} that the SM approach gives the exact kinetics $C_A(t)$ in one
special case, when $A$ particles are immobile, i.e., in the so-called scavenger problem.
The time dependent rate Eq.(\ref{eq:Kt}) has the following universal asymptotic dependence \cite{Rice}:
\begin{equation}
   k(t) \approx k \left(1 + \frac {R_D} {\sqrt{\pi Dt}}\right)
\label{eq:Kasymp}
\end{equation}
As a result, the long-term kinetics in the SM approach also differ from a simple exponential dependence:
  \begin{equation}
   C_A(t) \approx C_A(0)  \exp\left[-kC_B \left(t +2 R_D \sqrt{\frac t{\pi D}}\right)\right]
\label{eq:Asymp}
\end{equation}
Thus, for the scavenger problem, the solution of the equation (\ref{eq:DET}) shows non-exponential kinetics of $C_A(t)$ at short times (static quenching) $\quad t\lesssim {R_D^2}/ D$.

In the case of immobile $B$-particles (the trapping problem), Eq.(\ref{eq:Asymp}) accurately describes kinetics on an intermediate time-scale only.
At large times, the kinetics takes a different form, due to fluctuations in the initial positions of the particles $B$ (fluctuation asymptotics):
\cite{BalagurovVaks, ZeldovichCP1978, ProcacciaJCP1975}:
 \begin{equation}
   C_A(t) \approx C_A(0)  C_B^{1/5} (Dt)^{3/10}  \exp \left[- \lambda  C_B^{2/5} (Dt)^{3/5}\right]
\label{eq:Balagurov}
\end{equation}
where $\lambda$ is a numerical coefficient.

In the general case of the SM approach, the kinetics depends on the sum $D_A+D_B$, so it should describe both trapping and scavenger problems in the corresponding limits.  Thus, as noted in Refs. \cite{BalagurovVaks,ZeldovichCP1978}  the reaction rate approach is applicable for times when the fluctuation asymptotics Eq.(\ref{eq:Balagurov}) is much smaller than Eq.(\ref{eq:Asymp})
\begin{equation}
t\ll   \left(\sqrt{ \xi} kC_B\right)^{-1}
\label{eq:SMTimeLimit}
\end{equation}
Notably, both Eq.(\ref{eq:Balagurov}) and Eq.(\ref{eq:SMTimeLimit}) are only valid when the initial distribution of reactants is uncorrelated.

Another important example of the fluctuation asymptotics is the reaction $A+B\longrightarrow C+D$ with equal initial concentrations of the $A$ and $B$ particles $C_A(0)=C_B(0)$ and diffusion coefficients $D_A=D_B$.
The time dependence of the concentrations within the SM approach is given by the kinetic equation
\begin{equation}
\frac d{dt} C_A(t) = -k(t) C^2_A(t)
\label{eq:AB}
\end{equation}
Using Eqs.(\ref{eq:AB}) and (\ref{eq:Kasymp}) we get the asymptotic behavior of the concentration:
\begin{equation}
   C_A(t) \approx \left[k\left(t +2 R_D \sqrt{\frac t{\pi D}}\right)\right]^{-1}
\label{eq:ABasymp}
\end{equation}
However, as was predicted by  Ovchinnikov and Zeldovich \cite{ZeldovichCP1978}
and subsequently confirmed by other authors \cite{WilczekJCP83,CardyJSP1995,BensonJCompPhys2014}  fluctuations in the initial positions of the particles cause a different asymptotics:
\begin{equation}
C_{A}(t) \sim  \sqrt{C_A(0)}(Dt)^{-3/4}
\label{eq:Zeldovich}
\end{equation}
The time dependence $t^{-3/4}$  of the concentration was subsequently observed experimentally \cite{KopelmanPRE2004}.

The presence of initial correlations in pairs gives rise to another class of chemical processes, geminate reactions, whose kinetics are not adequately described by LMA.
When the reactants $A$ and $B$ are generated as a result of the molecule fragmentation, they appear in pairs ($ C_A(0)= C_B(0)$) with a short average distance $\bar r $ between them
$$\bar r^3 C_A (0)\ll 1$$
A common way to describe the kinetics of a geminate reaction is to use a pair distribution function $p_{AB}(r,t)$, which obeys the following equation:
\begin{equation}
    \begin{gathered}
        \pdv{}{t} p_{AB}(r, t) =  D \nabla \left(\nabla p_{AB}(r, t) \left[\nabla U(r)\right]p_{AB}(r, t) \right) -\\- W(r)p_{AB}(r, t)
    \end{gathered}
    \label{eq:m}
\end{equation}
 with the initial and boundary  conditions
 \begin{eqnarray*}
 \begin{gathered}
     p_{AB}(r,0)=p_0(r); \qquad \lim_{r\rightarrow\infty}p_{AB}(r, t) = 0; \\ \left(\nabla p_{AB}(r, t)+ \left[\nabla U(r)\right]p_{AB}(r, t) \right)\bigg|_{r=b}=0.
 \end{gathered}
 \end{eqnarray*}
Here $p_0(r)$ is the initial pair distribution function, normalized by
$$\int p_0(r)\,d^3r=C_A(0)=C_B(0).$$
 The time dependence of the reactant concentration can be found as
 \begin{equation}
 C_A(t)=C_B(t)=\int p_{AB}(r,t)\,\dd^3 r
 \label{eq:Cgem}
 \end{equation}
Geminate pairs cannot be considered isolated when their separation exceeds the average distance between particles. Thus, Eq.(\ref{eq:m}) is valid for times
\begin{equation}
t\ll \frac 1 {C_A^{2/3}(0)D}
\label{eq:tgem}
\end{equation}

The generation of geminate pairs can also occur as a result of a bimolecular reaction, for example, the electron transfer from an excited donor molecule to an acceptor molecule in the bulk. This problem can be considered using a combination of the SM approach and geminate kinetics within the so-called unified theory (UT), independently proposed by Burshtein \cite{BurshteinCPL1992} and Dorfman and Fayer \cite{DorfmanFayer}.

Early attempts to go beyond focusing on just a pair of reactants as in SM and to construct a many-particle theory were based on the use of the Kirkwood \cite{Kirkwd} superposition approximation (SA)  \cite{WF,Monch,Waite,Kapral},  or its generalizations \cite{KarplusJCP1997,AgSzabo,MolKaiz,KK,LeeJCP1999I,LeeJCP1999II,LeeJCP1999III,LeeJCP2000}.
The SA being applied to reaction $A+B  \rightarrow C+B$ gives the SM kinetic equations (\ref{eq:DET}-\ref{eq:n}). However, applying the SA to more complex reaction networks was not as successful.
For example, when applied to a multistage reactive system, SA produced results inconsistent with simple physical considerations \cite{DoktorovRRDCP2012}.

A more systematic description of many-particle reactive systems used theoretical approaches borrowed from solid-state theory, namely the averaged $T$-matrix approximation, which is widely used in the theory of disordered systems (cf. Refs. \cite{Lifshitz,Elliot,Ziman}). This approach was explicitly used to describe a diffusion-influenced reaction kinetics by Bixon and Zwanzig \cite{ZwanzigJCP1981}. However, an earlier approach developed by Sakun \cite{SakunPhysicaA1975} and Doktorov \cite{DoktorovPhysicaA1978} turned out to be equivalent to the averaged $T$-matrix approximation.
The integro-differential kinetic equations for concentrations reported in Ref. \onlinecite{DoktorovPhysicaA1978} are known as the integral encounter theory (IET). In the simplest case of a bimolecular reaction $A+B\rightarrow C+B$, the IET kinetic equation has the following form:
\begin{equation}
\frac d{dt} C_A(t) = - C_B \int\limits_{-0}^t \Sigma(\tau) C_A(t-\tau)\,d\tau
\label{eq:IET}
\end{equation}
where the memory function $\Sigma(\tau)$ is expressed in terms of the time-dependent reaction constant Eq.(\ref{eq:Kt}):
$$ \Sigma(\tau) = k(0)\delta(\tau) + \frac d {d\tau} k(\tau)$$

IET can be formulated in a convenient matrix form and therefore applied to an arbitrary set of bimolecular reactions. It was successfully applied for several case studies where LMA (\ref{eq:RDE}) does not hold \cite{LukzenCP1986,LukzenJCP1995,FrantsuzovJCP1997,FrantsuzovJCP1997s,FrantsuzovJL1998,FrantsuzovCPL1998,FrantsuzovJCP1998}, including the geminate pair recombination generated in the bulk reaction  (see also the review Ref. \onlinecite{BurshteinACP2004} and references therein).

However, the time interval for IET applicability proved to be very limited.
It was found \cite{ZwanzigJCP1981,KipriyanovCP1994} that Eq.(\ref{eq:IET}) gives a power-law asymptotic dependence of $A$ particle concentration  on time  instead of an exponential one (\ref{eq:Asymp}):
\begin{equation}
C_A(t)\approx C_A(0) \sqrt{\frac 3 {4\pi} \xi} \left(kC_B t\right)^{-3/2}
\label{eq:IETas}
\end{equation}
As a result, in this case, IET is applicable in a very short time interval  \cite{KipriyanovCP1994}:
\begin{equation}
t\ll  |\ln \xi|  \left( kC_B\right)^{-1}
\label{eq:ETTimeLimit}
\end{equation}
The reason is that the approximation of the averaged $T$-matrix used in the calculation of the integral kernel only accounts for the dynamics of the reacting pair.

To overcome this shortcoming of IET, theoretical approaches were developed based on hierarchical equations for reduced distribution functions, i.e., analogues of the BBGKY hierarchies for reactive systems \cite{Balescu}. For the $A+B\rightarrow C+B$ reaction, with a regular expansion on a scaling parameter, Gopich and collaborators derived an integro-differential kinetic equation with a modified kernel \cite{GopichPhysicaA1998}. This approach is called the modified encounter theory (MET).
The same result was obtained by applying Faddeev's three-body theory method to the closure of the kinetic equations hierarchy \cite{IgoshinPhysicaA1999,IgoshinCP1999}.
 As was shown in Refs. \cite{GopichJCP1996,GopichPhysicaA1998,IgoshinPhysicaA1999, IgoshinCP1999,GopichJCP1999}, the physical essence of
MET is that, in addition to the reaction between the particles forming a
reacting pair, it also takes into account the reaction between the pair particles and the 3rd particle from the bulk that is assumed to be uncorrelated with the pair. In other words, MET corresponds to the mean-field correction for the reaction of the pair's particles with the homogeneous bulk environment.  It was shown in \cite{KipriyanovCP1995} that, in some special cases where exact solutions are available, MET agrees with them for a time period considerably longer than IET.

In a previous work, some of the authors formulated a matrix generalization of MET for an arbitrary set of bimolecular reactions \cite{FrantsuzovCPL2000}. Furthermore, these equations can be reformulated in a purely differential form for the kinetic equations of IET/MET  for concentrations and pair distribution functions \cite{FrantsuzovCPL2000}.  In this form, the numerical solution of the kinetic equations for IET/MET is no more complex than for LMA/SM.  A  matrix generalization of MET is also being developed within the conventional integro-differential approach \cite{IvanovJCP2001I,IvanovJCP2001II,IvanovJCP2001III}. A similar approach to MET is used in the works
of Yang, Lee, and Shin (YLS) \cite{YLS,YLS2,YLS3}. They make a different correction using a relatively complicated iterative self-consistent calculation procedure. Nevertheless, YLS and MET agree well within the limits of their applicability and
accuracy (cf. Ref. \onlinecite{GopichJCP1999}).

However, numerical simulation of the reaction $A+B\leftrightarrow C+D$ \cite{AgmonJCP2003,PopovJCP2003} showed that the MET method cannot reproduce its asymptotic kinetics.
Using the self-consistent relaxation time approximation (SCRTA), Gopich and Szabo \cite{GopichSzaboJCP2002} obtained kinetic equations for concentrations and pair distribution functions. The asymptotic time behavior of the kinetics of this reaction within the SCRTA \cite{GopichSzaboJCP2002,GopichSzaboCP2002} coincides with the results of numerical simulations \cite{AgmonJCP2003,PopovJCP2003}.
 It should be noted that, although the SCRTA kinetic equations obtained for the reaction $A+B\leftrightarrow C+D$ resemble the MET kinetic equations \cite{FrantsuzovCPL2000}, they nevertheless contain some additional mean-field correction terms. Gopich and Szabo also suggested a form of the SCRTA kinetic equations for the general case of the multistage system of $A_i + A_k \leftrightarrow A_j + A_l$  reactions \cite{GopichSzaboJCP2002} and later for the system of the     $A_i+A_k\leftrightarrow A_j$ reactions \cite{GopichSzaboJPCB2018}. Another notable attempt to generalize the MET for specific reactions(the so-called generalized encounter theory) accounted for additional mean-field corrections \cite{DoktorovJCP2010,KipriyanovJCP2010}. However, the resulting kinetic equations for the reaction $A+B\rightarrow C+D$ obtained in Ref. \onlinecite{KipriyanovJCP2010} failed to reproduce asymptotics Eq.(\ref{eq:Zeldovich}) in the case of equal initial concentrations.

The many-particle nature of chemical reactions has inspired Doi \cite{DoiJPA1976,DoiJPA1976s} and independently Zeldovich and Ovchinnikov  \cite{ZeldovichJETP1978} to apply the quantum field theory approach to describe their kinetics. Later, this approach was applied to the description of several case cases  \cite{YashinJSP1985,CardyJSP1995,NaumannJCP1993,RudavetsJPA1993, TauberJPA2005,LucivjanskyTMP2011,LucivjanskyTMP2011s,HanaiJSP2023}. The QFT approach can study many-particle systems in various dimensions, including non-integer ones. The results indicate that, for any bimolecular diffusion-influenced reaction system, the upper critical dimension is $d_c=2$  \cite{CardyJSP1995,TauberJPA2005}. Thus, for any dimension larger than 2, including the 3D case considered in this paper, there is no need for a renormalization procedure.  With the QFT approach to the reaction $A+A \rightarrow B+B$, Doi \cite{DoiJPA1976s} reproduced Smoluchowski's expression Eq.(\ref{RD}) for the steady-state reaction rate under absorption boundary conditions. Mikhailov and Yashin \cite{YashinJSP1985} determine how the rate depends on the diffusion coefficient in the case of partially absorbing boundary conditions.
The power-law asymptotics Eq.(\ref{eq:Zeldovich}) for the reaction $A+B\rightarrow C+D$ was also reproduced within the quantum field theory approach \cite{CardyJSP1995,LeeCardyJSP1995-er,TauberJPA2005}.
Some examples of multistage reaction systems were also considered  \cite{TauberJPA2005}, but the QFT approach has not yet been generalized for an arbitrary set of reactions.

\section{Model statement and second quantization formalism}
\label{sec:SecondQuantization}

In this paper, we will consider a many-particle reactive system  formalized in the
following terms:
\begin{enumerate}
\item  $N$ kinds of structureless reacting particles $A_i$,
       $i=1\dots N$, diffuse freely in inert media each with a diffusion coefficient $D_i$.
       For particles with internal structure, different internal states are treated as distinct particles.
\item  The particles may undergo internal transitions
       $A_i\to A_j$ with rate constants of $q_{ji}$.
\item  These particles can react pairwise
       $A_i+A_k\to A_j+A_l$ with a reaction
       rate constant of $w_{jl,ik}(|\vec r|)$ depending on  the
       interparticle distance $\vec r$.
\item  There is interaction potential  $U_{ik}(|\vec r|)$ (in units
       of $kT$) in pairs $A_i+A_k$.
\end{enumerate}
The current state of the reactive system  can be described
by specifying the number of particles of a given kind, $\a_i$,
and their coordinates $\{\vec r\}_i \equiv
\{\vec r^{\,i}_1,\dots\vec r^{\,i}_{\a_i}\}$.
The probabilities of the specific realizations of the reactive system are then measured
by the many-particle distribution functions (MPDF)at time $t$
$f_{\a_1,\dots,\a_N}
\left(\{\vec r\}_1,\dots,\{\vec r\}_{N},t\right)$,
which satisfy the normalization condition
\begin{equation}
\label{eq:Norm}
\sum_{\a_1,\ldots,\a_N=0}^{\infty} \int \{\dd^3 r\}^N_{i} f_{\a_1,\a_2,\dots,\a_N}\left(
   \{\vec r\}_1,\dots,\{\vec r\}_{N},t\right) = 1
\end{equation}
where $\{\dd^3 r\}_{i}^N\equiv\prod_{i=1}^N
\prod_{\beta=1}^{\a_i}d^3 r_{\beta}^{i}$. The distribution functions $f$ are symmetrical under the permutations of the particles of the same kind. However, these particles are still distinguishable in the sense of classical objects. The definition of
MPDF and Eq.(\eq{Norm}) are only meaningful when there is a finite number of particles $\a_i$. Nevertheless, the resulting equations remain valid in the thermodynamic limit, enabling the description of infinite systems \cite{Balescu}.

The time evolution of the reactive system is governed by a hierarchical set of equations for MPDFs that describe the spatial diffusion of particles and their chemical transformations.
Such hierarchical sets of equations can be formulated using a classical analog of second quantization. This approach was previously employed in Refs. \cite{DoiJPA1976,DoiJPA1976s,YashinJSP1985,CardyJSP1995} for single reactions. In the present work, we generalize this approach
 involving an arbitrary number of chemical reactions. To this end, we introduce the empty state $\ket{0}$, corresponding to a solvent
containing no particles. Each $A_i$ particle is generated in the
position $\vec r$ by the creation operator $\Psi^\dag_i(\vec r)$ and removed from
the position $\vec r$ with  the annihilation operator $\Psi_i(\vec r)$. These operators obey the following  commutation relations:
\begin{equation}
\label{eq:CommRel}
\begin{gathered}
    \left[\Psi_i(\vec r),\Psi^\dag_j(\vec r^{\,\prime})\right] =
  \delta_{ij}\delta\left(\vec r-\vec r^{\,\prime}\right),\\
\left[\Psi_i(\vec r),\Psi_j(\vec r^{\,\prime})\right]=0,\qquad
\left[\Psi^\dag_i(\vec r),\Psi^\dag_j(\vec r^{\,\prime})\right]=0
\end{gathered}
\end{equation}
Note that the empty state $\ket{0}$ possesses the following useful properties:
\begin{equation}
\label{eq:ZeroPr}
\Psi_i(\vec r)\ket{0}\,\,=0,\qquad\bra{0}\,\Psi^\dag_i(\vec r)=0,\qquad \bra{0} 0\rangle~=1
\end{equation}
Eq.\eq{ZeroPr} means that the annihilation of a particle
in an empty state yields {a zero vector}. The last equation defines
the conjugate empty state $\bra{0}\,$.

We may now describe the state of the reactive system by the generalized
distribution function
\begin{widetext}
\begin{equation}
\label{eq:PsiDef} \ket{\Phi(t)}~~=
\sum_{\a_1,\ldots,\a_N=0}^{\infty}
    \int \{\dd^3 r\}^N_{i}f_{\a_1,\a_2,\dots,\a_N}\left(
         \{\vec r\}_1,\{\vec r\}_2,\dots,\{\vec r\}_{N},t\right)
         \prod_{i=1}^N \prod_{\beta=1}^{\alpha_i}
         \Psi^\dag_i\left(\vec r_{\beta}^{~i}\right)\ket{0}
\end{equation}
\end{widetext}
In this expression, the action of $\Psi^\dag$ operators on an empty
state $\ket{0}$ produces a set of $N$ kinds of particles with
$\a_i$ particles of each kind located in the positions
$\vec r_{\beta}^{~i}$. Since spatial coordinates in Eq. (\ref{eq:PsiDef}) are integrated out,  $\ket{\Phi(t)}$ is a vector of probabilities to find the reactive system in the states with a given
amount of particles of each kind $\{\a_1,\a_2,\dots,\a_N\}$.

The dynamics of the reactive system can now be viewed as the time evolution of
the components of $\ket{\Phi(t)}$ is described by the following
kinetic equation \cite{DoiJPA1976,ZeldovichJETP1978,DoiJPA1976s,YashinJSP1985}:
\begin{equation}
\label{eq:EqPhi} \frac d{dt}\ket{\Phi(t)}\,\,=\,\widehat{\cal
L}\ket{\Phi(t)}\,\,=\,
 \left(\widehat{\cal L}_0+\widehat{\cal V}\right)\ket{\Phi(t)}
\end{equation}
In this expression, $\widehat{\cal L}_0$ describes the
diffusion and monomolecular transformation of particles:
\begin{equation}
\label{eq:L0} \widehat{\cal L}_0 = \sum_{i,j}\int
\Psi^\dag_i(\vec r)\widehat L^0_{ij}
                        \Psi_j(\vec r)\,d^3 r, \qquad
\widehat L^0_{ij} = \delta_{ij}D_i\Delta+Q_{ij}
\end{equation}
where $\widehat Q$ is the relaxation-excitation matrix describing the internal conversion of the particles:
\begin{equation}
\label{eq:Q} Q_{i,j}=q_{ij} -
\delta_{ij}\sum_k q_{ki},\qquad (q_{ii}=0)
\end{equation}
Operator $\widehat{\cal V}$,
\begin{widetext}
\begin{equation}
\label{eq:V}
\widehat{\cal V} = \frac{1}{2} \sum_{i,j,l,m}
    \int \Psi^\dag_i(\vec r)\Psi^\dag_j(\vec r^{\,\prime})
    \widehat V_{ij,lm} (\vec r,\vec r^{\,\prime})
    \Psi_l(\vec r)\Psi_m(\vec r^{\,\prime})\,d^3 r\,d^3 r^{\,\prime}
\end{equation}
\end{widetext}
stands for the force-chemical interaction given by the interaction operator
\begin{equation}
\label{eq:Vik}
\begin{gathered}
\widehat V_{ik,lm}\left(\vec r_1,\vec r_2\right) =
  W_{ik,lm}\left(\vec r_1-\vec r_2\right) +\\+
  \delta_{il}\delta_{km}\left(D_i\nabla_1-D_k\nabla_2\right)
  \left(\nabla U_{ik}(\vec r_1-\vec r_2)\right)
\end{gathered}
\end{equation}
where
\begin{equation}
\label{eq:W}
\begin{gathered}
    W_{ik,jl}(r) = w_{ik,jl}(r) -
    \delta_{ij}\delta_{kl}{\sum_{j'l'}w_{j'l',ik}(r)},\\
    (w_{ik,ik}=0)
\end{gathered}
\end{equation}

It is easy to see that the interaction operator has the following property
\begin{equation}
 \sum_{kj} \int \int \widehat V_{kj,lm}(\vec
r_1,\vec r_1^{\,\prime})h(\vec r_1,\vec r_1^{\,\prime}) \,d^3r_1\,d^3r_1'=0
 \label{eq:Vint}
\end{equation}
where $h(\vec r_1,\vec r_1^{\,\prime})$ is an arbitrary function. The first term in the interaction (\ref{eq:Vik})
will be zero after the summation over the indices. The second one contains derivatives with respect to the coordinates, which means it vanishes after coordinate integration.

First, we assume that the particles' positions are initially ($t=0$) uncorrelated, i.e., the probability to find an $A_i$ particle in position $\vec r $ does not depend on the other particles. This assumption is, generally speaking, not compatible with the presence of force interactions and solvent structure. We use it here initially for simplification purposes. As shown below, the resulting kinetic equations can be generalized to cases with initial correlations.

Without correlations, the initial MFDF is of the form
\begin{equation}
\label{eq:EqPhi0}
f_{\a_1,\a_2,\dots,\a_N}\left(\{\vec r\}_1,\{\vec r\}_2,\dots,
\{\vec r\}_{N},t=0\right) =
\prod_{i=1}^N\prod_{\beta_i=1}^{\a_i} f_i(\vec r^{\,i}_{\beta_i})
\end{equation}
where initial single-particle probability density functions $f_i(\vec r)$ are normalized as:
\begin{equation}
\label{f_inorm}
\int f_i(\vec r)\, d^3 r = 1
\end{equation}
Then the initial condition for Eq.\eq{EqPhi}
takes the form
\begin{equation}
\label{eq:InPhi} \ket{\Phi(0)}\,\,=\,\prod_{i=1}^N
  \left( \int f_i(\vec r)\Psi^\dag_i(\vec r)\, d^3 r\right)^{\a_i}\ket{0}
\end{equation}

In order to find the particle concentrations $C_i(\vec r,t)$, the following relation may be derived from properties Eq.\eq{CommRel}
and \eq{ZeroPr}:
\begin{equation}
\label{eq:ConcRel}
\begin{gathered}
    \Psi^\dag_i(\vec r)\Psi_i(\vec r)
\prod_\a \Psi^\dag_i \left(\vec r^{~i}_\a\right)\ket{0}\,\,=\\=\,
\left(\sum_\a \delta\left(\vec r-\vec
r_\a^{~i}\right)\right) \prod_\a \Psi^\dag_i\left(\vec
r^{~i}_\a\right)\ket{0}
\end{gathered}
\end{equation}
As can be seen from this expression,
$\Psi^\dag_i(\vec r)\Psi_i(\vec r)$ is the local
density operator. The concentration of particles $A_i$ can therefore
be found as the average of $\Psi^\dag_i(\vec r)\Psi_i(\vec r)$
\cite{DoiJPA1976,DoiJPA1976s}:
\begin{widetext}
    \begin{eqnarray}
    \label{eq:LocalC}
            C_i(\vec r,t) &\equiv&
  \sum_{\a_1,\ldots,\a_N=0}^{\infty}
  \int \{\dd^3 r\}^N_{i}\left(
     \sum_{\alpha=1}^{\alpha_i}
           \delta(\vec r-\vec r^{~i}_{\alpha})\right)
     f_{\a_1,\a_2,\dots,\a_N}
      \left(\{\vec r\}_1,\{\vec r\}_2,\dots,\{\vec r\}_{N},t\right)= \nonumber\\
      &=&\,\,\bra{\Upsilon}\,\Psi^\dag_i(\vec r)\Psi_i(\vec r)\ket{\Phi(t)}
    \end{eqnarray}
\end{widetext}
where \cite{DoiJPA1976,DoiJPA1976s,ZeldovichJETP1978,YashinJSP1985}
\begin{equation}
\label{eq:BarPhi}
\begin{gathered}
    \bra{\Upsilon}\, =\,\,\bra{0}\,\exp\left(
  \sum_{i=1}^N \int \Psi_i\left(\vec r\right)\, d^3 r \right) =\\=\,\,
    \bra{0}\,\sum_{\alpha=0}^{\infty} \frac{1}{\a!}\left(
            \sum_{i=1}^N\int \Psi_i(\vec r) d^3 r\right)^{\a}
\end{gathered}
\end{equation}
The state $\bra{\Upsilon}\,$ has the following important properties
\cite{DoiJPA1976,DoiJPA1976s,ZeldovichJETP1978,YashinJSP1985}:
\begin{subequations}
\label{eq:UpsilonProps}
\begin{eqnarray}
\label{eq:Norm1}
\bra{\Upsilon}\,\ket{\Phi(t)}\,\,=1\\
\label{eq:PhiPsi}
\bra{\Upsilon}\,\Psi_i^\dag(\vec r)=\,\,\bra{\Upsilon}\,
\end{eqnarray}
\end{subequations}
which may be derived from Eqs.(\ref{eq:Norm}-\ref{eq:ZeroPr}). Applying Eq.\eq{PhiPsi}
to Eq.\eq{LocalC}, results in
\begin{equation}
\label{eq:LocalConc}
C_i(\vec r,t) = \,\,\bra{\Upsilon}\,\Psi_i(\vec r)\ket{\Phi(t)}
\end{equation}

It is possible to obtain the pair distribution function
$P_{ij}(\vec r_1,\vec r_2,t)$ that measures the probability
to find a pair of particles $A_i$ and $A_j$ at positions $\vec r_1$ and
$\vec r_2$, correspondingly. It is found as \cite{DoiJPA1976,ZeldovichJETP1978}
\begin{equation}
\label{eq:Pair}
P_{ij}(\vec r_1,\vec r_2,t) = \,\,
    \bra{\Upsilon}\,\Psi_i(\vec r_1)\Psi_j(\vec r_2)\ket{\Phi(t)}
\end{equation}

Finally, to scale the theory to the limit of large volumes and numbers of particles (the thermodynamic limit), we define the initial concentrations $C^0_i(\vec r)$ distributed in infinite space. Let's select the area of volume $V$.
 Then, the mean number of particles within this volume is
 \begin{equation}
    \alpha_i = \int_V \dd^3 r\ C^0_i(\vec r);
\end{equation}

Defining  the functions $f_i(\vec r)$ within the volume $V$ as
\begin{equation}
f_i(\vec r) = \frac {C^0_i(\vec r)}{\alpha_i}
\label{eq:fidef}
\end{equation}

we get from Eqs.\eq{LocalConc} and \eq{Pair}
$$ C_i(\vec r,0)=C^0_i(\vec r); \qquad P_{ij}(\vec r_1,\vec r_2,0) = C^0_i(\vec r_1)C^0_j(\vec r_1)$$

The thermodynamic limit is reached by increasing the volume $V$ to infinity.

\section{Diagrammatic approach to Master Equation}
\label{sec:DiarammaticApproach}

For the convenience of further discussion, we now need to
represent Eqs.\eq{LocalConc} and \eq{Pair} in a slightly different
form. Note that Eq.\eq{EqPhi} may be rewritten in the alternative form (cf. Ref. \onlinecite{DoiJPA1976s})
\begin{equation}
\label{eq:EqTPhi} \frac {d}{dt}\ket{\Phi^\bullet(t)}\,\,=
   \widehat{\cal V}^\bullet(t)\ket{\Phi^\bullet(t)}
\end{equation}
where the bullet sign denotes the interaction representation
\begin{equation}
\label{eq:IntPhiV} \begin{gathered}
    \ket{\Phi^\bullet(t)}\,\,=
  \widehat{\cal G}(t)^{-1}\ket{\Phi(t)}, \\
\widehat{\cal V}^\bullet(t) =
  \widehat{\cal G}(t)^{-1}\widehat{\cal V}\widehat{\cal G}(t)
\end{gathered}
\end{equation}
and $\widehat{\cal G}(t)$ is a Green function of a free diffusion without interactions. It obeys the following equation ($\widehat{\cal I}$
is the unit operator):
\begin{equation}
\label{eq:IntG} \frac {d}{dt}\widehat{\cal G}(t) =
    \widehat{\cal L}_0\widehat{\cal G}(t), \qquad
\widehat{\cal G}(0) = \widehat{\cal I}
\end{equation}

Equation \eq{EqTPhi} has a formal solution
\begin{widetext}
\begin{subequations}
\label{eq:FormalSolution}
\begin{eqnarray}
\label{eq:FSol} \ket{\Phi^\bullet(t)}\,\,=
  \left(1+\int_0^t \widehat{\cal V}^\bullet(\tau)d\tau+
    \int_0^t d\tau \int_0^\tau\,d\tau' \,
    \widehat{\cal V}^\bullet(\tau)
    \widehat{\cal V}^\bullet(\tau')d\tau d\tau' + \cdots
  \right)\ket{\Phi^\bullet(0)}\,\, \\
\label{eq:Texp} = {\cal T}\exp\left(
     \int_0^t\widehat{\cal V}^\bullet(\tau)d\tau
             \right)\ket{\Phi^\bullet(0)}
\end{eqnarray}
\end{subequations}
\end{widetext}
where $\ket{\Phi^\bullet(0)}\,\,=\ket{\Phi(0)}$, as follows from Eqs.\eq{IntPhiV} and \eq{IntG}. Here, we introduce the time ordering
operator ${\cal T}$ that acts on the products
of time-dependent operators, ordering them by decreasing their
time argument (cf. Ref. \onlinecite{Landau}). Using Eq.\eq{Texp}, one can represent the
concentration $C_i(\vec r,t)$ \eq{LocalConc} and the pair distribution function
$P_{ij}(\vec r_1,\vec r_2,t)$ \eq{EqPhi} as follows:
\begin{equation}
\label{eq:EqCInt}
C_i(\vec r,t) = \,\,\bra{\Upsilon}\,{\cal T}\Psi_i^\bullet(\vec r,t)
\exp\left(\int_0^t\widehat{\cal V}^\bullet(\tau)d\tau\right)
  \ket{\Phi^\bullet(0)}
\end{equation}
\begin{equation}
\label{eq:EqPInt}
\begin{split}
&P_{ij}(\vec r_1,\vec r_2,t) =\\
 &=\,
  \bra{\Upsilon}\,{\cal T}\Psi_i^\bullet(\vec r_1,t)
  \Psi_j^\bullet(\vec r_2,t)\exp
  \left(\int_0^t\widehat{\cal V}^\bullet(\tau)d\tau\right)
  \ket{\Phi^\bullet(0)}
  \end{split}
\end{equation}\\

where $\Psi_i^\bullet(\vec r,t)$ is the annihilation operator
$\Psi_i(\vec r)$ in the interaction representation:
\begin{equation}
\label{eq:PsiInt}
\begin{split}
    \Psi_i^\bullet(\vec r,t) = \widehat{\cal G}(t)^{-1}\Psi_i(\vec r)
\widehat{\cal G}(t),\\ \Psi_i^{\dagger\bullet}(\vec r,t) = \widehat{\cal G}(t)^{-1}\Psi_i^\dagger
(\vec r) \widehat{\cal G}(t).
\end{split}
\end{equation}

Using Eqs.(\ref{eq:InPhi},\ref{eq:BarPhi},\ref{eq:PsiInt}) in Eqs.(\ref{eq:EqCInt},\ref{eq:EqPInt}), one can represent both $C_i$ and $P_{ij}$
as $\bra{0}\,\ldots\ket{0}$ averages of infinite sums:\\

\begin{widetext}
    \begin{equation}
\label{eq:CAsAverage}
C_i(\vec r,t) = \,\, \bra{0}\,{\cal T}\exp\left(\sum_{j=1}^n
\int \Psi_j^\bullet(\vec r_f,t)\,d^3 r_f\right) \Psi_i^\bullet(\vec r,t)
\exp\left(\int_0^t\widehat{\cal V}^\bullet(\tau)d\tau\right)
\prod_{k=1}^N \left( \int f_k(\vec r_0)\Psi^{\dag\bullet}_k(\vec r_0,0)\,
 d^3 r_0\right)^{\a_k}\ket{0}
\end{equation}
\begin{equation}
\label{eq:PAsAverage}\begin{split}
   & P_{ij}(\vec r_1,\vec r_2,t) =\\
    &= \bra{0}\,{\cal T}\exp\left(\sum_{l=1}^n
\int \Psi_l^\bullet(\vec r_f,t)\,d^3 r_f\right) \Psi_i^\bullet
(\vec r_1,t)\Psi_j^\bullet(\vec r_2,t)
\exp\left(\int_0^t\widehat{\cal V}^\bullet(\tau)d\tau\right)
\prod_{k=1}^N \left( \int f_k(\vec r_0)\Psi^{\dag\bullet}_k(\vec
r_0,0)\, d^3 r_0\right)^{\a_k}\ket{0}
\end{split}
\end{equation}
\end{widetext}

Each term of the expansion series for the exponents contains the product of several $\Psi_i^\bullet(\vec r,t)$ and $\Psi_i^{\dagger\bullet}$ operators. In Eqs. (\ref{eq:CAsAverage}) and (\ref{eq:PAsAverage}), each term can be averaged using the
well-known Wick's theorem \cite{Landau}, which allows one to represent
any expectation value $\bra{0}\,\ldots\ket{0}$ as a sum of pair expectation products,
containing only two operators. The mean values
$\bra{0}\,{\cal T}\Psi_i^\bullet(\vec r,t)\Psi_k^\bullet(\vec r^{\,\prime},t')\ket{0}$ and
$\bra{0}\,{\cal T}\Psi^{\dag\bullet}_i(\vec r,t)\Psi^{\dag\bullet}_k(\vec r^{\,\prime},t')\ket{0}$
yield zero because of conditions (\ref{eq:ZeroPr}). As obtained
from Eq.(\ref{eq:ZeroPr}), the term
$\bra{0}\,{\cal T}\Psi^\bullet_i(\vec r,t)\Psi^{\dag\bullet}_k(\vec r^{\,\prime},t')\ket{0}$
yields zero if $t<t'$. For $t\ge t'$, Eqs.(\ref{eq:PsiInt})  gives
\begin{equation}
\label{eq:Green}
\begin{gathered}
\bra{0}\, {\cal T}\Psi^\bullet_i(\vec r,t)\Psi^{\dag\bullet}_k(\vec r^{\,\prime},t') \ket{0} =\\=
\bra{0}\, {\cal T}\Psi_i(\vec r) \exp\left(\widehat {\cal L}_0 (t-t')\right) \Psi^\dag_k(\vec r^{\,\prime}) \ket{0} \equiv
G^0_{ik}(\vec r,t,\vec r^{\,\prime},t')
\end{gathered}
\end{equation}
It follows from Eqs.(\ref{eq:IntG}), (\ref{eq:PsiInt}),
 and (\ref{eq:Green}) that
$G^0_{ik}(\vec r,t,\vec r^{\,\prime},t')$ obeys the following equation
\begin{equation}
\label{EqGreen}
\frac{\partial}{\partial t} G^0_{ik}(\vec r,t,\vec r^{\,\prime},t') =
 \sum_{l}\widehat L^0_{il}  G^0_{lk}(\vec r,t,\vec r^{\,\prime},t')
\end{equation}
where $\widehat L^0_{il}(t)$ is given by Eq.(\ref{eq:L0}).
$G^0_{ik}(\vec r,t,\vec r^{\,\prime},t')$ is therefore the single particle Green's function.
Eq.(\ref{EqGreen}) can be written in the matrix form
\begin{equation}
\label{EqGreenM}
\frac{\partial}{\partial t} \widehat G^0(\vec r,t,\vec r^{\,\prime},t') = \widehat L_0  \widehat G^0(\vec r,t,\vec r^{\,\prime},t')
\end{equation}
where
$$\widehat L_0=\widehat D\Delta+\widehat Q$$
and elements of the matrix $\widehat D$ defined as
$$D_{ik}=D_i\delta_{ik}$$
 As follows from Eq.(\ref{eq:Green}),
$\widehat G_{ik}(\vec r,t,\vec r^{\,\prime},t')=0$ if $t<t'$. The initial conditions
for Eq.(\ref{EqGreenM}) are defined at $t=t'$ and can also be derived from
Eq.(\ref{eq:Green}):
\begin{equation}
\label{G0Cond}
\widehat G^0(\vec r,t,\vec r^{\,\prime},t)=\widehat I \delta(\vec r-\vec r^{\,\prime})
\end{equation}
Physically, Green's function $G^0_{ik}(\vec r,t,\vec r^{\,\prime},t')$ represents the
probability density to find a particle in state $A_i$ in position
$\vec r$ in the time moment $t$ if it started from state $A_k$ in position
$\vec r^{\,\prime}$ in the time moment $t'$. From Eqs. (\ref{EqGreen}), (\ref{eq:L0}), {(\ref{eq:Q})} and (\ref{G0Cond}) we can obtain that $\widehat G_{ik}$ conserves the total probability:
\begin{equation}
\label{G0sum}
\sum_i \int G^0_{ik}(\vec r,t,\vec r^{\,\prime},t') \,d^3 r^{\,\prime} = 1, \quad
 \mbox{   for    } t\ge t'
\end{equation}

Expansion of the ${\cal T}$-exponent in Eq.(\ref{eq:CAsAverage}) gives the concentration series
\begin{equation}
\label{CSeries}
C_i(t)=C^{(0)}_i(t)+C^{(1)}_i(t)+C^{(2)}_i(t)+\cdots
\end{equation}
Each term  {$C^{(n)}_i(t)$} in this expansion can be expressed through Green's
functions $G^0_{ik}(\vec r,t,\vec r^{\,\prime},t')$ and the initial pair distribution
functions $f_i(\vec r)$ (cf. Eq.(\ref{f_inorm})). Expression (\ref{CSeries}) is essentially a perturbation series on the interparticle interaction.
Analytic expressions for $C^{(n)}_i(t)$ are impractical.
To obtain tractable expressions for the particle concentrations
$C_i(t)$, we employ the diagram technique, similar to that of Feynman
\cite{Landau} (cf. Supplementary Note I). Similar techniques were also
used in Refs. \cite{DoiJPA1976,DoiJPA1976s,YashinJSP1985}.

We introduce the following notations:

\begin{itemize}
\item
The cross represents concentration $C_i(\vec r,t)$
\begin{center}
\begin{picture}(1,0.5)
\put(0.2,0.4){\line(1,0){.3}} \put(0.5,0.4){\circle*{0.1}}
\put(0.0,0.2){\line(1,1){0.4}}
\put(0.0,0.6){\line(1,-1){0.4}}
\put(0.5,0.6){\makebox(0,0)[b]{$i$}}
\put(.6,0){\makebox(0,0)[b]{$rt$}}
\end{picture}
\end{center}
\item  The initial concentrations $C^0_i(\vec r)=\alpha_i f_i(\vec r)$ are
represented as a cross and circle
\begin{equation}\label{Cinitdiag}
\begin{picture}(1,0.5)
\put(0.2,0.4){\line(1,0){.3}}
\put(0.5,0.4){\circle*{0.1}}
\put(0.0,0.2){\line(1,1){0.4}}
\put(0.0,0.6){\line(1,-1){0.4}}
\put(0.2,0.4){\circle{0.2}}
\put(0.5,0.6){\makebox(0,0)[b]{$i$}}
\put(.6,0){\makebox(0,0)[b]{$r$}}
\end{picture}
\end{equation}
\item The solid horizontal line represents Green's function $G^0_{ik}(\vec r,t,\vec r^{\,\prime},t')$
\begin{center}
\begin{picture}(3,0.75)
\put(0,0.4){\circle*{0.1}}
\put(0.1,0.6){\makebox(0,0)[b]{$k$}}
\put(0,0.0){\makebox(0,0)[b]{$r't'$}}
\put(3,0.0){\makebox(0,0)[b]{$rt$}}
\put(0,0.4){\line(1,0){3}}
\put(2.9,0.6){\makebox(0,0)[b]{$i$}}
\put(3,0.4){\circle*{0.1}}
\end{picture}
\end{center}
\item
The dashed vertical line between two points with different spatial coordinates and equal
time coordinates corresponds to the interaction operator
$\widehat V_{ik,lm}(\vec r,\vec r^{\,\prime})$
\begin{center}
\begin{picture}(1,1.125)(0,-0.25)
\put(0.5,0){\circle*{0.1}}
\put(0,0){\line(1,0){1}}
\put(0.5,0.8){\circle*{0.1}}
\put(0,0.8){\line(1,0){1}}
\multiput(0.5,0.8)(0,-0.125){7}{\line(0,-1){0.1}}
\put(0.9,0.1){\makebox(0,0)[b]{$i$}}
\put(0.5,-0.4){\makebox(0,0)[b]{$r\,t$}}
\put(0.5,1){\makebox(0,0)[b]{$r'$}}
\put(0.9,0.5){\makebox(0,0)[b]{$k$}}
\put(0.1,0.1){\makebox(0,0)[b]{$l$}}
\put(0.1,0.5){\makebox(0,0)[b]{$m$}}
\end{picture}
\end{center}
\item
We assume summation over all unspecified indices and integration over all unspecified coordinates/times.
The time integral corresponding to the rightmost vertical line has limits from 0 to $t$.  The time integrals corresponding to the remaining vertical lines have limits from 0 to the integration time of the integral corresponding to the next vertical line to the right.
\end{itemize}

Using this notation, one can represent the particle concentration $C_i(r,t)$,
Eq.(\ref{CSeries}), as an infinite sum of diagrams. The diagrams of up
to the third order of interaction are given below
\begin{widetext}
$$
\begin{picture}
(1,0.5)
\put(0.2,0.){\line(1,0){.3}}
\put(0.5,0){\circle*{0.1}}
\put(0.0,-0.2){\line(1,1){0.4}}
\put(0.0,0.2){\line(1,-1){0.4}}
\put(0.5,0.2){\makebox(0,0)[b]{$i$}}
\put(.6,-0.4){\makebox(0,0)[b]{$rt$}}
\end{picture}
\quad=\quad
\begin{picture}(1,1)
\put(1,0){\circle*{0.1}}
\put(0.2,0.){\line(1,0){.8}}
\put(0.5,0){\circle*{0.1}}
\put(0.0,-0.2){\line(1,1){0.4}}
\put(0.0,0.2){\line(1,-1){0.4}}
\put(0.2,0){\circle{0.2}}
\put(0.9,0.2){\makebox(0,0)[b]{$i$}}
\put(1.0,-0.4){\makebox(0,0)[b]{$rt$}}
\end{picture}\quad+\quad
\begin{picture}(1.5,1)
\put(0.0,0.3){\line(1,1){0.4}}
\put(0.0,0.7){\line(1,-1){0.4}}
\put(0.2,0.5){\circle{0.2}}
\put(0.0,-0.2){\line(1,1){0.4}}
\put(0.0,0.2){\line(1,-1){0.4}}
\put(0.2,0){\circle{0.2}}
\put(0.2,0){\line(1,0){1.3}}
\put(0.2,0.5){\line(1,0){1}}
\multiput(0.5,0.5)(0.5,0){2}{\circle*{0.1}}
\multiput(0.5,0)(0.5,0){3}{\circle*{0.1}}
\multiput(1,0.5)(0,-0.125){4}{\line(0,-1){0.1}}
\put(1.4,0.2){\makebox(0,0)[b]{$i$}}
\put(1.5,-0.4){\makebox(0,0)[b]{$rt$}}
\end{picture}\quad +\quad
\begin{picture}(2,1)
\put(0.0,0.3){\line(1,1){0.4}}
\put(0.0,0.7){\line(1,-1){0.4}}
\put(0.2,0.5){\circle{0.2}}
\put(0.0,-0.2){\line(1,1){0.4}}
\put(0.0,0.2){\line(1,-1){0.4}}
\put(0.2,0.){\circle{0.2}}
\put(0.2,0){\line(1,0){1.8}}
\put(0.2,0.5){\line(1,0){1.5}}
\multiput(0.5,0.5)(0.5,0){3}{\circle*{0.1}}
\multiput(0.5,0)(0.5,0){4}{\circle*{0.1}}
\multiput(1,0.5)(0,-0.125){4}{\line(0,-1){0.1}}
\multiput(1.5,0.5)(0,-0.125){4}{\line(0,-1){0.1}}
\put(1.9,0.2){\makebox(0,0)[b]{$i$}}
\put(2,-0.4){\makebox(0,0)[b]{$rt$}}
\end{picture}\quad +\quad
\begin{picture}(2,1)
\put(0.0,0.3){\line(1,1){0.4}}
\put(0.0,0.7){\line(1,-1){0.4}}
\put(0.2,0.5){\circle{0.2}}
\put(0.0,-0.2){\line(1,1){0.4}}
\put(0.0,0.2){\line(1,-1){0.4}}
\put(0.2,0.){\circle{0.2}}
\put(0.,0.8){\line(1,1){0.4}}
\put(0.,1.2){\line(1,-1){0.4}}
\put(0.2,0){\line(1,0){1.8}}
\put(0.2,1){\circle{0.2}}
\put(0.2,0.5){\line(1,0){1.5}}
\put(0.2,1){\line(1,0){1}}
\put(1.5,0.5){\circle*{0.1}}
\put(2,0){\circle*{0.1}}
\put(0.5,1){\circle*{0.1}}
\put(1,1){\circle*{0.1}}
\put(0.5,0.5){\circle*{0.1}}
\multiput(0.5,0)(0.5,0){4}{\circle*{0.1}}
\multiput(1,1)(0,-0.125){8}{\line(0,-1){0.09}}
\multiput(1.5,0.5)(0,-0.125){4}{\line(0,-1){0.09}}
\put(1.9,0.2){\makebox(0,0)[b]{$i$}}
\put(2,-0.4){\makebox(0,0)[b]{$rt$}}
\end{picture}$$

$$\quad +\quad
\begin{picture}(2.5,1)
\put(0.0,0.3){\line(1,1){0.4}}
\put(0.0,0.7){\line(1,-1){0.4}}
\put(0.2,0.5){\circle{0.2}}
\put(0.0,-0.2){\line(1,1){0.4}}
\put(0.0,0.2){\line(1,-1){0.4}}
\put(0.2,0.){\circle{0.2}}
\put(0.2,0){\line(1,0){2.3}}
\put(0.2,0.5){\line(1,0){2}}
\multiput(0.5,0.5)(0.5,0){4}{\circle*{0.1}}
\multiput(0.5,0)(0.5,0){5}{\circle*{0.1}}
\multiput(1,0.5)(0,-0.125){4}{\line(0,-1){0.1}}
\multiput(1.5,0.5)(0,-0.125){4}{\line(0,-1){0.1}}
\multiput(2,0.5)(0,-0.125){4}{\line(0,-1){0.1}}
\put(2.4,0.2){\makebox(0,0)[b]{$i$}}
\put(2.5,-0.4){\makebox(0,0)[b]{$rt$}}
\end{picture}\quad +\quad
\begin{picture}(2.5,1)
\put(0.0,0.3){\line(1,1){0.4}}
\put(0.0,0.7){\line(1,-1){0.4}}
\put(0.2,0.5){\circle{0.2}}
\put(0.0,-0.2){\line(1,1){0.4}}
\put(0.0,0.2){\line(1,-1){0.4}}
\put(0.2,0.){\circle{0.2}}
\put(0.,0.8){\line(1,1){0.4}}
\put(0.,1.2){\line(1,-1){0.4}}
\put(0.2,1){\circle{0.2}}
\put(0.2,0){\line(1,0){2.3}}
\put(0.2,0.5){\line(1,0){2}}
\put(0.2,1){\line(1,0){1}}
\put(1.5,0.5){\circle*{0.1}}
\put(2,0){\circle*{0.1}}
\put(0.5,1){\circle*{0.1}}
\put(1,1){\circle*{0.1}}
\put(0.5,0.5){\circle*{0.1}}
\multiput(0.5,0)(0.5,0){5}{\circle*{0.1}}
\put(2,0.5){\circle*{0.1}}
\multiput(1,1)(0,-0.125){8}{\line(0,-1){0.09}}
\multiput(1.5,0.5)(0,-0.125){4}{\line(0,-1){0.09}}
\multiput(2,0.5)(0,-0.125){4}{\line(0,-1){0.09}}
\put(2.4,0.2){\makebox(0,0)[b]{$i$}}
\put(2.5,-0.4){\makebox(0,0)[b]{$rt$}}
\end{picture}\quad +\quad
\begin{picture}(2.5,1)
\put(0.0,0.3){\line(1,1){0.4}}
\put(0.0,0.7){\line(1,-1){0.4}}
\put(0.2,0.5){\circle{0.2}}
\put(0.0,-0.2){\line(1,1){0.4}}
\put(0.0,0.2){\line(1,-1){0.4}}
\put(0.2,0.){\circle{0.2}}
\put(0.,0.8){\line(1,1){0.4}}
\put(0.,1.2){\line(1,-1){0.4}}
\put(0.2,1){\circle{0.2}}
\put(0.2,0){\line(1,0){2.3}}
\put(0.2,0.5){\line(1,0){2}}
\put(0.2,1){\line(1,0){1.5}}
\put(1.,0.5){\circle*{0.1}}
\put(2,0){\circle*{0.1}}
\put(0.5,1){\circle*{0.1}}
\put(1.5,1){\circle*{0.1}}
\put(0.5,0.5){\circle*{0.1}}
\multiput(0.5,0)(0.5,0){5}{\circle*{0.1}}
\put(2,0.5){\circle*{0.1}}
\multiput(1.5,1)(0,-0.125){8}{\line(0,-1){0.09}}
\multiput(1,0.5)(0,-0.125){4}{\line(0,-1){0.09}}
\multiput(2,0.5)(0,-0.125){4}{\line(0,-1){0.09}}
\put(2.4,0.2){\makebox(0,0)[b]{$i$}}
\put(2.5,-0.4){\makebox(0,0)[b]{$rt$}}
\end{picture}\quad +\quad
\begin{picture}(2.5,1)
\put(0.0,0.3){\line(1,1){0.4}}
\put(0.0,0.7){\line(1,-1){0.4}}
\put(0.2,0.5){\circle{0.2}}
\put(0.0,-0.2){\line(1,1){0.4}}
\put(0.0,0.2){\line(1,-1){0.4}}
\put(0.2,0.){\circle{0.2}}
\put(0.,0.8){\line(1,1){0.4}}
\put(0.,1.2){\line(1,-1){0.4}}
\put(0.2,1){\circle{0.2}}
\put(0.2,0){\line(1,0){2.3}}
\put(0.2,0.5){\line(1,0){1.5}}
\put(0.2,1){\line(1,0){2}}
\put(1.,0.5){\circle*{0.1}}
\put(2,0){\circle*{0.1}}
\put(0.5,1){\circle*{0.1}}
\put(2,1){\circle*{0.1}}
\put(0.5,0.5){\circle*{0.1}}
\multiput(0.5,0)(0.5,0){5}{\circle*{0.1}}
\put(1.5,0.5){\circle*{0.1}}
\multiput(2,1)(0,-0.125){8}{\line(0,-1){0.09}}
\multiput(1,0.5)(0,-0.125){4}{\line(0,-1){0.09}}
\multiput(1.5,0.5)(0,-0.125){4}{\line(0,-1){0.09}}
\put(2.4,0.2){\makebox(0,0)[b]{$i$}}
\put(2.5,-0.4){\makebox(0,0)[b]{$rt$}}
\end{picture} $$

\begin{equation} \quad +\quad
\begin{picture}(2.5,1)
\put(0.0,0.3){\line(1,1){0.4}}
\put(0.0,0.7){\line(1,-1){0.4}}
\put(0.2,0.5){\circle{0.2}}
\put(0.0,-0.2){\line(1,1){0.4}}
\put(0.0,0.2){\line(1,-1){0.4}}
\put(0.2,0.){\circle{0.2}}
\put(0.,0.8){\line(1,1){0.4}}
\put(0.,1.2){\line(1,-1){0.4}}
\put(0.2,1){\circle{0.2}}
\put(0.2,0){\line(1,0){2.3}}
\put(0.2,0.5){\line(1,0){2}}
\put(0.2,1){\line(1,0){1.5}}
\put(1.,0.5){\circle*{0.1}}
\put(2,0){\circle*{0.1}}
\put(0.5,1){\circle*{0.1}}
\put(1.5,1){\circle*{0.1}}
\put(0.5,0.5){\circle*{0.1}}
\multiput(0.5,0)(0.5,0){2}{\circle*{0.1}}
\put(2,0.5){\circle*{0.1}}
\multiput(2,0)(0.5,0){2}{\circle*{0.1}}
\put(2,0.5){\circle*{0.1}}
\put(1.5,0.5){\circle*{0.1}}
\multiput(1.5,1)(0,-0.125){4}{\line(0,-1){0.09}}
\multiput(1,0.5)(0,-0.125){4}{\line(0,-1){0.09}}
\multiput(2,0.5)(0,-0.125){4}{\line(0,-1){0.09}}
\put(2.4,0.2){\makebox(0,0)[b]{$i$}}
\put(2.5,-0.4){\makebox(0,0)[b]{$rt$}}
\end{picture}\quad +\quad
\begin{picture}(2.5,1)
\put(0.0,0.3){\line(1,1){0.4}}
\put(0.0,0.7){\line(1,-1){0.4}}
\put(0.2,0.5){\circle{0.2}}
\put(0.0,-0.2){\line(1,1){0.4}}
\put(0.0,0.2){\line(1,-1){0.4}}
\put(0.2,0.){\circle{0.2}}
\put(0.,0.8){\line(1,1){0.4}}
\put(0.,1.2){\line(1,-1){0.4}}
\put(0.2,1){\circle{0.2}}
\put(0.2,0){\line(1,0){2.3}}
\put(0.2,0.5){\line(1,0){2}}
\put(0.2,1){\line(1,0){1}}
\put(1.,0.5){\circle*{0.1}}
\put(2,0){\circle*{0.1}}
\put(0.5,1){\circle*{0.1}}
\put(0.5,0.5){\circle*{0.1}}
\put(0.5,0){\circle*{0.1}}
\put(2,0.5){\circle*{0.1}}
\put(1,1){\circle*{0.1}}
\multiput(1.5,0)(0.5,0){3}{\circle*{0.1}}
\put(2,0.5){\circle*{0.1}}
\put(1.5,0.5){\circle*{0.1}}
\multiput(1.,1)(0,-0.125){4}{\line(0,-1){0.09}}
\multiput(1.5,0.5)(0,-0.125){4}{\line(0,-1){0.09}}
\multiput(2,0.5)(0,-0.125){4}{\line(0,-1){0.09}}
\put(2.4,0.2){\makebox(0,0)[b]{$i$}}
\put(2.5,-0.4){\makebox(0,0)[b]{$rt$}}
\end{picture}\quad +\quad
\begin{picture}(2.5,1)
\put(0.0,0.3){\line(1,1){0.4}}
\put(0.0,0.7){\line(1,-1){0.4}}
\put(0.2,0.5){\circle{0.2}}
\put(0.0,-0.2){\line(1,1){0.4}}
\put(0.0,0.2){\line(1,-1){0.4}}
\put(0.2,0.){\circle{0.2}}
\put(0.,0.8){\line(1,1){0.4}}
\put(0.,1.2){\line(1,-1){0.4}}
\put(0.2,1){\circle{0.2}}
\put(0.2,0){\line(1,0){2.3}}
\put(0.2,0.5){\line(1,0){1.5}}
\put(0.2,1){\line(1,0){2}}
\put(1.,0.5){\circle*{0.1}}
\put(2,0){\circle*{0.1}}
\put(0.5,1){\circle*{0.1}}
\put(2,1){\circle*{0.1}}
\put(0.5,0.5){\circle*{0.1}}
\multiput(0.5,0)(0.5,0){2}{\circle*{0.1}}
\put(1.5,0.5){\circle*{0.1}}
\multiput(2,0)(0.5,0){2}{\circle*{0.1}}
\put(1.5,1){\circle*{0.1}}
\multiput(2,1)(0,-0.125){8}{\line(0,-1){0.09}}
\multiput(1,0.5)(0,-0.125){4}{\line(0,-1){0.09}}
\multiput(1.5,1)(0,-0.125){4}{\line(0,-1){0.09}}
\put(2.4,0.2){\makebox(0,0)[b]{$i$}}
\put(2.5,-0.4){\makebox(0,0)[b]{$rt$}}
\end{picture}
\quad +\quad
\begin{picture}(2.5,2)
\put(0.0,0.3){\line(1,1){0.4}}
\put(0.0,0.7){\line(1,-1){0.4}}
\put(0.2,0.5){\circle{0.2}}
\put(0.0,-0.2){\line(1,1){0.4}}
\put(0.0,0.2){\line(1,-1){0.4}}
\put(0.2,0.){\circle{0.2}}
\put(0.,0.8){\line(1,1){0.4}}
\put(0.,1.2){\line(1,-1){0.4}}
\put(0.2,1){\circle{0.2}}
\put(0.,1.3){\line(1,1){0.4}}
\put(0.,1.7){\line(1,-1){0.4}}
\put(0.2,1.5){\circle{0.2}}
\put(0.2,0){\line(1,0){2.3}}
\put(0.2,0.5){\line(1,0){1}}
\put(0.2,1){\line(1,0){1.5}}
\put(0.2,1.5){\line(1,0){2}}
\put(1.,0.5){\circle*{0.1}}
\put(2,0){\circle*{0.1}}
\put(0.5,1){\circle*{0.1}}
\put(1.5,1){\circle*{0.1}}
\put(0.5,0.5){\circle*{0.1}}
\multiput(0.5,0)(0.5,0){5}{\circle*{0.1}}
\put(0.5,1.5){\circle*{0.1}}
\put(2,1.5){\circle*{0.1}}
\multiput(1.5,1)(0,-0.125){8}{\line(0,-1){0.09}}
\multiput(1,0.5)(0,-0.125){4}{\line(0,-1){0.09}}
\multiput(2,0.5)(0,-0.125){4}{\line(0,-1){0.09}}
\multiput(2,1.5)(0,-0.125){12}{\line(0,-1){0.09}}
\put(2.4,0.2){\makebox(0,0)[b]{$i$}}
\put(2.5,-0.4){\makebox(0,0)[b]{$rt$}}
\end{picture}\quad +\quad\cdots
\label{DCSeries}
\end{equation}
\mbox{}\\
\end{widetext}
All diagrams in Eq.(\ref{DCSeries}) have the following features in common.
First, all of them lead to the particle in the state $A_i$ at the point
$(\vec r,t)$, i.e., their rightmost lines correspond to the Green's function
$G_{i*}(\vec r,t,\vec *,*)$ ($*$ stands for the unspecified
integration/summation variables). Second, all diagrams are \emph{connected},
i.e., all points of each diagram are connected by a continuous path.
Third, all diagrams are \emph{topologically different}. Topologically different diagrams can not be transformed into each other by means of
reshaping, without changing their connectivity, and therefore, they represent
different analytical expressions. Eq.(\ref{DCSeries}) includes all possible
diagrams with these described properties. The zero- and first-order terms in
Eq.(\ref{CSeries}) and (\ref{DCSeries}) correspond to the following
expressions
\begin{widetext}
\begin{equation}
C^{(0)}_i(\vec r,t) =
  \sum_j \int G^0_{ij}(\vec r,t,\vec r_0,0) C^0_j(\vec r_0)\,d^3 r_0
\label{1stTerm}
\end{equation}
\begin{eqnarray}\nonumber
C^{(1)}_i(\vec r,t)& = &\sum_{n n'} \sum_{kjlm}   \int_0^t\,dt_1
\int  d^3 r_1\,d^3 r^{\,\prime}_1\,d^3 r_0\,d^3 r^{\,\prime}_0
 G^0_{ik}(\vec r,t,\vec r_1,t_1)
\widehat V_{kj,lm}(\vec r_1,\vec r^{\,\prime}_1)\\
\label{2ndTerm} &\times& G^0_{ln}(\vec r_1,t_1,\vec r_0,0)
G^0_{m n'}(\vec r^{\,\prime}_1,t_1,\vec r^{\,\prime}_0,0)C^0_n(\vec r_0) C^0_{n'}(\vec r^{\,\prime}_0)
\end{eqnarray}\\

Analytical expressions for the higher-order terms can be written in a similar
way.

The pair distribution function (\ref{eq:PAsAverage})
$P_{ik}(\vec r_1,\vec r_2,t)$ can also be represented as a sum of
diagrams leading to points $(i\vec r_1 t)$ and $(k\vec r_2 t)$

\begin{equation}
\label{eq:PDiagram}
P_{ik}(\vec r_1,\vec r_2,t)\quad=\quad
\begin{picture}(1.5,1.5)
\put(0.0,0.8){\line(1,1){0.4}}
\put(0.0,1.2){\line(1,-1){0.4}}
\put(0.2,1){\circle{0.2}}
\put(0.0,-0.2){\line(1,1){0.4}}
\put(0.0,0.2){\line(1,-1){0.4}}
\put(0.2,0){\circle{0.2}}
\put(0.2,0){\line(1,0){0.8}}
\put(0.2,1){\line(1,0){0.8}}
\multiput(0.5,1)(0.5,0){2}{\circle*{0.1}}
\multiput(0.5,0)(0.5,0){2}{\circle*{0.1}}
\put(.9,0.2){\makebox(0,0)[b]{$i$}}
\put(.9,1.2){\makebox(0,0)[b]{$k$}}
\put(1.,-0.4){\makebox(0,0)[b]{$r_1t$}}
\put(1.,0.6){\makebox(0,0)[b]{$r_2t$}}
\end{picture}\quad +\quad
\begin{picture}(2,1.5)
\put(0.0,0.8){\line(1,1){0.4}}
\put(0.0,1.2){\line(1,-1){0.4}}
\put(0.0,-0.2){\line(1,1){0.4}}
\put(0.0,0.2){\line(1,-1){0.4}}
\put(0.2,1){\circle{0.2}}
\put(0.2,0){\circle{0.2}}
\put(0.2,0){\line(1,0){1.3}}
\put(0.2,1){\line(1,0){1.3}}
\multiput(0.5,1)(0.5,0){3}{\circle*{0.1}}
\multiput(0.5,0)(0.5,0){3}{\circle*{0.1}}
\multiput(1,1)(0,-0.125){8}{\line(0,-1){0.09}}
\put(1.4,0.2){\makebox(0,0)[b]{$i$}}
\put(1.4,1.2){\makebox(0,0)[b]{$k$}}
\put(1.5,-0.4){\makebox(0,0)[b]{$r_1t$}}
\put(1.5,0.6){\makebox(0,0)[b]{$r_2t$}}
\end{picture}\quad +\quad
\begin{picture}(2,1.5)
\put(0.0,0.8){\line(1,1){0.4}}
\put(0.0,1.2){\line(1,-1){0.4}}
\put(0.2,1){\circle{0.2}}
\put(0.0,-0.2){\line(1,1){0.4}}
\put(0.0,0.2){\line(1,-1){0.4}}
\put(0.2,0){\circle{0.2}}
\put(0.,0.3){\line(1,1){0.4}}
\put(0.,0.7){\line(1,-1){0.4}}
\put(0.2,0){\line(1,0){1.3}}
\put(0.2,0.5){\circle{0.2}}
\put(0.2,0.5){\line(1,0){1.1}}
\put(0.2,1){\line(1,0){1.3}}
\put(1.5,1){\circle*{0.1}}
\put(0.5,1){\circle*{0.1}}
\put(0.5,0.5){\circle*{0.1}}
\put(1,0.5){\circle*{0.1}}
\multiput(0.5,0)(0.5,0){3}{\circle*{0.1}}
\multiput(1,0.5)(0,-0.125){4}{\line(0,-1){0.09}}
\put(1.4,0.2){\makebox(0,0)[b]{$i$}}
\put(1.4,1.2){\makebox(0,0)[b]{$k$}}
\put(1.5,-0.4){\makebox(0,0)[b]{$r_1t$}}
\put(1.5,0.6){\makebox(0,0)[b]{$r_2t$}}
\end{picture}\quad +\quad \cdots
\end{equation}
\mbox{}\\

\end{widetext}
In this expression, all diagrams lead to the pair of states $A_k$ and $A_i$.
Eq.(\ref{eq:PDiagram}) contains the connected and unconnected
diagrams. As follows from our diagram definitions, unconnected
diagrams correspond to the product of their connected parts (sub-diagrams),
each of which leads either to state $A_k$ or $A_i$. We can rearrange the
unconnected diagrams in Eq.(\ref{eq:PDiagram}) such that the sub-diagrams leading
to state $A_i$ are extracted as common multiples. Comparison with Eq.(\ref{DCSeries}) shows that the summation of unconnected diagrams
reduces to the product of particle concentrations
$C_i(\vec r_1,t) C_k(\vec r_2,t)$. Finally, the pair distribution function
can be recast in the following form
\begin{equation}
\label{eq:PairFunction}
P_{ik}(\vec r_1,\vec r_2,t)=C_i(\vec r_1,t) C_k(\vec r_2,t) +
p_{ik}(\vec r_1,\vec r_2,t)
\end{equation}
where  $p_{ik}(\vec r_1,\vec r_2,t)$ is the pair correlation function. This function is obtained
as a sum of all connected diagrams starting with two or more $C_j(r,0)$
($j=1\ldots M)$ and leading to $A_i$ at $\vec r_1\ t$ and $A_k$ at
$\vec r_2\ t$:
\begin{widetext}
\begin{equation}
p_{ik}(\vec r_1,\vec r_2,t)\equiv
\begin{picture}(1.,1)
\put(0.5,1){\circle*{0.1}}
\put(0.5,0){\circle*{0.1}}
\put(0.5,0.5){\oval(1,1)[l]}
\put(.4,0.2){\makebox(0,0)[b]{$i$}}
\put(.4,1.2){\makebox(0,0)[b]{$k$}}
\put(.5,-0.4){\makebox(0,0)[b]{$r_1t$}}
\put(.5,0.6){\makebox(0,0)[b]{$r_2t$}}
\end{picture}\quad =\quad
\begin{picture}(2,1.5)
\put(0.0,0.8){\line(1,1){0.4}}
\put(0.0,1.2){\line(1,-1){0.4}}
\put(0.2,1){\circle{0.2}}
\put(0.0,-0.2){\line(1,1){0.4}}
\put(0.0,0.2){\line(1,-1){0.4}}
\put(0.2,0){\circle{0.2}}
\put(0.2,0){\line(1,0){1.3}}
\put(0.2,1){\line(1,0){1.3}}
\multiput(0.5,1)(0.5,0){3}{\circle*{0.1}}
\multiput(0.5,0)(0.5,0){3}{\circle*{0.1}}
\multiput(1,1)(0,-0.125){8}{\line(0,-1){0.09}}
\put(1.4,0.2){\makebox(0,0)[b]{$i$}}
\put(1.4,1.2){\makebox(0,0)[b]{$k$}}
\put(1.5,-0.4){\makebox(0,0)[b]{$r_1t$}}
\put(1.5,0.6){\makebox(0,0)[b]{$r_2t$}}
\end{picture}\quad +\quad
\begin{picture}(2.5,1.5)
\put(0.0,0.8){\line(1,1){0.4}}
\put(0.0,1.2){\line(1,-1){0.4}}
\put(0.2,1){\circle{0.2}}
\put(0.0,-0.2){\line(1,1){0.4}}
\put(0.0,0.2){\line(1,-1){0.4}}
\put(0.2,0){\circle{0.2}}
\put(0.2,0){\line(1,0){1.8}}
\put(0.2,1){\line(1,0){1.8}}
\multiput(0.5,1)(0.5,0){4}{\circle*{0.1}}
\multiput(0.5,0)(0.5,0){4}{\circle*{0.1}}
\multiput(1,1)(0,-0.125){8}{\line(0,-1){0.09}}
\multiput(1.5,1)(0,-0.125){8}{\line(0,-1){0.09}}
\put(1.9,0.2){\makebox(0,0)[b]{$i$}}
\put(1.9,1.2){\makebox(0,0)[b]{$k$}}
\put(2,-0.4){\makebox(0,0)[b]{$r_1t$}}
\put(2,0.6){\makebox(0,0)[b]{$r_2t$}}
\end{picture}\quad +\quad
\begin{picture}(2.5,1.5)
\put(0.0,0.8){\line(1,1){0.4}}
\put(0.0,1.2){\line(1,-1){0.4}}
\put(0.2,1){\circle{0.2}}
\put(0.0,-0.2){\line(1,1){0.4}}
\put(0.0,0.2){\line(1,-1){0.4}}
\put(0.2,0){\circle{0.2}}
\put(0.,0.3){\line(1,1){0.4}}
\put(0.,0.7){\line(1,-1){0.4}}
\put(0.2,0){\line(1,0){1.8}}
\put(0.2,0.5){\circle{0.2}}
\put(0.2,0.5){\line(1,0){1.1}}
\put(0.2,1){\line(1,0){1.8}}
\put(1.5,1){\circle*{0.1}}
\put(0.5,1){\circle*{0.1}}
\put(0.5,0.5){\circle*{0.1}}
\put(1,0.5){\circle*{0.1}}
\put(2,1){\circle*{0.1}}
\multiput(0.5,0)(0.5,0){4}{\circle*{0.1}}
\multiput(1,0.5)(0,-0.125){4}{\line(0,-1){0.09}}
\multiput(1.5,1)(0,-0.125){8}{\line(0,-1){0.09}}
\put(1.9,0.2){\makebox(0,0)[b]{$i$}}
\put(1.9,1.2){\makebox(0,0)[b]{$k$}}
\put(2,-0.4){\makebox(0,0)[b]{$r_1t$}}
\put(2,0.6){\makebox(0,0)[b]{$r_2t$}}
\end{picture}\quad +\quad \cdots
\label{pDiag}
\end{equation}
\mbox{}\\
Using this result, one can derive the following \emph{exact} relation:
\begin{equation}
\label{Svyaz'CiP}
\begin{picture}(1,0.5)
\put(0.2,0.){\line(1,0){.3}}
\put(0.5,0){\circle*{0.1}}
\put(0.0,-0.2){\line(1,1){0.4}}
\put(0.0,0.2){\line(1,-1){0.4}}
\put(0.5,0.2){\makebox(0,0)[b]{$i$}}
\put(.6,-0.4){\makebox(0,0)[b]{$rt$}}
\end{picture}
\quad=\quad
\begin{picture}(1,1)
\put(1,0){\circle*{0.1}}
\put(0.2,0.){\line(1,0){.8}}
\put(0.5,0){\circle*{0.1}}
\put(0.0,-0.2){\line(1,1){0.4}}
\put(0.0,0.2){\line(1,-1){0.4}}
\put(0.2,0.){\circle{0.2}}
\put(.9,0.2){\makebox(0,0)[b]{$i$}}
\put(1.0,-0.4){\makebox(0,0)[b]{$rt$}}
\end{picture}\quad+\quad
\begin{picture}(1.,1)
\put(0.0,0.8){\line(1,1){0.4}}
\put(0.0,1.2){\line(1,-1){0.4}}
\put(0.0,-0.2){\line(1,1){0.4}}
\put(0.0,0.2){\line(1,-1){0.4}}
\put(0.2,0){\line(1,0){0.8}}
\put(0.2,1){\line(1,0){0.6}}
\put(0.5,1){\circle*{0.1}}
\multiput(0.5,0)(0.5,0){2}{\circle*{0.1}}
\multiput(0.5,1)(0,-0.125){8}{\line(0,-1){0.1}}
\put(.9,0.2){\makebox(0,0)[b]{$i$}}
\put(1.,-0.4){\makebox(0,0)[b]{$rt$}}
\end{picture}\quad +\quad
\begin{picture}(1.,1)
\put(0.5,0){\line(1,0){0.5}}
\put(0.5,1){\line(1,0){0.3}}
\put(0.5,0.5){\oval(1,1)[l]}
\put(0.5,1){\circle*{0.1}}
\multiput(0.5,0)(0.5,0){2}{\circle*{0.1}}
\multiput(0.5,1)(0,-0.125){8}{\line(0,-1){0.1}}
\put(.9,0.2){\makebox(0,0)[b]{$i$}}
\put(1.,-0.4){\makebox(0,0)[b]{$rt$}}
\end{picture}
\end{equation}

\end{widetext}
The derivation is based on the fact that all diagrams in Eq.(\ref{DCSeries}), except the first one, have $X=
\begin{picture}(1,1)
\put(0.2,0.8){\circle*{0.1}}
\put(0.0,0.8){\line(1,0){0.4}}
\multiput(0.225,0.8)(0,-0.125){4}{\line(0,-1){0.1}}
\put(0.2,0.3){\circle*{0.1}}
\put(0.0,0.3){\line(1,0){0.7}}
\put(0.7,0.3){\circle*{0.1}}
\put(0.7,0.5){\makebox(0,0)[b]{$i$}}
\put(0.7,-0.1){\makebox(0,0)[b]{$rt$}}
\end{picture}$
as their rightmost element. This element represents the product of Green's
function and interaction operator with the corresponding summation of the
omitted indices and integration over the omitted coordinates and time variables. This element can be extracted as a common multiplier of the sum of diagrams identical to that in Eq.(\ref{eq:PDiagram}). Substitution
of Eqs.(\ref{eq:PairFunction}) and (\ref{pDiag}), appended from the right
with $X$, results in Eq.(\ref{Svyaz'CiP}). In this equation, the first term
represents the evolution of initial conditions in the absence of
interparticle interactions, the second and third terms represent the
contributions from the concentration product and pair correlation function in Eq.(\ref{eq:PairFunction}), respectively. The analytical expression for
(\ref{Svyaz'CiP}) is as follows:
\begin{widetext}
\begin{equation}
\label{CiP}
\begin{gathered}
C_i(\vec r,t)=\sum_k \int G^0_{ik}(\vec r,t,\vec r^{\,\prime},0) C^0_k(\vec r^{\,\prime})\,
d^3 r^{\,\prime} + \sum_{klmn} \int_0^t \,dt'
\int G^0_{ik}(\vec r,t,\vec r_1,t') \widehat V_{klmn}(\vec r_1,\vec r_2)
P_{mn}(\vec r_1,\vec r_2,t') \,d^3 r_1\,d^3 r_2
\end{gathered}
\end{equation}

\end{widetext}
The time differentiation of both sides of this equation, using
Eq.(\ref{EqGreen}), gives
\begin{equation}
\label{EqCi}
\partial_t C_i(\vec r,t)= \sum_j  \widehat L^0_{ij}
C_j(\vec r,t)+ R_i(\vec r,t)
\end{equation}
where $R_i(\vec r,t)$ is a collision integral representing the
reaction rate at point $\vec r$ at time $t$
\begin{equation}
\label{EqRi}
R_i(\vec r,t) = \sum_{{jlm}}\int\widehat{V}_{ij,lm}(\vec r,\vec r^{\,\prime})
  P_{lm}(\vec r,\vec r^{\,\prime},t)\,d^3 r^{\,\prime}
\end{equation}
This equation can also be derived directly from Eqs.(\ref{eq:LocalConc}),
(\ref{eq:Pair}) and ({\ref{eq:EqPhi}) (cf. Refs. \cite{DoiJPA1976,DoiJPA1976s}). In fact, it
represents the first equation of the BBGKY hierarchy for our reacting system.
Eq.(\ref{EqRi}) has the following diagrammatic representation:
\begin{equation}
\label{DiagR_i}
R_i(\vec r,t) =
\begin{picture}(1.,1.25)(0,-0.125)
\put(0.0,0.8){\line(1,1){0.4}} \put(0.0,1.2){\line(1,-1){0.4}}
\put(0.0,-0.2){\line(1,1){0.4}} \put(0.0,0.2){\line(1,-1){0.4}}
\put(0.2,0){\line(1,0){0.8}} \put(0.2,1){\line(1,0){0.8}}
\put(0.7,1){\circle*{0.1}} \put(0.7,0){\circle*{0.1}}
\multiput(0.7,1)(0,-0.125){8}{\line(0,-1){0.1}}
\put(.9,0.2){\makebox(0,0)[b]{$i$}}
\put(0.7,-0.4){\makebox(0,0)[b]{$rt$}}
\end{picture}\quad + \quad
\begin{picture}(1.,1.25)(0,-0.125)
\put(0.5,0){\line(1,0){0.3}} \put(0.5,1){\line(1,0){0.3}}
\put(0.5,0.5){\oval(1,1)[l]} \put(0.5,1){\circle*{0.1}}
\put(0.5,0){\circle*{0.1}}
\multiput(0.5,1)(0,-0.125){8}{\line(0,-1){0.1}}
\put(.7,0.2){\makebox(0,0)[b]{$i$}}
\put(0.5,-0.4){\makebox(0,0)[b]{$rt$}}
\end{picture}
\end{equation}

Eq.(\ref{EqCi}) gives the algorithm for the calculation of
$C_i(r,t)$. It includes the calculation of the collision integral
(\ref{EqRi}), which is not tractable without making approximations.
We shall consider these approximations in the next Sections. We will present the diagrammatic expression for $p_{ik}$ in an irreducible form by using partial summation, similar
to that used for the derivation of Eq.(\ref{Svyaz'CiP}). In Eq.(\ref{pDiag}), we will consider diagrams which contain s subdiagram, ${\cal D}_1$, connected to the the rest of the diagram,
${\cal D}_2$ only by the interaction with state $A_*$ at $(\vec *,*)$
(here $*$ stands for the unspecified integration/summation
variables), such that erasing the line corresponding to this interaction
would break the diagram into two disconnected parts, one of which
(${\cal D}_1$) leads to state $A_*$ at $(\vec *,*)$. By extracting ${\cal D}_2$  as common multipliers and using
Eq.(\ref{DCSeries}), the sum over the corresponding ${\cal D}_1$  can
be reduced to concentrations $C_*(\vec *,*)$. Recursively repeating the
described procedure, one can express $p_{ik}(\vec r_1, \vec r_2,t)$
via concentrations at time $t=*$ (instead of those at $t=0$):
\begin{widetext}
\begin{equation}
\label{IrrP}
p_{ik}(\vec r_1, \vec r_2,t) \quad\equiv\quad
\begin{picture}(1.,1)
\put(0.5,1){\circle*{0.1}}
\put(0.5,0){\circle*{0.1}}
\put(0.5,0.5){\oval(1,1)[l]}
\put(.4,0.2){\makebox(0,0)[b]{$i$}}
\put(.4,1.2){\makebox(0,0)[b]{$k$}}
\put(.5,-0.4){\makebox(0,0)[b]{$r_1t$}}
\put(.5,0.6){\makebox(0,0)[b]{$r_2t$}}
\end{picture}=~
\begin{picture}(1.25,1.5)
\put(0,0.8){\line(1,1){0.4}}
\put(0.,1.2){\line(1,-1){0.4}}
\put(0.,-0.2){\line(1,1){0.4}}
\put(0,0.2){\line(1,-1){0.4}}
\put(0.2,0){\line(1,0){0.3}}
\put(0.2,1){\line(1,0){0.3}}
\put(0.5,0){\circle*{0.1}}
\put(1.2,0.){\circle*{0.1}}
\put(0.5,0.0){\line(1,0){0.7}}
\put(0.5,1){\circle*{0.1}}
\put(1.2,1){\circle*{0.1}}
\put(0.5,1.0){\line(1,0){0.7}}
\multiput(0.5,1)(0,-0.125){8}{\line(0,-1){0.09}}
\put(1.1,0.2){\makebox(0,0)[b]{$i$}}
\put(1.1,1.2){\makebox(0,0)[b]{$k$}}
\put(1.2,0.6){\makebox(0,0)[b]{$r_2t$}}
\put(1.2,-0.4){\makebox(0,0)[b]{$r_1t$}}
\end{picture}~~+~
\begin{picture}(1.5,1.5)
\put(0,0.8){\line(1,1){0.4}}
\put(0.,1.2){\line(1,-1){0.4}}
\put(0.,-0.2){\line(1,1){0.4}}
\put(0,0.2){\line(1,-1){0.4}}
\put(0.2,0){\line(1,0){0.3}}
\put(0.2,1){\line(1,0){0.3}}
\put(0.5,0){\circle*{0.1}}
\put(1.5,0){\circle*{0.1}}
\put(1.0,0){\circle*{0.1}}
\put(0.5,0.0){\line(1,0){1}}
\put(0.5,1){\circle*{0.1}}
\put(1.5,1){\circle*{0.1}}
\put(1,1){\circle*{0.1}}
\put(0.5,1.0){\line(1,0){1}}
\multiput(0.5,1)(0,-0.125){8}{\line(0,-1){0.09}}
\multiput(1,1)(0,-0.125){8}{\line(0,-1){0.09}}
\put(1.4,0.2){\makebox(0,0)[b]{$i$}}
\put(1.4,1.2){\makebox(0,0)[b]{$k$}}
\put(1.5,-0.4){\makebox(0,0)[b]{$r_1t$}}
\put(1.5,0.6){\makebox(0,0)[b]{$r_2t$}}
\end{picture}~~+~
\begin{picture}(2,1)
\put(0,0.8){\line(1,1){0.4}}
\put(0.,1.2){\line(1,-1){0.4}}
\put(0.,-0.2){\line(1,1){0.4}}
\put(0,0.2){\line(1,-1){0.4}}
\put(0.2,0){\line(1,0){0.3}}
\put(0.2,1){\line(1,0){0.3}}
\put(0.5,1){\circle*{0.1}}
\put(2,1){\circle*{0.1}}
\put(0.5,0.){\circle*{0.1}}
\put(2,0.){\circle*{0.1}}
\put(0.5,1.0){\line(1,0){1.5}}
\put(0.5,0.0){\line(1,0){1.5}}
\multiput(1,1)(0.5,0){2}{\circle*{0.1}}
\multiput(1,0)(0.5,0){2}{\circle*{0.1}}
\multiput(0.5,1)(0,-0.125){8}{\line(0,-1){0.09}}
\multiput(1,1)(0,-0.125){8}{\line(0,-1){0.09}}
\multiput(1.5,1)(0,-0.125){8}{\line(0,-1){0.09}}
\put(1.9,0.2){\makebox(0,0)[b]{$i$}}
\put(1.9,1.2){\makebox(0,0)[b]{$k$}}
\put(2,-0.4){\makebox(0,0)[b]{$r_1t$}}
\put(2,0.6){\makebox(0,0)[b]{$r_2t$}}
\end{picture}~~+~\dots~+~
\begin{picture}(2,1)
\put(0,0.8){\line(1,1){0.4}}
\put(0.,1.2){\line(1,-1){0.4}}
\put(0.,-0.2){\line(1,1){0.4}}
\put(0,0.2){\line(1,-1){0.4}}
\put(0.2,0){\line(1,0){0.3}}
\put(0.2,1){\line(1,0){0.3}}
\put(0.5,1){\circle*{0.1}}
\put(0.5,0.){\circle*{0.1}}
\put(0.5,1.0){\line(1,0){1}}
\put(0.5,0.0){\line(1,0){1}}
\multiput(1,1)(0.5,0){2}{\circle*{0.1}}
\put(1.5,0){\circle*{0.1}}
\put(0.5,1.3){\line(1,1){0.4}}
\put(0.5,1.7){\line(1,-1){0.4}}
\put(0.7,1.5){\line(1,0){0.5}}
\put(1.,1.5){\circle*{0.1}}
\multiput(0.5,1)(0,-0.125){8}{\line(0,-1){0.09}}
\multiput(1,1.5)(0,-0.125){4}{\line(0,-1){0.09}}
\put(1.4,0.2){\makebox(0,0)[b]{$i$}}
\put(1.4,1.2){\makebox(0,0)[b]{$k$}}
\put(1.5,-0.4){\makebox(0,0)[b]{$r_1t$}}
\put(1.5,0.6){\makebox(0,0)[b]{$r_2t$}}
\end{picture}~~+~\cdots
\end{equation}
\mbox{}\\

The substitution of this series into  Eq.(\ref{EqRi}) gives the diagrammatic expansion
of the collision integral $R_i$:
\begin{equation}
\label{IntRi}
R_i(\vec r,t)\quad = \quad
\begin{picture}(1.,1)
\put(0.0,0.8){\line(1,1){0.4}}
\put(0.0,1.2){\line(1,-1){0.4}}
\put(0.0,-0.2){\line(1,1){0.4}}
\put(0.0,0.2){\line(1,-1){0.4}}
\put(0.2,0){\line(1,0){0.8}} \put(0.2,1){\line(1,0){0.8}}
\put(0.7,1){\circle*{0.1}}
\put(0.7,0){\circle*{0.1}}
\multiput(0.7,1)(0,-0.125){8}{\line(0,-1){0.1}}
\put(.9,0.2){\makebox(0,0)[b]{$i$}}
\put(0.7,-0.4){\makebox(0,0)[b]{$rt$}}
\end{picture}~~+~
\begin{picture}(1.5,1.5)
\put(0,0.8){\line(1,1){0.4}}
\put(0.,1.2){\line(1,-1){0.4}}
\put(0.,-0.2){\line(1,1){0.4}}
\put(0,0.2){\line(1,-1){0.4}}
\put(0.2,0){\line(1,0){0.3}}
\put(0.2,1){\line(1,0){0.3}}
\put(0.5,0){\circle*{0.1}}
\put(1.25,0.){\circle*{0.1}}
\put(0.5,0.0){\line(1,0){1}}
\put(0.5,1){\circle*{0.1}}
\put(1.25,1){\circle*{0.1}}
\put(0.5,1.0){\line(1,0){1}}
\multiput(0.5,1)(0,-0.125){8}{\line(0,-1){0.09}}
\multiput(1.25,1)(0,-0.125){8}{\line(0,-1){0.09}}
\put(1.45,0.2){\makebox(0,0)[b]{$i$}}
\put(1.25,-0.4){\makebox(0,0)[b]{$rt$}}
\end{picture}~~+~
\begin{picture}(1.75,1.5)
\put(0,0.8){\line(1,1){0.4}}
\put(0.,1.2){\line(1,-1){0.4}}
\put(0.,-0.2){\line(1,1){0.4}}
\put(0,0.2){\line(1,-1){0.4}}
\put(0.2,0){\line(1,0){0.3}}
\put(0.2,1){\line(1,0){0.3}}
\put(0.5,0){\circle*{0.1}}
\put(1.5,0){\circle*{0.1}}
\put(1.0,0){\circle*{0.1}}
\put(0.5,0.0){\line(1,0){1.3}}
\put(0.5,1){\circle*{0.1}}
\put(1.5,1){\circle*{0.1}}
\put(1,1){\circle*{0.1}}
\put(0.5,1.0){\line(1,0){1.3}}
\multiput(0.5,1)(0,-0.125){8}{\line(0,-1){0.09}}
\multiput(1,1)(0,-0.125){8}{\line(0,-1){0.09}}
\multiput(1.5,1)(0,-0.125){8}{\line(0,-1){0.09}}
\put(1.7,0.2){\makebox(0,0)[b]{$i$}}
\put(1.5,-0.4){\makebox(0,0)[b]{$rt$}}
\end{picture}~~+~
\begin{picture}(2.25,1)
\put(0,0.8){\line(1,1){0.4}}
\put(0.,1.2){\line(1,-1){0.4}}
\put(0.,-0.2){\line(1,1){0.4}}
\put(0,0.2){\line(1,-1){0.4}}
\put(0.2,0){\line(1,0){0.3}}
\put(0.2,1){\line(1,0){0.3}}
\put(0.5,1){\circle*{0.1}}
\put(2,1){\circle*{0.1}}
\put(0.5,0.){\circle*{0.1}}
\put(2,0.){\circle*{0.1}}
\put(0.5,1.0){\line(1,0){1.8}}
\put(0.5,0.0){\line(1,0){1.8}}
\multiput(1,1)(0.5,0){2}{\circle*{0.1}}
\multiput(1,0)(0.5,0){2}{\circle*{0.1}}
\multiput(0.5,1)(0,-0.125){8}{\line(0,-1){0.09}}
\multiput(1,1)(0,-0.125){8}{\line(0,-1){0.09}}
\multiput(1.5,1)(0,-0.125){8}{\line(0,-1){0.09}}
\multiput(2,1)(0,-0.125){8}{\line(0,-1){0.09}}
\put(2.2,0.2){\makebox(0,0)[b]{$i$}}
\put(2,-0.4){\makebox(0,0)[b]{$rt$}}
\end{picture}~~+~\dots~+~
\begin{picture}(1.75,1)
\put(0,0.8){\line(1,1){0.4}}
\put(0.,1.2){\line(1,-1){0.4}}
\put(0.,-0.2){\line(1,1){0.4}}
\put(0,0.2){\line(1,-1){0.4}}
\put(0.2,0){\line(1,0){0.3}}
\put(0.2,1){\line(1,0){0.3}}
\put(0.5,1){\circle*{0.1}}
\put(0.5,0.){\circle*{0.1}}
\put(0.5,1.0){\line(1,0){1.3}}
\put(0.5,0.0){\line(1,0){1.3}}
\multiput(1,1)(0.5,0){2}{\circle*{0.1}}
\put(1.5,0){\circle*{0.1}}
\put(0.5,1.3){\line(1,1){0.4}}
\put(0.5,1.7){\line(1,-1){0.4}}
\put(0.7,1.5){\line(1,0){0.6}}
\put(1.,1.5){\circle*{0.1}}
\multiput(0.5,1)(0,-0.125){8}{\line(0,-1){0.09}}
\multiput(1,1.5)(0,-0.125){4}{\line(0,-1){0.09}}
\multiput(1.5,1)(0,-0.125){8}{\line(0,-1){0.09}}
\put(1.7,0.2){\makebox(0,0)[b]{$i$}}
\put(1.5,-0.4){\makebox(0,0)[b]{$rt$}}
\end{picture}~~+~\cdots
\end{equation}
\mbox{}
\end{widetext}

The diagrams in Eqs.(\ref{IrrP}) and (\ref{IntRi}) can be classified
by their order in the concentration expansion. This order is defined by
the number of crosses the diagram starts with. We will denote the
order as a superscript in parentheses:
\begin{equation}
\label{eq:COrder}
R_i(\vec r,t) = R^{(2)}_i(\vec r,t) + R^{(3)}_i(\vec r,t) + \dots
+ R^{(n)}_i(\vec r,t) + \dots
\end{equation}

We will interchangeably use these expansions for both the
collision integral and the pair distribution functions. These series
will serve as a starting point for making further approximations
to obtain the set of kinetic equations. The first
approximation that we consider corresponds to integral encounter theory (IET).

\section{Integral encounter theory }
\label{sec:EncounterTheory}

The IET  corresponds to the lowest (second) order  approximation
in the concentration series of the collision integral Eq.(\ref{IntRi}).
Leaving only the second-order terms in concentration results in the following series:
\newpage

\begin{widetext}
\begin{equation}
\label{IntRi2}
R^{(2)}_i(\vec r,t)\quad = \quad
\begin{picture}(1.,1)
\put(0.0,0.8){\line(1,1){0.4}}
\put(0.0,1.2){\line(1,-1){0.4}}
\put(0.0,-0.2){\line(1,1){0.4}}
\put(0.0,0.2){\line(1,-1){0.4}}
\put(0.2,0){\line(1,0){0.8}} \put(0.2,1){\line(1,0){0.8}}
\put(0.7,1){\circle*{0.1}}
\put(0.7,0){\circle*{0.1}}
\multiput(0.7,1)(0,-0.125){8}{\line(0,-1){0.1}}
\put(.9,0.2){\makebox(0,0)[b]{$i$}}
\put(0.7,-0.4){\makebox(0,0)[b]{$rt$}}
\end{picture}~~+~
\begin{picture}(1.5,1.5)
\put(0,0.8){\line(1,1){0.4}}
\put(0.,1.2){\line(1,-1){0.4}}
\put(0.,-0.2){\line(1,1){0.4}}
\put(0,0.2){\line(1,-1){0.4}}
\put(0.2,0){\line(1,0){0.3}}
\put(0.2,1){\line(1,0){0.3}}
\put(0.5,0){\circle*{0.1}}
\put(1.25,0.){\circle*{0.1}}
\put(0.5,0.0){\line(1,0){1}}
\put(0.5,1){\circle*{0.1}}
\put(1.25,1){\circle*{0.1}}
\put(0.5,1.0){\line(1,0){1}}
\multiput(0.5,1)(0,-0.125){8}{\line(0,-1){0.09}}
\multiput(1.25,1)(0,-0.125){8}{\line(0,-1){0.09}}
\put(1.45,0.2){\makebox(0,0)[b]{$i$}}
\put(1.25,-0.4){\makebox(0,0)[b]{$rt$}}
\end{picture}~~+~
\begin{picture}(1.75,1.5)
\put(0,0.8){\line(1,1){0.4}}
\put(0.,1.2){\line(1,-1){0.4}}
\put(0.,-0.2){\line(1,1){0.4}}
\put(0,0.2){\line(1,-1){0.4}}
\put(0.2,0){\line(1,0){0.3}}
\put(0.2,1){\line(1,0){0.3}}
\put(0.5,0){\circle*{0.1}}
\put(1.5,0){\circle*{0.1}}
\put(1.0,0){\circle*{0.1}}
\put(0.5,0.0){\line(1,0){1.3}}
\put(0.5,1){\circle*{0.1}}
\put(1.5,1){\circle*{0.1}}
\put(1,1){\circle*{0.1}}
\put(0.5,1.0){\line(1,0){1.3}}
\multiput(0.5,1)(0,-0.125){8}{\line(0,-1){0.09}}
\multiput(1,1)(0,-0.125){8}{\line(0,-1){0.09}}
\multiput(1.5,1)(0,-0.125){8}{\line(0,-1){0.09}}
\put(1.7,0.2){\makebox(0,0)[b]{$i$}}
\put(1.5,-0.4){\makebox(0,0)[b]{$rt$}}
\end{picture}~~+~
\begin{picture}(2.25,1)
\put(0,0.8){\line(1,1){0.4}}
\put(0.,1.2){\line(1,-1){0.4}}
\put(0.,-0.2){\line(1,1){0.4}}
\put(0,0.2){\line(1,-1){0.4}}
\put(0.2,0){\line(1,0){0.3}}
\put(0.2,1){\line(1,0){0.3}}
\put(0.5,1){\circle*{0.1}}
\put(2,1){\circle*{0.1}}
\put(0.5,0.){\circle*{0.1}}
\put(2,0.){\circle*{0.1}}
\put(0.5,1.0){\line(1,0){1.8}}
\put(0.5,0.0){\line(1,0){1.8}}
\multiput(1,1)(0.5,0){2}{\circle*{0.1}}
\multiput(1,0)(0.5,0){2}{\circle*{0.1}}
\multiput(0.5,1)(0,-0.125){8}{\line(0,-1){0.09}}
\multiput(1,1)(0,-0.125){8}{\line(0,-1){0.09}}
\multiput(1.5,1)(0,-0.125){8}{\line(0,-1){0.09}}
\multiput(2,1)(0,-0.125){8}{\line(0,-1){0.09}}
\put(2.2,0.2){\makebox(0,0)[b]{$i$}}
\put(2,-0.4){\makebox(0,0)[b]{$rt$}}
\end{picture}~+~\cdots
\end{equation}
\mbox{}\\

This diagram can be rewritten as
\begin{equation}
\label{R2ndorder}
R^{(2)}_i(\vec r,t)
\quad= \quad
\begin{picture}(2,1.125)
\put(-0.2,0.6){\line(1,1){0.4}} \put(-0.2,1.0){\line(1,-1){0.4}}
\put(-0.2,-0.2){\line(1,1){0.4}} \put(-0.2,0.2){\line(1,-1){0.4}}
\put(0.4,0){\circle*{0.1}} \put(1.1,0){\circle*{0.1}}
\multiput(0.4,0)(0,0.1){7}{\line(4,1){0.7}}
\put(0.4,0.8){\circle*{0.1}} \put(1.1,0.8){\circle*{0.1}}
\put(0,0.8){\line(1,0){1.3}} \put(0,0.){\line(1,0){1.3}}
\put(0.4,0.8){\line(0,-1){0.8}} \put(1.1,0.8){\line(0,-1){0.8}}
\put(1.3,0.1){\makebox(0,0)[b]{$i$}}
\put(1.1,-0.4){\makebox(0,0)[b]{$rt$}}
\end{picture} \quad \mbox{, where} \qquad
\begin{picture}(1.5,1.125)
\put(0.4,0){\circle*{0.1}}
\put(1.1,0){\circle*{0.1}}
\multiput(0.4,0)(0,0.1){7}{\line(4,1){0.7}}
\put(0.4,0.8){\circle*{0.1}}
\put(1.1,0.8){\circle*{0.1}}
\put(0,0.8){\line(1,0){1.5}}
\put(0,0.){\line(1,0){1.5}}
\put(0.4,0.8){\line(0,-1){0.8}}
\put(1.1,0.8){\line(0,-1){0.8}}
\put(1.4,0.1){\makebox(0,0)[b]{$i$}}
\put(1.1,-0.4){\makebox(0,0)[b]{$r_1t$}}
\put(1.1,1){\makebox(0,0)[b]{$r_2$}}
\put(1.4,0.5){\makebox(0,0)[b]{$k$}}
\put(0.1,0.1){\makebox(0,0)[b]{$l$}}
\put(0.4,-0.4){\makebox(0,0)[b]{$r_1't'$}}
\put(0.4,1){\makebox(0,0)[b]{$r_2'$}}
\put(0.1,0.5){\makebox(0,0)[b]{$m$}}
\end{picture} \quad \equiv \quad
T_{ik,lm}(\vec r_1,\vec r_2,t,\vec r^{\,\prime}_1,\vec r^{\,\prime}_2,t')
\end{equation}
\mbox{}\\
is the pair $T$-matrix, defined as the sum of all ladder \cite{Balescu} diagrams:
\begin{equation}
\begin{gathered}
\label{TmatrixDiagram}
\begin{picture}(1.5,1.125)
\put(0.4,0){\circle*{0.1}}
\put(1.1,0){\circle*{0.1}}
\multiput(0.4,0)(0,0.1){7}{\line(4,1){0.7}}
\put(0.4,0.8){\circle*{0.1}}
\put(1.1,0.8){\circle*{0.1}}
\put(0,0.8){\line(1,0){1.5}}
\put(0,0.){\line(1,0){1.5}}
\put(0.4,0.8){\line(0,-1){0.8}}
\put(1.1,0.8){\line(0,-1){0.8}}
\put(1.4,0.1){\makebox(0,0)[b]{$i$}}
\put(1.1,-0.4){\makebox(0,0)[b]{$r_1t$}}
\put(1.1,1){\makebox(0,0)[b]{$r_2$}}
\put(1.4,0.5){\makebox(0,0)[b]{$k$}}
\put(0.1,0.1){\makebox(0,0)[b]{$l$}}
\put(0.4,-0.4){\makebox(0,0)[b]{$r_1't'$}}
\put(0.4,1){\makebox(0,0)[b]{$r_2'$}}
\put(0.1,0.5){\makebox(0,0)[b]{$m$}}
\end{picture}
\quad = \quad
\begin{picture}(1,1.125)
\put(0.5,0){\circle*{0.1}}
\put(0,0){\line(1,0){1}}
\put(0.5,0.8){\circle*{0.1}}
\put(0,0.8){\line(1,0){1}}
\multiput(0.5,0.8)(0,-0.125){7}{\line(0,-1){0.1}}
\put(0.9,0.1){\makebox(0,0)[b]{$i$}}
\put(0.5,-0.4){\makebox(0,0)[b]{$r_1t$}}
\put(0.5,1){\makebox(0,0)[b]{$r_2$}}
\put(0.9,0.5){\makebox(0,0)[b]{$k$}}
\put(0.1,0.1){\makebox(0,0)[b]{$l$}}
\put(0.1,0.5){\makebox(0,0)[b]{$m$}}
\end{picture}\quad \times \delta(t-t')\delta(\vec r_1-\vec r^{\,\prime}_1)
\delta(\vec r_2-\vec r^{\,\prime}_2)
 + \quad
\begin{picture}(1.5,1.125)
\put(0,0){\line(1,0){1.5}}
\put(0,0.8){\line(1,0){1.5}}
\multiput(0.4,0.8)(0.7,0){2}{\circle*{0.1}}
\multiput(0.4,0)(0.7,0){2}{\circle*{0.1}}
\multiput(0.4,0.8)(0,-0.125){7}{\line(0,-1){0.1}}
\multiput(1.1,0.8)(0,-0.125){7}{\line(0,-1){0.1}}
\put(1.4,0.1){\makebox(0,0)[b]{$i$}}
\put(1.1,-0.4){\makebox(0,0)[b]{$r_1t$}}
\put(1.1,1){\makebox(0,0)[b]{$r_2$}}
\put(1.4,0.5){\makebox(0,0)[b]{$k$}}
\put(0.1,0.1){\makebox(0,0)[b]{$l$}}
\put(0.4,-0.4){\makebox(0,0)[b]{$r_1't'$}}
\put(0.4,1){\makebox(0,0)[b]{$r_2'$}}
\put(0.1,0.5){\makebox(0,0)[b]{$m$}}
\end{picture}\quad +\quad
\begin{picture}(2,1.125)
\put(0,0){\line(1,0){2}}
\put(0,0.8){\line(1,0){2}}
\multiput(0.5,0.8)(0.5,0){3}{\circle*{0.1}}
\multiput(0.5,0)(0.5,0){3}{\circle*{0.1}}
\multiput(0.5,0.8)(0,-0.125){7}{\line(0,-1){0.1}}
\multiput(1,0.8)(0,-0.125){7}{\line(0,-1){0.1}}
\multiput(1.5,0.8)(0,-0.125){7}{\line(0,-1){0.1}}
\put(1.9,0.1){\makebox(0,0)[b]{$i$}}
\put(1.5,-0.4){\makebox(0,0)[b]{$r_1t$}}
\put(1.5,1){\makebox(0,0)[b]{$r_2$}}
\put(1.9,0.5){\makebox(0,0)[b]{$k$}}
\put(0.1,0.1){\makebox(0,0)[b]{$l$}}
\put(0.5,-0.4){\makebox(0,0)[b]{$r_1't'$}}
\put(0.5,1){\makebox(0,0)[b]{$r_2'$}}
\put(0.1,0.5){\makebox(0,0)[b]{$m$}}
\end{picture}+\quad \cdots
\end{gathered}
\end{equation}
\mbox{}\\

Substituting Eq.(\ref{R2ndorder}) for Eq.(\ref{EqCi}), we get the master equation of IET:
\begin{equation}
\label{eq:MasterETi}
\frac \partial{\partial t} C_i(\vec r,t) =
  \sum_j \widehat L^0_{ij} C_j(\vec r,t) +
  \sum_{klm}\int\limits_0^t\,dt' \int d^3 r^{\,\prime}\,d^3 r_1\,
  d^3 r^{\,\prime}_1
   T_{ik,lm} (\vec r,\vec r_1,t,\vec r^{\,\prime},\vec r^{\,\prime}_1,t')
   C_l(\vec r^{\,\prime},t')C_m(\vec r^{\,\prime}_1,t')
\end{equation}
This equation can be rewritten in the vector form
\begin{equation}
\label{eq:MasterET}
\frac \partial{\partial t}\vec C(\vec r,t) =
  \widehat L_0\vec C(\vec r,t) +
  \mbox{Tr}_2\int\limits_0^t\,dt' \int d^3 r^{\,\prime}\,d^3 r_1\,
  d^3 r^{\,\prime}_1
  \widehat{\bf T} (\vec r,\vec r_1,t,\vec r^{\,\prime},\vec r^{\,\prime}_1,t')
  \left(\vec C(\vec r^{\,\prime},t')\otimes\vec C(\vec r^{\,\prime}_1,t')\right)
\end{equation}

 The matrices in two-index (pair) space are bolded, acting on the matrices in the one-index (single particle) space,
  $\otimes$ is a direct product, and $\mbox{Tr}_2$ denotes summation over the
second particle $(\mbox{Tr}_2\widehat{ A})_i = \sum_j{ A}_{ij}$.

The pair $T$-matrix (\ref{TmatrixDiagram}) can be expressed as

\begin{eqnarray}
\label{Tmatrix}
 \widehat {\bf T}(\vec r_1,\vec r_2,t,\vec r^{\,\prime}_1,\vec r^{\,\prime}_2,t')
 =\widehat {\bf V}(\vec r_1,\vec r_2)\delta (t-t')
\delta(\vec r_1-\vec r^{\,\prime}_1)\delta(\vec r_2-\vec r^{\,\prime}_2)+
 \widehat {\bf V}(\vec r_1,\vec r_2)
 \widehat {\bf G}(\vec r_1,\vec r_2,t,\vec r^{\,\prime}_1,\vec r^{\,\prime}_2,t')
 \widehat {\bf V}(\vec r^{\,\prime}_1,\vec r^{\,\prime}_2)
\end{eqnarray}
where  $\widehat {\bf G}$ is the pair Green's function (PGF) which obeys the following equation

\begin{equation}
\label{GET}
\frac \partial{\partial t}
  \widehat {\bf G}(\vec r_1,\vec r_2,t,\vec r^{\,\prime}_1,\vec r^{\,\prime}_2,t')
 = \left( \widehat {\bf L} +
  \widehat {\bf Q} + \widehat {\bf V}(\vec r_1,\vec r_2) \right)
   \widehat {\bf G}(\vec r_1,\vec r_2,t,\vec r^{\,\prime}_1,\vec r^{\,\prime}_2,t')
\end{equation}

with the initial condition
$$\widehat {\bf G}(\vec r_1,\vec r_2,t,\vec r^{\,\prime}_1,\vec r^{\,\prime}_2,t) =
  \widehat {\bf I}\delta(\vec r_1-\vec r^{\,\prime}_1)
    \delta(\vec r_2-\vec r^{\,\prime}_2)$$
where
$$\widehat{\bf Q} = \widehat Q\otimes \widehat I +
                      \widehat I\otimes\widehat Q \quad, \mbox{and} \quad \widehat { L}_{ik,lm}= \delta_{il}\delta_{km}
(D_i\Delta_1 +D_j\Delta_2)$$

Eq.(\ref{eq:MasterET}) implies that the integral encounter theory is equivalent to the averaged $T$-matrix approximation (ATA). ATA formalism is convenient for the description of diffusion-influenced
reactions (Ref.\cite{DoiJPA1976,DoiJPA1976s}), particularly because the $T$-matrix remains finite
even in the limit of infinite reactivity or interaction potential (black
sphere approximation or hardcore repulsion).

As follows from Eq.(\ref{DiagR_i}), the pair correlation function in this approximation is

\begin{equation}
\label{eq:pET}
p_{ik}(\vec r_1,\vec r_2,t) \quad = \quad
\begin{picture}(1.5,1.5)
\put(0,0.8){\line(1,1){0.4}}
\put(0.,1.2){\line(1,-1){0.4}}
\put(0.,-0.2){\line(1,1){0.4}}
\put(0,0.2){\line(1,-1){0.4}}
\put(0.2,0){\line(1,0){0.3}}
\put(0.2,1){\line(1,0){0.3}}
\put(0.5,0){\circle*{0.1}}
\put(1.5,0.){\circle*{0.1}}
\put(0.5,0.0){\line(1,0){1}}
\put(0.5,1){\circle*{0.1}}
\put(1.5,1){\circle*{0.1}}
\put(0.5,1.0){\line(1,0){1}}
\multiput(0.5,1)(0,-0.125){8}{\line(0,-1){0.09}}
\put(1.4,0.2){\makebox(0,0)[b]{$i$}}
\put(1.4,1.2){\makebox(0,0)[b]{$k$}}
\put(1.5,0.6){\makebox(0,0)[b]{$r_2t$}}
\put(1.5,-0.4){\makebox(0,0)[b]{$r_1t$}}
\end{picture}\quad + \quad
\begin{picture}(1.5,1.5)
\put(0,0.8){\line(1,1){0.4}}
\put(0.,1.2){\line(1,-1){0.4}}
\put(0.,-0.2){\line(1,1){0.4}}
\put(0,0.2){\line(1,-1){0.4}}
\put(0.2,0){\line(1,0){0.3}}
\put(0.2,1){\line(1,0){0.3}}
\put(0.5,0){\circle*{0.1}}
\put(1.5,0){\circle*{0.1}}
\put(1.0,0){\circle*{0.1}}
\put(0.5,0.0){\line(1,0){1}}
\put(0.5,1){\circle*{0.1}}
\put(1.5,1){\circle*{0.1}}
\put(1,1){\circle*{0.1}}
\put(0.5,1.0){\line(1,0){1}}
\multiput(0.5,1)(0,-0.125){8}{\line(0,-1){0.09}}
\multiput(1,1)(0,-0.125){8}{\line(0,-1){0.09}}
\put(1.4,0.2){\makebox(0,0)[b]{$i$}}
\put(1.4,1.2){\makebox(0,0)[b]{$k$}}
\put(1.5,-0.4){\makebox(0,0)[b]{$r_1t$}}
\put(1.5,0.6){\makebox(0,0)[b]{$r_2t$}}
\end{picture}\quad + \quad
\begin{picture}(2,1)
\put(0,0.8){\line(1,1){0.4}}
\put(0.,1.2){\line(1,-1){0.4}}
\put(0.,-0.2){\line(1,1){0.4}}
\put(0,0.2){\line(1,-1){0.4}}
\put(0.2,0){\line(1,0){0.3}}
\put(0.2,1){\line(1,0){0.3}}
\put(0.5,1){\circle*{0.1}}
\put(2,1){\circle*{0.1}}
\put(0.5,0.){\circle*{0.1}}
\put(2,0.){\circle*{0.1}}
\put(0.5,1.0){\line(1,0){1.5}}
\put(0.5,0.0){\line(1,0){1.5}}
\multiput(1,1)(0.5,0){2}{\circle*{0.1}}
\multiput(1,0)(0.5,0){2}{\circle*{0.1}}
\multiput(0.5,1)(0,-0.125){8}{\line(0,-1){0.09}}
\multiput(1,1)(0,-0.125){8}{\line(0,-1){0.09}}
\multiput(1.5,1)(0,-0.125){8}{\line(0,-1){0.09}}
\put(1.9,0.2){\makebox(0,0)[b]{$i$}}
\put(1.9,1.2){\makebox(0,0)[b]{$k$}}
\put(2,-0.4){\makebox(0,0)[b]{$r_1t$}}
\put(2,0.6){\makebox(0,0)[b]{$r_2t$}}
\end{picture}\quad +\quad \cdots
\end{equation}
\mbox{}\\
It can be expressed in terms of the pair Green's function Eq.(\ref{GET})
\begin{equation}
\label{pP}
p_{ik}(\vec r_1,\vec r_2,t) =
\sum_{lmns} \int_0^t\,dt'\int    G_{ik,lm}(\vec r_1,\vec r_2,t,\vec r^{\,\prime}_1,\vec r^{\,\prime}_2,t')
  \widehat { V}_{lm,ns}(\vec r^{\,\prime}_1,\vec r^{\,\prime}_2) C_n(\vec r^{\,\prime}_1,t') C_s(\vec r^{\,\prime}_2,t') \,d^3 r^{\,\prime}_1\,d^3 r^{\,\prime}_2
\end{equation}
or in the matrix form
\begin{equation}
\label{pPvec}
\widehat{ p}(\vec r_1,\vec r_2,t) =
 \int_0^t\,dt'\int  \widehat {\bf G}(\vec r_1,\vec r_2,t,\vec r^{\,\prime}_1,\vec r^{\,\prime}_2,t')
  \widehat {\bf V}(\vec r^{\,\prime}_1,\vec r^{\,\prime}_2)
 \left(\vec C(\vec r^{\,\prime}_1,t')\otimes\vec C(\vec r^{\,\prime}_2,t')\right) \,d^3 r^{\,\prime}_1\,d^3 r^{\,\prime}_2
\end{equation}

Making use of Eqs.~(\ref{GET}), one can rewrite this equation in the equivalent differential form
(cf. Ref. \onlinecite{FrantsuzovCPL2000})
\begin{equation}
\label{Diffp}
\frac \partial {\partial t}\widehat{ p}
(\vec r_1,\vec r_2,t) =
\left(  \widehat{\bf L}+\widehat {\bf Q} +\widehat{\bf V}
(\vec r_1,\vec r_2)\right) \widehat{ p}(\vec r_1,\vec r_2,t) +
\widehat{\bf V}(\vec r_1,\vec r_2)\left(\vec C(\vec r_1,t)\otimes\vec C(\vec r_2,t)\right)
\end{equation}
Together with Eq.(\ref{EqCi}), which can also be written in the vector form
\begin{equation}
\label{EqCvec}
\partial_t \vec C(\vec r,t)= \widehat L_0
\vec C(\vec r,t)+ \mbox{Tr}_2 \int \widehat{\bf V}(\vec r,\vec r_1)\left(\vec C(\vec r,t)\otimes\vec C(\vec r_1,t)+ \widehat{ p}
(\vec r,\vec r_1,t) \right) \, d^3 r_1
\end{equation}
and represents the differential formulation of IET. It is sufficient
and convenient for many practical calculations (cf. Ref. \onlinecite{FrantsuzovCPL2000}).
\end{widetext}

\section{Expansion in terms of small parameter $\alpha$ }
\label{sec:ScaleMod}

The integral encounter theory represents only the second-order term in the concentration
expansion of the collision integral. Therefore, IET is only applicable when the reactant concentration is very low. A direct attempt to make corrections by the inclusion of higher-order terms
from Eq.(\ref{IntRi}) results in severe time limitations
because these terms diverge at $t\to\infty$. This divergence is illustrated in Supplementary Note II for the following third-order diagram from
Eq.(\ref{IntRi}):
\begin{equation}
\label{eq:DivergingDiagram}
\begin{picture}(2,1.5)
\put(0,0.8){\line(1,1){0.4}} \put(0.,1.2){\line(1,-1){0.4}}
\put(0.,-0.2){\line(1,1){0.4}} \put(0,0.2){\line(1,-1){0.4}}
\put(0.2,0){\line(1,0){0.3}} \put(0.2,1){\line(1,0){0.3}}
\put(0.5,1){\circle*{0.1}} \put(1.5,1){\circle*{0.1}}
\put(0.5,0.){\circle*{0.1}} \put(1.5,0.){\circle*{0.1}}
\put(0.5,1.0){\line(1,0){1.2}} \put(0.5,0.0){\line(1,0){1.2}}
\put(1,1){\circle*{0.1}} \put(0.5,1.3){\line(1,1){0.4}}
\put(0.5,1.7){\line(1,-1){0.4}} \put(0.7,1.5){\line(1,0){0.6}}
\put(1.,1.5){\circle*{0.1}}
\multiput(0.5,1)(0,-0.125){8}{\line(0,-1){0.09}}
\multiput(1,1.5)(0,-0.125){4}{\line(0,-1){0.09}}
\multiput(1.5,1)(0,-0.125){8}{\line(0,-1){0.09}}
\put(1.7,0.2){\makebox(0,0)[b]{$i$}}
\put(1.5,-0.4){\makebox(0,0)[b]{$rt$}}\end{picture}
\end{equation}
\mbox{}\\
which diverges as $\sqrt{t}$ at $t\to\infty$. Higher-order expansion terms
diverge even faster. Consequently, the standard concentration expansion fails to converge uniformly over all timescales.


A uniformly convergent approximation of the collision integral can be obtained by choosing a dimensionless expansion variable.
For this purpose, we will use the parameter $\alpha$, defined as
$$\alpha = \sqrt \xi$$
where $\xi$ is a generalization of the parameter Eq.(\ref{xiAB}) to a multistage reactive system
\begin{equation}
 \xi  =  \max_{ijkl} \max_{rt}\, \left\{\frac {k^3_{ij,kl}}{48 \pi^2 (D_k+D_l)^3} C_l(r,t)\right\}
 \label{xi}
\end{equation}
The concentrations $C_i(\vec r, t)$ can be presented in the following  form
\begin{eqnarray}
      C_i(\vec r, t) = \alpha^2\underline{C}_i( \underline {\vec r},  \underline t)
      \label{eq:Scaling}
\end{eqnarray}
where $t$ and $r$ scale as follows to conserve the form of the FCK equation:
$$\underline t = \alpha^2 t, \quad  \underline {\vec r} = \alpha \vec r.$$
Here
$\underline{C}_i (\vec r, t)$ is critical kinetics on which
\begin{equation}
 \max_{ijkl} \max_{rt}\, \left\{\frac {k^3_{ij,kl}}{48 \pi^2 (D_k+D_l)^3} \underline {C}_l(\vec r,t)\right\}\sim 1
 \nonumber
\end{equation}
and, therefore, all the terms of the perturbation series (diagrams) are of the same order of magnitude. The representation  (\ref{eq:Scaling}) has a very useful property: if  Eq.(\ref{eq:Scaling}) is a solution of  Eq.~(\ref{eq:RDE}) at some value $\alpha$, then it is a solution at any other value \cite{GopichPhysicaA1998}. Thus, equation (\ref{eq:Scaling}) gives the correct asymptotics in the limit of infinitely dilute solutions $\alpha\rightarrow 0$, since equation (\ref{eq:RDE}) is exact in this limit.
To keep constant steady-state reaction rates,  the distance-dependent reaction rate and interaction potential must be  written in real coordinates (instead of critical ones):
\begin{equation}
    \widehat {\bf W} (\vec r )= \widehat { \underline {\bf W}} (\vec r),\quad  \widehat {\bf U} (\vec r )= \widehat { \underline {\bf U}} (\vec r)
\end{equation}


For further consideration, it will be helpful to represent the collision integral $R_i(\vec r, t)$ (\ref{IntRi}) purely in terms of the pair $T$-matrices ($T$-matrix representation).
This procedure can be achieved by partial summation of diagrams in Eq.(\ref{IntRi}),
with all consecutive interactions summed into a single pair $T$-matrix.
Therefore, one can use a simple "rule of thumb" for the transition from the interaction to a $T$-matrix diagrammatic representation: the diagrams
consisting of two or more consecutive interactions are discarded, and in the remaining diagrams, the interaction lines are replaced by $T$-matrices.
The application of this rule to the diagrammatic representation for
$R^{(2)}_i(\vec r,t)$ (cf. Eq.(\ref{eq:COrder})) gives only one diagram, shown in Eq.(\ref{R2ndorder}).
The corresponding expression for $R^{(3)}_i(\vec r,t)$
is obtained similarly, resulting in an infinite sum of
the pair $T$-matrix diagrams:
\begin{widetext}
\begin{eqnarray}
\nonumber
R^{(3)}_i(\vec r,t)\quad =&  \quad
\begin{picture}(3.2,1.5)
\multiput(1.6,0.5)(0,0.1){5}{\line(4,1){0.4}}
\multiput(2.4,0)(0,0.1){5}{\line(4,1){0.4}}
\multiput(0.8,0)(0,0.1){5}{\line(4,1){0.4}}
\put(0.8,0){\circle*{0.1}}
\put(1.2,0){\circle*{0.1}}
\put(0.8,0.5){\circle*{0.1}}
\put(1.2,0.5){\circle*{0.1}}
\put(1.6,0.5){\circle*{0.1}}
\put(2,0.5){\circle*{0.1}}
\put(1.6,1){\circle*{0.1}}
\put(2,1){\circle*{0.1}}
\put(2.4,0.5){\circle*{0.1}}
\put(2.8,0.5){\circle*{0.1}}
\put(2.4,0){\circle*{0.1}}
\put(2.8,0){\circle*{0.1}}
\put(0.8,0.5){\line(0,-1){0.5}}
\put(1.6,1){\line(0,-1){0.5}}
\put(2,1){\line(0,-1){0.5}}
\put(1.2,0.5){\line(0,-1){0.5}}
\put(2.4,0.5){\line(0,-1){0.5}}
\put(2.8,0.5){\line(0,-1){0.5}}
\put(0.2,0.3){\line(1,1){0.4}}
\put(0.2,0.7){\line(1,-1){0.4}}
\put(0.2,-0.2){\line(1,1){0.4}}
\put(0.2,0.2){\line(1,-1){0.4}}
\put(1,0.8){\line(1,1){0.4}}
\put(1,1.2){\line(1,-1){0.4}}
\put(1.2,1){\line(1,0){1}}
\put(0.4,0.5){\line(1,0){2.6}}
\put(0.4,0.){\line(1,0){2.6}}
\put(3.0,0.1){\makebox(0,0)[b]{$i$}}
\put(2.8,-0.4){\makebox(0,0)[b]{$rt$}}
\end{picture}  \quad &+ \quad
\begin{picture}(3.6,1.5)
\put(0.4,0){\circle*{0.1}}
\put(0.8,0){\circle*{0.1}}
\put(0.4,0.5){\circle*{0.1}}
\put(0.8,0.5){\circle*{0.1}}
\multiput(0.4,0)(0,0.1){5}{\line(4,1){0.4}}
\put(0.8,0.5){\line(0,-1){0.5}}
\put(0.4,0.5){\line(0,-1){0.5}}
\put(2,0.5){\circle*{0.1}}
\put(2.4,0.5){\circle*{0.1}}
\put(2,1){\circle*{0.1}}
\put(2.4,1){\circle*{0.1}}
\multiput(2,0.5)(0,0.1){5}{\line(4,1){0.4}}
\put(2,1){\line(0,-1){0.5}}
\put(2.4,1){\line(0,-1){0.5}}
\put(2.8,0){\circle*{0.1}}
\put(3.2,0){\circle*{0.1}}
\put(2.8,0.5){\circle*{0.1}}
\put(3.2,0.5){\circle*{0.1}}
\multiput(2.8,0)(0,0.1){5}{\line(4,1){0.4}}
\put(2.8,0.5){\line(0,-1){0.5}}
\put(3.2,0.5){\line(0,-1){0.5}}
\put(-0.2,0.3){\line(1,1){0.4}}
\put(-0.2,0.7){\line(1,-1){0.4}}
\put(-0.2,-0.2){\line(1,1){0.4}}
\put(-0.2,0.2){\line(1,-1){0.4}}
\put(1.4,0.3){\line(1,1){0.4}}
\put(1.4,0.7){\line(1,-1){0.4}}
\put(0.0,0.5){\line(1,0){1}}
\put(0.,0.){\line(1,0){3.4}}
\put(1.5,1){\line(1,0){1.1}}
\put(1.0,0.5){\line(1,1){0.5}}
\put(1.6,0.5){\line(1,0){1.8}}
\put(3.4,0.1){\makebox(0,0)[b]{$i$}}
\put(3.2,-0.4){\makebox(0,0)[b]{$rt$}}
\end{picture}\\
 \quad +&   \quad
\begin{picture}(4.6,1.5)
\multiput(1.6,0.5)(0,0.1){5}{\line(4,1){0.4}}
\multiput(3.2,0.5)(0,0.1){5}{\line(4,1){0.4}}
\multiput(2.4,0)(0,0.1){5}{\line(4,1){0.4}}
\multiput(0.8,0)(0,0.1){5}{\line(4,1){0.4}}
\put(0.8,0){\circle*{0.1}}
\put(1.2,0){\circle*{0.1}}
\put(0.8,0.5){\circle*{0.1}}
\put(1.2,0.5){\circle*{0.1}}
\put(1.6,0.5){\circle*{0.1}}
\put(2,0.5){\circle*{0.1}}
\put(1.6,1){\circle*{0.1}}
\put(2,1){\circle*{0.1}}
\put(2.4,0.5){\circle*{0.1}}
\put(2.8,0.5){\circle*{0.1}}
\put(2.4,0){\circle*{0.1}}
\put(2.8,0){\circle*{0.1}}
\put(3.2,0.5){\circle*{0.1}}
\put(3.6,0.5){\circle*{0.1}}
\put(3.2,1){\circle*{0.1}}
\put(3.6,1){\circle*{0.1}}
\put(0.8,0.5){\line(0,-1){0.5}}
\put(1.6,1){\line(0,-1){0.5}}
\put(2,1){\line(0,-1){0.5}}
\put(1.2,0.5){\line(0,-1){0.5}}
\put(2.4,0.5){\line(0,-1){0.5}}
\put(2.8,0.5){\line(0,-1){0.5}}
\put(3.2,1){\line(0,-1){0.5}}
\put(3.6,1){\line(0,-1){0.5}}
\put(0.2,0.3){\line(1,1){0.4}}
\put(0.2,0.7){\line(1,-1){0.4}}
\put(0.2,-0.2){\line(1,1){0.4}}
\put(0.2,0.2){\line(1,-1){0.4}}
\put(1,0.8){\line(1,1){0.4}}
\put(1,1.2){\line(1,-1){0.4}}
\put(1.2,1){\line(1,0){2.6}}
\put(0.4,0.5){\line(1,0){3.4}}
\put(0.4,0.){\line(1,0){2.6}}
\put(3.8,0.6){\makebox(0,0)[b]{$i$}}
\put(3.6,0.1){\makebox(0,0)[b]{$rt$}}
\end{picture}
\quad &+ \quad
\begin{picture}(4.6,1.5)
\put(0.4,0){\circle*{0.1}}
\put(0.8,0){\circle*{0.1}}
\put(0.4,0.5){\circle*{0.1}}
\put(0.8,0.5){\circle*{0.1}}
\multiput(0.4,0)(0,0.1){5}{\line(4,1){0.4}}
\put(0.8,0.5){\line(0,-1){0.5}}
\put(0.4,0.5){\line(0,-1){0.5}}
\put(2,0.5){\circle*{0.1}}
\put(2.4,0.5){\circle*{0.1}}
\put(2,1){\circle*{0.1}}
\put(2.4,1){\circle*{0.1}}
\multiput(2,0.5)(0,0.1){5}{\line(4,1){0.4}}
\put(2,1){\line(0,-1){0.5}}
\put(2.4,1){\line(0,-1){0.5}}
\put(2.8,0){\circle*{0.1}}
\put(3.2,0){\circle*{0.1}}
\put(2.8,0.5){\circle*{0.1}}
\put(3.2,0.5){\circle*{0.1}}
\multiput(2.8,0)(0,0.1){5}{\line(4,1){0.4}}
\put(2.8,0.5){\line(0,-1){0.5}}
\put(3.2,0.5){\line(0,-1){0.5}}
\put(3.8,0){\circle*{0.1}}
\put(4.2,0){\circle*{0.1}}
\put(3.8,0.5){\circle*{0.1}}
\put(4.2,0.5){\circle*{0.1}}
\multiput(3.8,0)(0,0.1){5}{\line(4,1){0.4}}
\put(3.8,0.5){\line(0,-1){0.5}}
\put(4.2,0.5){\line(0,-1){0.5}}
\put(-0.2,0.3){\line(1,1){0.4}}
\put(-0.2,0.7){\line(1,-1){0.4}}
\put(-0.2,-0.2){\line(1,1){0.4}}
\put(-0.2,0.2){\line(1,-1){0.4}}
\put(1.4,0.3){\line(1,1){0.4}}
\put(1.4,0.7){\line(1,-1){0.4}}
\put(0.0,0.5){\line(1,0){1}}
\put(0.,0.){\line(1,0){4.4}}
\put(3.8,0.5){\line(1,0){0.6}}
\put(1.5,1){\line(1,0){1.8}}
\put(1.0,0.5){\line(1,1){0.5}}
\put(1.6,0.5){\line(1,0){1.8}}
\put(3.8,0.5){\line(-1,1){0.5}}
\put(4.4,0.1){\makebox(0,0)[b]{$i$}}
\put(4.2,-0.4){\makebox(0,0)[b]{$rt$}}
\end{picture}\\
 \quad +& \quad
\begin{picture}(4.6,1.5)
\multiput(1.6,0.5)(0,0.1){5}{\line(4,1){0.4}}
\multiput(3.2,0.5)(0,0.1){5}{\line(4,1){0.4}}
\multiput(2.4,0)(0,0.1){5}{\line(4,1){0.4}}
\multiput(0.8,0)(0,0.1){5}{\line(4,1){0.4}}
\multiput(4,0)(0,0.1){5}{\line(4,1){0.4}}
\put(0.8,0){\circle*{0.1}}
\put(1.2,0){\circle*{0.1}}
\put(0.8,0.5){\circle*{0.1}}
\put(1.2,0.5){\circle*{0.1}}
\put(1.6,0.5){\circle*{0.1}}
\put(2,0.5){\circle*{0.1}}
\put(1.6,1){\circle*{0.1}}
\put(2,1){\circle*{0.1}}
\put(2.4,0.5){\circle*{0.1}}
\put(2.8,0.5){\circle*{0.1}}
\put(2.4,0){\circle*{0.1}}
\put(2.8,0){\circle*{0.1}}
\put(3.2,0.5){\circle*{0.1}}
\put(3.6,0.5){\circle*{0.1}}
\put(3.2,1){\circle*{0.1}}
\put(3.6,1){\circle*{0.1}}
\put(4,0.0){\circle*{0.1}}
\put(4.,0.5){\circle*{0.1}}
\put(4.4,.0){\circle*{0.1}}
\put(4.4,0.5){\circle*{0.1}}
\put(0.8,0.5){\line(0,-1){0.5}}
\put(1.6,1){\line(0,-1){0.5}}
\put(2,1){\line(0,-1){0.5}}
\put(1.2,0.5){\line(0,-1){0.5}}
\put(2.4,0.5){\line(0,-1){0.5}}
\put(2.8,0.5){\line(0,-1){0.5}}
\put(3.2,1){\line(0,-1){0.5}}
\put(3.6,1){\line(0,-1){0.5}}
\put(4.,0.5){\line(0,-1){0.5}}
\put(4.4,0.5){\line(0,-1){0.5}}
\put(0.2,0.3){\line(1,1){0.4}}
\put(0.2,0.7){\line(1,-1){0.4}}
\put(0.2,-0.2){\line(1,1){0.4}}
\put(0.2,0.2){\line(1,-1){0.4}}
\put(1,0.8){\line(1,1){0.4}}
\put(1,1.2){\line(1,-1){0.4}}
\put(1.2,1){\line(1,0){2.6}}
\put(0.4,0.5){\line(1,0){4.2}}
\put(0.4,0.){\line(1,0){4.2}}
\put(4.6,0.1){\makebox(0,0)[b]{$i$}}
\put(4.4,-0.4){\makebox(0,0)[b]{$rt$}}
\end{picture}
\quad&+  \quad \cdots
\label{R3}
\end{eqnarray}
\mbox{}\\

Substituting the concentrations in the form Eq.(\ref{eq:Scaling}) and the variables $\vec r$ and $t$ in the form $\alpha^{-1} \underline{\vec r}$ and $\alpha^{-2}\underline t$, respectively, into the given diagram, we find the order of this diagram in $\alpha$.
As shown in Additional Note III, in the limit $\alpha\to 0$ the Green's function has the following order in $\alpha$:
\begin{equation}
\label{eq:ScalingG}
 G^0_{ik}(\alpha^{-1} \underline{\vec r},\alpha^{-2} \underline t,\alpha^{-1} \underline{\vec r}_1, \alpha^{-2} \underline t_1) \sim \alpha^3 + O(\alpha^5)
\end{equation}

The pair $T$-matrix has the following order in $\alpha$ (see Supplementary Note IV for the details):

\begin{equation}
\label{Tpoint2}
    \widehat{\bf T}(\vec r,\vec r^{\,\prime},   t,\vec r_1,\vec r_1^{\,\prime}, t_1) =
    \alpha^{11} \widehat {\bf K}\delta(\underline {\vec r}- \underline {\vec r}_1)\delta(\underline {\vec r}^{\,\prime}-\underline {\vec r}_1)
  \delta(\underline {\vec r}_1^{\,\prime}-\underline {\vec r}_1)\delta(\underline t-\underline t_1)+ O(\alpha^{12})
\end{equation}
where $\widehat{\mathbf K}$ is the matrix of LMA rate constants (cf. Eq.(\ref{eq:RDE})) defined as:
$$ K_{ij;kl}=k_{ij;kl} \text{ , for}\, \{ij\} \neq \{kl\};\qquad
 K_{ij;ij}=-\sum_{kl} k_{kl;ij}$$
$\widehat{\mathbf K}$ represents the following integral of the pair $T$-matrix:
\begin{equation}
\label{DefK}
\widehat {\bf K} = \int\limits_0^{\infty}\, d\tau \int\!\int\!\int \widehat {\bf T}(\vec r,\vec r+\vec r_2,t ,\vec r +\vec r_1^{\,\prime}, \vec r+ \vec r_2^{\,\prime},t-\tau)\,
 d^3 r_2\,d^3 r_1'\,d^3 r_2'
\end{equation}

\end{widetext}
Details of the $\widehat{\mathbf K}$ matrix calculation are given in Supplementary Note V.
When a $T$-matrix appears in a diagram, it is integrated over four spatial and two temporal variables. Using the factor powers of $\alpha$ found in Eqs.(\ref{eq:Scaling},\ref{Tpoint2}), the leading order of a $T$-matrix in diagrams is
calculated as
$$\alpha^{11}\alpha^{4\times (-3)}\alpha^{2\times(-2)}=\alpha^{-5}$$
Summarizing all of the above, one can formulate the following rules for  calculating the order of $\alpha$ for an arbitrary diagram:
\begin{itemize}
\item Each concentration cross contributes $\alpha^2$ \item Each Green's function line contributes  $\alpha^3+O(\alpha^5)$
\item Each T-matrix contributes $\alpha^{-5}+O(\alpha^{-4})$ \item The $\alpha$-order of the whole diagram
is a product of contributions from all components times $\alpha^5$. The latter factor arises because there is no space-time integration at the diagram's final point.
\end{itemize}

The $\alpha$-expansion of the collision integral $R_i(\vec r,t)$ may be represented
similarly to Eq.(\ref{eq:COrder}):
\begin{equation}
\label{eq:Ralpha}
R_i(\vec r,t) = R^{[4]}_i(\vec r,t) + R^{[5]}_i(\vec r,t) + \dots +
                R^{[n]}_i(\vec r,t) + \dots
\end{equation}
where the superscript $[n]$ denotes the $n$-th order in $\alpha$. It can now be verified that the collision integral $R^{(2)}_i(\vec r,t)$ (\ref{R2ndorder}) has an order of
$\alpha^4+O(\alpha^5)$.
 It follows from Eq.(\ref{Tpoint2})  that the fourth order term in $\alpha$ has the following form:
\begin{equation}
\begin{gathered}
    R_i^{[4]}(\vec r,t)=  \sum\limits_{jkl} K_{ij,kl}C_k(\vec r,\ t)  C_l(\vec r, t) =\\=\alpha^4 \sum\limits_{jkl} K_{ij,kl}\underline {C}_k(\underline{ \vec r},\underline t) \underline {C}_l(\underline{\vec r},\underline t)
\end{gathered}
 \label{R4}
\end{equation}
Eq.(\ref{EqCi}) with the collision integral Eq.(\ref{R4}) gives the diffusion-reaction equations  Eq.(\ref{eq:RDE}). Therefore, Eq.(\ref{eq:RDE})
can be obtained within IET using the following approximation of the pair $T$-matrix:
\begin{equation}
\label{DefpT}
 \begin{gathered}
     T_{ik,lm}(\vec r_1,\vec r_2,t,\vec r_1^{\,\prime},\vec r_2^{\,\prime},t') =\\=
   K_{ik,lm}\delta(\vec r_2-\vec r_1)
   \delta(\vec r_1^{\,\prime}-\vec r_1)\delta(\vec r_2^{\,\prime}-\vec r_1)\delta(t-t')
 \end{gathered}
\end{equation}
 which is sometimes called a point approximation \cite{GopichPhysicaA1998} because the $T$-matrix is nonzero only at a spatial-temporal point.
 The IET Master  equation Eq.(\ref{eq:MasterET}) contains all terms of the $\alpha$-expansion up to the order of
$\alpha^4$, but only some terms of the order of $\alpha^5$ and higher. Authors of Ref. \onlinecite{GopichPhysicaA1998} have shown
that the kinetics of the $A+B\rightarrow C+B$ reaction can be correctly described only if all $\alpha$-terms up to the order of $\alpha^5$
are taken into consideration.
For example, the diagram (\ref{eq:DivergingDiagram}), included in $R^{(3)}_i(\vec r,t)$, has the fifth order in $\alpha$. Therefore, we aim to modify the IET by including all diagrams in the series $\alpha^n$, where $n\le 5$.

 It is easy to find that all possible diagrams up to order $\alpha^5$ are contained in the expression
 \begin{widetext}
$$R^{[4]}_i(\vec r,t) + R^{[5]}_i(\vec r,t) + O\left(\alpha^6\right) \quad =$$
\begin{equation}
\begin{picture}(1.2,1.125)
\put(-0.2,0.6){\line(1,1){0.4}} \put(-0.2,1.0){\line(1,-1){0.4}} \put(-0.2,-0.2){\line(1,1){0.4}} \put(-0.2,0.2){\line(1,-1){0.4}}
\put(0.4,0){\circle*{0.1}} \put(1.1,0){\circle*{0.1}} \multiput(0.4,0)(0,0.1){7}{\line(4,1){0.7}} \put(0.4,0.8){\circle*{0.1}}
\put(1.1,0.8){\circle*{0.1}} \put(0,0.8){\line(1,0){1.3}} \put(0,0.){\line(1,0){1.3}} \put(0.4,0.8){\line(0,-1){0.8}}
\put(1.1,0.8){\line(0,-1){0.8}} \put(1.3,0.1){\makebox(0,0)[b]{$i$}} \put(1.1,-0.4){\makebox(0,0)[b]{$rt$}}
\end{picture}\quad+\quad
\begin{picture}(3,1.125)
\put(-0.2,0.6){\line(1,1){0.4}} \put(-0.2,1.0){\line(1,-1){0.4}}
\put(-0.2,-0.2){\line(1,1){0.4}} \put(-0.2,0.2){\line(1,-1){0.4}}
\put(0.4,0.8){\circle*{0.1}} \put(1.1,0.8){\circle*{0.1}}

\put(2,0){\circle*{0.1}} \put(2.7,0){\circle*{0.1}}
\multiput(2,0)(0,0.1){7}{\line(4,1){0.7}}
\put(2,0.8){\circle*{0.1}} \put(2.7,0.8){\circle*{0.1}}

\put(0.4,0){\circle*{0.1}} \put(1.1,0){\circle*{0.1}}
\multiput(0.4,0)(0,0.1){7}{\line(4,1){0.7}}

\put(0.4,0.8){\circle*{0.1}} \put(1.1,0.8){\circle*{0.1}}
\put(0,0.8){\line(1,0){2.9}} \put(0,0.){\line(1,0){2.9}}
\put(2,0.8){\circle*{0.1}} \put(2,0){\circle*{0.1}}
\put(2,0.8){\line(0,-1){0.8}}\put(2.7,0.8){\line(0,-1){0.8}}
\put(0.4,0.8){\line(0,-1){0.8}}
\put(1.1,0.8){\line(0,-1){0.8}}

\put(1.4,-0.15){\line(1,0){0.3}} \put(1.4,-0.15){\line(0,1){0.3}}
\put(1.4,0.15){\line(1,0){0.3}} \put(1.7,-0.15){\line(0,1){0.3}}

\put(2.9,0.1){\makebox(0,0)[b]{$i$}}
\put(2.7,-0.4){\makebox(0,0)[b]{$rt$}}
\end{picture}\quad +\quad
\begin{picture}(2.9,1.125)
\put(-0.2,0.6){\line(1,1){0.4}} \put(-0.2,1.0){\line(1,-1){0.4}}
\put(-0.2,-0.2){\line(1,1){0.4}} \put(-0.2,0.2){\line(1,-1){0.4}}
\put(0.4,0.8){\circle*{0.1}} \put(1.1,0.8){\circle*{0.1}}

\put(2,0){\circle*{0.1}} \put(2.7,0){\circle*{0.1}}
\multiput(2,0)(0,0.1){7}{\line(4,1){0.7}}
\put(2,0.8){\circle*{0.1}} \put(2.7,0.8){\circle*{0.1}}

\put(0.4,0){\circle*{0.1}} \put(1.1,0){\circle*{0.1}}
\multiput(0.4,0)(0,0.1){7}{\line(4,1){0.7}}

\put(0.4,0.8){\circle*{0.1}} \put(1.1,0.8){\circle*{0.1}}
\put(0,0.8){\line(1,0){2.9}} \put(0,0.){\line(1,0){2.9}}
\put(2,0.8){\circle*{0.1}} \put(2,0){\circle*{0.1}}
\put(2,0.8){\line(0,-1){0.8}}\put(2.7,0.8){\line(0,-1){0.8}}
\put(0.4,0.8){\line(0,-1){0.8}}
\put(1.1,0.8){\line(0,-1){0.8}}

\put(1.4,0.65){\line(1,0){0.3}} \put(1.4,0.65){\line(0,1){0.3}}
\put(1.4,0.95){\line(1,0){0.3}} \put(1.7,0.65){\line(0,1){0.3}}

\put(2.9,0.1){\makebox(0,0)[b]{$i$}}
\put(2.7,-0.4){\makebox(0,0)[b]{$rt$}}
\end{picture}\quad +\quad
\begin{picture}(2.9,1.125)
\put(-0.2,0.6){\line(1,1){0.4}} \put(-0.2,1.0){\line(1,-1){0.4}}
\put(-0.2,-0.2){\line(1,1){0.4}} \put(-0.2,0.2){\line(1,-1){0.4}}
\put(0.4,0.8){\circle*{0.1}} \put(1.1,0.8){\circle*{0.1}}

\put(2,0){\circle*{0.1}} \put(2.7,0){\circle*{0.1}}
\multiput(2,0)(0,0.1){7}{\line(4,1){0.7}}
\put(2,0.8){\circle*{0.1}} \put(2.7,0.8){\circle*{0.1}}

\put(0.4,0){\circle*{0.1}} \put(1.1,0){\circle*{0.1}}
\multiput(0.4,0)(0,0.1){7}{\line(4,1){0.7}}

\put(0.4,0.8){\circle*{0.1}} \put(1.1,0.8){\circle*{0.1}}
\put(0,0.8){\line(1,0){2.9}}
\put(0,0.){\line(1,0){2.9}}
\put(2,0.8){\circle*{0.1}} \put(2,0){\circle*{0.1}}
\put(2,0.8){\line(0,-1){0.8}}\put(2.7,0.8){\line(0,-1){0.8}}
\put(0.4,0.8){\line(0,-1){0.8}}
\put(1.1,0.8){\line(0,-1){0.8}}

\put(1.4,-0.15){\line(1,0){0.3}} \put(1.4,-0.15){\line(0,1){0.3}}
\put(1.4,0.15){\line(1,0){0.3}} \put(1.7,-0.15){\line(0,1){0.3}}
\put(1.4,0.65){\line(1,0){0.3}} \put(1.4,0.65){\line(0,1){0.3}}
\put(1.4,0.95){\line(1,0){0.3}} \put(1.7,0.65){\line(0,1){0.3}}

\put(2.9,0.1){\makebox(0,0)[b]{$i$}}
\put(2.7,-0.4){\makebox(0,0)[b]{$rt$}}
\end{picture}
\label{R45}
\end{equation}
\mbox{}\\

where the horizontal line carrying the square represents the "Green's function with interactions", which is described by the following diagrammatic series

$$
\begin{picture}(1,1)
\put(0,0.){\circle*{0.1}} \put(0,0.0){\line(1,0){1}}
\put(1,0.){\circle*{0.1}} \put(.9,0.2){\makebox(0,0)[b]{$i$}}
\put(0.1,0.2){\makebox(0,0)[b]{$k$}}
\put(1,-0.4){\makebox(0,0)[b]{$rt$}}
\put(0.,-0.4){\makebox(0,0)[b]{$r't'$}}
\put(0.35,-0.15){\line(1,0){0.3}}
\put(0.35,-0.15){\line(0,1){0.3}} \put(0.35,0.15){\line(1,0){0.3}}
\put(0.65,-0.15){\line(0,1){0.3}}
\end{picture} \quad =\quad
\begin{picture}(2,1)
\put(0.8,0){\circle*{0.1}} \put(1.2,0){\circle*{0.1}}
\multiput(0.8,0)(0,0.1){5}{\line(4,1){0.4}}
\put(0.8,0.5){\circle*{0.1}} \put(1.2,0.5){\circle*{0.1}}
\put(0.4,0.5){\line(1,0){1.2}} \put(0.2,0.3){\line(1,1){0.4}}
\put(0.2,0.7){\line(1,-1){0.4}} \put(0,0.){\line(1,0){2}}
\put(0.8,0.5){\line(0,-1){0.5}} \put(1.2,0.5){\line(0,-1){0.5}}
\put(0.,0.){\circle*{0.1}} \put(2,0.0){\circle*{0.1}}
\put(1.9,0.2){\makebox(0,0)[b]{$i$}}
\put(0.,0.2){\makebox(0,0)[b]{$k$}}
\put(2,-0.4){\makebox(0,0)[b]{$rt$}}
\put(0.,-0.4){\makebox(0,0)[b]{$r't'$}}
\end{picture}\quad+\quad
\begin{picture}(2,1)
\put(0.8,0){\circle*{0.1}} \put(1.2,0){\circle*{0.1}}
\multiput(0.8,0)(0,0.1){5}{\line(4,1){0.4}}
\put(0.8,0.5){\circle*{0.1}} \put(1.2,0.5){\circle*{0.1}}
\put(0.4,0.5){\line(1,0){1.1}} \put(0.2,0.3){\line(1,1){0.4}}
\put(0.2,0.7){\line(1,-1){0.4}} \put(0,0.){\line(1,0){1.5}}
\put(0.8,0.5){\line(0,-1){0.5}} \put(1.2,0.5){\line(0,-1){0.5}}
\put(1.5,0.5){\line(1,-1){0.5}} \put(0.,0.){\circle*{0.1}}
\put(2,0.0){\circle*{0.1}} \put(1.9,0.3){\makebox(0,0)[b]{$i$}}
\put(0.,0.2){\makebox(0,0)[b]{$k$}}
\put(2,-0.4){\makebox(0,0)[b]{$rt$}}
\put(0.,-0.4){\makebox(0,0)[b]{$r't'$}}
\end{picture}\quad+\quad
\begin{picture}(4,1)
\put(0.8,0){\circle*{0.1}} \put(1.2,0){\circle*{0.1}}
\multiput(0.8,0)(0,0.1){5}{\line(4,1){0.4}}
\put(0.8,0.5){\circle*{0.1}} \put(1.2,0.5){\circle*{0.1}}
\put(0.4,0.5){\line(1,0){1.2}} \put(0.2,0.3){\line(1,1){0.4}}
\put(0.2,0.7){\line(1,-1){0.4}} \put(0,0.){\line(1,0){4}}
\put(0.8,0.5){\line(0,-1){0.5}} \put(1.2,0.5){\line(0,-1){0.5}}
\put(0.,0.){\circle*{0.1}} \put(4,0.0){\circle*{0.1}}
\put(2.8,0){\circle*{0.1}} \put(3.2,0){\circle*{0.1}}
\multiput(2.8,0)(0,0.1){5}{\line(4,1){0.4}}
\put(2.8,0.5){\circle*{0.1}} \put(3.2,0.5){\circle*{0.1}}
\put(2.4,0.5){\line(1,0){1.2}} \put(2.2,0.3){\line(1,1){0.4}}
\put(2.2,0.7){\line(1,-1){0.4}} \put(2.8,0.5){\line(0,-1){0.5}}
\put(3.2,0.5){\line(0,-1){0.5}}
\put(3.9,0.2){\makebox(0,0)[b]{$i$}}
\put(4,-0.4){\makebox(0,0)[b]{$rt$}}
\put(0.,0.2){\makebox(0,0)[b]{$k$}}
\put(0.,-0.4){\makebox(0,0)[b]{$r't'$}}
\end{picture}\quad+\quad$$
\begin{equation}
\label{Blc}
\begin{picture}(4,1)
\put(0.8,0){\circle*{0.1}} \put(1.2,0){\circle*{0.1}}
\multiput(0.8,0)(0,0.1){5}{\line(4,1){0.4}}
\put(0.8,0.5){\circle*{0.1}} \put(1.2,0.5){\circle*{0.1}}
\put(0.4,0.5){\line(1,0){1.2}} \put(0.2,0.3){\line(1,1){0.4}}
\put(0.2,0.7){\line(1,-1){0.4}} \put(0,0.){\line(1,0){3.5}}
\put(0.8,0.5){\line(0,-1){0.5}} \put(1.2,0.5){\line(0,-1){0.5}}
\put(0.,0.){\circle*{0.1}} \put(4,0.0){\circle*{0.1}}
\put(2.8,0){\circle*{0.1}} \put(3.2,0){\circle*{0.1}}
\multiput(2.8,0)(0,0.1){5}{\line(4,1){0.4}}
\put(2.8,0.5){\circle*{0.1}} \put(3.2,0.5){\circle*{0.1}}
\put(2.4,0.5){\line(1,0){1.1}} \put(2.2,0.3){\line(1,1){0.4}}
\put(2.2,0.7){\line(1,-1){0.4}} \put(2.8,0.5){\line(0,-1){0.5}}
\put(3.2,0.5){\line(0,-1){0.5}}\put(3.5,0.5){\line(1,-1){0.5}}
\put(3.9,0.3){\makebox(0,0)[b]{$i$}}
\put(4,-0.4){\makebox(0,0)[b]{$rt$}}
\put(0.,0.2){\makebox(0,0)[b]{$k$}}
\put(0.,-0.4){\makebox(0,0)[b]{$r't'$}}
\end{picture}\quad+\quad
\begin{picture}(4,1)
\put(0.8,0){\circle*{0.1}} \put(1.2,0){\circle*{0.1}}
\multiput(0.8,0)(0,0.1){5}{\line(4,1){0.4}}
\put(0.8,0.5){\circle*{0.1}} \put(1.2,0.5){\circle*{0.1}}
\put(0.4,0.5){\line(1,0){1.1}} \put(0.2,0.3){\line(1,1){0.4}}
\put(0.2,0.7){\line(1,-1){0.4}} \put(2,0.){\line(1,0){2}}
\put(0,0.){\line(1,0){1.5}} \put(1.5,0.5){\line(1,-1){0.5}}
\put(0.8,0.5){\line(0,-1){0.5}} \put(1.2,0.5){\line(0,-1){0.5}}
\put(0.,0.){\circle*{0.1}} \put(4,0.0){\circle*{0.1}}
\put(2.8,0){\circle*{0.1}} \put(3.2,0){\circle*{0.1}}
\multiput(2.8,0)(0,0.1){5}{\line(4,1){0.4}}
\put(2.8,0.5){\circle*{0.1}} \put(3.2,0.5){\circle*{0.1}}
\put(2.4,0.5){\line(1,0){1.2}} \put(2.2,0.3){\line(1,1){0.4}}
\put(2.2,0.7){\line(1,-1){0.4}} \put(2.8,0.5){\line(0,-1){0.5}}
\put(3.2,0.5){\line(0,-1){0.5}}
\put(3.9,0.2){\makebox(0,0)[b]{$i$}}
\put(0.,0.2){\makebox(0,0)[b]{$k$}}
\put(4,-0.4){\makebox(0,0)[b]{$rt$}}
\put(0.,-0.4){\makebox(0,0)[b]{$r't'$}}
\end{picture}\quad+\quad
\begin{picture}(4,1)
\put(0.8,0){\circle*{0.1}} \put(1.2,0){\circle*{0.1}}
\multiput(0.8,0)(0,0.1){5}{\line(4,1){0.4}}
\put(0.8,0.5){\circle*{0.1}} \put(1.2,0.5){\circle*{0.1}}
\put(0.4,0.5){\line(1,0){1.1}} \put(0.2,0.3){\line(1,1){0.4}}
\put(0.2,0.7){\line(1,-1){0.4}} \put(2,0.){\line(1,0){1.5}}
\put(0,0.){\line(1,0){1.5}} \put(1.5,0.5){\line(1,-1){0.5}}
\put(0.8,0.5){\line(0,-1){0.5}} \put(1.2,0.5){\line(0,-1){0.5}}
\put(0.,0.){\circle*{0.1}} \put(4,0.0){\circle*{0.1}}
\put(2.8,0){\circle*{0.1}} \put(3.2,0){\circle*{0.1}}
\multiput(2.8,0)(0,0.1){5}{\line(4,1){0.4}}
\put(2.8,0.5){\circle*{0.1}} \put(3.2,0.5){\circle*{0.1}}
\put(2.4,0.5){\line(1,0){1.1}} \put(2.2,0.3){\line(1,1){0.4}}
\put(2.2,0.7){\line(1,-1){0.4}} \put(2.8,0.5){\line(0,-1){0.5}}
\put(3.2,0.5){\line(0,-1){0.5}} \put(3.5,0.5){\line(1,-1){0.5}}
\put(3.9,0.3){\makebox(0,0)[b]{$i$}}
\put(0.,0.2){\makebox(0,0)[b]{$k$}}
\put(4,-0.4){\makebox(0,0)[b]{$rt$}}
\put(0.,-0.4){\makebox(0,0)[b]{$r't'$}}
\end{picture}\quad+\quad \cdots
\end{equation}
\mbox{}\\

In this series, all $T$-matrices can be approximated by the first nonzero terms of their $\alpha$-expansion (point approximation)  without affecting the contributions of the $\alpha^5$  order. This observation will be used to simplify the resulting equations.

While it is sufficient to leave only the first two terms in the expansion Eq.(\ref{eq:Ralpha}), the resulting
analytical expressions are somewhat cumbersome.  Certain
higher-order terms may be added to Eq.(\ref{R45}) without a reduction in accuracy to simplify the final expressions. This procedure,
described in Supplementary Note VI, results in the following approximation for the collision integral:

\begin{equation}
\label{CMETR} R^{CM}_i(\vec r,t)=\quad
\begin{picture}(1.,1.125)
\put(0.0,0.8){\line(1,1){0.4}} \put(0.0,1.2){\line(1,-1){0.4}} \put(0.0,-0.2){\line(1,1){0.4}}
\put(0.0,0.2){\line(1,-1){0.4}} \put(0.2,0){\line(1,0){0.8}} \put(0.2,1){\line(1,0){0.8}}
\put(0.7,1){\circle*{0.1}} \put(0.7,0){\circle*{0.1}}
\multiput(0.7,1)(0,-0.125){8}{\line(0,-1){0.1}}
\put(.9,0.2){\makebox(0,0)[b]{$i$}} \put(0.7,-0.4){\makebox(0,0)[b]{$rt$}}
\end{picture}\quad +\quad
\begin{picture}(1.7,1.125)
\put(0,0.8){\line(1,1){0.4}} \put(0.,1.2){\line(1,-1){0.4}} \put(0.,-0.2){\line(1,1){0.4}}
\put(0,0.2){\line(1,-1){0.4}} \put(0.2,0){\line(1,0){0.3}} \put(0.2,1){\line(1,0){0.3}}
\put(0.5,0){\circle*{0.1}} \put(1.5,0.){\circle*{0.1}} \put(0.5,0.05){\line(1,0){1}}
\put(0.5,1.){\circle*{0.1}}
\put(0.5,-0.05){\line(1,0){1}}  \put(1.5,1){\circle*{0.1}}
\put(0.5,0.95){\line(1,0){1}}\put(0.5,1.05){\line(1,0){1}}
\multiput(0.5,1)(0,-0.125){8}{\line(0,-1){0.09}}
\multiput(1.5,1)(0,-0.125){8}{\line(0,-1){0.09}}
\put(1.5,1){\line(1,0){0.2}}\put(1.5,0){\line(1,0){0.2}}
\put(1.7,0.2){\makebox(0,0)[b]{$i$}}
\put(1.5,-0.4){\makebox(0,0)[b]{$rt$}}
\end{picture}\quad + \quad
\begin{picture}(1.5,1.125)
\put(0,0.8){\line(1,1){0.4}} \put(0.,1.2){\line(1,-1){0.4}} \put(0.,-0.2){\line(1,1){0.4}}
\put(0,0.2){\line(1,-1){0.4}} \put(0.2,0){\line(1,0){0.3}} \put(0.2,1){\line(1,0){0.3}}
\put(0.5,0){\circle*{0.1}} \put(1.5,0){\circle*{0.1}} \put(1.0,0){\circle*{0.1}}
\put(0.5,-0.05){\line(1,0){1}} \put(0.5,0.05){\line(1,0){1}}
\put(0.5,1){\circle*{0.1}} \put(1.5,1){\circle*{0.1}} \put(1,1){\circle*{0.1}}
\put(0.5,.95){\line(1,0){1}}\put(0.5,1.05){\line(1,0){1}}
\multiput(0.5,1)(0,-0.125){8}{\line(0,-1){0.09}} \multiput(1,1)(0,-0.125){8}{\line(0,-1){0.09}}
\multiput(1.5,1)(0,-0.125){8}{\line(0,-1){0.09}} \put(1.5,1){\line(1,0){0.2}}
\put(1.5,0){\line(1,0){0.2}}
\put(1.7,0.2){\makebox(0,0)[b]{$i$}}
\put(1.5,-0.4){\makebox(0,0)[b]{$rt$}}
\end{picture}\quad + \quad
\begin{picture}(2.2,1.125)
\put(0,0.8){\line(1,1){0.4}} \put(0.,1.2){\line(1,-1){0.4}} \put(0.,-0.2){\line(1,1){0.4}}
\put(0,0.2){\line(1,-1){0.4}} \put(0.2,0){\line(1,0){0.3}} \put(0.2,1){\line(1,0){0.3}}
\put(0.5,1){\circle*{0.1}} \put(2,1){\circle*{0.1}} \put(0.5,0.){\circle*{0.1}}
\put(2,0.){\circle*{0.1}}
\put(0.5,1.05){\line(1,0){1.5}} \put(0.5,0.05){\line(1,0){1.5}} \put(0.5,.95){\line(1,0){1.5}}
\put(0.5,-0.05){\line(1,0){1.5}} \multiput(1,1)(0.5,0){2}{\circle*{0.1}}
\multiput(1,0)(0.5,0){2}{\circle*{0.1}}
\multiput(0.5,1)(0,-0.125){8}{\line(0,-1){0.09}} \multiput(1,1)(0,-0.125){8}{\line(0,-1){0.09}}
\multiput(1.5,1)(0,-0.125){8}{\line(0,-1){0.09}} \multiput(2,1)(0,-0.125){8}{\line(0,-1){0.09}}
\put(2,1){\line(1,0){0.2}}\put(2,0){\line(1,0){0.2}} \put(2.2,0.2){\makebox(0,0)[b]{$i$}}
\put(2,-0.4){\makebox(0,0)[b]{$rt$}}
\end{picture}\quad +\quad \cdots
\end{equation}
\mbox{}\\
where we introduce an ``effective'' Green's function $G^E_{ik}(\vec r,t,\vec r^{\,\prime},t')$
\begin{equation}
\begin{picture}(1,0.5)
\put(0,0.1){\circle*{0.1}} \put(0,0.05){\line(1,0){1}}
\put(0,0.15){\line(1,0){1}} \put(1,0.1){\circle*{0.1}} \put(.9,0.3){\makebox(0,0)[b]{$i$}}
\put(0.1,0.3){\makebox(0,0)[b]{$k$}} \put(1,-0.4){\makebox(0,0)[b]{$rt$}}
\put(0.,-0.4){\makebox(0,0)[b]{$r't'$}}
\end{picture} \quad =\quad
\begin{picture}(1,0.5)
\put(0,0.){\circle*{0.1}} \put(0,0.0){\line(1,0){1}} \put(1,0.){\circle*{0.1}}
\put(.9,0.2){\makebox(0,0)[b]{$i$}} \put(0.1,0.2){\makebox(0,0)[b]{$k$}} \put(1,-0.4){\makebox(0,0)[b]{$rt$}}
\put(0.,-0.4){\makebox(0,0)[b]{$r't'$}}
\end{picture} \quad +\quad
\begin{picture}(1,0.5)
\put(0,0.){\circle*{0.1}} \put(0,0.0){\line(1,0){1}} \put(1,0.){\circle*{0.1}}
\put(.9,0.2){\makebox(0,0)[b]{$i$}} \put(0.1,0.2){\makebox(0,0)[b]{$k$}} \put(1,-0.4){\makebox(0,0)[b]{$rt$}}
\put(0.,-0.4){\makebox(0,0)[b]{$r't'$}} \put(0.35,-0.15){\line(1,0){0.3}} \put(0.35,-0.15){\line(0,1){0.3}}
\put(0.35,0.15){\line(1,0){0.3}} \put(0.65,-0.15){\line(0,1){0.3}}
\end{picture}
\label{Gdress}
\end{equation}
\mbox{}\\[-3pt]
As may be derived from Eqs.(\ref{Blc}) and (\ref{Gdress}), the ``effective'' Green's function
obeys the Dyson equation \cite{Landau}
\begin{equation}
\label{Dyson}
\begin{picture}(1,0.75)
\put(0,0.1){\circle*{0.1}} \put(0,0.05){\line(1,0){1}}
\put(0,0.15){\line(1,0){1}} \put(1,0.1){\circle*{0.1}}
\put(.9,0.3){\makebox(0,0)[b]{$i$}}
\put(0.1,0.3){\makebox(0,0)[b]{$k$}}
\put(1,-0.4){\makebox(0,0)[b]{$rt$}}
\put(0.,-0.4){\makebox(0,0)[b]{$r't'$}}
\end{picture} \quad =\quad
\begin{picture}(1,0.75)
\put(0,0.){\circle*{0.1}} \put(0,0.0){\line(1,0){1}}
\put(1,0.){\circle*{0.1}} \put(.9,0.2){\makebox(0,0)[b]{$i$}}
\put(0.1,0.2){\makebox(0,0)[b]{$k$}}
\put(1,-0.4){\makebox(0,0)[b]{$rt$}}
\put(0.,-0.4){\makebox(0,0)[b]{$r't'$}}
\end{picture} \quad +\quad
\begin{picture}(2.2,0.75)
\put(1,0){\circle*{0.1}} \put(1.4,0){\circle*{0.1}}
\multiput(1,0)(0,0.1){5}{\line(4,1){0.4}}
\put(1,0.5){\circle*{0.1}} \put(1.4,0.5){\circle*{0.1}}
\put(0.,0.05){\line(1,0){1}}
\put(0.,-0.05){\line(1,0){1}}\put(0.6,0.5){\line(1,0){1.2}}
\put(0.4,0.3){\line(1,1){0.4}} \put(0.4,0.7){\line(1,-1){0.4}}
\put(1,0.){\line(1,0){1.2}} \put(1,0.5){\line(0,-1){0.5}}
\put(1.4,0.5){\line(0,-1){0.5}} \put(0.,0.){\circle*{0.1}}
\put(2.2,0.0){\circle*{0.1}} \put(2.1,0.2){\makebox(0,0)[b]{$i$}}
\put(0.,0.2){\makebox(0,0)[b]{$k$}}
\put(2.2,-0.4){\makebox(0,0)[b]{$rt$}}
\put(0.,-0.4){\makebox(0,0)[b]{$r't'$}}
\end{picture}\quad+\quad
\begin{picture}(2.2,0.75)
\put(1,0){\circle*{0.1}} \put(1.4,0){\circle*{0.1}}
\multiput(1,0)(0,0.1){5}{\line(4,1){0.4}}
\put(1,0.5){\circle*{0.1}} \put(1.4,0.5){\circle*{0.1}}
\put(0.6,0.5){\line(1,0){1.1}} \put(0.4,0.3){\line(1,1){0.4}}
\put(0.4,0.7){\line(1,-1){0.4}}\put(1,0){\line(1,0){0.6}}
\put(0,0.05){\line(1,0){1.}}\put(0,-0.05){\line(1,0){1.}}
\put(1,0.5){\line(0,-1){0.5}} \put(1.4,0.5){\line(0,-1){0.5}}
\put(1.7,0.5){\line(1,-1){0.5}} \put(0.,0.){\circle*{0.1}}
\put(2.2,0.0){\circle*{0.1}} \put(2.1,0.3){\makebox(0,0)[b]{$i$}}
\put(0.,0.2){\makebox(0,0)[b]{$k$}}
\put(2.2,-0.4){\makebox(0,0)[b]{$rt$}}
\put(0.,-0.4){\makebox(0,0)[b]{$r't'$}}
\end{picture}
\end{equation}
\mbox{}\\
Using a point approximation for the $T$-matrices (see discussion after Eq.(\ref{Blc})),
equation (\ref{Dyson}) may be rewritten in the following analytic form
\begin{equation}
G^E_{ik}(\vec r,t,\vec r^{\,\prime},t') = G^0_{ik}(\vec r,t,\vec r^{\,\prime},t') +
  \int\sum_{jlms} G^0_{ij}(\vec r,t,\vec r_1,t_1)[K_{js,lm}+K_{js,lm}]
  G^E_{lk}(\vec r_1,t_1,\vec r^{\,\prime},t') C_m(\vec r_1',t_1) \,d^3r_1
\end{equation}

This equation is analogous to the Galitsky approximation in solid-state theory
(see, e.g., Ref. \onlinecite{Mattuck}). Taking the time derivatives of both its parts, we obtain
\begin{equation}
\label{eq:GreenApp}
\frac{d}{dt} G^E_{ik}(\vec r,t,\vec r^{\,\prime},t') =
 \sum_j\widehat L^0_{ij} G^E_{jk}(\vec r,t,\vec r^{\,\prime},t') +\sum_{jlm}
 [K_{ij,lm}+K_{ji,lm}] G^E_{lk}(\vec r,t,\vec r^{\,\prime},t')  C_m(\vec r,t)
\end{equation}
or in matrix notation
\begin{equation}
\label{eq:GreenCMET}
    \frac{d}{dt} \widehat G^{E}(\vec r,t,\vec r^{\,\prime},t') =
\widehat L_0 \widehat G^{E}(\vec r,t,\vec r^{\,\prime},t') +\widehat S(\vec r,t)
 \widehat G^{E}(\vec r,t,\vec r^{\,\prime},t')
\end{equation}
where
\begin{equation}
S_{il}(\vec r,t)=
\sum_{jm} [K_{ij,lm}+K_{ji,lm}]  C_m(\vec r,t)
\label{def:S}
\end{equation}

\section{Equations for the pair distribution function}
\label{sec:PDFeq}

Comparison of Eqs.(\ref{DiagR_i}) and (\ref{CMETR}) gives the following diagrammatic
expression for the pair distribution function, which includes all terms of the $\alpha$-expansion
up to the order of $\alpha^5$:

\begin{equation}
\label{no3-enc}
\begin{picture}(0.5,1.25)
\put(0.5,0.5){\oval(1,1)[l]} \put(0.5,0){\circle*{0.1}}
\put(0.5,1){\circle*{0.1}}
\put(.4,0.1){\makebox(0,0)[b]{i}}
\put(.5,-0.4){\makebox(0,0)[b]{$r_1t$}}
\put(.5,0.6){\makebox(0,0)[b]{$r_2t$}}
\put(.4,1.2){\makebox(0,0)[b]{k}}
\end{picture}\quad =
\quad
\begin{picture}(1.5,1.25)
\put(0,0.8){\line(1,1){0.4}} \put(0.,1.2){\line(1,-1){0.4}}
\put(0.,-0.2){\line(1,1){0.4}} \put(0,0.2){\line(1,-1){0.4}}
\put(0.2,0){\line(1,0){0.3}} \put(0.2,1){\line(1,0){0.3}}
\put(0.5,0){\circle*{0.1}} \put(1.5,0.){\circle*{0.1}}
\put(0.5,0.05){\line(1,0){1}} \put(0.5,1.){\circle*{0.1}}
\put(0.5,-0.05){\line(1,0){1}}  \put(1.5,1){\circle*{0.1}}
\put(0.5,0.95){\line(1,0){1}}\put(0.5,1.05){\line(1,0){1}}
\multiput(0.5,1)(0,-0.125){8}{\line(0,-1){0.09}}
\put(1.4,0.2){\makebox(0,0)[b]{$i$}}
\put(1.4,1.2){\makebox(0,0)[b]{$k$}}
\put(1.5,0.6){\makebox(0,0)[b]{$r_2t$}}
\put(1.5,-0.4){\makebox(0,0)[b]{$r_1t$}}
\end{picture}\quad + \quad
\begin{picture}(1.5,1.25)
\put(0,0.8){\line(1,1){0.4}} \put(0.,1.2){\line(1,-1){0.4}}
\put(0.,-0.2){\line(1,1){0.4}} \put(0,0.2){\line(1,-1){0.4}}
\put(0.2,0){\line(1,0){0.3}} \put(0.2,1){\line(1,0){0.3}}
\put(0.5,0){\circle*{0.1}} \put(1.5,0){\circle*{0.1}}
\put(1.0,0){\circle*{0.1}}
\put(0.5,-0.05){\line(1,0){1}} \put(0.5,0.05){\line(1,0){1}}
\put(0.5,1){\circle*{0.1}} \put(1.5,1){\circle*{0.1}}
\put(1,1){\circle*{0.1}}
\put(0.5,.95){\line(1,0){1}}\put(0.5,1.05){\line(1,0){1}}
\multiput(0.5,1)(0,-0.125){8}{\line(0,-1){0.09}}
\multiput(1,1)(0,-0.125){8}{\line(0,-1){0.09}}
\put(1.4,0.2){\makebox(0,0)[b]{$i$}}
\put(1.4,1.2){\makebox(0,0)[b]{$k$}}
\put(1.5,-0.4){\makebox(0,0)[b]{$r_1t$}}
\put(1.5,0.6){\makebox(0,0)[b]{$r_2t$}}
\end{picture}\quad + \quad
\begin{picture}(2,1.25)
\put(0,0.8){\line(1,1){0.4}} \put(0.,1.2){\line(1,-1){0.4}}
\put(0.,-0.2){\line(1,1){0.4}} \put(0,0.2){\line(1,-1){0.4}}
\put(0.2,0){\line(1,0){0.3}} \put(0.2,1){\line(1,0){0.3}}
\put(0.5,1){\circle*{0.1}} \put(2,1){\circle*{0.1}}
\put(0.5,0.){\circle*{0.1}} \put(2,0.){\circle*{0.1}}
\put(0.5,1.05){\line(1,0){1.5}} \put(0.5,0.05){\line(1,0){1.5}}
\put(0.5,.95){\line(1,0){1.5}} \put(0.5,-0.05){\line(1,0){1.5}}
\multiput(1,1)(0.5,0){2}{\circle*{0.1}}
\multiput(1,0)(0.5,0){2}{\circle*{0.1}}
\multiput(0.5,1)(0,-0.125){8}{\line(0,-1){0.09}}
\multiput(1,1)(0,-0.125){8}{\line(0,-1){0.09}}
\multiput(1.5,1)(0,-0.125){8}{\line(0,-1){0.09}}
\put(1.9,0.2){\makebox(0,0)[b]{$i$}}
\put(1.9,1.2){\makebox(0,0)[b]{$k$}}
\put(2,-0.4){\makebox(0,0)[b]{$r_1t$}}
\put(2,0.6){\makebox(0,0)[b]{$r_2t$}}
\end{picture}\quad +\quad \cdots
\quad = \quad
\begin{picture}(1.5,1.25)
\put(0,0.8){\line(1,1){0.4}} \put(0.,1.2){\line(1,-1){0.4}}
\put(0.,-0.2){\line(1,1){0.4}}
\put(0,0.2){\line(1,-1){0.4}} \put(0.2,0){\line(1,0){0.3}}
\put(0.2,1){\line(1,0){0.3}} \put(0.5,0){\circle*{0.1}}
\put(1.5,0.){\circle*{0.1}} \put(0.5,-0.05){\line(1,0){1}}
\put(0.5,0.05){\line(1,0){1}} \put(0.5,1){\circle*{0.1}}
\put(1.5,1){\circle*{0.1}} \put(0.5,1.05){\line(1,0){1}}
\put(0.5,0.95){\line(1,0){1}}
\multiput(0.5,1)(0,-0.25){4}{\line(0,-1){0.15}}
\put(1.4,1.2){\makebox(0,0)[b]{$k$}}
\put(1.5,0.6){\makebox(0,0)[b]{$r_2t$}}
\put(1.4,.2){\makebox(0,0)[b]{$i$}}
\put(1.5,-0.4){\makebox(0,0)[b]{$r_1t$}}
\end{picture}\quad + \quad
\begin{picture}(1.5,1.25)
\put(0.5,0.5){\oval(1,1)[l]} \put(0.5,0){\circle*{0.1}}
\put(1.5,0.){\circle*{0.1}} \put(0.5,-0.05){\line(1,0){1}}
\put(0.5,0.05){\line(1,0){1}} \put(0.5,1){\circle*{0.1}}
\put(1.5,1){\circle*{0.1}} \put(0.5,1.05){\line(1,0){1}}
\put(0.5,0.95){\line(1,0){1}}
\multiput(0.5,1)(0,-0.25){4}{\line(0,-1){0.15}}
\put(1.4,1.2){\makebox(0,0)[b]{$k$}}
\put(1.5,0.6){\makebox(0,0)[b]{$r_2t$}}
\put(1.4,.2){\makebox(0,0)[b]{$i$}}
\put(1.5,-0.4){\makebox(0,0)[b]{$r_1t$}}
\end{picture}
\end{equation}
\mbox{}\\
The analytical form of Eq.(\ref{no3-enc}) reads
\begin{equation}
\label{pPCM} p_{ik}(\vec r_1,\vec r_2,t) = \sum_{lmns} \int_0^t\,dt'\int  G^{E}_{il}(\vec r_1,t,\vec r_1^{\,\prime},t' )
   G^{E}_{km}(\vec r_2,t,\vec r_2^{\,\prime},t' )
             \widehat V_{lm,ns}(\vec r_1^{\,\prime},\vec r_2^{\,\prime})
              P_{ns}(\vec r_1^{\,\prime},\vec r_2^{\,\prime},t')\,d^3r_1'\,d^3r_2'
\end{equation}
Making use of Eqs.(\ref{eq:PairFunction}) and (\ref{eq:GreenCMET}), one can rewrite this equation in the equivalent differential form
\begin{equation}
\label{DiffpCM}
\frac \partial {\partial t}\widehat{ p}(\vec r_1,\vec r_2,t)=
 \left(  \widehat{\bf L}+\widehat {\bf Q}+\widehat{\bf V}(\vec r_1,\vec r_2) +
 \widehat S(\vec r_1,t)\otimes \widehat I +\widehat I \otimes \widehat S(\vec r_2,t)
 \right)\widehat{ p}(\vec r_1,\vec r_2,t)+
 \widehat{\bf V}(\vec r_1,\vec r_2)
 \left(\vec C(\vec r_1,t)\otimes \vec C(\vec r_2,t)\right)
\end{equation}

\end{widetext}
where the pair correlation function satisfies the following boundary condition
\begin{equation}
\label{eq:Infcon}
\lim_{|\vec r_1-\vec r_2|\to \infty}\widehat{ p}(\vec r_1,\vec r_2,t)=0
\end{equation}
Having the pair distribution function as a solution to the above equation, the concentration kinetics can be calculated as a solution to Eqs.~(\ref{EqCi}) and (\ref{EqRi}). The latter may be brought into matrix form as
follows (see also Eq.(\ref{eq:PairFunction})):
\begin{equation}
\frac \partial{\partial t}\vec C(\vec r,t) =
    \widehat L_0 \vec C(\vec r,t) +
    \mbox{Tr}_2\int \widehat{\bf V}(\vec r,\vec r^{\,\prime})
    \widehat{ P}(\vec r,\vec r^{\,\prime},t)\,d^3r
    \label{eq:MasterCMET}
\end{equation}
Eq.(\ref{DiffpCM}) and Eq.(\ref{eq:MasterCMET}) give the pair distribution function
formulation for the theory that includes a complete set of terms of $\alpha$-expansion of the collision integral
up to the order of $\alpha^5$.
These equations are the core of the differential formulation of the modified encounter theory  \cite{FrantsuzovCPL2000,IvanovJCP2001I,IvanovJCP2001II,IvanovJCP2001III}, with one significant correction.

Dyson equation for the "effective" Green's function  in the MET  has the following form:
\begin{equation}
\label{eq:METdiags}
\begin{picture}(1,0.75)
\put(0,0.1){\circle*{0.1}} \put(0,0.05){\line(1,0){1}}
\put(0,0.15){\line(1,0){1}} \put(1,0.1){\circle*{0.1}}
\put(.9,0.3){\makebox(0,0)[b]{$i$}}
\put(0.1,0.3){\makebox(0,0)[b]{$k$}}
\put(1,-0.4){\makebox(0,0)[b]{$rt$}}
\put(0.,-0.4){\makebox(0,0)[b]{$r't'$}}
\end{picture} \quad =\quad
\begin{picture}(1,0.75)
\put(0,0.){\circle*{0.1}} \put(0,0.0){\line(1,0){1}}
\put(1,0.){\circle*{0.1}} \put(.9,0.2){\makebox(0,0)[b]{$i$}}
\put(0.1,0.2){\makebox(0,0)[b]{$k$}}
\put(1,-0.4){\makebox(0,0)[b]{$rt$}}
\put(0.,-0.4){\makebox(0,0)[b]{$r't'$}}
\end{picture} \quad +\quad
\begin{picture}(2.2,0.75)
\put(1,0){\circle*{0.1}} \put(1.4,0){\circle*{0.1}}
\multiput(1,0)(0,0.1){5}{\line(4,1){0.4}}
\put(1,0.5){\circle*{0.1}} \put(1.4,0.5){\circle*{0.1}}
\put(0.,0.05){\line(1,0){1}}
\put(0.,-0.05){\line(1,0){1}}\put(0.6,0.5){\line(1,0){1.2}}
\put(0.4,0.3){\line(1,1){0.4}} \put(0.4,0.7){\line(1,-1){0.4}}
\put(1,0.){\line(1,0){1.2}} \put(1,0.5){\line(0,-1){0.5}}
\put(1.4,0.5){\line(0,-1){0.5}} \put(0.,0.){\circle*{0.1}}
\put(2.2,0.0){\circle*{0.1}} \put(2.1,0.2){\makebox(0,0)[b]{$i$}}
\put(0.,0.2){\makebox(0,0)[b]{$k$}}
\put(2.2,-0.4){\makebox(0,0)[b]{$rt$}}
\put(0.,-0.4){\makebox(0,0)[b]{$r't'$}}
\end{picture}
\end{equation}
\mbox{}\\
The last term means that the reaction of a propagating particle with an arbitrary particle in the bulk results in the replacement of the original particle by the reaction product.
The Dyson equation we obtained Eq.(\ref{Dyson}) includes the additional term:
\begin{equation}
\label{eq:newterm}
\begin{picture}(2.2,0.75)
\put(1,0){\circle*{0.1}} \put(1.4,0){\circle*{0.1}}
\multiput(1,0)(0,0.1){5}{\line(4,1){0.4}}
\put(1,0.5){\circle*{0.1}} \put(1.4,0.5){\circle*{0.1}}
\put(0.6,0.5){\line(1,0){1.1}} \put(0.4,0.3){\line(1,1){0.4}}
\put(0.4,0.7){\line(1,-1){0.4}}\put(1,0){\line(1,0){0.6}}
\put(0,0.05){\line(1,0){1.}}\put(0,-0.05){\line(1,0){1.}}
\put(1,0.5){\line(0,-1){0.5}} \put(1.4,0.5){\line(0,-1){0.5}}
\put(1.7,0.5){\line(1,-1){0.5}} \put(0.,0.){\circle*{0.1}}
\put(2.2,0.0){\circle*{0.1}} \put(2.1,0.3){\makebox(0,0)[b]{$i$}}
\put(0.,0.2){\makebox(0,0)[b]{$k$}}
\put(2.2,-0.4){\makebox(0,0)[b]{$rt$}}
\put(0.,-0.4){\makebox(0,0)[b]{$r't'$}}
\end{picture}
\end{equation}\\
which takes into account that the original particle can also be replaced by the second reaction product as a result of the bulk reaction.

 We therefore refer to our equations as the complete modified encounter theory (CMET). As has
been mentioned in the Theoretical Background Section, the MET corresponds to the mean-field
approximation for the reaction between the pair's particles and the isotropic
environment. CMET provides a complete mean-field approximation, available within the same
order of $\alpha$-expansion, which accounts for the effect of secondary
correlations between the environment and the pair's particles, developing in
the course of reaction.

Eq.(\ref{pPCM}) implies zero initial conditions
$p_{ik}(\vec r_1,\vec r_2,0)=0$.
However, with a minor modification of the above-described diagrammatic technique, it is possible to generalize the results for arbitrary initial pair correlations. This modification leaves
Eqs.(\ref{DiffpCM},\ref{eq:MasterCMET}) unchanged, but non-zero initial
conditions can be introduced:
\begin{equation}
\label{eq:arbIC}
p_{ik}(\vec r_1,\vec r_2,0)=p^0_{ik}(\vec r_1,\vec r_2)
\end{equation}

\section{Spatially homogeneous concentrations}

For systems with spatially homogeneous initial conditions, homogeneity is maintained throughout the temporal evolution of the reaction.
 $$C_i(\vec r,t)=C_i(t)$$
 In this case, the resulting kinetic equations can be further simplified. The pair distribution function depends only on the relative position of the reactants $\vec r = \vec r_{2}- \vec {r}_{1} $.   Invoking spherical symmetry, the pair distribution function becomes a function solely of the interparticle distance $r=|\vec{r}|$, allowing for a significant reduction of Eqs. (\ref{DiffpCM}) and (\ref{eq:MasterCMET}):
\begin{equation}
\label{eq:MasterCi}
\frac d{d t}\vec C(t) =
\mbox{Tr}_2\int_0^\infty \widehat{\bf V}( r)\widehat{ P}( r,t)\,d^3r +
      \widehat Q\vec C( t)
\end{equation}
\begin{equation}
\label{DiffpC}
 \begin{gathered}
     \frac \partial {\partial t}\widehat{ p}(r,t)=
 \left(  \widehat{\bf D}\Delta+\widehat {\bf Q} +
 \widehat{\bf V}(r) +\widehat {\bf S}(t) \right)
 \widehat{ p}(r,t)+\\+\widehat{\bf V}(r)\left(\vec C(t)\otimes \vec C(t)\right)
 \end{gathered}
\end{equation}
where
\begin{eqnarray}
    \begin{gathered}
        \widehat{P}(r,t)=\widehat{ p}(r,t)+ \vec C(t)\otimes \vec C(t), \\
  \widehat {\bf S}(t)=\widehat S(t)\otimes \widehat I +
  \widehat I \otimes \widehat S(t), \\ {D}_{ik,lm}=
  \delta_{il}\delta_{km} (D_i+D_k)
    \end{gathered}
    \label{PSD}
\end{eqnarray}
and the generalized interaction operator $\widehat {\bf V}$ (cf. Eq.(\ref{eq:Vik}))
acts as follows
\begin{equation}
\label{eq:SphericalV}
\widehat {\bf V}(r) \widehat {p}(r,t) =
  \widehat {\bf W}(r)\widehat {p}(r,t) +
  \widehat {\bf D}\nabla\left([\nabla\widehat{\bf U}(r)]\widehat{ p}(r,t)\right)
\end{equation}
with ${ U}_{ik,lm}(r)=\delta_{il}\delta_{km}U_{ik}(r)$.
The operator $\widehat S(t)$ is defined as
\begin{equation}
    S_{il}(t)=\sum_{jm} [K_{ij,lm}+K_{ji,lm}]  C_m(t)
    \label{S_CMET}
\end{equation}
Note that the general MET equations formulated in Ref. \onlinecite{FrantsuzovCPL2000} coincides with the Eqs.(\ref{DiffpC}-\ref{eq:SphericalV}) when $\widehat S(t)$ is defined as
\begin{equation}
    S_{il}(t)=\sum_{jm} K_{ij,lm} C_m(t)
    \label{S_MET}
\end{equation}

Substituting
Eq.(\ref{eq:SphericalV}) into Eq.(\ref{eq:MasterCi}), we can obtain
\begin{equation}
\label{eq:MasterCW}
\frac d{d t}\vec C(t) =
\mbox{Tr}_2\int_0^\infty \widehat {\bf W}( r)\widehat{ P}(r,t)\,d^3r +
      \widehat Q\vec C(t)
\end{equation}

The spherically-symmetric pair correlation function $\widehat {p}(r,t)$ is
subject to the upper and lower boundary conditions. The former follows
directly from Eq.(\ref{eq:Infcon}):
\begin{equation}
\label{eq:UpperBC}
\lim_{r\to\infty}\widehat{ p}(r,t) = 0, \quad \mbox{or} \quad
\lim_{r\to\infty}\widehat{ P}(r,t) = \vec C(t)\otimes \vec C(t)
\end{equation}
This condition means that correlations will be lost as the pair separates to large distances.
The lower boundary condition comes as a result of the continuity of
$\widehat{ p}(\vec r_1,\vec r_2,t)$ at $\vec r_1\to\vec r_2$:
$$\left.\nabla \widehat{ P}(r,t)\right|_{r=0}=0$$
which can also be formulated as a null flux through point $r=0$. However,
in realistic chemical systems, particles cannot approach each other closer than a certain contact distance $b$. This constraint is incorporated into the CMET framework via a hard-sphere (infinite) or steep repulsive potential at $r \le b$:
\begin{equation}
\label{eq:hardcoreU}
\widehat {\bf U}(r)= \widehat { \bf U}_h \Theta(b-r) + \widehat { \bf U}_r(r) \qquad
\mbox{at} \quad \widehat{ \bf U}_h\to \infty
\end{equation}
In addition, zero reactivity within the contact sphere is assumed
\begin{equation}
\label{eq:hardcoreW}
\widehat {\bf W}(r)= \left(1-\Theta(b-r)\right)\widehat {\bf W}_r(r)
\end{equation}
and the corresponding initial condition as allowed by Eq.(\ref{eq:arbIC}):
\begin{equation}
\label{eq:hardcoreIC}
\widehat{ P}(r,t=0)=
 \left(1-\Theta(b-r)\right)\left(\vec C(0)\otimes \vec C(0)\right)
\end{equation}
Provided that Eqs.(\ref{eq:hardcoreU}-\ref{eq:hardcoreIC}) hold, the pair
distribution function $\widehat{ P}(r,t)$ remains zero inside the sphere with a
radius of $b$, as does the flux through radius $b$:
\begin{equation}
\label{eq:MasterBC}
\left.\left\{ \nabla \widehat{ P}(r,t) +
  [\nabla \widehat {\bf U}(r)] \widehat{ P}(r,t)\right\}\right|_{r=b}=0
\end{equation}
which is essentially the lower boundary condition for Eq.(\ref{DiffpC}).
Since $\widehat{ P}(r<b,t)=0$, the lower integration limit in
Eqs.(\ref{eq:MasterCi},\ref{eq:MasterCW}) is actually $b$.

\section{Materials and Methods}
\label{sec:MaterialsMethods}

Based on the results of Ref.~[\onlinecite{FrantsuzovCPL2000}], we developed a computer program, {\bf Tegro}, for computing the kinetics of an arbitrary set of reactions under the assumption of spatially homogeneous concentrations {and no initial pair correlations}.
The program implements the kinetic equations (\ref{eq:MasterCW}) and (\ref{DiffpC}-\ref{eq:SphericalV}) subject to the boundary conditions (\ref{eq:UpperBC}) and (\ref{eq:MasterBC}) in the CMET (option “Complete MET”), MET (option “Modified ET”), and IET (option “Encounter theory”) formulations, where  $\widehat S(t)$ is defined by Eq. (\ref{S_CMET}), by Eq. (\ref{S_MET}), and $\hat{S} (t)=0$, respectively.
{\bf Tegro} is available from the GitLab repository at \texttt{https://gitlab.com/krissinel/tegro} under the CC~BY--NC~4.0 licence. The repository also contains configuration files required to reproduce the calculations presented in Figs.~\ref{fig1}-\ref{fig:Zeldovich} and~\ref{fig:comparison} in the following Sections.

\section{Case studies and numerical comparisons}
\label{sec:PartSys}

In this Section, we present results from computational studies aimed at testing CMET in cases with known concentration kinetics or asymptotic limits.

\subsection{Irreversible quenching of excitation }
\label{subs:smol}

The first example we consider is the reaction of quenching the molecule $D$ excitation    by an energy transfer to the acceptor molecule $A$
\begin{equation}
D^*  + A\rightarrow D + A
\label{Ds}
\end{equation}
Specifically, we consider the case where the diffusion coefficient $D^*$ is equal to zero. In this case, as discussed in the Theoretical Background Section, the SM approach Eqs.~(\ref{eq:DET}-\ref{eq:n})  result in the exact kinetics $C_{D^*}(t)$ \cite{AllBlum,Szabo}.
The solution of Eq.(\ref{eq:DET}) is
\begin{equation}
    C_{D^*}(t) = C_{D^*}(0)\exp\left(-C_A\int\limits_0^tk(\tau)\dd \tau\right),
    \label{PCD*}
\end{equation}
where $k(t)$ is given by the Eqs.(\ref{eq:Kt}-\ref{eq:m}). Here, we numerically calculate $k(t)$ with Spherically Symmetric Diffusion Problem (SSDP)  application \cite{KrissinelSSDP} using the following distance-dependent energy transfer rate (F\"{o}rster formula):
\begin{equation}
W(r)=W_c\left(\frac b r\right)^6
\label{Forster}
\end{equation}

In this case, the CMET equation (\ref{eq:MasterCW}) has to be written only for the concentration $D^*$. The equations (\ref{DiffpC}-\ref{S_CMET}) have to be written for the distribution functions of the pairs $D^* - A$ and $D^* - D^*$.

The kinetics of the $D^*$ concentration calculated within  CMET, the SM approach, and formal chemical kinetics are shown in Fig.~\ref{fig1}. Numerical results confirm that CMET accurately reproduces the SM solution within its valid temporal domain. Kinetics for both CMET and the SM approach are different from the exponential dependence of LMA:
\begin{equation}
    C_{D^*}(t) = C_{D^*}(0)\exp(-C_A kt), \quad k = \lim_{t\rightarrow \infty}k(t).
    \label{eq:exp}
\end{equation}
\begin{figure}[h]
    \centering
    \includegraphics[width=1\linewidth]{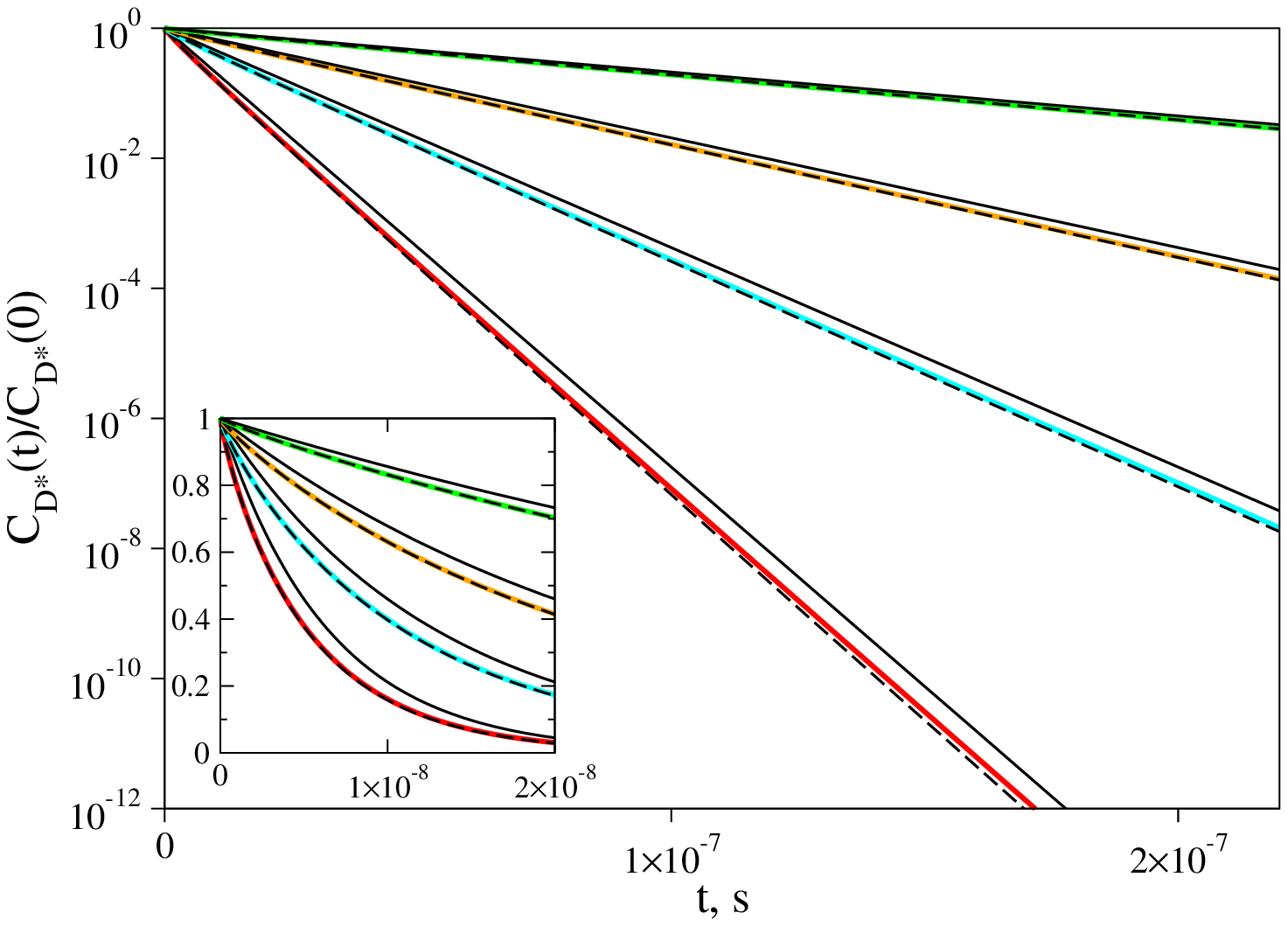}
  \caption{Time dependence of the concentration $C_{D^*}(t)$ for the reaction $D^*  + A\rightarrow D + A$, calculated using CMET Eqs.(\ref{eq:MasterCW}) and (\ref{DiffpC}-\ref{S_CMET}) (color lines), the SM approach Eqs.(\ref{PCD*}) and (\ref{eq:DET}-\ref{eq:n}) (dashed lines), and formal kinetics Eq.(\ref{eq:exp}) (thin black lines)  at various concentrations of $A$ particles. The inset gives a more detailed view of the initial stage of the reaction.  The parameters of the reaction are:  $D_{D}=0,\, D_A = 5\times 10^{6}\ \text{cm}^2/\text{s}$, the distance dependent reaction rate  $W(r)$ is given by Eq.(\ref{Forster}) where {$W_c = 10^{10}\ \text{s}^{-1}$}, the contact radius is $b= 5\AA$. Concentration of $A$ is $C_A= 0.1\, M$ (red), $C_A= 0.05\, M$ (cyan), $C_A= 0.025\, M$ (orange), $C_A= 0.01\, M$ (green). }
    \label{fig1}
\end{figure}

\subsection{Stationary quenching of luminescence}
\label{subs:sternvolmer}

The next example represents a more complex system:
$$ \begin{array}{lll}
&D^* & + A\rightarrow D + A\\
&&\\
q(t)&\uparrow\downarrow&  1 /{\tau}\\
&&\\
&D&
\end{array}
$$
The particle $D$ is excited to the excited state $D^*$ (e.g., by light radiation) with the excitation rate  $q(t)$. The excited state can relax by emitting a photon at a rate of $1/\tau$ or can be quenched by an energy transfer to the acceptor $A$. Let us consider the case when the excitation is weak $C_{D^*}\ll C_{D}$.
In the case of pulsed excitation $q(t)=P\delta(t)$, the kinetics of $D^*(t)$ can be described by the SM kinetic equation:
\begin{equation}
    \frac d{dt}C_{D^*}(t) = -\frac{1}{\tau}C_{D^*}(t) - k(t)C_A C_{D^*}(t)
\end{equation}
with the initial condition
$C_{D^*}(0)=P C_D$.

This equation could be solved explicitly as
\begin{equation}
   C_{D^*}(t) = P C_D \exp{-\frac{t}{\tau} - C_A\int\limits_0^tk(t')\dd t'}.
    \label{eq:singexc}
\end{equation}
This is result is exact when $D^\ast$ is immobile.
For an arbitrary $q(t)$ dependence, the concentration $C_{D^*}(t)$ can be expressed   as
\begin{equation}
    \begin{gathered}
        C_{D^*}(t) =\\= C_D \int\limits_{-\infty}^t q(t')\exp{-\frac{t-t'}{\tau} - C_A\int\limits_0^{t-t'}k(t'')\dd t''}\dd t'
    \end{gathered}
\end{equation}

When the constant excitation is switched on at $t=0$, i.e., $q(t)=q\Theta(t)$, where $\Theta(t)$ is the Heaviside function, Eq. (10.7) reduces to:
\begin{equation}
    C_{D^*}(t) = qC_D \int\limits_0^t\exp{-\frac{t'}{\tau} - C_A\int\limits_0^{t'}k(t'')\dd t''}\dd t'
    \label{CDt}
\end{equation}

Under continuous-wave excitation, $D^\ast$  evolves toward a steady-state concentration, given by the $t\rightarrow\infty$ limit of Eq. (\ref{CDt}) as follows:
\begin{equation}
C_{D^*}=qC_D \int\limits_0^{\infty}\exp{-\frac{t}{\tau} - C_A\int\limits_0^{t}k(t')\dd t'}\dd t'
\end{equation}

Then, the luminescence quantum yield under continuous wave excitation is estimated as

\begin{equation}
    \begin{gathered}
        \phi(C_A)\equiv \frac {C_{D^*}}{\tau qC_D}=\\=\frac1 {\tau} \int\limits_0^t\exp{-\frac{t'}{\tau} - C_A\int\limits_0^{t'}k(t'')\dd t''}\dd t'
    \end{gathered}
    \label{phi}
 \end{equation}
Expanding the integrand in series on $C_A$, we get the Stern - Volmer dependence at a low acceptor concentration (see, for example, Ref. \onlinecite{Rice}):
\begin{equation}
 [\phi(C_A)]^{-1} = 1 + C_A\Tilde k(1/\tau),\quad \Tilde k(s) = \int\limits_0^\infty k(t) e^{-st}\dd t
 \label{Stern}
 \end{equation}
\begin{figure}[h]
\centering
  		\includegraphics[width=1\linewidth]{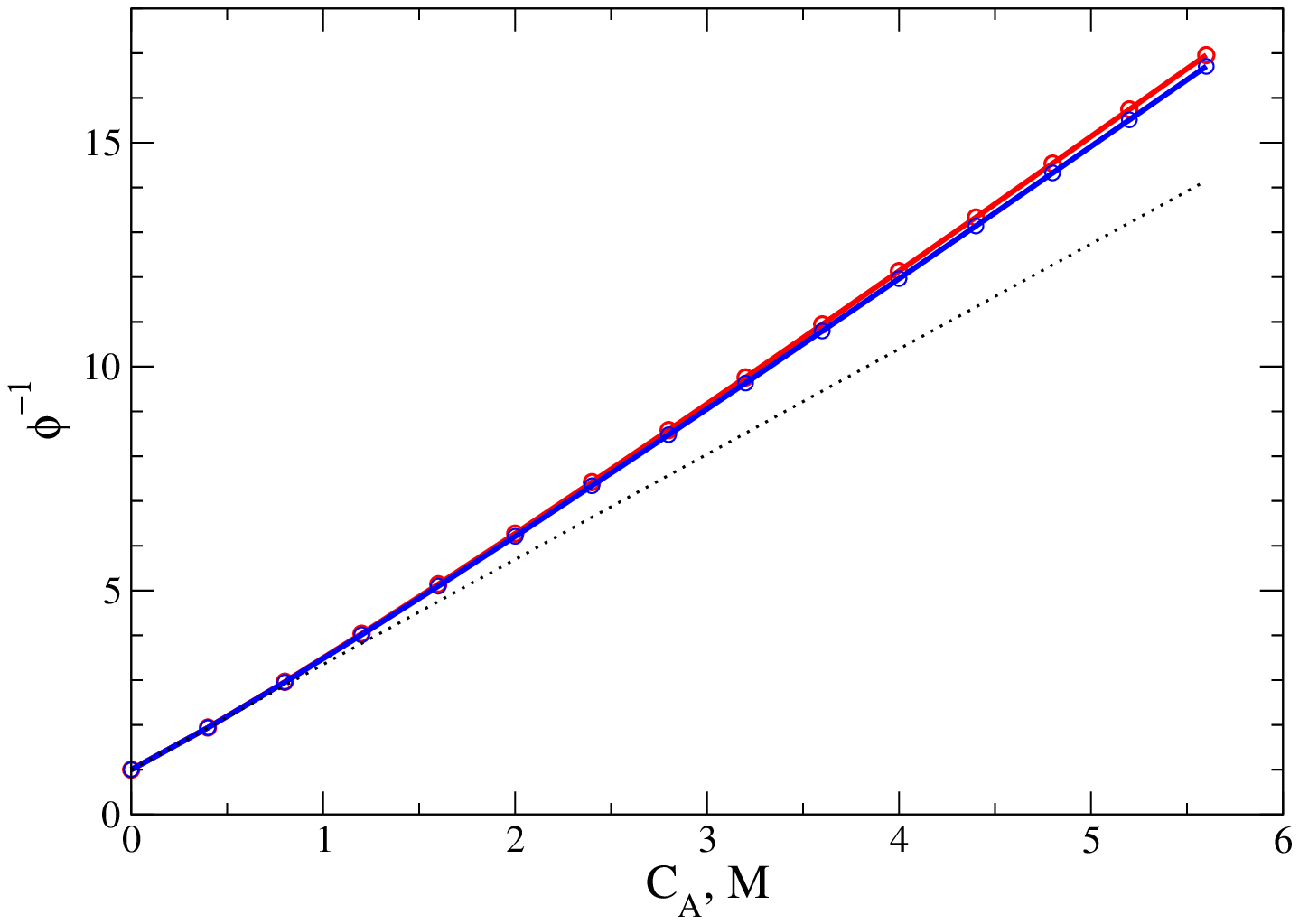}
   		\includegraphics[width=1\linewidth]{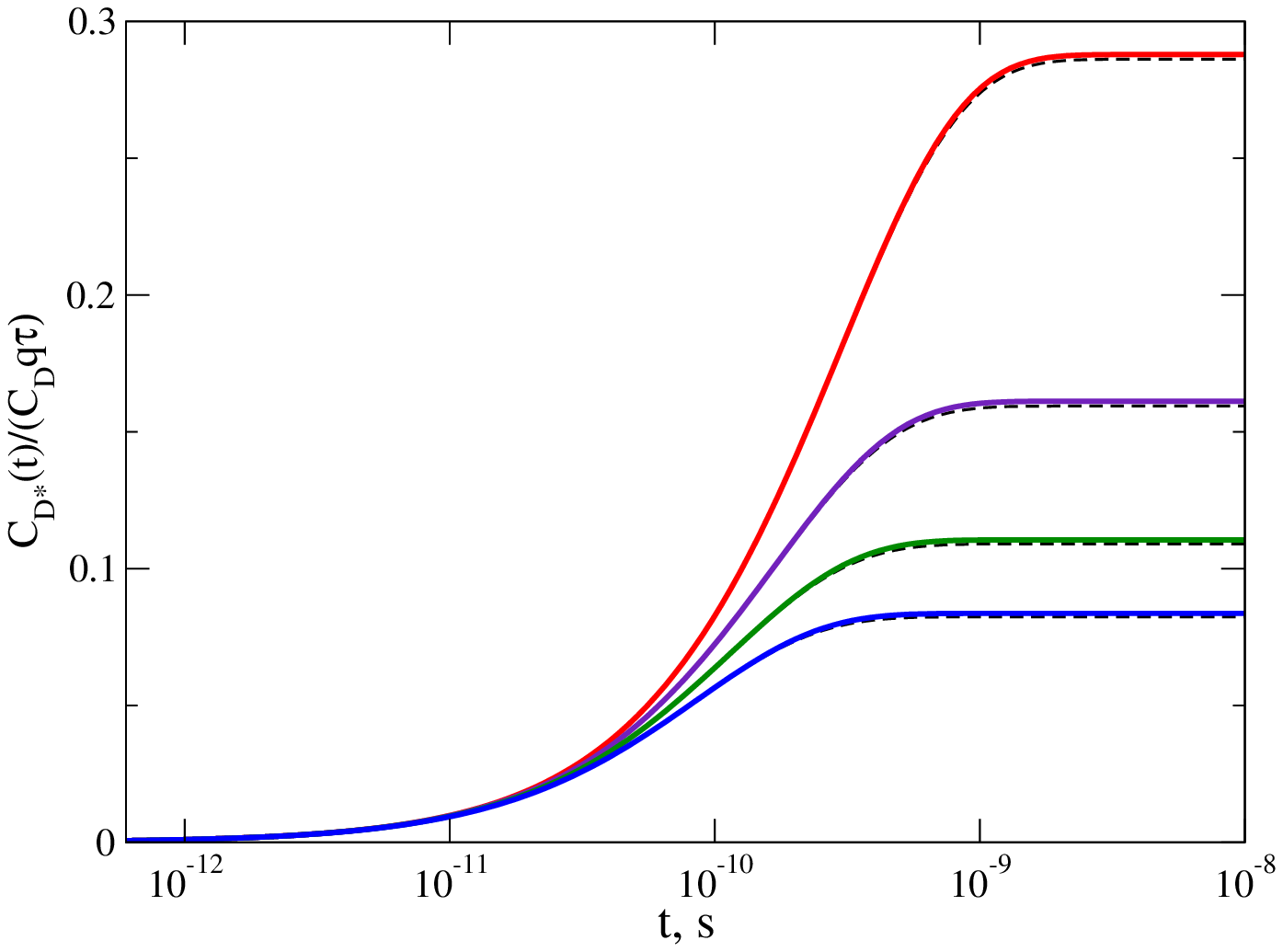}
        \caption{(top)
      Comparison of Stern–Volmer plots predicted by CMET (blue circles) and the SM model (Eq. \ref{phi}; red circles), showing the inverse luminescence quantum yield as a function of acceptor concentration under continuous-wave excitation. The dotted line represents the linear Stern-Volmer dependence Eq.(\ref{Stern}).
        (bottom) Time dependence of the  $D^*$ concentration under CW excitation for different $A$ concentrations calculated within CMET  Eqs.(\ref{eq:MasterCW}) and (\ref{DiffpC}-\ref{S_CMET}) (solid lines) and the SM approach Eq.(\ref{CDt}) (dashed lines) with colors indicating the concentration of $A$ particle:
        $1$ M (red),  $2$ M (indigo), $3$ M (green), and $4$ M (blue). The remaining parameters of the reaction are: $\tau = 10^{-9}\ \text{s}$, $D_A = 5\cdot 10^{-6}\ \text{cm}^2/\text{s}$, { $D_D = 0$ }, { $D_{D*} = 0$ }, $q = 10^7 \ \text{s}^{-1}$,    the distance dependent reaction rate  $W(r)$ is given by Eq.(\ref{Forster}), where $W_{{c}} = 10^{10}\ \text{s}^{-1}$, the contact radius is $b= 5\AA$.}
        \label{fig2}
\end{figure}
As shown in Fig. \ref{fig2}, CMET correctly reconstructs the time dependence of the concentration under continuous excitation and
the acceptor concentration dependence of the inverse luminescence quantum yield $\phi^{-1}$ up to $5$  M.

\subsection{Electron transfer and consequent geminate recombination}
\label{subs:burshtein}

Another representative example is the electron transfer from an excited particle $D^*$ to the acceptor $A$ in the bulk, generating a geminate pair $D^+$ and $A^-$. The $D^+$ and $A^-$ particles can then recombine via reverse electron transfer during geminate kinetics or can separate and subsequently recombine in a bulk reaction.

\begin{equation}
 \begin{array}{ccc}
D^*  + A\rightarrow& [D^+ \dots A^-]& \rightarrow D + A\\
\\
&\downarrow& \\
\\
&D^+ + A^-& \rightarrow D + A
\end{array}
\label{Dorf}
\end{equation}

The theory describing the generation of the geminate pairs in the bulk reaction and geminate kinetics was suggested independently by Burshtein \cite{BurshteinCPL1992} and Dorfman and Fayer \cite{DorfmanFayer}.
In this theory, the concentration kinetics of $D^*$ obey the SM kinetic equation:

\begin{eqnarray}
    \pdv{}{t}C_{D^*}(t) = -k_i(t) C_AC_{D^*}(t)
    \label{CD*UT}
\end{eqnarray}

with a time-dependent rate constant $k_i(t)$
\begin{eqnarray}
    \begin{gathered}
        k_i(t) = \int W_i(r)n(r, t)\dd^3 r,\\ \pdv{}{t}n(r, t) = D \Delta n(r, t) - W_i(r)n(r, t).
    \end{gathered}
\end{eqnarray}
where $W_i(r)$ is the electron transfer rate
$$W_i(r)=W_i^0\exp(-2(r-b)/L)$$
 The ion concentration $C_{D+}(t)$ can be found by integrating the geminate pair distribution function $p(r,t)$ Eq.(\ref{eq:Cgem}):

\begin{eqnarray}
    C_{D+}(t) = \int p(r,t)    \dd^3 r
    \label{CD+}
\end{eqnarray}

where  $p(r, t)$ obeys the following equation
\begin{equation}
    \begin{gathered}
        \pdv{t} p(r, t) =  D \nabla \left(\nabla p(r, t)+ \left[\nabla U(r)\right]p(r, t) \right) -\\- W_r(r)p(r, t)+ W_{{i}}(r) C_AC_{D^*}(t)n(r,t)
    \end{gathered}
    \label{eq:p}
\end{equation}
where $U(r)$ is the electrostatic potential energy and $W_r(r)$ is the back electron transfer rate
 $$W_r(t)=W_r^0\exp(-2(r-b)/L)$$
Since the theory utilizes both the SM and geminate kinetic equations,  it was called the unified theory (UT) in Ref. \onlinecite{BurshteinACP2004}.
This theory gives an exact solution for kinetics as long as the ion pairs $D^+ \dots A^-$ can be considered as geminate, i.e., condition Eq.(\ref{eq:tgem}) is satisfied.
At longer times, the pairs are separated, and recombination of ions in the volume will be observed.
Note that in the limit $D^*(0)\rightarrow 0 $ the characteristic time of applicability of UT tends to infinity.

Burshtein and Frantsuzov \cite{FrantsuzovJCP1997,FrantsuzovJCP1997s} showed that the kinetics of reaction Eq.(\ref{Dorf}), including the geminate recombination of ions and their bulk recombination, can be correctly described using IET.
However, as discussed above, the time interval where IET is applicable is too short. The transition to CMET provides a significant increase in this time interval.
 Fig~\ref{fig:UT}  illustrates this by comparing the kinetics obtained by solving the CMET and UT kinetic equations.
{Due to the irreversibility of the first stage in both theories}, the kinetics of the $D^*$ concentration are  {nearly} identical, as shown in subsection A.
The time dependence of the {$D^+$} concentration in UT ({dashed} line) shows an initial rise, caused by the generation of the geminate pairs in a forward reaction.
It then decays due to the geminate recombination,  reaching a plateau when all unrecombined pairs have separated.
The kinetics of $D^*$ calculated in CMET (solid lines) coincide with the kinetics calculated in UT at short times, but show the decay of the latter due to the recombination of ions in the bulk.
As the initial concentration $D^*(0)$ decreases, the kinetics of $D^+$ calculated in the CMET approaches the kinetics calculated in UT.

Thus, CMET quantitatively describes the kinetics of multistage reactions, including geminate and bulk ion recombination.

\begin{figure}[h]
    \centering
    \includegraphics[width=1\linewidth]{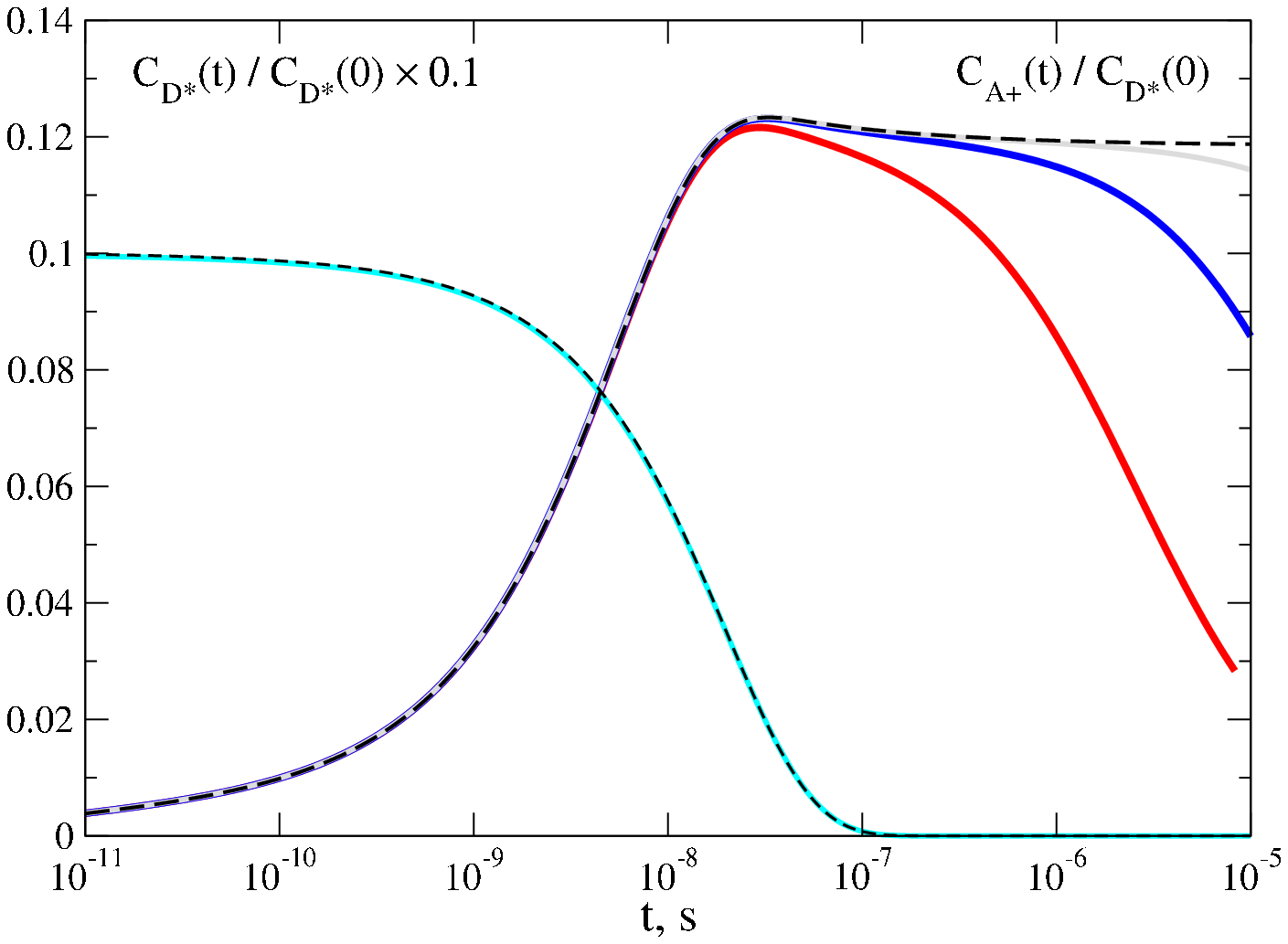}
    \caption{Time dependence of the $D^*$ concentration for the multistage reaction Eq.(\ref{Dorf}), calculated using UT Eq.(\ref{CD*UT})) (dashed line) and CMET Eqs.(\ref{eq:MasterCW}) and (\ref{DiffpC}-\ref{S_CMET}) (cyan line). These lines are multiplied by the factor of $0.1$ for clarity.
    The time dependence of the $D^+$ concentration, calculated using UT Eqs.(\ref{CD+}-\ref{eq:p}) (dashed line) and CMET with the initial concentration $C_{D^*}(0)$ being equal to $10^{-3}\ \text{M}$ (red line),  $10^{-4}\ \text{M}$ (blue line), and  $10^{-5}\ \text{M}$ ({gray} line). The parameters of the reactions are $W_i^0=10^{12} s^{-1} , W_r^0=10^{11} \text{s}^{-1}, L=1\AA, b=5 \AA, D_A=5\times10^{-6}\ \text{cm}^2/\text{s}, C_A=10^{-2}\ \text{M} $.}
    \label{fig:UT}
\end{figure}

\subsection{Reversible reaction $A+B\leftrightarrow C+D$ }
\label{subs:gopichszabo}

As a final example, consider the reversible reaction $A+B\leftrightarrow C+D$, in which the rate of the forward reaction $W_f(r)$ is related to the rate of the reverse reaction $W_b(r)$ by the equilibrium constant
$$ W_f(r)= K_e W_b(r)$$
At long times, the concentrations of $A$, $B$, $C$ and $D$ reach equilibrium values $C^e_A$, $C^e_B$, $C^e_C$ and $C^e_D$, respectively, which obey the following relation
$$ K_eC^e_A C^e_B= C^e_C C^e_D$$
The kinetics of the reaction are described by the relaxation function
\begin{eqnarray*}
    \begin{gathered}
        R(t)=\frac{C_A(t)-C^e_A}{C_A(0)-C^e_A} = \frac{C_B(t)-C^e_B}{C_B(0)-C^e_B}=\\=\frac{C_C(t)-C^e_C}{C_C(0)-C^e_C}=\frac{C_D(t)-C^e_D}{C_D(0)-C^e_D}
    \end{gathered}
\end{eqnarray*}\\

There is an exact analytic result for the long-term (fluctuation) asymptotics for this reaction, obtained by Gopich and Szabo \cite{GopichSzaboCP2002}.
 This result can be written in a simpler form  in the following special cases \cite{GopichSzaboCP2002}:\\

 \begin{widetext}
1. Pseudo-first order limit $C_A^e,\ C_C^e \rightarrow 0$ :

\begin{equation}
        R_A(t) = \frac{K_eC_B^e}{(K_eC_B^e+C_D^e)^2} \left(4\pi(D_D + \mathcal D)t\right)^{-3/2}
        + \frac{K_eC_D^e}{(K_eC_B^e+C^e_D)^2} \left(4\pi(D_B + \mathcal D)t\right)^{-3/2}
    \label{eq:GS0}
\end{equation}\\

where 
$$\mathcal D =\frac{C_D^eD_A + K_eC_B^eD_C}{K_eC_B^e + C_D^e}$$\\
2. Equal diffusion constants $D_A = D_C,\ D_B = D_D$:

\begin{equation}
\begin{split}
   R_A(t) = &\frac{2(K_e-1)^2K_eC_A^eC_B^e}{(K_e(C_A^e+C_B^e)+C_C^e+C_D^e)^3} \left(8\pi \mathcal D t\right)^{-3/2}\\
   + &\frac{K_e(C_B^e+C_D^e)}{(K_e(C_A^e+C_B^e)+C_C^e+C_D^e)^2} \left(4\pi(D_A+ \mathcal D) t\right)^{-3/2}\\
    + &\frac{K_e(C_A^e+C_C^e)}{(K_e(C_A^e+C_B^e)+C_C^e+C_D^e)^2} \left(4\pi(D_B+ \mathcal D) t\right)^{-3/2}
\end{split}
\label{eq:GS1}
\end{equation}\\
where
$$\mathcal D =\frac{D_A(K_eC^e_A+C^e_C) + D_B(K_e C_B^e+C_D^e)}{K_e(C_A^e+C_B^e) + C_C^e+C_D^e}$$\\
3. Equal diffusion constants $D_A = D_B, \ D_C = D_D$:\\

\begin{equation}
    \begin{split}
      R_A(t) = &\frac{2K_eC_A^eC_B^e}{(K_e(C_A^e+C_B^e)+C_C^e+C_D^e)(C_A^e+C_B^e)^2} \left(8\pi  D_A t\right)^{-3/2}\\
           + &\frac{2C_C^eC_D^e}{(K_e(C_A^e+C_B^e)+C_C^e+C_D^e)(C_C^e+C_D^e)^2} \left(8\pi  D_C t\right)^{-3/2}\\
   + &\frac{2K_eC_A^eC_B^e\left(K_e((C_A^e)^2+(C_B^e)^2)-(C_C^e)^2-(C_D^e)^2\right)^2}{(K_e(C_A^e+C_B^e)+C_C^e+C_D^e)^3(C_A^e+C_B^e)^2(C_C^e+C_D^e)^2} \left(8\pi \mathcal D t\right)^{-3/2}\\
    + &\frac{K_e(C_C^e+C_D^e)(C_A^e-C_B^e)^2}{(K_e(C_A^e+C_B^e)+C_C^e+C_D^e)^2(C_A^e+C_B^e)^2} \left(4\pi(D_A+ \mathcal D) t\right)^{-3/2}\\
    + &\frac{K_e(C_A^e+C_B^e)(C_C^e-C_D^e)^2}{(K_e(C_A^e+C_B^e)+C_C^e+C_D^e)^2(C_C^e+C_D^e)^2} \left(4\pi(D_C+ \mathcal D) t\right)^{-3/2}
    \end{split}
    \label{eq:GS2}
\end{equation} 

where
$$\mathcal D =\frac{D_CK_e(C^e_A+C^e_B) + D_A( C_C^e+C_D^e)}{K_e(C_A^e+C_B^e) + C_C^e+C_D^e}$$\\

\end{widetext}

The time dependence of the relaxation function $R_A(t)$, calculated within the CMET for these three cases, is shown in Fig. \ref{fig:GopichSzabo}, with the parameters shown in Table \ref{t:pars}.
In this case, the CMET equations (\ref{eq:MasterCW}) have to be written for the concentrations of  $A$, $B$, $C$, and $D$ particles. The equations (\ref{DiffpC}-\ref{S_CMET}) have to be written for the distribution functions of the pairs $A$-$B$,  $C$-$D$, $A$-$D$,  $B$-$C$,  $A$-$A$,  $B$-$B$, $A$-$C$,  $B$-$D$, $C$-$C$, and $D$-$D$.

The relaxation function at short times can be described by LMA equations, when at large times it shows fluctuation asymptotics proportional to $t^{-3/2}$.
In all three cases, the asymptotic aligns precisely with the exact theoretical predictions of  Eqs.(\ref{eq:GS0}-\ref{eq:GS2}).

\begin{table*}
\begin{ruledtabular}
    \begin{tabular}{|l|c|c|c|c|c|c|}
        \hline & $C_A(0), M$ & $C_B(0), M$ & $D_A, \text{cm}^2/\text{s}$& ${D_B}, \text{cm}^2/\text{s}$ & ${D_C},  \text{cm}^2/\text{s}$& $D_D, \text{cm}^2/\text{s}$\\ \hline
         \text{Case 1 purple line} & $10^{-5}$ & $10^{-3}$ & $5\times 10^{-6}$ &  $5\times 10^{-6}$ & $5\times 10^{-6}$ & $5\times 10^{-6}$ \\ \hline
         \text{Case 2 blue line} &  $10^{-3}$ & $10^{-3}$ & $5\times 10^{-5}$ &$5\times 10^{-5}$ &$5\times 10^{-6}$ &$5\times 10^{-6}$\\\hline
         \text{Case 3 green line} & $10^{-3}$ & $10^{-3}$ & $5\times 10^{-5}$ &$5\times 10^{-6}$ & $5\times 10^{-5}$ & $5\times 10^{-6}$\\\hline

    \end{tabular}
    \caption{Parameters of reaction  $A+B\leftrightarrow C+D$  for Fig.~\ref{fig:GopichSzabo}.}
    \label{t:pars}

\end{ruledtabular}
\end{table*}

\begin{figure}[t]
    \centering
    \includegraphics[width=1\linewidth]{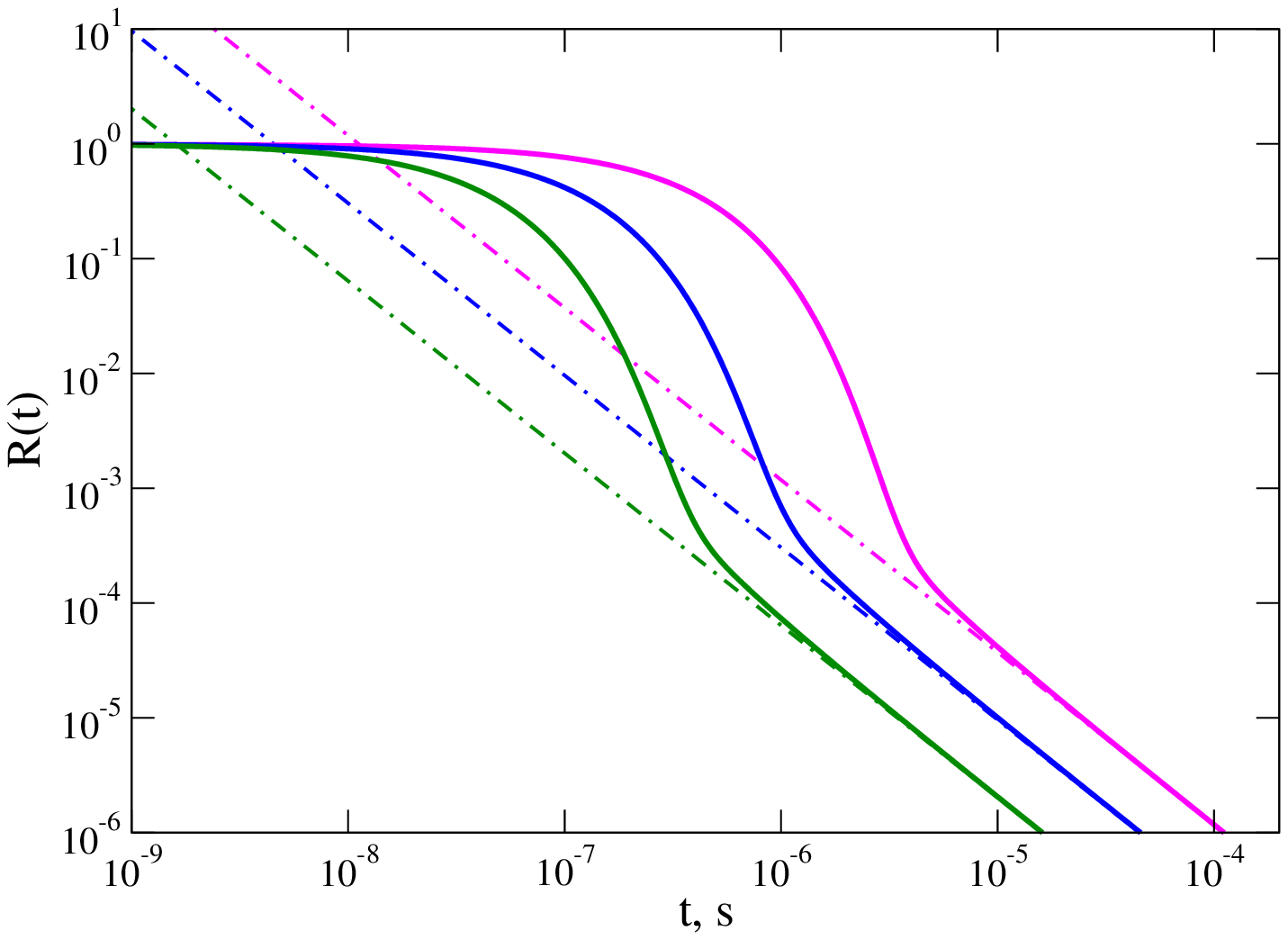}
    \caption{Time dependence of the relaxation functions for $A+B\leftrightarrow C+D$ reaction, calculated within CMET Eqs.(\ref{eq:MasterCW}) and (\ref{DiffpC}-\ref{S_CMET}) (solid lines).  The asymptotic dependencies Eqs.(\ref{eq:GS0}-\ref{eq:GS2}) are shown by dash-dotted lines. The distance dependent reaction {rates are} given by Eq.(\ref{Cont}), where $W_{\delta}= 10^{10}\ \text{s}^{-1}$. The equilibrium constant is  $K_e = 1$. The initial concentrations and diffusion coefficients for different colors are given in Table \ref{t:pars}.}
    \label{fig:GopichSzabo}
\end{figure}

Figure \ref{fig:Agmon} shows a comparison of the kinetics of the reaction $A + B \leftrightarrow C + D$, obtained by the CMET method, with the results of the simulations of Agmon and Popov \cite{AgmonJCP2003}. The modeling method used in Ref. \onlinecite{AgmonJCP2003} allows us to obtain accurate kinetics for the reaction in the pseudo-first order limit $C_A,C_C \ll C_B,C_D$. From Fig.\ref{fig:Agmon}, it can be concluded that the kinetics coincide within the statistical error of the simulation.

\begin{figure}[h]
    \centering
    \includegraphics[width=1\linewidth]{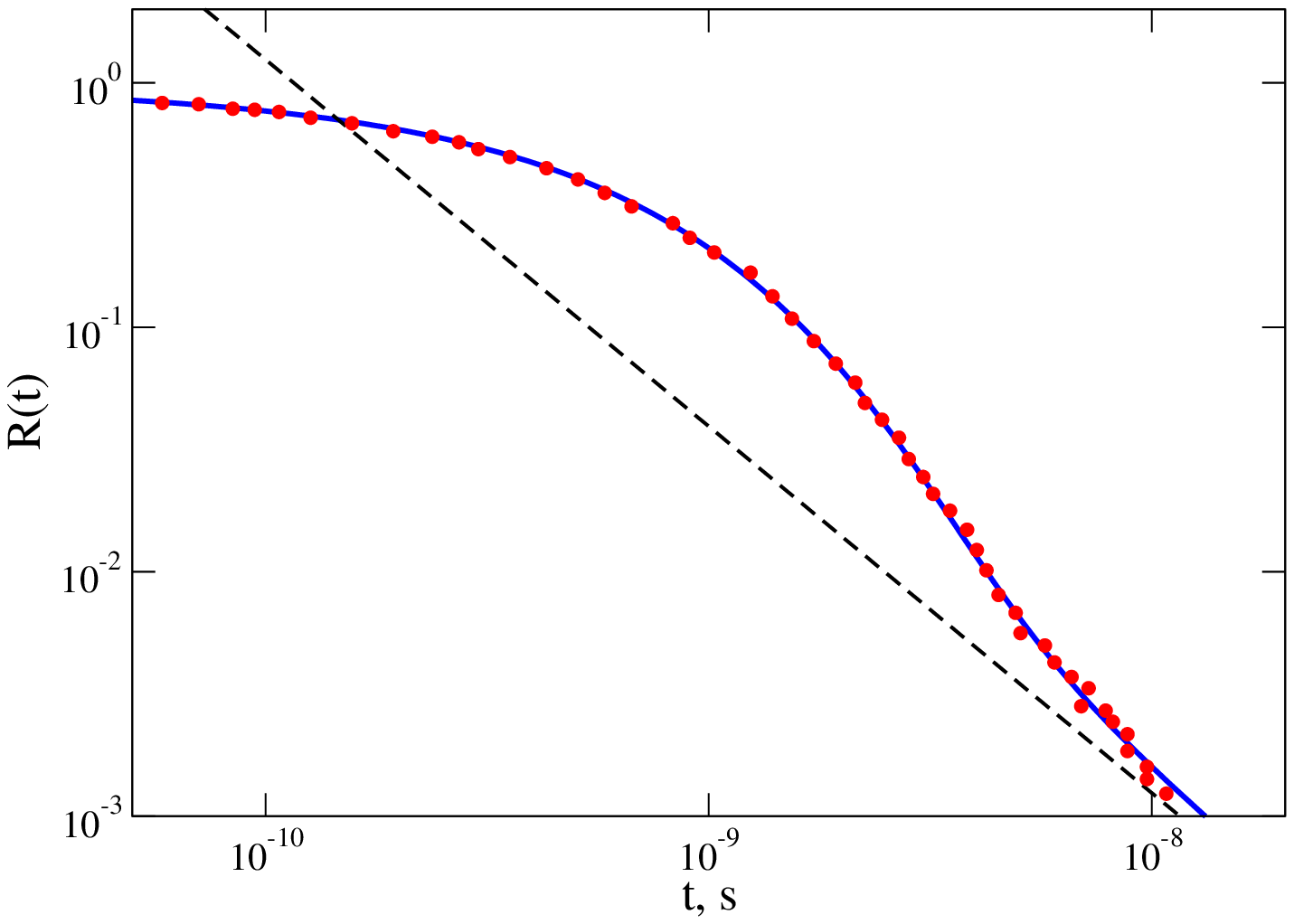}
    \caption{
Time dependence of the relaxation function for  $A + B \leftrightarrow C + D$ reaction, calculated within CMET (\ref{eq:MasterCW}) and (\ref{DiffpC}-\ref{S_CMET}) (blue line) and by numerical simulations \cite{AgmonJCP2003} (red  dots).  The parameters of the reaction are: $C_A(0) = {10^{-10}}\ \text{M}$, $C_B(0) = C_D(0) = {0.167}\ \text{M}$, $C_C(0)=0$, $D_A=D_B=D_C=D_D = {10^{-5}}\ \text{cm}^2/\text{s}$, the distance dependent reaction {rates  are} given by Eq.(\ref{Cont}), where $W_{\delta} = {10^{12}}\ \text{\AA}/\text{s}$.  The equilibrium constant is  $K_e = 1$.  The dotted line represents the Gopich - Szabo asymptotics given by  Eq. (\ref{eq:GS0}).}
    \label{fig:Agmon}
\end{figure}

Let us consider the case of equal initial concentrations of $A$ and $B$ particles $C_A(0)=C_B(0)=C_0$  and  $C_C(0)=C_D(0)=0$.
Fig.~\ref{fig:Zeldovich} shows the CMET kinetic curve $C_A(t)$ for different values of $K_e$. {We assumed, both $W_f(r)$ and $W_b(r)$ to be of contact reaction form:
\begin{equation}
   W_f(r)= W_{\delta}\delta\left(r - b\right),\ W_b(r)= \frac{W_{\delta}}{K_e}\delta\left(r - b\right)
    \label{Cont}
\end{equation} }

As can be concluded from Fig.~\ref{fig:Zeldovich}, with an increase in $K_e$ the $C_A(t)$ dependence approaches the CMET kinetic curve of the irreversible reaction $A+B \rightarrow C+D$.
{This kinetics is } almost identical to the solution of the following kinetic equation {obtained by Pasteur, Bolster, and  Benson \cite{BensonJCompPhys2014} to describe an irreversible reaction}:
\begin{equation}
\begin{gathered}
    \frac{d}{dt}C_{A}(t) = -kC_A^2 \\+ k\frac {C_0}{2(8\pi Dt)^{3/2}} \left[1-\exp\left(-4k\int\limits_0^{t}C_A(t')\,dt'\right)\right],
\end{gathered}
\label{eq:Benson}
\end{equation}\\
where $k$ is a steady-state rate constant defined by Eqs.(\ref{eq:Kt}-\ref{RD}).

The asymptotic behavior of the solution of Eq.(\ref{eq:Benson}) has a power-law form:
\begin{equation}
C_{A}(t) = \frac {\sqrt{C_0}}{\sqrt{2}}(8\pi Dt)^{-3/4}
\label{eq:Zeld}
\end{equation}
Such fluctuation asymptotics of the irreversible reaction kinetics was predicted by Ovchinnikov and Zeldovich \cite{ZeldovichCP1978} and subsequently confirmed by other authors \cite{WilczekJCP83,CardyJSP1995,LeeCardyJSP1995-er,TauberJPA2005}.
It is important to note that the results in Refs. \onlinecite{ZeldovichCP1978,WilczekJCP83,CardyJSP1995,LeeCardyJSP1995-er,TauberJPA2005} differ from Eq.(\ref{eq:Zeld}) by a constant factor. The source of this difference will be subject of our future studies.

\begin{figure}[t]
    \centering
    \includegraphics[width=1\linewidth]{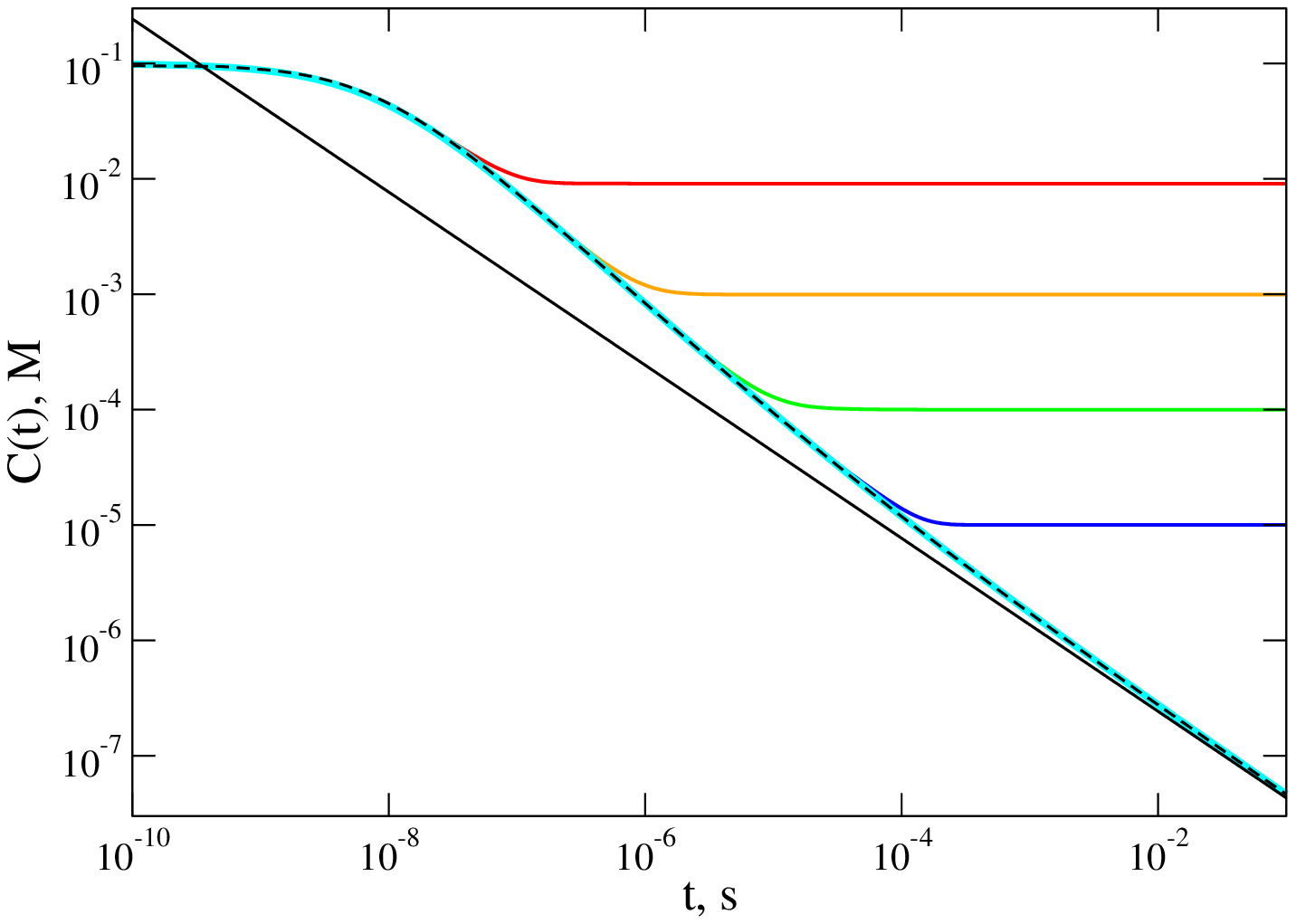}
    \caption{Time dependence of the $A$ particle concentration  for $A+B\leftrightarrow C+D$ reaction, calculated within CMET Eqs.(\ref{eq:MasterCW}) and (\ref{DiffpC}-\ref{S_CMET})  (solid lines with the colors corresponding to the $K_e = {10}$ for the red, $K_e = {10^{3}}$ for the orange, $K_e = {10^{5}}$ for the green, $K_e = {10^{7}}$ for the blue line). Initial parameters of the reaction are: $C_A(0) = C_B(0) = 0.1\ \text{M}$, $D = 5\times 10^{-6}\ \text{cm}^2/\text{s}$,  the distance dependent reaction {rates are} given by Eq.(\ref{Cont}), where $W_{\delta}= 10^{10}\ \text{\AA}/\text{s}$. The cyan line represents the $A$ particle concentration of the irreversible reaction $A+B\rightarrow C+D$, calculated within the CMET.   The dashed black line represents a solution of Eq. (\ref{eq:Benson}). The solid black line represents the asymptotic dependence Eq.(\ref{eq:Zeld}). }
    \label{fig:Zeldovich}
\end{figure}

\section{Discussion}
\label{sec:Discussion}
In the present work, the second quantization formalism is applied for the first time to systematically derive kinetic equations for multistage diffusion-influenced reactions in a general set of unimolecular and bimolecular reactive systems.
As a convenient tool for visualizing calculations, a corresponding diagrammatic technique was developed. In section \ref{sec:EncounterTheory}, we show that this technique in the second order of the expansion of the collision integral in concentration reproduces the general equations of the IET, i.e, Eqs.(\ref{Diffp}-\ref{EqCvec}). Since the IET has a limited range of applicability, a regular series expansion of the collision integral in terms of  parameter $\alpha$ was developed.
Taking into account the first two non-vanishing terms of the expansion, proportional to $\alpha^4$ and $\alpha^5$ as well as some additional higher order terms, we obtained kinetic equations for the concentrations and pair distribution functions.
The resulting equations are similar to the equations of the modified encounter theory (MET) \cite{FrantsuzovCPL2000,IvanovJCP2001I}, but contain additional terms.
Therefore, we called these equations the complete modified encounter theory (CMET).

To illustrate the differences in the IET, MET, and CMET approaches, we use a reaction $A+B\rightarrow C+D$.
For this example, the master equations (\ref{eq:MasterCMET}) for the concentrations in all these theories have the following form:
\begin{widetext}
\begin{eqnarray} \label{Ca}
\frac {\partial}{\partial t} C_{A}(\vec r,t) &=& D_A\Delta C_{A}(\vec r,t)  -\int W(\vec r-\vec r_1) \left(p_{AB}(\vec r,\vec r_2,t)+ C_A(\vec r,t) C_B(\vec r_2,t)\right) \,d^3 r_2
\\  \label{Cb}
\frac {\partial}{\partial t} C_{B}(\vec r,t) &=& D_B\Delta C_{B}(\vec r,t)  -\int W(\vec r-\vec r_1) \left(p_{AB}(\vec r_2,\vec r,t)+ C_A(\vec r_2,t) C_B(\vec r,t)\right) \,d^3 r_2
\\  \label{Cc}
\frac {\partial}{\partial t} C_{C}(\vec r,t) &=& D_C\Delta C_{C}(\vec r,t)  +\int W(\vec r-\vec r_1) \left(p_{AB}(\vec r,\vec r_2,t)+ C_A(\vec r,t) C_B(\vec r_2,t)\right) \,d^3 r_2
\\
\frac {\partial}{\partial t} C_{D}(\vec r,t) &=& D_D\Delta C_{D}(\vec r,t)  +\int W(\vec r-\vec r_1) \left(p_{AB}(\vec r_2,\vec r,t)+ C_A(\vec r_2,t) C_B(\vec r,t)\right) \,d^3 r_2
\label{Cd}
\end{eqnarray}

Equations for the pair distribution functions Eq.(\ref{DiffpCM}) are formulated below with the following notation:
 $$\vec \rho=\vec r_2-\vec r_1,\quad (1)\equiv (\vec r_1,t),\quad  (2)\equiv (\vec r_2,t), \quad (12)\equiv (\vec r_1,\vec r_2,t), \quad  (21)\equiv (\vec r_2,\vec r_1,t)$$

\begin{eqnarray}
\frac {\partial}{\partial t} p_{AB}(12)&=&(D_A \Delta_1+D_B\Delta_2) p_{AB}(12)  - W(\vec \rho) p_{AB}(12)- \underline {\underline{W(\vec \rho)C_A(1) C_B(2)}}
\nonumber\\
 &&  {\color{red}- K\, p_{AB}(12)C_B(1) -\underline{ K\, p_{AB}(12) C_A(2)}}{\color{blue} - K\, p_{AA}(12)C_B(2) - K\, p_{BB}(12) C_A(1)}
\label{eq:pAB_CMET}\\
\frac {\partial}{\partial t} p_{CD}(12)&=&(D_C \Delta_1+D_D\Delta_2) p_{CD}(12)  + W(\vec \rho) p_{AB}(12)+ \underline {\underline{ W(\vec \rho) C_A(1) C_B(2)} }
\nonumber\\
 && {\color{red}+ K\, p_{AD}(12)C_B(1)+ K\, p_{BC}(21) C_A(2) } {\color{blue} + K\, p_{BD}(12) C_A(1) + K\, p_{AC}(21)C_B(2)}
\label{eq:pCD_CMET}\\
\frac {\partial}{\partial t} p_{AD}(12)&=&(D_A \Delta_1+D_D\Delta_2) p_{AD}(12)
\nonumber\\
 &&{\color{red} -  K\, p_{AD}(12)C_B(1) + \underline{K\, p_{AB}(12)C_A(2)}} {\color{blue}  + K\, p_{AA}(12)C_B(2)}
\label{eq:pAD_CMET}\\
\frac {\partial}{\partial t} p_{BC}(12)&=&(D_B \Delta_1+D_C \Delta_2) p_{AC}(12)
\nonumber\\
&&{\color{red}-  K\, p_{BC}(12)C_A(1)
 + K\, p_{AB}(21)C_B(2)}  {\color{blue} +  K\, p_{BB}(12)C_A(2) }
\label{eq:pBC_CMET}\\
\frac {\partial}{\partial t} p_{AA}(12)&=&\left(D_A \Delta_1+D_A\Delta_2\right)p_{AA}(12)
\nonumber\\
&& {\color{red} -  K\, p_{AA}(12)C_B(1) - K\, p_{AA}(12)C_B(2)}
  {\color{blue} - K\, p_{AB}(21)C_A(1)-\underline {K\, p_{AB}(12)C_A(2)}}
\label{eq:pAA_CMET}\\
\frac {\partial}{\partial t} p_{BB}(12)&=&(D_B \Delta_1+D_B\Delta_2)p_{BB}(12)
\nonumber\\
&& {\color{red}  -  K\, p_{BB}(12)C_A(1) - K\, p_{BB}(12)C_A(2)}
 {\color{blue} - K\, p_{AB}(12)C_B(1)-K\, p_{AB}(21)C_B(1)}
\label{eq:pBB_CMET}\\
\frac {\partial}{\partial t} p_{AC}(12)&=&(D_A \Delta_1+D_C \Delta_2) p_{AC}(12)
\nonumber\\
&& {\color{red} -  K\, p_{AC}(12)C_B(1)  +  K\, p_{AA}(12)C_B(2)}   {\color{blue} + \underline{K\, p_{AB}(12)C_A(2)}}
\label{eq:pAC_CMET}\\
\frac {\partial}{\partial t} p_{BD}(12)&=&(D_B \Delta_1+D_D\Delta_2) p_{BD}(12)
\nonumber\\
&&  {\color{red} -  K\, p_{BD}(12)C_A(1)
 + K\, p_{BB}(12)C_A(2) } {\color{blue} + K\, p_{AB}(21)C_B(2)}
\label{eq:pBD_CMET}\\
\frac {\partial}{\partial t} p_{CC}(12)&=&(D_C \Delta_1+D_C\Delta_2)p_{CC}(12)
\nonumber\\
&&  {\color{red} +  K\, p_{AC}(12)C_B(1)+ K\, p_{AC}(21)C_B(2) }
  {\color{blue}+ K\, p_{BC}(12)C_A(1) +  K\, p_{BC}(21)C_A(1)}
\label{eq:pCC_CMET}\\
\frac {\partial}{\partial t} p_{DD}(12)&=&(D_D \Delta_1+D_D\Delta_2)p_{DD}(12)
\nonumber\\
&&  {\color{red}+ K\, p_{BD}(12)C_A(1)+ K\, p_{BD}(21)C_A(2)}
 {\color{blue}+  K\, p_{AD}(12)C_B(1) + K\, p_{AD}(21)C_B(2)}
\label{eq:pDD_CMET}
\end{eqnarray}

\end{widetext}
where $K=K_{CD,AB}$ is the element of $\hat K$ matrix defined in Eq.(\ref{DefK}).  The value of $K$ in the case of an irreversible reaction $A+B\rightarrow C+D$ is equal to the steady-state rate constant $k$, determined by Eqs.(\ref{eq:Kt}-\ref{RD}) \cite{IgoshinCP1999}.

Terms appearing in both the MET and CMET but not in the IET are highlighted in red; terms appearing only in the CMET  are highlighted in blue. With no initial correlations for IET (no colored terms present), it is easy to see that only two equations, Eqs.(\ref{eq:pAB_CMET}-\ref{eq:pCD_CMET}), give non-zero solutions for pair distribution functions.
Thus, in the IET, one needs to solve only two coupled equations (\ref{eq:pAB_CMET}) and (\ref{Ca}) to obtain concentrations dynamics.
In the MET, the number of equations giving non-zero solutions for pair distribution functions increases up to four Eqs.(\ref{eq:pAB_CMET}-\ref{eq:pBC_CMET}) but still only equations (\ref{eq:pAB_CMET}) and (\ref{Ca})  are needed to compute the concentration dynamics. For CMET, all the pair correlations are non-zero, but only solutions of  (\ref{eq:pAB_CMET}), (\ref{eq:pAA_CMET}), and ((\ref{eq:pBB_CMET}) are needed to compute $p_{AB}$ which is then used in equation (\ref{Ca}). For the reversible reaction $A+B\leftrightarrow C+D$, all pair correlations need to be computed (not shown).

To understand the physical meaning of the difference refer to Fig.~\ref{fig:IET} and Fig.~\ref{fig:MET}.
Fig.~\ref{fig:IET} illustrates the processes accounted for in the IET:  Fig.~\ref{fig:IET}a depicts two uncorrelated particles $A$ and $B$ approaching each other in the bulk and the elementary act of reaction (thick black arrow). As a result of the reaction, particles $A$ and $B$ disappear, and a geminate pair of particles $C$ and $D$ appears at the same positions blue arrows show the correspondences).
The pair of particles $C$ and $D$  then diffuse away from each other.
Fig.~\ref{fig:IET}b shows how this process is described by the IET. Initially, there is no correlation between particles $A$ and $B$ (no line connecting them).
 As a result of the reaction, a positive correlation arises between particles $C$ and $D$, described by the doubly underlined term in the equation (\ref{eq:pCD_CMET}), as well as a negative correlation between particles $A$ and $B$, described by the doubly underlined term in the equation (\ref{eq:pAB_CMET}).
As follows from Eq.(\ref{eq:pAB_CMET}) without MET and CMET terms, the time evolution of the negative pair distribution function $A$ and $B$ generated as a result of the interaction of uncorrelated particles in the bulk is described by the equation of geminate kinetics.
This means that in the IET such a correlation in pairs, spreading over large distances by diffusion, exists indefinitely.
This explains the short time interval during which the IET is applicable.

\begin{figure}[h]
    \centering
    \includegraphics[width=1\linewidth]{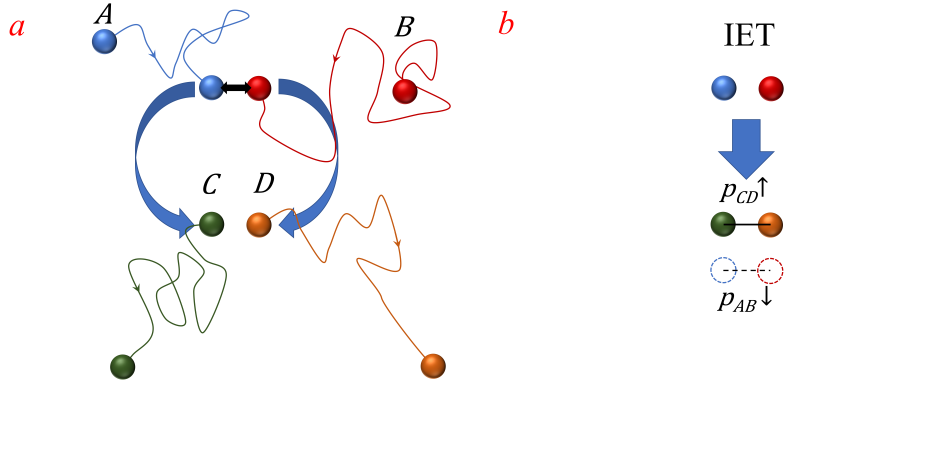}
    \caption{Illustration of the IET action for the reaction $A+B\rightarrow C+D$.
    (a) Schematic picture of the elementary act of reaction.  Two uncorrelated particles $A$ and $B$ approaching each other in the bulk.  As a result of the reaction (thick black arrow), particles $A$ and $B$ disappear, and a geminate pair of particles $C$ and $D$ appears at the same positions blue arrows show the correspondences). The pair of particles $C$ and $D$  then diffuse away from each other.
(b)  Description of this process within the IET. Initially, there is no correlation between particles $A$ and $B$ (no line connecting them).
 As a result of the reaction, a positive correlation arises between particles $C$ and $D$, described by the doubly underlined term in the equation (\ref{eq:pCD_CMET}), as well as a negative correlation between particles $A$ and $B$, described by the doubly underlined term in the equation (\ref{eq:pAB_CMET}).
    }
    \label{fig:IET}
\end{figure}

Fig.~\ref{fig:MET} illustrates the processes included in the MET and CMET.
Fig.~\ref{fig:MET}a shows a geminate pair of particles $A$ and $B$ coming apart with a particle $B$ approaching and reacting with another (uncorrelated) particle $A$ in the bulk. As a result of the reaction (blue arrows show the correspondence), particles $A$ and $B$ disappear, and a geminate pair of particles $C$ and $D$ appears at their corresponding  positions. Meanwhile, the original particle $A$ remains unchanged.
Fig.~\ref{fig:MET}b shows how this process is described by the MET and CMET.
The initial correlation between particles $A$ and $B$ is shown by the black line. For illustrative purposes, we assume a positive correlation. It is maintained when the particles separate to a distance $\rho$ significantly larger than the effective reaction radius.
In the MET, as a result of the reaction between the particle $B$ and the uncorrelated (bulk) particle $A$, a positive correlation appears at a distance $\rho$ between particle $A$ and the newly-appeared particle $D$. This is mathematically described by  the underlined  term in Eq.(\ref{eq:pAD_CMET}). Meanwhile, the existing correlation between particle $A$ and particle $B$ disappears, as described by the underlined term in Eq. (\ref{eq:pAB_CMET}).

The CMET additionally takes into account the appearance of a positive correlation at distance $\rho$ between particle $A$ and the particle $C$ located near $D$  (formally, given the point approximation (\ref{Tpoint2}), at the same positions as $D$) when the bulk reaction occurs. This effect is described by the underlined term in Eq.(\ref{eq:pAC_CMET}). The same bulk reaction leads to the appearance of a negative correlation between particle $A$ and the disappeared particle $A$, as described by the underlined term in Eq.(\ref{eq:pAA_CMET}).
\begin{figure}[ht]
    \centering
    \includegraphics[width=1\linewidth]{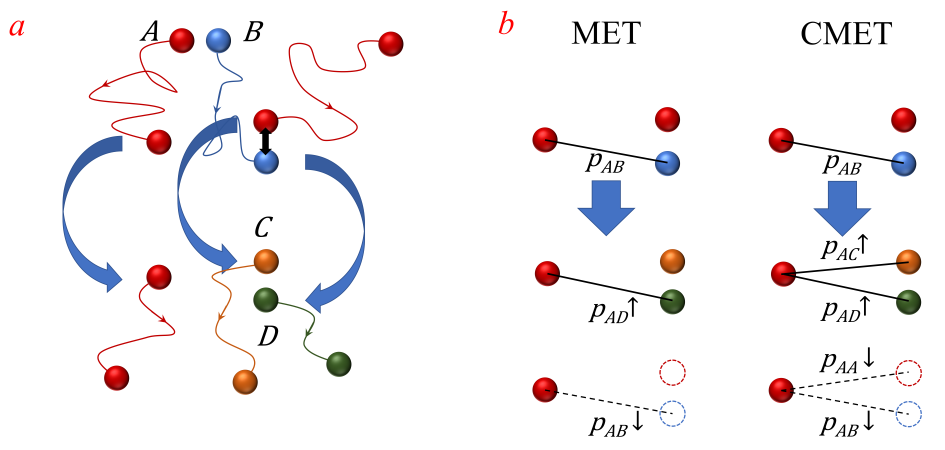}
    \caption{{Illustration of the MET and the CMET action for the reaction $A+B\rightarrow C+D$.
  (a) Schematic representation of an elementary act of reaction of particle B, which forms a geminate pair with particle A, with another (uncorrelated) particle A in the bulk.  (b) Description of this process within the MET and CMET.    }}
    \label{fig:MET}
\end{figure}

 Now,  based on this, we can interpret the underlying physical basis of the theories:

 \begin{enumerate}
     \item IET formulates the correlation dynamics of the given pair of reactants, taking into account their diffusion, force-chemical interactions, and interconversion of pair correlations due to monomolecular reactions. No reactions of particles in a given pair with other particles are included.
     \item However, even in dilute solutions, the probability of conversion of a given particle to a different one due to bimolecular reaction in bulk can be comparable to the internal conversion and, therefore, needs to be taken into account. MET accounts for the contributions of these processes onto correlations, focusing only on the direct interconversion of the particles in the reacting pair, i.e., for a pair dynamic for particles $A_i$ and $A_j$, we need to account for their effective conversion $i\to l$ and $j\to n$  due to reactions via $\widehat V_{ij, ln}$ with bulk particles, which would result in correlation interconversion.
     \item CMET accounts for the fact that, as a result of these bulk reactions, another molecule appeared or disappeared in the vicinity of the original one, and thus another correlation interconversion should be accounted for.
 \end{enumerate}

 For the full schematics of the processes contributing to pair-correlation dynamics/interconversion in IET, MET, and CMET in the general case, the reader is referred to Fig. 1 in the Supplementary Materials.

The critical importance of the additional terms in the CMET is shown in Fig.~\ref{fig:comparison}, where the modeling results of the kinetics of the reaction $A + B \leftrightarrow C + D$ within the IET, MET, and CMET
{as well as the kinetics of the reaction  $A + B \leftrightarrow D + C$ within the MET} are compared  for a spatially homogeneous case.

From Fig.~\ref{fig:comparison}, we conclude that the IET shows a $t^{-3/2}$ dependence of the relaxation function at long times, but with incorrect amplitude. The CMET, on the other hand, as we already demonstrated in Fig. \ref{fig:GopichSzabo}, correctly reproduces the exact asymptotics Eq.(\ref{eq:GS1}).
{ The MET leads to exponential kinetics.  It is noteworthy that the MET gives different kinetics for the reactions $A+B\leftrightarrow\ C+D$   and $A+B\leftrightarrow\ D+C$.  This is due to the fact that MET is not invariant with respect to the permutation of reaction products, unlike IET and CMET. The non-invariance of MET with respect to the permutation of $C$ and $D$ can be seen, for example, in equations (\ref{eq:pAB_CMET}–\ref{eq:pDD_CMET}). }
\begin{figure}
    \centering
    \includegraphics[width=1\linewidth]{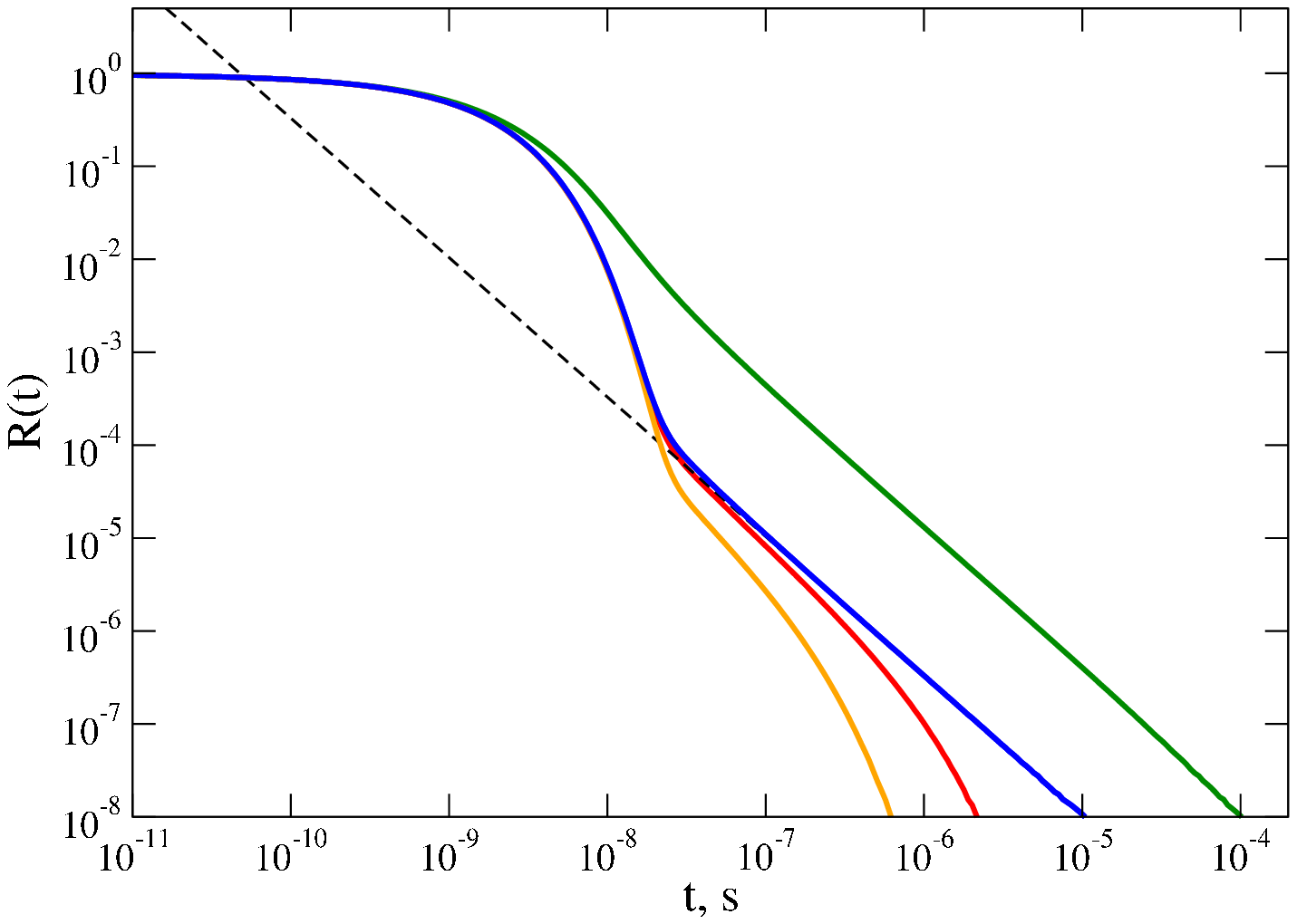}
    \caption{{ Time dependence of the relaxation function obtained by
     IET (green line), MET (red line) and CMET (blue line)  for $A + B \leftrightarrow C + D$ reaction and MET (orange line) for $A + B \leftrightarrow D + C$ reaction. }
     The dotted line represents the Gopich - Szabo asymptote given by Eq. (\ref{eq:GS1}). The parameters of the reaction are: $C_A(0) = 0.01\ \text{M}$, $C_B(0) = C_D(0) = 0.1\ \text{M}$, $C_C(0)=0$, $D_A=D_B=D_C=D_D = 5\times 10^{-6}\ \text{cm}^2/\text{s}$, the distance dependent reaction {rates  are} given by Eq.(\ref{Cont}), where $W_{\delta} = 10^{11}\ \text{\AA}/\text{s}$.  The equilibrium constant is  $K_e = 1$.}
    \label{fig:comparison}
\end{figure}

{
It is important to discuss the relations between CMET and self-consistent relaxation time approximation (SCRTA), suggested in Ref.  \onlinecite{GopichSzaboJCP2002}.  Detailed description of the SCRTA equations was only made for elementary reversible reactions  $A+B\leftrightarrow\ C+D$  and $A+B\leftrightarrow\ C$. Gopich and Szabo \cite{GopichSzaboJCP2002} also suggested (without derivations) a generalization of the SCRTA kinetic equations for a multistage reactive system. The structure of these generalized equations is formally the same as CMET equations (\ref{eq:MasterCi}-\ref{S_CMET}) for spatially homogeneous concentrations, with two important limitations:
\begin{enumerate}
\item 	SCRTA equations are formulated only for contact reactions, whereas CMET can consider reactions at a distance;
\item
	It is suggested in Ref. 95 to use the self-consistent or steady-state reaction rates for the calculation of the ${\hat{K}}_0\left(t\right)$ matrix which is an analog of our $\hat{S}\left(t\right)$ matrix Eq. (\ref{S_CMET}).  But the definition of the steady-state reaction rates as well as the self-consistent reaction rates for the multistage reactive system is not clear. The problems of the steady-state and the self-consistent reaction rates definition are discussed in the Supplementary Note VII using the example of the bulk electron transfer reaction and consequent geminate recombination. In contrast to SCRTA, CMET can be readily applied for this example with well-defined elements of the  $\hat{K}$ matrix Eq. (\ref{DefK}).
\end{enumerate}
}


\section{Summary}

This work provides the first systematic derivation of kinetic equations for general multistage reaction networks using the second-quantization formalism of quantum field theory. The resulting complete modified encounter theory framework addresses the limitations of the integral encounter theory and modified encounter theory without increasing their computational complexity. Formulated as a set of coupled differential equations, the complete modified encounter theory provides a versatile, computationally efficient approach to the first-principles kinetic modeling of arbitrary second-order reactive systems in liquid solutions.
As such, it can be used for kinetic modeling of chemical systems with coupled first and second-order reactions beyond the assumption of pairwise interactions.

{The diverse case studies presented in Section \ref{sec:PartSys} demonstrate that CMET provides a robust, unified framework capable of encompassing the full spectrum of kinetic regimes in multistage diffusion-influenced reactions. By rigorously accounting for necessary pair correlations, the theory successfully captures the transition from initial geminate recombination to non-stationary and stationary bulk kinetics. Crucially, CMET resolves the inaccuracies of previous encounter theories by accurately reproducing the long-time fluctuation asymptotics that arise in both reversible and irreversible systems.}

\begin{acknowledgments}
R.V.L. and P.A.F. (Voevodsky Institute of Chemical Kinetics and Combustion SB RAS) acknowledge the core funding from the Russian Federal Ministry of Science and Higher Education {(FWGF-2026-0007)}. O.A.I. acknowledges funding support from the Welch Foundation (Grant C-1995)  and from  the Center for Theoretical Biological Physics sponsored by the National Science Foundation (PHY-2019745)
. E.B.K.\ would like to thank the Alexander von Humboldt Foundation, the Weizmann Institute of Science, and the Fritz Haber Center of the Hebrew University of Jerusalem for their support during 1997--2000, when the present study was conceived.
\end{acknowledgments}


\bibliography{refs}

@book{Landau,
    author = {V. B. Berestetskii and E. M Lifshitz and L. P. Pitaevskii},
    title = {Quantum Electrodinamics},
    publisher = {Pergamon press},
    year = {1982}
}

@article{GopichSzaboJCP2002,
    author = {Irina V. Gopich and Attila Szabo},
    journal = {J. Chem. Phys.},
    year = {2002},
    volume = {117},
    pages = {507}
}

@book{Rice,
  title={Diffusion-limited Reactions},
  author={S.A. Rice},
  isbn={9780444416315},
 publisher = {\emph{Comprehensive Chemical Kinetics, v.25,\/} Elsevier, Amsterdam},
  url={https://books.google.ru/books?id=SdIQzwEACAAJ},
  year={1985}
}

@article{TauberJPA2005,
doi = {10.1088/0305-4470/38/17/R01},
url = {https://doi.org/10.1088/0305-4470/38/17/R01},
year = {2005},
month = {apr},
publisher = {},
volume = {38},
number = {17},
pages = {R79},
author = {Täuber, Uwe C and Howard, Martin and Vollmayr-Lee, Benjamin P},
title = {Applications of field-theoretic renormalization group methods to reaction–diffusion problems},
journal = {Journal of Physics A: Mathematical and General}
}

@book{Zhabotinsky,
    author = {A. M. Zhabotinsky},
    title = {Concentrational Autooscillations [in Russian]},
    publisher = {Science, Moscow},
    year = {1974},
    month = {Jule},
    day = {15}
}

@book{ehrenfest2014conceptual,
  title={The Conceptual Foundations of the Statistical Approach in Mechanics},
  author={Ehrenfest, P. and Ehrenfest, T.},
  isbn={9780486163147},
  series={Dover Books on Physics},
  url={https://books.google.ru/books?id=g0xoBQAAQBAJ},
  year={2014},
  publisher={Dover Publications}
}

@article{Murphy_2024, title={Breakdown of Boltzmann-type models for the alignment of self-propelled rods}, volume={376}, ISSN={0025-5564}, url={http://dx.doi.org/10.1016/j.mbs.2024.109266}, DOI={10.1016/j.mbs.2024.109266}, journal={Mathematical Biosciences}, publisher={Elsevier BV}, author={Murphy, Patrick and Perepelitsa, Misha and Timofeyev, Ilya and Lieber-Kotz, Matan and Islas, Brandon and Igoshin, Oleg A.}, year={2024}, month=oct, pages={109266} }

@misc{lu2025,
      title={Spectral BBGKY: a scalable scheme for nonlinear Boltzmann and correlation kinetics},
      author={Xingjian Lu and Shuzhe Shi},
      year={2025},
      eprint={2507.14243},
      archivePrefix={arXiv},
      primaryClass={nucl-th},
      url={https://arxiv.org/abs/2507.14243},
}

@article{Forster1974,
    author = {Dieter Forster},
    title = {Properties of the kinetic memory function in classical fluids},
    journal = {Physical Review A},
    year = {1974},
    volume = {9},
    number = {2},
    pages = {943-956}
}

@article{Harris_1988, title={The Nature of Simple Photodissociation Reactions in Liquids on Ultrafast Time Scales}, volume={39}, ISSN={1545-1593}, url={http://dx.doi.org/10.1146/annurev.pc.39.100188.002013}, DOI={10.1146/annurev.pc.39.100188.002013}, number={1}, journal={Annual Review of Physical Chemistry}, publisher={Annual Reviews}, author={Harris, A L and Brown, J K and Harris, C B}, year={1988}, month=oct, pages={341–366} }

@article{Zeng2018,
  title = {Adsorptive Separation of Fructose and Glucose by Metal-Organic Frameworks: Equilibrium,  Kinetic,  Thermodynamic,  and Adsorption Mechanism Studies},
  volume = {57},
  ISSN = {1520-5045},
  url = {http://dx.doi.org/10.1021/acs.iecr.8b00435},
  DOI = {10.1021/acs.iecr.8b00435},
  number = {28},
  journal = {Industrial Engineering Chemistry Research},
  publisher = {American Chemical Society (ACS)},
  author = {Zeng,  Zhouliangzi and Lyu,  Jiafei and Bai,  Peng and Guo,  Xianghai},
  year = {2018},
  month = jun,
  pages = {9200-9209}
}

@article{Ma2021,
  title = {Adsorptive capture of perrhenate (ReO4-) from simulated wastewater by cationic 2D-MOF BUC-17},
  volume = {202},
  ISSN = {0277-5387},
  url = {http://dx.doi.org/10.1016/j.poly.2021.115218},
  DOI = {10.1016/j.poly.2021.115218},
  journal = {Polyhedron},
  publisher = {Elsevier BV},
  author = {Ma,  Jing and Wang,  Chong-Chen and Zhao,  Zi-Xuan and Wang,  Peng and Li,  Jun-Jiao and Wang,  Fu-Xue},
  year = {2021},
  month = jul,
  pages = {115218}
}

@article{Klann2011,
  title = {Agent-based simulation of reactions in the crowded and structured intracellular environment: Influence of mobility and location of the reactants},
  volume = {5},
  ISSN = {1752-0509},
  url = {http://dx.doi.org/10.1186/1752-0509-5-71},
  DOI = {10.1186/1752-0509-5-71},
  number = {1},
  journal = {BMC Systems Biology},
  publisher = {Springer Science and Business Media LLC},
  author = {Klann,  Michael T and Lapin,  Alexei and Reuss,  Matthias},
  year = {2011},
  pages = {71}
}

@article{Lee2016,
  title = {Single-molecule imaging reveals modulation of cell wall synthesis dynamics in live bacterial cells},
  volume = {7},
  ISSN = {2041-1723},
  url = {http://dx.doi.org/10.1038/ncomms13170},
  DOI = {10.1038/ncomms13170},
  number = {1},
  pages={ 13170},
  journal = {Nature Communications},
  publisher = {Springer Science and Business Media LLC},
  author = {Lee,  Timothy K. and Meng,  Kevin and Shi,  Handuo and Huang,  Kerwyn Casey},
  year = {2016},
  month = oct
}

@article{Sm,
    author = {M. von Smoluchowski},
    title = {Versucheiner Mathematischen Theorie der Koagulations Kinetic Kolloider Lousungen.},
    journal = {Z. Phys. Chem.},
    pages = {129-168},
    volume = {92},
    year = {1917}
}

@article{CK,
    author = {F.C. Collins and G.E. Kimball},
    title = {Diffusion-Controlled Reaction Rates},
    journal = {J. Colloid. Sci.},
    year = {1949},
    volume = {4},
    pages = {425}
}

@article{KMR,
    author = {  S. F. Kilin and M. S. Mikhelashvili and I. M. Rozman},
    title = {Transfer of Electronic Excitation Energy in Liquid Solutions},
    journal = {Opt. Spectrosc.},
    year = {1964},
    volume = {16},
    pages = {576}
}

@article{TunBag,
    author = {N. N. Tunitsky and H. S. Bagdasaryan},
    journal = {Opt.  Spectrosc.},
    year = {1963},
    volume = {15},
    pages = {303}
}

@article{WF,
  title = {General theory of diffusion-controlled reactions},
  volume = {58},
  ISSN = {1089-7690},
  url = {http://dx.doi.org/10.1063/1.1679757},
  DOI = {10.1063/1.1679757},
  number = {9},
  journal = {J. Chem. Phys.},
  publisher = {AIP Publishing},
  author = {Gerald Wilemski   and Marshall Fixman  },
  year = {1973},
  month = may,
  pages = {4009–4019}
}

@article{AllBlum,
  title = {On the direct energy transfer to moving acceptors},
  volume = {72},
  ISSN = {1089-7690},
  url = {http://dx.doi.org/10.1063/1.439703},
  DOI = {10.1063/1.439703},
  number = {8},
  journal = {J. Chem. Phys.},
  publisher = {AIP Publishing},
  author = {K. Allinger   and  A. Blumen },
  year = {1980},
  month = apr,
  pages = {4608–4619}
}

@article{Szabo,
  title = {Theory of diffusion-influenced fluorescence quenching},
  volume = {93},
  ISSN = {1541-5740},
  url = {http://dx.doi.org/10.1021/j100356a011},
  DOI = {10.1021/j100356a011},
  number = {19},
  journal = {J.  Phys. Chem.},
  publisher = {American Chemical Society (ACS)},
  author = {Attila Szabo  },
  year = {1989},
  month = sep,
  pages = {6929–6939}
}

@article{Monch,
    author = {L. Monchick and J. L. Magee and A. H. Samuel},
    journal = {J. Chem. Phys.},
    year = {1957},
    volume = {26},
    pages = {935}
}

@article{Waite,
    author = {T. R. Waite},
    journal = {Phys. Rev.},
    year = {1957},
    volume = {107},
    pages = {463}
}

@article{Kapral,
    author = {R. Kapral},
    journal = {Adv. Chem. Phys.},
    year = {1981},
    volume = {48},
    pages = {71}
}

@article{KarplusJCP1997,
    author = {S. Lee and M. Karplus},
    journal = {J. Chem. Phys.},
    year = {1987},
    volume = {86},
    pages = {1883}
}

@article{AgSzabo,
    author = {N. Agmon and A. Szabo},
    journal = {J. Chem. Phys.},
    year = {1990},
    volume = {92},
    pages = {5270}
}

@article{MolKaiz,
    author = {A. Molski and J. Kaizer},
    journal = {J. Chem. Phys.},
    year = {1992},
    volume = {96},
    pages = {1391}
}

@book{KK,
    author = {E. Kotomin and V. Kuzovkov},
    title = {Modern aspects of diffusion-controlled reactions.
                   Cooperative phenommena in bimolecular processes},
    publisher = {\emph{Comprehensive Chemical Kinetics, v.34,\/}, Elsevier, Amsterdam},
    year = {1996}
}

@article{IgoshinPhysicaA1999,
    author = {A. A. Kipriyanov and O. A. Igoshin and A. B. Doktorov},
    journal = {Physica A},
    year = {1999},
    volume = {268},
    pages = {567}
}

@article{IgoshinCP1999,
    author = {O. A. Igoshin and A. A. Kipriyanov  and A. B. Doktorov},
    journal = {Chem. Phys.},
    year = {1999},
    volume = {244},
    pages = {371-385}
}

@article{IvanovJCP2001I,
    author = { K. L. Ivanov and N. N. Lukzen and A. B. Doktorov and  A. I. Burshtein},
    journal = {J. Chem. Phys.},
    year = {2001},
    volume = {114},
    pages = {1754}
}

@article{IvanovJCP2001II,
    author = { K. L. Ivanov and N. N. Lukzen and A. B. Doktorov and  A. I. Burshtein},
    journal = {J. Chem. Phys.},
    year = {2001},
    volume = {114},
    pages = {1763}
}

@article{IvanovJCP2001III,
    author = { K. L. Ivanov and N. N. Lukzen and A. B. Doktorov and  A. I. Burshtein},
    journal = {J. Chem. Phys.},
    year = {2001},
    volume = {114},
    pages = {5682}
}

@book{Balescu,
    author = {R. Balescu},
    title = {Equilibrium and Non-equilibrium Statistical Mechanics},
    publisher = {Wiley \& Sons, New York},
    year = {1975}
}

@article{SakunPhysicaA1975,
    author = {V. P. Sakun},
    journal = {Physica A},
    year = {1975},
    volume = {80},
    pages = {128}
}

@article{DoiJPA1976,
    author = {M. Doi},
    journal = {J. Phys. A: Math. Gen.},
    year = {1976},
    volume = {9},
    pages = {1465}
}

@article{DoiJPA1976s,
    author = {M. Doi},
    journal = {J. Phys. A: Math. Gen.},
    year = {1976},
    volume = {9},
    pages = {1479}
}

@article{DoktorovPhysicaA1978,
    author = {A. B. Doktorov},
    journal = {Physica A},
    year = {1978},
    volume = {90},
    pages = {109}
}

@article{ZwanzigJCP1981,
    author = {M. Bixon and R. Zwanzig},
    journal = {J. Chem. Phys.},
    year = {1981},
    volume = {75},
    pages = {2354}
}

@article{GopichJCP1996,
    author = {I. V. Gopich and A. B. Doktorov},
    journal = {J. Chem. Phys.},
    year = {1996},
    volume = {105},
    pages = {2320}
}

@article{GopichPhysicaA1998,
    author = {A. A. Kipriyanov and I. V. Gopich and A. B. Doktorov},
    journal = {Physica A},
    year = {1998},
    volume = {255},
    pages = {347}
}

@article{YLS,
    author = {M. Yang and S. Lee and K. J. Shin},
    journal = {J. Chem. Phys.},
    year = {1998},
    volume = {108},
    pages = {117}
}

@article{YLS2,
    author = {M. Yang and S. Lee and K. J. Shin},
    journal = {J. Chem. Phys.},
    year = {1998},
    volume = {108},
    pages = {8557}
}

@article{YLS3,
    author = {M. Yang and S. Lee and K. J. Shin},
    journal = {J. Chem. Phys.},
    year = {1998},
    volume = {108},
    pages = {9069}
}

@article{GopichJCP1999,
    author = {I. V. Gopich and A. A. Kipriyanov and A. B. Doktorov},
    journal = {J. Chem. Phys.},
    year = {1999},
    volume = {110},
    pages = {22}
}

@article{LukzenCP1986,
    author = {N.N. Lukzen and A. B. Doktorov and A. I. Burshtein},
    journal = {Chem. Phys.},
    year = {1986},
    volume = {102},
    pages = {149}
}

@article{KipriyanovCP1994,
    author = {A. A. Kipriyanov and I. V. Gopich and A. B. Doktorov},
    journal = {Chem. Phys.},
    year = {1994},
    volume = {187},
    pages = {241}
}

@article{LukzenJCP1995,
    author = {A. I. Burshtein and N. N. Lukzen},
    journal = {J. Chem. Phys.},
    year = {1995},
    volume = {103},
    pages = {9631}
}

@book{Lifshitz,
    author = {I. M. Lifshitz and S. A. Gredescool and L. A. Pastur},
    title = {The introduction to the theory of disordered systems},
    publisher = {Willey \& Sons, New York},
    year = {1988}
}

@article{Elliot,
    author = {R. J. Elliot and J. A. Krumhansl and P. L. Leath},
    journal = {Rev. Mod. Phys.},
    year = {1974},
    volume = {46},
    pages = {465}
}

@book{Ziman,
    author = {J. M. Ziman},
    title = {Models of disorder},
    publisher = {Cambridge University, Cambridge},
    year = {1979}
}

@article{Kirkwd,
    author = {J. G. J. Kirkwood},
    journal = {Chem. Phys.},
    year = {1935},
    volume = {3},
    pages = {300}
}

@article{KipriyanovCP1995,
    author = {A. A. Kipriyanov and I. V. Gopich and A. B. Doktorov},
    journal = {Chem. Phys.},
    year = {1995},
    volume = {191},
    pages = {101}
}

@article{FrantsuzovCPL2000,
    author = {P. A. Frantsuzov and O. A. Igoshin and E. B. Krissinel},
    journal = {Chem. Phys. Lett.},
    year = {2000},
    volume = {317},
    pages = {481}
}

@article{YashinJSP1985,
    author = {A. S. Mikhailov and V. V. Yashin},
    journal = {J. Stat. Phys.},
    year = {1985},
    volume = {38},
    pages = {347}
}

@article{CardyJSP1995,
    author = {B. P. Lee and  J. Cardy},
    journal = {J. Stat. Phys.},
    year = {1995},
    volume = {80},
    pages = {971}}

@book{Mattuck,
    author = {R. D. Mattuck},
    title = {A guide to Feynman diagrams in the many-body problem},
    publisher = {Dover, New York},
    year = {1992}
}

@article{ZeldovichCP1978,
    author = {A. A. Ovchinnikov and Ya. B. Zheldovich},
    journal = {Chem. Phys.},
    year = {1978},
    volume = {28},
    pages = {215}
}

@article{WilczekJCP83,
    author = {D. Toussant and F. Wilczek},
    journal = {J. Chem. Phys.},
    year = {1983},
    volume = {78},
    pages = {2642}
}

@article{KrissinelSSDP,
    author = {Evgenii B. Krissinel' and Noam Agmon},
    journal = {J. Comput. Chem.},
    year = {1996},
    volume = {17(9)},
    pages = {1085}
}

@article{BalagurovVaks,
    author = {B. Ya. Balagurov and V. G. Vaks},
    journal = {Zh. Eksp. Teor. Fiz.},
    year = {1973},
    volume = {65},
    pages = {1939}
}

@article{LeeCardyJSP1995-er,
    author = {Benjamin P. Lee and John Cardy},
    journal = {J. Stat. Phys.},
    year = {1995},
    volume = {80},
    pages = {971}
}

@article{DorfmanFayer,
    author = {R. C. Dorfman and M. D. Fayer},
    journal = {J. Chem. Phys.},
    year = {1992},
    volume = {96},
    pages = {7410}
}

@article{GopichSzaboCP2002,
    author = {Irina V. Gopich and Attila Szabo},
    journal = {Chem. Phys.},
    year = {2002},
    volume = {284},
    pages = {91}
}

@article{GopichSzaboJPCB2018,
    author = {Irina V. Gopich and Attila Szabo},
    journal = {J. Phys. Chem. B},
    year = {2018},
    volume = {122},
    pages = {11338 - 11354}
}

@article{BurshteinACP2004,
    author = {A. I. Burshtein},
    journal = {Adv. Chem. Phys.},
    year = {2004},
    volume = {129},
    pages = {105}
}

@article{BurshteinCPL1992,
    author = {A. I. Burshtein},
    journal = {Chem. Phys. Let.},
    year = {1992},
    volume = {194},
    pages = {247}
}

@article{FrantsuzovJCP1997s,
    author = {A. I. Burshtein and P. A. Frantsuzov},
    journal = {J. Chem. Phys.},
    year = {1997},
    volume = {107},
    pages = {2872-2880}
}

@article{FrantsuzovJCP1997,
    author = {A. I. Burshtein and P. A. Frantsuzov},
    journal = {J. Chem. Phys.},
    year = {1997},
    volume = {106},
    pages = {3948-3955}
}

@article{FrantsuzovJL1998,
    author = {A. I. Burshtein and P. A. Frantsuzov},
    journal = {J. Lumin.},
    year = {1998},
    volume = {78},
    pages = {33-52}
}

@article{FrantsuzovCPL1998,
    author = {A. I. Burshtein I. V. Gopih and P. A. Frantsuzov},
    journal = {Chem. Phys. Lett. },
    year = {1998},
    volume = {298},
    pages = {60-66}
}

@article{FrantsuzovJCP1998,
    author = {P. A. Frantsuzov and A. I. Burshtein },
    journal = {J. Chem. Phys.},
    year = {1998},
    volume = {109},
    pages = {5957-5962}
}

@article{ZeldovichJETP1978,
    author = { Ya. B. Zheldovich and A. A. Ovchinnikov},
    journal = {Sov. Phys. JETP},
    year = {1978},
    volume = {47},
    pages = {829-834}
}

@article{NaumannJCP1993,
    author = {W. Naumann and N.V. Shokhirev and A. Szabo},
    journal = {J. Chem. Phys.},
    year = {1993},
    volume = {98},
    pages = {2353-2365}
}

@article{RudavetsJPA1993,
    author = {M. Doi},
    journal = {J. Phys. A: Math. Gen.},
    year = {1993},
    volume = {26},
    pages = {5313-5337}
}

@article{GlasserRMP1998,
    author = {Daniel C. Mattis and M. Lawrence Glasser},
    journal = {Rev. Mod. Phys.},
    year = {1998},
    volume = {70},
    pages = {979-1001}
}

@article{HanaiJSP2023,
    author = {Yuji Hirono and Ryo Hanai},
    journal = {J. Stat. Phys.},
    year = {2023},
    volume = {190},
    pages = {86}
}

@article{LucivjanskyTMP2011,
    author = {M. Hnatich and J. Honkonen  and T. Lu\v{c}ivjansk\'{y}},
    journal = {Theor. Math. Phys.},
    year = {2011},
    volume = {169},
    pages = {1481–1488}
}

@article{LucivjanskyTMP2011s,
    author = {M. Hnatich and J. Honkonen  and T. T. Lu\v{c}ivjansk\'{y}},
    journal = {Theor. Math. Phys.},
    year = {2011},
    volume = {169},
    pages = {1489–1498}
}

@article{LeeJCP1999I,
    author = {Jaeyoung Sung and Sangyoub Lee},
    journal = {J. Chem. Phys.},
    year = {1999},
    volume = {111},
    pages = {796}
}

@article{LeeJCP1999II,
    author = {Jaeyoung Sung and Sangyoub Lee},
    journal = {J. Chem. Phys.},
    year = {1999},
    volume = {111},
    pages = {804}
}

@article{LeeJCP1999III,
    author = {Jaeyoung Sung and Sangyoub Lee},
    journal = {J. Chem. Phys.},
    year = {1999},
    volume = {111},
    pages = {10159}
}

@article{LeeJCP2000,
    author = {Jaeyoung Sung and Sangyoub Lee},
    journal = {J. Chem. Phys.},
    year = {2000},
    volume = {112},
    pages = {2128}
}

@article{DoktorovRRDCP2012,
    author = {A. B. Doktorov},
    journal = {Recent Res. Devel. Chem. Physics},
    year = {2012},
    volume = {6},
    pages = {135–192}
}

@article{BensonJCompPhys2014,
title = {Connecting the dots: Semi-analytical and random walk numerical solutions of the diffusion–reaction equation with stochastic initial conditions},
journal = {Journal of Computational Physics},
volume = {263},
pages = {91-112},
year = {2014},
issn = {0021-9991},
doi = {https://doi.org/10.1016/j.jcp.2014.01.020},
url = {https://www.sciencedirect.com/science/article/pii/S0021999114000473},
author = {Amir Paster and Diogo Bolster and David A. Benson}
}

@article{ProcacciaJCP1975,
  title = {Rotational relaxation: An analytic solution of the master equation with applications to HCl},
  volume = {63},
  ISSN = {1089-7690},
  url = {http://dx.doi.org/10.1063/1.431720},
  DOI = {10.1063/1.431720},
  number = {7},
  journal = {The Journal of Chemical Physics},
  publisher = {AIP Publishing},
  author = {Procaccia,  I. and Shimoni,  Y. and Levine,  R. D.},
  year = {1975},
  month = oct,
  pages = {3181–3182}
}

@article{DoktorovJCP2010,
    author = {A. B. Doktorov and A. A. Kipriyanov and A. A. Kipriyanov },
    journal = {J. Chem.Phys.},
    year = {2010},
    volume = {132},
    pages = {204502}
}

@article{KipriyanovJCP2010,
    author = {A. A. Kipriyanov and A. A. Kipriyanov and A. B. Doktorov},
    journal = {J. Chem.Phys.},
    year = {2010},
    volume = {133},
    pages = {174508}
}

@misc{HempelArXiv2024,
      title={The Potential of Geminate Pairs in Lead Halide Perovskite revealed via Time-resolved Photoluminescence},
      author={Hannes Hempel and Martin Stolterfoht and Orestis Karalis and Thomas Unold},
      year={2024},
      eprint={2409.06382},
      archivePrefix={arXiv},
      primaryClass={cond-mat.mtrl-sci},
      url={https://arxiv.org/abs/2409.06382}
}

@article{GulbinasAEM2017,
    author = {Ram\={u}nas Augulis and  Marius Franckevi\v{c}ius and Vytautas Abramavi\v{c}ius and Darius Abramavi\v{c}ius and Shaik Mohammed Zakeeruddin ang Michael Gr\"{a}tzel and Vidmantas Gulbinas},
    journal = {Adv. Energy Mater},
    year = {2017},
    volume = {1},
    pages = {1700405}
}

@article{GoldsmithJPCC2017,
author = {Manger, Lydia H. and Rowley, Matthew B. and Fu, Yongping and Foote, Alexander K. and Rea, Morgan T. and Wood, Sharla L. and Jin, Song and Wright, John C. and Goldsmith, Randall H.},
title = {Global Analysis of Perovskite Photophysics Reveals Importance of Geminate Pathways},
journal = {The Journal of Physical Chemistry C},
volume = {121},
number = {2},
pages = {1062-1071},
year = {2017},
doi = {10.1021/acs.jpcc.6b11547},
URL = {        https://doi.org/10.1021/acs.jpcc.6b11547 },


}

@article{KosterNatureComm2017,
    author = {Davide Bartesaghi and Irene del Carmen P\'{e}rez and Juliane Kniepert and  Steffen Roland and Mathieu Turbiez and Dieter Neher and L. Jan Anton Koster},
    journal = {Nat. Commun.},
    year = {2015},
    volume = {6},
    pages = {7083}
}

@book{Seeger,
  title={Semiconductor physics: An Introduction},
  author={K. Seeger},
  isbn={ 978-3-642-06023-6},
 publisher = {Springer-Verlag Berlin Heidelberg GmbH},
  year={2004}
}

@article{DebyeTES1942,
author = {P. Debye},
title = {Reaction rates in ionic solutions},
journal = {Trans. Electrochem. Soc.},
volume = {82},
pages = { 265—272},
year = {1942},
doi = {10.1149/1.3071413}

}

@book{Salikhov,
  title={Spin Polarization and Magnetic Effects in Radical Reactions},
  author={K. Salikhovand Y. Molin and R. Sagdeev and A. Buchachenko},
 publisher = {Elsevier Science, Amsterdam},
  year={1984}
}

@article{GrzybowskiiACIEE2010,
author = {S. Soh and  M. Byrska and K. Kandere-Grzybowska and B. A. Grzybowski},
title = {Reaction-diffusion systems in intracellular molecular transport and control},
journal = {Angew. Chem. Int. Ed. Engl.},
volume = {49},
pages = {4170–4198},
year = {2010},
doi = {10.1002/anie.200905513}

}

@article{KolmogorovBMU1937,
author = {A. Kolmogorov and I. Petrovskii and N. Piskunov},
title = {A study of the diffusion equation with increase in the amount of substance, and its application to a biological problem},
journal = {Bull. Moscow Univ., Math. Mech.},
volume = {1},
pages = {1–25},
year = {1937}

}

@article{FisherAE1937,
author = {R. A. Fisher},
title = {The Wave of Advance of Advantageous Genes},
journal = {Ann.  Eug.},
volume = {7},
pages = {355},
year = {1937}

}

@article{ZeldovichAP1938,
author = {Y. B. Zeldovich and D. A. Frank-Kamenetsky},
title = {The Wave of Advance of Advantageous Genes},
journal = {Acta Physicochim.},
volume = {9},
pages = {341},
year = {1938}

}

@article{TuringPTRSL1952,
author = {A. M. Turing },
title = {The chemical basis of morphogenesis},
journal = {Philos. Trans. R. Soc. B },
volume = {237},
pages = {37-72},
year = {1952}

}

@article{AgmonJCP2003,
    author = {A. V. Popov and N. Agmon},
    journal = {J. Chem.Phys.},
    year = {2003},
    volume = {118},
    pages = {11057}
}

@article{PopovJCP2003,
    author = { N. Agmon and A. V. Popov},
    journal = {J. Chem.Phys.},
    year = {2003},
    volume = {119},
    pages = {6680}
}

@article{KopelmanPRE2004,
    author = { E. Monson and R. Kopelman},
    journal = {Phys. Rev. E},
    year = {2004},
    volume = {69},
    pages = {021103}
}

\end{document}



\title{Quantum field theory approach for multi-stage chemical kinetics in liquids.\\ Supplementary materials}

\author{Roman~V.~Li$^a$,~ Oleg~A.~Igoshin$^b$,~Evgeny~B.~Krissinel$^c$, ~ Pavel~A.~Frantsuzov$^a$}
\address{$^a$Voevodsky Institute of Chemical Kinetics and Combustion, Novosibirsk, 630090, Russian Federation\\
$^b$Department of Bioengineering, Rice University, Houston, TX 77005, USA\\
$^c$Science and Technology Facilities Council,  Swindon, SN2 1SZ, United Kingdom}

\date{\today }

\vspace{12pt}

\maketitle

\section*{Supplementary Note I.   Diagram construction}

The concentration series Eq. (4.15) in the main article

\begin{equation}
\label{CSeries}
C_i(t)=C^{(0)}_i(t)+C^{(1)}_i(t)+C^{(2)}_i(t)+C^{(3)}_i(t)+\cdots
\end{equation}
 is obtained by the concentration expansion
of the following $T$-exponent (Eq.(4.8) in the main article)

\begin{equation}
\label{eq:CAsAverage}
C_i(\vec r,t) = \,\, \bra{0}\,{\cal T}\exp\left(\sum_{j=1}^n
\int \Psi_j^\bullet(\vec r_f,t)\,d^3\vec r_f\right) \Psi_i^\bullet(\vec r,t)
\exp\left(\int_0^t\widehat{\cal V}^\bullet(\tau)d\tau\right)
\prod_{k=1}^N \left( \int f_k(\vec r_0)\Psi^{\dag\bullet}_k(\vec r_0,0)\,
 d^3\vec r_0\right)^{\a_k}\ket{0}
\end{equation}

 We will first consider the lowest-order term of the expansion $C^{(0)}_i(\vec r,t)$, which corresponds to
the zero-order term of the $T$-exponent series,
which is $1$. Substituting $1$ for
$\exp\left(\int_0^t\widehat{\cal V}^\bullet(\tau)d\tau\right)$ in Eq.(\ref{eq:CAsAverage}), we obtain
\begin{equation}
\label{C00}
C^{(0)}_i(\vec r,t) = \bra{0}\, {\cal T}
 \exp\left(\sum_{k=1}^N\int \Psi_k^\bullet(\vec r_f,t) d^3r_f\right)
 \Psi_i^\bullet(\vec r,t)
 \prod_{j=1}^N \left( \int f_j(\vec r_0)\Psi^{\dagger\bullet}_j(\vec r_0,0)\,
 d^3r_0\right)^{\alpha_j}\ket{0}
\end{equation}
This expression can be expanded into the sum of expectation values
$\bra{0}\,{\cal T}\prod_i \Psi_i^\bullet(\vec r_i,t_i)
\prod_j\Psi^{\dagger\bullet}_j(\vec r_j,t_j)\ket{0}$, each of which can be reduced
to simpler terms $\bra{0}\,{\cal T}\Psi_i^\bullet\Psi^\bullet_j\ket{0}$,
$\bra{0}\,{\cal T}\Psi_i^\bullet\Psi^{\dagger\bullet}_j\ket{0}$ and
$\bra{0}\,{\cal T}\Psi_i^{\dagger\bullet}\Psi^{\dagger\bullet}_j\ket{0}$ with the help of Wick's theorem \cite{Landau}:\\
{\it The vacuum expectation value of the product of any number of operators $\Psi^\bullet$ and $\Psi^{\dagger\bullet}$ is equal to the sum of the products of all possible expectation values of these operators taken in pairs. In each pair, the factors must be placed in the same order as in the original product.}

As was mentioned in Section VI in the main article,
\begin{eqnarray}
\label{eq:ZeroPair}
\bra{0}\,{\cal T}\Psi_i^\bullet(\vec r_1,t_1)\Psi^{\bullet}_k(\vec r_2,t_2)\ket{0} &=&
  \bra{0}\,{\cal T}\Psi_i^{\dagger\bullet}(\vec r_1,t_1)
    \Psi^{\dagger\bullet}_k(\vec r_2,t_2)\ket{0} = 0\\
\label{eq:GFPair}
\bra{0}\,{\cal T} \Psi_i^\bullet(\vec r_1,t_1)
  \Psi^{\dagger\bullet}_k(\vec r_2,t_2)\ket{0} &=&
  G^0_{ik}(\vec r_1,t_1,\vec r_2,t_2)
\end{eqnarray}
Making use of the Wick's theorem and Eqs. (\ref{eq:ZeroPair}) and (\ref{eq:GFPair}),
one can represent $C^{(0)}_i(\vec r,t)$, Eq.(\ref{C00}), as well as all other
$C^{(k)}_i(\vec r,t)$ from Eq.(\ref{eq:CAsAverage}), in terms of Green functions.
This technique is exemplified in the following:
$$ \bra{0}\,{\cal T} \Psi_i^\bullet(\vec r_1,t_1)
\Psi_j^\bullet(\vec r_2,t_2)\Psi^{\dagger\bullet}_k(\vec r_3,t_3)
\Psi^{\dagger\bullet}_l(\vec r_4,t_4)\ket{0}=$$
$$\bra{0}\,{\cal T} \Psi_i^\bullet(\vec r_1,t_1)\Psi^{\dagger\bullet}_k
(\vec r_3,t_3)\ket{0} \bra{0}\,{\cal T} \Psi_j^\bullet(\vec r_2,t_2)
\Psi^{\dagger\bullet}_l(\vec r_4,t_4)\ket{0}+\bra{0}\,{\cal T}\Psi_i^\bullet(\vec r_1,t_1)\Psi^{\dagger\bullet}_l
(\vec r_4,t_4)\ket{0}
\bra{0}\,{\cal T} \Psi_j^\bullet(\vec r_2,t_2)\Psi^{\dagger\bullet}_k(\vec r_3,t_3)\ket{0}=$$
$$G^0_{ik}(\vec r_1,t_1,\vec r_3,t_3)G^0_{jl}(\vec r_2,t_2,\vec r_4,t_4)+
G^0_{il}(\vec r_1,t_1,\vec r_4,t_4)G^0_{jk}(\vec r_2,t_2,\vec r_3,t_3)$$

It follows from Wick's theorem that if an expression contains an unequal number of creation and annihilation
operators, its vacuum expectation is zero. Eq.(\ref{C00}) contains $M=\sum_i \alpha_i$ creation operators, i.e., the initial number of particles in the system. Only the $(M-1)$-th term of the exponent expansion in Eq.(\ref{C00})
contains the same number of annihilation operators. Eq.(\ref{C00}) can therefore be rewritten as follows:
\begin{equation}
\label{C0}
 C^0_i(\vec r,t)= \bra{0}\, {\cal T} \frac 1 {(M-1)!}
 \left(\sum_{k=1}^N\int \Psi_k^\bullet(\vec r_f,t) d^3r_f\right)^{M-1}
 \Psi_i^\bullet(\vec r,t)
 \prod_{j=1}^N \left( \int f_j(\vec r_0)\Psi^{\dagger\bullet}_j(\vec r_0,0)
 \, d^3r_0\right)^{\alpha_j}\ket{0}
\end{equation}

Eq.(\ref{C0}) contains a vacuum average of the product of $M$ creation and $M$ annihilation operators. For this
vacuum average,  Wick's theorem gives the sum of $M!$ terms (all possible permutations of pairs). Each term is a
product of $M$ Green's functions corresponding to a given $\bra{0}\,T \Psi^\bullet\Psi^{\dagger\bullet}\ket{0}$ pair. All of
them can be represented in a graphical form
$$
\begin{picture}(7,8)
\multiput(0.5,0.5)(0,0.5){2}{\circle*{0.1}} \put(0.5,7.5){\circle*{0.1}}
\multiput(0.5,1.5)(0,0.25){3}{\circle*{0.03}}\multiput(0.5,2.5)(0,0.5){2}{\circle*{0.1}}

 \multiput(0.5,3.5)(0,0.25){3}{\circle*{0.03}}
\put(0.5,4.5){\circle*{0.1}}
 \multiput(0.5,5)(0,0.25){3}{\circle*{0.03}}
\put(0.5,6){\circle*{0.1}}
 \multiput(0.5,6.5)(0,0.25){3}{\circle*{0.03}}

  \put(-0.2,1.5){\makebox(0,0)[r]{$\Psi^{\dagger\bullet}_1$}}

  \put(0.1,0.5){\line(0,1){2}} \put(0.1,3){\line(0,1){1.5}}
\put(-0.2,3.5){\makebox(0,0)[r]{$\Psi^{\dagger\bullet}_2$}}

 \put(0.1,6){\line(0,1){1.5}}
\put(-0.2,6.5){\makebox(0,0)[r]{$\Psi^{\dagger\bullet}_N$}}

\put(5.5,0.5){\circle*{0.1}}\multiput(5.5,1)(0,0.5){2}{\circle{0.1}}\multiput(5.5,2)(0,0.25){7}{\circle*{0.03}}
\put(5.5,7.5){\circle{0.1}} \put(5.5,4){\circle{0.1}}

\put(5.5,4.5){\circle*{0.03}} \put(5.5,5){\circle{0.1}} \multiput(5.5,6)(0,0.25){5}{\circle*{0.03}}

\put(0.5,2.5) {\line(5,-2){5}} \put(0.5,0.5) {\line(5,1){5}} \put(0.5,1) {\line(1,0){5}}
  \put(0.5,3) {\line(5,1){5}} \put(0.5,4.5) {\line(5,3){5}} \put(0.5,6) {\line(5,-1){5}}
\put(0.5,7.5) {\line(5,-2){5}}  \put(5.5,5.5){\circle{0.1}}

\put(6.5,0.5){\makebox(0,0)[l]{$\Psi_i^\bullet(r,t)$}}

 \put(6.5,1.5){\makebox(0,0)[l]{$\Psi_\ast^\bullet(\ast,t)$}}
\end{picture}$$
The filled circles on the left correspond to the initial creation operators $\psi$ multiplied by the initial distributions $f_j(r_0)$. Initial coordinate $r_0$ integrations are assumed. For each kind of
particles $A_j$, we have $\alpha_j$ creation operators (circles). Therefore, the total number of circles to
the left is $M$. The only filled circle on the bottom right corresponds to the annihilation operator
$\Psi_i^\bullet(\vec r,t)$. The remaining empty circles to the right correspond to annihilation operators summed over their indices and integrated over their spatial variable. We represent those indices and variables
as stars.

Applying  Wick's theorem to Eq.(\ref{C0}) results in a sum of $M!$ terms. Each term is a product of $M$
expectation values of the form Eq.(\ref{eq:GFPair}), i.e. Green's function. We choose to represent each  Green's
function as a line that connects the left and right circles of the diagram. After summation and integration, the contribution
from each line connecting a filled circle on the left and an empty circle on the right gives the following factor
\begin{equation}
\sum_k\int\int G_{kj}(\vec r_f,t,\vec r_0,0)f_j(\vec r_0)\,d^3r_f\,d^3r_0=1
\end{equation}
The latter equality  follows from the following property of the Green's function (Eqs.(4.14) in the main article)
\begin{equation}
\label{G0sum}
\sum_i \int G^0_{ik}(\vec r,t,\vec r^{\,\prime},t') \,d^3\vec r^{\,\prime} = 1, \quad
 \mbox{   for    } t\ge t'
\end{equation}

 Therefore, the only non-unit factor in the product corresponds to the term represented by the line connecting the filled circle to the right with a
filled circle to the left.  As a result, one can drop all lines ending with an empty circle  and obtain diagram:
$$\begin{picture}(2,1) \put(2,0.5){\circle*{0.1}} \put(0,0.5){\line(1,0){2}} \put(0,0.5){\circle*{0.1}}
\put(1.9,0.7){\makebox(0,0)[b]{$i$}}\put(0.1,0.7){\makebox(0,0)[b]{$j$}}
\put(2.0,0.1){\makebox(0,0)[b]{$rt$}} \put(0,0.1){\makebox(0,0)[b]{$r_00$}}
\end{picture}$$
We chose this line to be a graphical representation of Green's function $G^0_{ij}(\vec r,t,\vec r_0,0)$ according to
Eq.(\ref{eq:GFPair}).

For each of the left circles, there are $(M-1)!$ identical contributions, corresponding to all possible
permutations of other pairs of circles. As such, the contribution of this term to Eq.(\ref{C0}) is
\begin{equation}
\label{f_j}
 \int G^0_{ij}(\vec r,t,\vec r_0,0) f_j(\vec r_0)\,d^3r_0
\end{equation}
There is $\alpha_j$ of these terms for each $j$. Summing all the contributions, we obtain
\begin{equation}
C^0_i(\vec r,t)=\sum_j \int G^0_{ij}(\vec r,t,\vec r_0,0)
 \alpha_j f_j(\vec r_0)\,d^3r_0
\end{equation}
Using the  following notation for the initial concentration
\begin{equation}
 C_j(\vec r,0)=\alpha_j f_j(\vec r)
 \label{eq:Cinf}
 \end{equation}
 we have
\begin{equation}
C^0_i(\vec r,t)=\sum_j \int G^0_{ij}(\vec r,t,\vec r_0,0) C_j(\vec r_0,0)\,d^3r_0
\end{equation}
Or in diagram notation (see Eq.(4.17)in the main article)
$$\begin{picture}(2,1) \put(2,0.5){\circle*{0.1}} \put(0.2,0.5){\line(1,0){1.8}} \put(0.5,0.5){\circle*{0.1}}
\put(0.0,0.3){\line(1,1){0.4}} \put(0.0,0.7){\line(1,-1){0.4}} \put(0.2,0.5){\circle{0.2}}
\put(1.9,0.7){\makebox(0,0)[b]{$i$}}
\put(2.0,0.1){\makebox(0,0)[b]{$rt$}}
\end{picture}$$

Calculating the contribution of the first-order (in interaction) term of Eq.(\ref{eq:CAsAverage}) can be done in a similar fashion. Replacing the $T$-exponent in Eq.(\ref{eq:CAsAverage}) by its first order expansion term according to
Eq.(4.4), we obtain
$$C^1_i(\vec r,t)=\bra{\Upsilon}\, {\cal T}\Psi^\bullet_i(\vec r,t)
\int_0^t \widehat{ \cal V}^\bullet(t_1) \,dt_1\ket{\Phi^\bullet(0)} = $$
$$\bra{0}\,{\cal T} \frac 1 {(M-1)!}(\sum_{n=1}^N\int \Psi^\bullet_n(\vec
r_f,t) d^3r_f)^{M-1} \Psi^\bullet_i(\vec r,t)$$
$$\frac 1 2  \sum_{kjlm} \int_0^t\,dt_1 \int \int \Psi^{\dagger\bullet}_k(\vec r_1,t_1)
\Psi^{\dagger\bullet}_j(\vec r_1^{\,\prime},t_1)\widehat V_{kj,lm}(\vec r_1,\vec r_1^{\,\prime}) \Psi^\bullet_l(\vec
r_1,t_1)\Psi^\bullet_m(\vec r_1^{\,\prime},t_1)\,d^3r_1\,d^3r_1'$$
\begin{equation}
\prod_{s=1}^N ( \int f_s(\vec r_0)\Psi^{\dagger\bullet}_s(\vec r_0,0)\, d^3r_0)^{\alpha_s}\ket{0}
\end{equation}

As in the previous case, we can represent all the operators as circles in a graph

$$
\begin{picture}(7,8)
\multiput(0.5,0.5)(0,0.5){2}{\circle*{0.1}} \put(0.5,7.5){\circle*{0.1}}
\multiput(0.5,1.5)(0,0.25){3}{\circle*{0.03}}\multiput(0.5,2.5)(0,0.5){2}{\circle*{0.1}}

 \multiput(0.5,3.5)(0,0.25){3}{\circle*{0.03}}
\put(0.5,4.5){\circle*{0.1}}
 \multiput(0.5,5)(0,0.25){3}{\circle*{0.03}}
\put(0.5,6){\circle*{0.1}}
 \multiput(0.5,6.5)(0,0.25){3}{\circle*{0.03}}

  \put(-0.2,1.5){\makebox(0,0)[r]{$\Psi^{\dagger\bullet}_1$}}

  \put(0.1,0.5){\line(0,1){2}} \put(0.1,3){\line(0,1){1.5}}
\put(-0.2,3.5){\makebox(0,0)[r]{$\Psi^{\dagger\bullet}_2$}}

 \put(0.1,6){\line(0,1){1.5}}
\put(-0.2,6.5){\makebox(0,0)[r]{$\Psi^{\dagger\bullet}_N$}}

\put(5.5,0.5){\circle*{0.1}}\multiput(5.5,1)(0,0.5){5}{\circle{0.1}}\multiput(5.5,3.5)(0,0.25){15}{\circle*{0.03}}
\put(5.5,7.5){\circle{0.1}}

\put(3,2){\circle*{0.1}}  \put(3,3){\circle*{0.1}} \multiput(3,3)(0,-0.125){8}{\line(0,-1){0.1}}
\put(3,1.65){\makebox(0,0)[b]{$r_1\,t_1$}} \put(3,3.2){\makebox(0,0)[b]{$r_1',t_1$}}

\put(6.5,0.5){\makebox(0,0)[l]{$\Psi_i^\bullet(r,t)$}}

 \put(6.5,1.5){\makebox(0,0)[l]{$\Psi_\ast^\bullet(\ast,t)$}}
\end{picture}$$

Two circles connected by a dashed line  correspond to a pair of operators $\Psi^{\dagger\bullet}_k(\vec
r_1,t_1) \Psi^{\bullet}_l(\vec r_1,t_1)$ and $\Psi^{\dagger\bullet}_j(\vec r_1^{\,\prime},t_1) \Psi^{\bullet}_m(\vec
r_1^{\,\prime},t_1)$ in the interaction operator.

 Application of Wick's theorem gives a sum of all the terms, that are represented by the set of lines connecting circles in the graph. Each of the circles $r_1$ and
$r_1'$ has to be connected to one circle on the right side and one circle on the left side of the graph. We can once again drop all lines connecting the filled circle on the left to the empty circle on the right.
The remaining terms correspond to three distinct topologies. In the first, both $r_1$ and $r_1'$ circles are connected to
 empty circles on the right side.

$$
\begin{picture}(6,2.5)
\multiput(0.5,0.5)(0,0.5){2}{\circle*{0.1}} \put(0.5,2){\circle*{0.1}}

\multiput(5.5,1)(0,1){2}{\circle{0.1}} \put(5.5,0.5){\circle*{0.1}}

\put(0.5,0.5){\line(1,0){5}} \put(0.5,1){\line(1,0){5}} \put(0.5,2){\line(1,0){5}}

\put(0.2,0.5){\makebox(0,0)[r]{$\Psi^{\dagger\bullet}_m$}}\put(0.2,1){\makebox(0,0)[r]{$\Psi^{\dagger\bullet}_s$}}
\put(0.2,2){\makebox(0,0)[r]{$\Psi^{\dagger\bullet}_{s'}$}} \put(6,0.5){\makebox(0,0)[l]{$\Psi_i^\bullet$}}

 \put(3,1){\circle*{0.1}}  \put(3,2){\circle*{0.1}}
\multiput(3,2)(0,-0.125){8}{\line(0,-1){0.1}} \put(3,0.65){\makebox(0,0)[b]{$r_1\,t_1$}}
\put(3,2.2){\makebox(0,0)[b]{$r_1',t_1$}}
\end{picture}$$

The corresponding term has the following form
$$\sum_{m} \int \bra{0}\,T\Psi^\bullet_i(\vec r,t)\Psi^{\dagger\bullet}_m(\vec r_0,0)\ket{0}\,d^3r_0$$
$$\times \frac 1 2 \sum_{nn'}\sum_{kjlm}\int_0^t\, dt_1 \int\int\int\int\int\int
\bra{0}\,T\Psi^\bullet_n(\vec r_f,t)\Psi^{\dagger\bullet}_k(\vec r_1,t_1)\ket{0}\bra{0}\,T
\Psi^\bullet_{n'}(\vec r_f^{\,\prime},t)\Psi^{\dagger\bullet}_j(\vec r_1^{\,\prime},t_1)\ket{0}$$
$$\cross\widehat V_{kj,lm}(\vec r_1,\vec r_1^{\,\prime}) \bra{0}\,T\Psi^\bullet_m(\vec r_1^{\,\prime},t_1)
\Psi^{\dagger\bullet}_{s'}(\vec r_0^{\,\prime},0)\ket{0}$$
$$\cross\bra{0}\,T\Psi^\bullet_l(\vec r_1,t_1)
\Psi^{\dagger\bullet}_s(\vec r_0,0)\ket{0} f_s(\vec r_0)f_{s'}(\vec r_0^{\,\prime})
\,d^3 r_f \,d^3r_f'\,d^3r_1\,d^3r_1'\,d^3r_0\,d^3r_0'$$
The first factor in this expression is equivalent to Eq.(\ref{f_j}), the second can be written in the other form:

$$\frac 1 2 \sum_{nn'}\sum_{kjlm} \int_0^t\,dt_1 \int \int \int \int \int \int G^0_{nk}(\vec r_f,t,\vec r_1,t_1)
G^0_{n'j}(\vec r_f^{\,\prime},t,\vec r_1^{\,\prime},t_1)$$
$$\cross \widehat V_{kj,lm}(\vec r_1,\vec r_1^{\,\prime})
 G^0_{ls}(\vec r_1,t_1,\vec r_0,0) G^0_{ms'}(\vec r_1^{\,\prime},t_1,\vec r_0^{\,\prime},0)
 f_{s}(\vec r_0,0)f_{s'}(\vec r_0^{\,\prime},0) d^3r_f\,d^3r_f'\,d^3r_1\,d^3r_1'\,d^3r_0\,d^3r_0$$

Using Eq.(\ref{G0sum}), it can be recast as

$$\frac 1 2 \int_0^t\,dt_1 \sum_{kjlm} \int \int \int \int \widehat V_{kj,lm}(\vec r_1,\vec r_1^{\,\prime})
 G^0_{ls}(\vec r_1,t_1,\vec r_0,0) G^0_{ms'}(\vec r_1^{\,\prime},t_1,\vec r_0^{\,\prime},0)
 f_{s}(\vec r_0,0)f_{s'}(\vec r_0^{\,\prime},0) \,d^3r_1\,d^3r_1'\,d^3r_0\,d^3r_0$$
This expression is equal to $0$ because of Eq.(3.11) in the main article
\begin{equation}
 \sum_{kj} \int \int \widehat V_{kj,lm}(\vec
r_1,\vec r_1^{\,\prime})f(\vec r_1,\vec r_1^{\,\prime}) \,d^3r_1\,d^3r_1'=0
 \label{eq:Vint}
\end{equation}

In the second topology, the circle  $(\vec r_1,t_1)$ is connected to  $(\vec r,t)$

$$
\begin{picture}(6,2)
\multiput(0.5,0.5)(0,1){2}{\circle*{0.1}}

\put(5.5,1.5){\circle{0.1}} \put(5.5,0.5){\circle*{0.1}}

\put(0.5,0.5){\line(1,0){5}} \put(0.5,1.5){\line(1,0){5}}

\put(0.2,0.5){\makebox(0,0)[r]{$\Psi^{\dagger\bullet}_s$}}
\put(0.2,1.5){\makebox(0,0)[r]{$\Psi^{\dagger\bullet}_{s'}$}}
\put(6,0.5){\makebox(0,0)[l]{$\Psi_i^\bullet(r,t)$}}

 \put(3,0.5){\circle*{0.1}}  \put(3,1.5){\circle*{0.1}}
\multiput(3,1.5)(0,-0.125){8}{\line(0,-1){0.1}} \put(3,0.15){\makebox(0,0)[b]{$r_1\,t_1$}}
\put(3,1.7){\makebox(0,0)[b]{$r_1',t$}}
\end{picture}$$

There are $M-1$ identical contributions, due to all possible empty circles. Each of them should
be multiplied by $(M-2)!$ due to all possible permutations of the other circle pairs. Thus, we can skip a
factor $1/(M-1)!$ in the corresponding expression

$$\frac 1 2 \sum_{n}\sum_{kjlm} \int_0^t\,dt_1 \int \int \int \int \int  G^0_{ik}(\vec r,t,\vec r_1,t_1)
G^0_{nj}(\vec r_f,t,\vec r_1^{\,\prime},t_1)$$
\begin{equation}
 \cross\widehat V_{kj,lm}(\vec r_1,\vec r_1^{\,\prime})
 G^0_{ls}(\vec r_1,t_1,\vec r_0,0) G^0_{ms'}(\vec r_1^{\,\prime},t_1,\vec r_0^{\,\prime},0)
 f_{s}(\vec r_0,0)f_{s'}(\vec r_0^{\,\prime},0) d^3r_f\,d^3r_1\,d^3r_1'\,d^3r_0\,d^3r_0
 \label{eq:Sec}
 \end{equation}

The topology of the third kind contains the connection between circle $(\vec r_1^{\,\prime},t_1)$  and $(r,t)$

$$
\begin{picture}(6,2)
\multiput(0.5,0.5)(0,1){2}{\circle*{0.1}}

\put(5.5,1.5){\circle{0.1}} \put(5.5,0.5){\circle*{0.1}}

\put(0.5,0.5){\line(1,0){5}} \put(0.5,1.5){\line(1,0){5}}

\put(0.2,0.5){\makebox(0,0)[r]{$\Psi^{\dagger\bullet}_{s'}$}}
\put(0.2,1.5){\makebox(0,0)[r]{$\Psi^{\dagger\bullet}_{s}$}}
\put(6,0.5){\makebox(0,0)[l]{$\Psi_i^\bullet(r,t)$}}

 \put(3,0.5){\circle*{0.1}}  \put(3,1.5){\circle*{0.1}}
\multiput(3,1.5)(0,-0.125){8}{\line(0,-1){0.1}} \put(3,0.15){\makebox(0,0)[b]{$r_1'\,t_1$}}
\put(3,1.7){\makebox(0,0)[b]{$r_1,t$}}
\end{picture}$$
This graph is topologically equivalent to the previous one. The corresponding term is equal to Eq.(\ref{eq:Sec}). The sum of two terms is twice as large. Using Eq.(\ref{G0sum}), we can skip one Green's function together with an
integration and summation, rewriting Eq. (\ref{eq:Sec}) as

$$ \sum_{kjlm} \int_0^t\,dt_1
\int \int \int \int  G^0_{ik}(\vec r,t,\vec r_1,t_1) \widehat V_{kj,lm}(\vec r_1,\vec r_1^{\,\prime}) G^0_{ls}(\vec
r_1,t_1,\vec r_0,0)$$
\begin{equation}
\cross G^0_{ms'}(\vec r_1^{\,\prime},t_1,\vec r_0^{\,\prime},0)f_s(\vec r_0)f_{s'}(\vec r_0^{\,\prime}) d^3r_1\,d^3r_1'\,d^3r_0\,d^3r_0'
\end{equation}

This corresponds to omitting the line connecting to the empty circle in the graph. For each pair of  indexes
$s$ and  $s'$ ( $s\neq s'$), there are $\alpha_s\alpha_{s'}$ equal terms in the sum. If  $s=s'$, there are
$\alpha_s(\alpha_{s}-1)$.  Collecting together all the terms and using Eq.(\ref{eq:Cinf}), we finally
get

$$C^{(1)}_i(\vec r,t)= \sum_{ss'} \sum_{kjlm} \int_0^t\,dt'
\int \int \int \int \int G^0_{ik}(\vec r,t,\vec r_1,t_1)
 \widehat V_{kj,lm}(\vec r_1,\vec r_1^{\,\prime})$$
\begin{equation}
\times G^0_{ls}(\vec r_1,t_1,\vec r_0,0) G^0_{ms'}(\vec r_1^{\,\prime},t_1,\vec r_0^{\,\prime},0)C_s(\vec r_0,0) C_{s'}(\vec
r_0^{\,\prime},0)\, d^3r_1\,d^3r_1'\,d^3r_0\,d^3r_0'
\end{equation}
where  we neglect the difference between $\alpha_s(\alpha_{s}-1)$ and $\alpha_s^2$
 in the thermodynamic limit $\alpha_s\to\infty$, $f_s\to 0$ when $C_s=\operatorname{const}$.
 This formula can be represented by the following diagram

$$\begin{picture}(2,1.5) \put(0.0,0.8){\line(1,1){0.4}} \put(0.0,1.2){\line(1,-1){0.4}}
\put(0.0,-0.2){\line(1,1){0.4}} \put(0.0,0.2){\line(1,-1){0.4}} \put(0.2,1){\circle{0.2}}
\put(0.2,0){\circle{0.2}} \put(0.2,0){\line(1,0){1.3}} \put(0.2,1){\line(1,0){1.0}}
\multiput(0.5,1)(0.5,0){2}{\circle*{0.1}} \multiput(0.5,0)(0.5,0){3}{\circle*{0.1}}
\multiput(1,1)(0,-0.125){8}{\line(0,-1){0.09}} \put(1.4,0.2){\makebox(0,0)[b]{$i$}}

\put(1.5,-0.4){\makebox(0,0)[b]{$rt$}}
\end{picture}$$

In a similar fashion, we can develop diagram representation for the $n$-th order term in the expansion of $T$-exponent in Eq.(\ref{eq:CAsAverage})

$$C^{(n)}_i(\vec r,t)=\bra{\Upsilon}\, {\cal T} \Psi^\bullet_i(\vec r,t)
\int_0^{t_2}\,dt_1 \int_0^{t_3}\,dt_2\cdots \int_0^t\,dt_n \widehat{ \cal V}^\bullet(t_n)\cdots\widehat{ \cal
V}^\bullet(t_2)\widehat{ \cal V}^\bullet(t_1) \ket{\Phi^\bullet(0)}$$
 The corresponding graph has the form

$$
\begin{picture}(9,8)
\multiput(0.5,0.5)(0,0.5){2}{\circle*{0.1}} \put(0.5,7.5){\circle*{0.1}}
\multiput(0.5,1.5)(0,0.25){3}{\circle*{0.03}}\multiput(0.5,2.5)(0,0.5){2}{\circle*{0.1}}

 \multiput(0.5,3.5)(0,0.25){3}{\circle*{0.03}}
\put(0.5,4.5){\circle*{0.1}}
 \multiput(0.5,5)(0,0.25){3}{\circle*{0.03}}
\put(0.5,6){\circle*{0.1}}
 \multiput(0.5,6.5)(0,0.25){3}{\circle*{0.03}}

  \put(-0.2,1.5){\makebox(0,0)[r]{$\Psi^{\dagger\bullet}_1$}}

  \put(0.1,0.5){\line(0,1){2}} \put(0.1,3){\line(0,1){1.5}}
\put(-0.2,3.5){\makebox(0,0)[r]{$\Psi^{\dagger\bullet}_2$}}

 \put(0.1,6){\line(0,1){1.5}}
\put(-0.2,6.5){\makebox(0,0)[r]{$\Psi^{\dagger\bullet}_N$}}

\put(8.5,0.5){\circle*{0.1}}\multiput(8.5,1)(0,0.5){5}{\circle{0.1}}\multiput(8.5,3.5)(0,0.25){15}{\circle*{0.03}}
\put(8.5,7.5){\circle{0.1}}

\put(2,3){\circle*{0.1}}  \put(2,3.8){\circle*{0.1}} \multiput(2,3.8)(0,-0.125){7}{\line(0,-1){0.1}}
\put(2,2.5){\makebox(0,0)[b]{$r_1\,t_1$}} \put(2,4.2){\makebox(0,0)[b]{$r_1',t_1$}}

\put(4,3){\circle*{0.1}}  \put(4,3.8){\circle*{0.1}} \multiput(4,3.8)(0,-0.125){7}{\line(0,-1){0.1}}
\put(4,2.5){\makebox(0,0)[b]{$r_2\,t_2$}} \put(4,4.2){\makebox(0,0)[b]{$r_2',t_2$}}

\multiput(5,3.4)(0.5,0){3}{\circle*{0.03}}

\put(7,3){\circle*{0.1}}  \put(7,3.8){\circle*{0.1}} \multiput(7,3.8)(0,-0.125){7}{\line(0,-1){0.1}}
\put(7,2.5){\makebox(0,0)[b]{$r_n\,t_n$}} \put(7,4.2){\makebox(0,0)[b]{$r_n',t_n$}}

\put(9.5,0.5){\makebox(0,0)[l]{$\Psi_i^\bullet(r,t)$}}

 \put(9.5,1.5){\makebox(0,0)[l]{$\Psi_\ast^\bullet(\ast,t)$}}
\end{picture}$$

Analyzing all possible connections in the same fashion according to Wick's theorem, we obtain the sum of
topologically distinct connected diagrams. The first few of them are given in Eq.(4.18) in the main article.

\section*{Supplementary Note II. Third order concentration diagram contribution estimation }
\label{supplementary:B}
The contribution of diagram (6.1) in the main article

\begin{equation}
\label{eq:DivergingDiagram}
\begin{picture}(2,1.5)
\put(0,0.8){\line(1,1){0.4}} \put(0.,1.2){\line(1,-1){0.4}}
\put(0.,-0.2){\line(1,1){0.4}} \put(0,0.2){\line(1,-1){0.4}}
\put(0.2,0){\line(1,0){0.3}} \put(0.2,1){\line(1,0){0.3}}
\put(0.5,1){\circle*{0.1}} \put(1.5,1){\circle*{0.1}}
\put(0.5,0.){\circle*{0.1}} \put(1.5,0.){\circle*{0.1}}
\put(0.5,1.0){\line(1,0){1.2}} \put(0.5,0.0){\line(1,0){1.2}}
\put(1,1){\circle*{0.1}} \put(0.5,1.3){\line(1,1){0.4}}
\put(0.5,1.7){\line(1,-1){0.4}} \put(0.7,1.5){\line(1,0){0.6}}
\put(1.,1.5){\circle*{0.1}}
\multiput(0.5,1)(0,-0.125){8}{\line(0,-1){0.09}}
\multiput(1,1.5)(0,-0.125){4}{\line(0,-1){0.09}}
\multiput(1.5,1)(0,-0.125){8}{\line(0,-1){0.09}}
\put(1.7,0.2){\makebox(0,0)[b]{$i$}}
\put(1.5,-0.4){\makebox(0,0)[b]{$rt$}}\end{picture}
\end{equation}
\mbox{}\\

is given by the following expression
  $$R^{(d)}_i(\vec r,t)=
\sum_{kk'ss'i'}\sum_{rr'll'mm'}\int_0^t dt_1 \int_0^{t_1}\,dt_2
\int \,d^3r_2\,d^3 r_3\, d^3r_4\,d^3 r_5d^3r_6\,
\widehat V_{ik,i'k'}(\vec r,\vec r_2) G^0_{i'l'}(\vec r,t,\vec r_3,t_1)$$
\begin{equation}
\times G^0_{k's}(\vec r_2,t,\vec r_4,t_2) V_{ss',rr'}(\vec r_4,\vec r_5) C_{r}(\vec r_5,t_2)
G^0_{r'm'}(\vec r_4,t_2,\vec r_6,t_1) \widehat V_{l'm',lm}(\vec r_3,\vec r_6)C_l(\vec r_3,t_1) C_m(\vec r_6,t_1)
\end{equation}
To simplify the estimation, we assume that $\widehat {\bf Q}=0$ and that all concentrations are homogeneous.
We also assume that the concentrations do not vary over time. Under these assumptions, integration over $\vec r_5$ is easy to perform
  $$R^{(d)}_i(\vec r,t)=
\sum_{kk'ss'i'}\sum_{rr'll'mm'}\int_0^t dt_1 \int_0^{t_1}\,dt_2
\int \,d^3r_2\,d^3 r_3\, d^3r_4\,d^3 r_5d^3r_6\,
\widehat V_{ik,i'k'}(\vec r,\vec r_2) G^0_{i'l'}(\vec r,t,\vec r_3,t_1)$$
\begin{equation}
\times G^0_{k's}(\vec r_2,t,\vec r_4,t_2) k^0_{ss',rr'} C_{r}
G^0_{r'm'}(\vec r_4,t_2,\vec r_6,t_1)  \widehat V_{l'm',lm}(\vec r_3,\vec r_6)C_l C_m
\end{equation}
where
$$k^0_{ss',rr'}=\int d^3r'\widehat V_{ss',rr'}(\vec r,\vec r')$$

The analytic expression for Green's function in case $Q_{ij}=0$ is given by

\begin{equation}
 G^0_{ik}(\vec r,t,\vec r_1,t_1)=\frac {\delta_{ik}}{(4\pi D_i(t-t_1))^{3/2}}\exp \left(-\frac {(\vec r-\vec r_1)^2}{4 D_i (t-t_1)}\right)
\end{equation}
For $t\to \infty$ particles that diffuse by a distance $(Dt)^{1/2}$ which is much bigger than reaction zone size.
Therefore, one can assume that the chemical reaction occurs only at contact:
\[
V_{ss',rr'}(\vec r,\vec r')=k^0_{ss',rr'}\delta(\vec r-\vec r')
\]
As a result
$$R^{(d)}_i(\vec r,t)\approx \sum_{kk'si'}\sum_{rr'lm}\int_0^t dt_1
\int_0^{t_1}\,dt_2 \int \,d^3 r_3\int \, d^3r_4\, k^0_{ik,i'k'}
G^0_{i'i'}(\vec r,t,\vec r_3,t_1)$$
\begin{equation}
 \times G^0_{k'k'}(\vec r,t,\vec r_4,t_2) k^0_{k's,rr'} C_{r}
G^0_{r'r'}(\vec r_4,t_2,\vec r_3,t_1)   k^0_{i'r',lm}C_l C_m
\label{IntPoint}
\end{equation}
The convolution of Green's function results is of the form
$$ \int \,d^3 r_3 \int \, d^3r_4\, G^0_{i'i'}(\vec r,t,\vec r_3,t_1)
G^0_{k'k'}(\vec r^{\,\prime},t,\vec r_4,t_2) G^0_{r'r'}(\vec r_4,t_2,\vec
r_3,t_1) =\frac {1}{(2\pi \Delta)^{3/2}}\exp \left(-\frac {(\vec
r-\vec r\, ')^2}{2 \Delta}\right)$$ where
$$\Delta=2 D_{i'}(t-t_1)+2 D_{k'}(t-t_2)+2 D_{r'}(t_2-t_1)$$
Using this result in Eq.(\ref{IntPoint}) with $\vec r=\vec r'$, we obtain
$$R^{(d)}_i(\vec r,t)\approx
\sum_{kk'si'}\sum_{rr'lm}
k^0_{ik,i'k'} k^0_{k's,rr'} C_{r}
 k^0_{i'r',lm}C_l C_m$$
\begin{equation}
\times \int_0^t dt_1 \int_0^{t_1}\,dt_2\, \frac 1 {[4\pi (D_{i'}(t-t_1)+D_{k'}(t-t_2)+D_{r'}(t_2-t_1)]^{3/2}}
\label{IntDiag}
\end{equation}
When $t\to\infty$, this integral diverges as $\sqrt{t}$.

\section*{Supplementary Note III. Green's function order in $\alpha$}
\label{supplementary:C}

The Green's function obeys the following equation (Eq.(4.12) in the main article)

\begin{equation}
    \label{G02}
    \pdv{}{t}\hat {G}^0(\vec r, t,\vec r^{\,\prime},t') = \hat D{\Delta \hat G}^0(\vec r, t,\vec r^{\,\prime},t') + \hat Q \hat {G}^0(\vec r, t,\vec r^{\,\prime},t')
\end{equation}
with the initial condition
$$\hat G^0(\vec r, t,\vec r^{\,\prime},t') = \hat I \delta(\vec r-\vec r^{\,\prime})$$
It is easy to conclude that the Green's function depends only on the time and coordinate differences: $\hat G^0(\vec r, t,\vec r^{\,\prime},t') = \hat G^0(\vec r-\vec r^{\,\prime},t-t')$. This function  obeys
\begin{equation}
    \label{G0}
    \pdv{}{t}\hat G^0(\vec r, t) = \hat D\Delta \hat G^0(\vec r, t) + \hat Q \hat G^0(\vec r, t)
\end{equation}
with the initial condition
$$\hat G^0(\vec r, 0) = \hat I \delta(\vec r)$$

Substituting $\vec r=\alpha^{-1}\vec {\underline r}$  and  $t=\alpha^{-2}\underline t$  into $\hat G^0$, we define:
$$\underline {\hat  G}^0(\underline {\vec r}, \underline t)=\hat G^0(\alpha^{-1}\vec {\underline r}, \alpha^{-2}{\underline t})$$
$\underline {\hat  G}^0(\underline {\vec r}, \underline t)$ obeys the following equation:
\begin{equation}
    \label{G0su}
     \pdv{}{\underline t} \underline {\hat G}^0(\vec {\underline r}, {\underline t}) = \hat D\underline \Delta \underline{\hat G}^0(\vec {\underline r}, {\underline t}) + \alpha^{-2}\hat Q {\hat G}^0(\vec {\underline r}, {\underline t})
\end{equation}
 with the initial condition
 $$\underline {\hat G}^0(\vec {\underline r}, 0) =\alpha^3 \hat I \delta(\vec {\underline r})$$

Fourier transform of $\underline {\hat G}^0(\vec {\underline r}, {\underline t})$ , defined as
$$\underline {\hat G}^0(\vec {\underline k}, {\underline t})= \int \underline {\hat G}^0(\vec {\underline r}, {\underline t})\exp(i\vec {\underline k}\vec {\underline r}) \,d^3\vec {\underline r},$$
obeys the equation
\begin{equation}
    \label{G0k}
    \pdv{}{\underline t}\underline {\hat G}^0(\vec {\underline k}, {\underline t}) = -\hat D |\vec {\underline k}|^2 \underline {\hat G}^0(\vec {\underline k}, {\underline t}) +  \alpha^{-2} \hat Q\underline {\hat G}^0(\vec {\underline k}, {\underline t})
\end{equation}
with the initial condition
$$\underline {\hat G}^0(\vec {\underline k}, 0) = \alpha^3 \hat I$$
The solution of Eq.(\ref{G0k}) can be expressed as follows:
\begin{equation}
\label{scaledsol}
    \underline {\hat G}^0(\vec {\underline k}, {\underline t}) =  \alpha^3 \exp(-\hat D |\vec {\underline k}|^2 {\underline t} + \alpha^{-2}\hat Q{\underline t})
\end{equation}

Obtaining the inverse Fourier transform of this expression for an arbitrary $\alpha$ value is cumbersome, because $\hat D$ and $\hat Q$ may not commute.
In the $\alpha\rightarrow 0$ limit the expression is dominated by the second term in the exponent.
Matrix $\hat Q$ can be diagonalized
$$\hat Q= \sum_{i=1}^N \mathbf {p}_i  Q_i \mathbf {l}_i$$
where $Q_i$ are the eigenvalues of $\hat Q$,   $\mathbf {p}_i$ and $ \mathbf {l}_i$ are  the right (column) and left (row) eigenvectors, defined as
$$ \hat Q \mathbf {p}_i =Q_i \mathbf {p}_i, \qquad    \mathbf {l}_i \hat Q =Q_i \mathbf {l}_i, \qquad \mathbf {l}_i \mathbf {p}_k = \delta_{ik}$$
 The exponent of $\hat Q$ can be expressed as
\begin{equation}
\label{exp}
    \exp(\alpha^{-2}\hat Q {\underline t}) = \sum\limits_{i=1}^N \mathbf {p}_i \exp(\alpha^{-2} Q_i {\underline t}) \mathbf {l}_i
\end{equation}
The matrix $\hat Q$ has at least one zero eigenvalue. Let's denote number of zero eigenvalues as $N_0\le N$.
Let's eigenvectors with zero eigenvalue have indexes between $1$ and  $ N_0$, and  eigenvectors with negative eigenvalues have indexes  large than $N_0$.
At  $\alpha\rightarrow 0$
\begin{equation}
\label{projector}
 \exp(\alpha^{-2}\hat Q {\underline t})  = \hat P +  \alpha^2 \delta({\underline t})  \hat R
\end{equation}
where $\hat P$ is the projection operator onto the subspace of vectors satisfying the condition   $\hat Q \mathbf{x} = 0$
$$\hat P=\sum\limits_{i=1}^{N_0}\mathbf {p}_i\mathbf {l}_i$$
and the operator $\hat R$ is defined as
$$ \hat R =\sum\limits_{j=N_0+1}^N\mathbf {p}_i \mathbf {l}_i \left|Q_j\right|^{-1}$$
  Thus, the asymptotic behaviour of the Eq. (\ref{scaledsol}) at  $\alpha\rightarrow 0$  is

\begin{equation}
    \label{newsol}
    \underline {\hat G}^0(\vec {\underline k}, {\underline t}) = \alpha^3 \exp\left(-\sum_{i=1}^{N_0}\sum_{k=1}^{N_0} \mathbf {p}_i D^p_{ik} \mathbf {l}_i |\vec {\underline k}|^2 {\underline t}\right) +  \alpha^5 \delta({\underline t})  \hat R
\end{equation}
where the elements of the matrix $\hat D^P$ with size $N_0\times N_0$ is defined as
 $$D^p_{ik} = \mathbf {l}_i \hat D \mathbf {p}_i$$

The inverse Fourier transformation of Eq.(\ref{newsol}) gives
\begin{equation}
  \hat G^0(  \alpha^{-1}\vec{\underline r}, \alpha^{-2} {\underline t}) = \frac{\alpha^3}{(4\pi {\underline t})^{3/2}|| \hat D^{p}||} \exp\left(-\hat D^{I}\frac{|\vec {\underline r}|^2}{4 {\underline t}}\right) + \frac{\alpha^5}{8\pi^3} \delta( {\underline t}) \delta(\vec {\underline r})  \hat R
\end{equation}
 $||\hat D^{p}||$ denotes  the determinant of the matrix $\hat D^p$ and $ \hat D^I$ is defined as
$$ \hat D^I=\sum_{i=1}^{N_0}\sum_{k=1}^{N_0}\mathbf {p}_i D^{I}_{ik}\mathbf {l}_i$$
where $D^{i}_{ik}$ denotes the elements of the matrix inverse to $D^p$.

\section*{Supplementary Note IV. The pair $T$-matrix  order in $\alpha$}

The pair $T$-matrix is defined by  Eq.(5.6) in the main article
\begin{eqnarray}
\label{Tm}
 \widehat {\bf T}(\vec r_1,\vec r_2,t,\vec r^{\,\prime}_1,\vec r^{\,\prime}_2,t')
 =\widehat {\bf V}(\vec r_1,\vec r_2)\delta (t-t')
\delta(\vec r_1-\vec r^{\,\prime}_1)\delta(\vec r_2-\vec r^{\,\prime}_2)+
 \widehat {\bf V}(\vec r_1,\vec r_2)
 \widehat {\bf G}(\vec r_1,\vec r_2,t,\vec r^{\,\prime}_1,\vec r^{\,\prime}_2,t')
 \widehat {\bf V}(\vec r^{\,\prime}_1,\vec r^{\,\prime}_2)
\end{eqnarray}

where the pair Green's function $\widehat {\bf G}$ obeys the following equation (Eq.(5.7) in the main article)
\begin{equation}
\label{GET}
\frac \partial{\partial t}
  \widehat {\bf G}(\vec r_1,\vec r_2,t,\vec r^{\,\prime}_1,\vec r^{\,\prime}_2,t')
 = ( \widehat {\bf L} +
  \widehat {\bf Q} + \widehat {\bf V}(\vec r_1,\vec r_2) )
   \widehat {\bf G}(\vec r_1,\vec r_2,t,\vec r^{\,\prime}_1,\vec r^{\,\prime}_2,t')
\end{equation}
with the initial condition
$$\widehat {\bf G}(\vec r_1,\vec r_2,t,\vec r^{\,\prime}_1,\vec r^{\,\prime}_2,t) =
  \widehat {\bf I}\delta(\vec r_1-\vec r^{\,\prime}_1)
    \delta(\vec r_2-\vec r^{\,\prime}_2)$$

The critical pair $T$-matrix is
\begin{eqnarray}
\label{Tms}
 \underline {\widehat {\bf T}}(\underline{\vec r}_1,\underline{\vec r}_2,\underline t,\underline{\vec r}_1^{\,\prime},\underline{\vec r}_2^{\,\prime},\underline t')
 =\widehat {\bf V}(\underline{\vec r}_1,\underline{\vec r}_2)\delta (\underline t-\underline t')
\delta(\underline{\vec r}_1-\underline{\vec r}_1^{\,\prime})\delta(\underline{\vec r}_2-\underline{\vec r}_2^{\,\prime})+
 \widehat {\bf V}(\underline{\vec r}_1,\underline{\vec r}_2)
 \underline {\widehat {\bf G}}(\underline{\vec r}_1,\underline{\vec r}_2,t,\underline{\vec r}_1^{\,\prime},\underline{\vec r}_2^{\,\prime},\underline t')
 \widehat {\bf V}(\underline{\vec r}_1^{\,\prime},\underline{\vec r}_2^{\,\prime})
\end{eqnarray}

The critical pair Green's function also obeys

\begin{equation}
\label{GETs}
  \frac \partial{\partial \underline t}
 \underline{ \widehat {\bf G}}(\underline{\vec r}_1,\underline{\vec r}_2, \underline t,\underline{\vec r}^{\,\prime}_1,\underline{\vec r}^{\,\prime}_2,\underline t')
 = ( \widehat {\bf L} +
 \widehat {\bf Q} + \widehat {\bf V}(\underline{\vec r}_1,\underline{\vec r}_2) )
   \underline{\widehat {\bf G}}(\underline{\vec r}_1,\underline{\vec r}_2,t,\underline{\vec r}^{\,\prime}_1,\underline{\vec r}^{\,\prime}_2,\underline t')
\end{equation}

with the initial condition
$$ \underline{ \widehat {\bf G}}(\underline{\vec r}_1,\underline{\vec r}_2,\underline t,\underline{\vec r}_1^{\,\prime},\underline{\vec r}_2^{\,\prime},\underline t) =
  \widehat {\bf I}\delta(\underline{\vec r}_1-\underline{\vec r}_1^{\,\prime})
    \delta(\underline{\vec r}_2-\underline{\vec r}_2^{\,\prime})$$

Comparing Eqs.(\ref{Tm}-\ref{GET}) and  Eqs.(\ref{Tms}-\ref{GETs}), one can conclude
$$\underline{ \widehat {\bf G}}({\vec r_1},{\vec r_2}, t,\vec r_1^{\,\prime},\vec r_2^{\,\prime}, t) =
\widehat {\bf G}({\vec r_1},{\vec r_2}, t,\vec r_1^{\,\prime},\vec r_2^{\,\prime}, t)$$
and
\begin{equation}
 \underline{\widehat {\bf T}}({\vec r_1},{\vec r_2}, t,\vec r_1^{\,\prime},\vec r_2^{\,\prime}, t')=
\widehat {\bf T}({\vec r_1},{\vec r_2},t,\vec r_1^{\,\prime},\vec r_2^{\,\prime}, t')
\label{Teqv}
\end{equation}

Due to the properties of the operators $\widehat {\bf L}$ and  $\widehat {\bf V}$, the  pair Green's function Eq.(\ref{GET})  has time and spatial translation symmetry
\begin{equation}
\widehat {\bf G} (  {\vec r}_1, {\vec r}_2,{ t}, {\vec r}_1^{\,\prime}, {\vec r}_2^{\,\prime} ,t') = \widehat {{\bf G}}(0, {\vec r}_2- {\vec r}_1,{t}-{ t}', {\vec r}_1^{\,\prime}-{\vec r}_1, {\vec r}_2^{\,\prime}-\vec r_1 ,0)
\label{trans}
\end{equation}

From this follows the same property for the  pair $T$-matrix:
\begin{equation}
 \widehat {\bf { T}}(  {\vec r}_1, {\vec r}_2,{ t}, {\vec r}_1^{\,\prime}, {\vec r}_2^{\,\prime} ,t')=\widehat {\bf { T}}(0, {\vec r}_2- {\vec r}_1,{t}-{ t}', {\vec r}_1^{\,\prime}-{\vec r}_1, {\vec r}_2^{\,\prime}-\vec r_1 ,0)
  \label{eq:Spat}
 \end{equation}

The  pair $T$-matrix appears in all the terms of the collision integral expansion together with coordinate and time
integration
\begin{equation}
{\widehat g}({\vec r}, t)=\int\limits_0^{ t} dt' \int  d^3{r_2}\int d^3{r_1}'\int d^3{r_2}'\,
 \widehat{\bf T}({\vec r},{\vec r_2}, t,{\vec r_1}',{\vec r_2}',t')
  {\widehat f}({\vec r_2},{\vec r_1}', {\vec r_2}', t')
\label{Act}
 \end{equation}
 where  $\widehat f$ can be expressed in terms of concentrations, Green's functions, and $T$-matrices.
 If we suppose that $\widehat f$ is the function of $m$-th order of the parameter $\alpha$

\begin{equation}
{\widehat f}({\vec r_2},\vec {r}_1^{\,\prime},\vec {r}_2^{\,\prime},{t'})=
\alpha ^m\widehat {\underline f}(\underline {\vec  r}_2,\underline {\vec r}_1^{\,\prime},\underline {\vec  r}_2^{\,\prime},{\underline t}')= \alpha^m \widehat {\underline f}
 (\alpha{\vec r_2},\alpha{\vec {r}_1^{\,\prime}},\alpha \vec {r}_2^{\,\prime},\alpha^2{t'})
 \label{eq:fscale}
 \end{equation}

The simplest example of this expression is the collision integral of the integral encounter theory (Eq.(5.2) in the main article)
where $\widehat f$ has the form:

\begin{equation}
f_{kl}(\vec r_2,\vec r^{\,\prime}_1,\vec r^{\,\prime}_2,t')= C_k(\vec r^{\,\prime}_1,t')C_l(\vec r^{\,\prime}_2,t')=\alpha^4 \underline C_k(\underline{\vec r}^{\,\prime}_1,\underline t')\underline C_l(\underline {\vec r}^{\,\prime}_2,\underline t')
 \label{fET}
  \end{equation}

Eq.(\ref{Act}) can be rewritten using Eq.(\ref{eq:Spat}) as
$${\widehat g}( {\vec r}, t)=\int\limits^{ t}_0\, d \tau \int  d^3{\rho}\int d^3{\rho_1}\int d^3{\rho_2}\,
 {\widehat{{\bf T}}}(0,{\vec \rho}, {\tau},{\vec \rho_1},  {\vec \rho_2},0)
   {\widehat f}({\vec r}+  {\vec\rho},{\vec r}+ {\vec\rho_1},{\vec r}+ {\vec\rho_2},t+ \tau)$$
where
$${\vec \rho}={\vec r_2}-{\vec r},\quad {\vec \rho_1}={\vec r_1}^{\,\prime}-{\vec r}, \quad    {\vec \rho_2}={\vec r_2}^{\,\prime}-{\vec r}, \quad  {\tau}=  t' - t$$
Eq.(\ref{eq:fscale}) and Eq.(\ref{Teqv}) allow us to express $ {\widehat g}$ in terms of the critical function $\widehat {\underline f}$ and critical $T$-matrix:

$${\widehat g}({\vec  r}, t)=\int\limits^{t}_0 d \tau \int  d^3{\rho}\int d^3{\rho_1}\int d^3{\rho_2}\,
{\widehat{ {\bf {\underline T}}}}(0,{\vec \rho}, {\tau}, {\vec \rho_1}, {\vec \rho_2},0)
\alpha^m  \widehat {\underline f}(\alpha  {\vec r}+\alpha  {\vec\rho},\alpha  {\vec r}+\alpha {\vec \rho_1},
  \alpha  {\vec r}+\alpha  {\vec \rho_2}, \alpha^2  t+\alpha^2{\tau})$$
Rewriting this expression as a function of the critical coordinate and time:
$${\widehat g}(\alpha^{-1} \vec {\underline r}, \alpha^{-2} {\underline t})=\int\limits^{\alpha^{-2}{\underline t}}_0 d\tau \int   d^3{\rho}\int d^3{\rho_1}\int  d^3{\rho_2}\,
 \widehat{\bf {\underline T}}(0,{\vec \rho},\tau,{\vec \rho_1},{\vec \rho_2},0)
\alpha^m \widehat {\underline f}(\vec {\underline r}+\alpha  {\vec\rho},\vec {\underline r}+\alpha  { \vec\rho_1},
  \vec {\underline r}+\alpha  {\vec \rho_2}, {\underline t}+\alpha^2\tau)$$
Expanding this expression in series at  $\alpha \rightarrow 0$ we get:
$${\widehat g}(\alpha^{-1} \vec {\underline r}, \alpha^{-2} {\underline t})=\alpha^m   \int\limits^{\infty}_0 d \tau \int  d^3 \rho \int d^3{\rho_1}\int d^3 {\rho_2}\,
{\widehat{\bf {\underline T}}}(0,{\vec \rho} \tau, {\vec \rho_1}, {\vec \rho_2},0)
 \widehat {\underline f}(\vec {\underline r},\vec {\underline r},\vec {\underline r}, {\underline t})\left(1 + O( \alpha) \right)$$
 or, presented in another form
  $${\widehat g}(\alpha^{-1} \vec {\underline r}, \alpha^{-2} {\underline t})= \alpha^m  \widehat{\bf { K}}\widehat {\underline f}(\vec {\underline r},\vec {\underline r},\vec {\underline r}, {\underline t})
  \left( 1+ O( \alpha) \right) $$
 where the matrix  $\widehat{\bf { K}}$ is defined as
 \begin{equation}
 \widehat{\bf { K}}= \int^{\infty}_0 d { \tau} \int  d^3 { \rho} \int d^3{{ \rho}_1}\int d^3 {{ \rho}_2}\,
 {\widehat{\bf T}}(0,{\vec {\rho}},\tau,{\vec { \rho}_1}, {\vec { \rho}_2},0 )
 \label{eq:K}
 \end{equation}
does not depend on $\alpha$.  Using Eq.(\ref{eq:fscale}), we get
$${\widehat g}({\vec  r}, t)=   \widehat{\bf { K}} {\widehat f}({\vec  r},{\vec  r},{\vec  r}, t)(1+ O( \alpha) ) $$

In the lowest order term of  $\alpha$ series,  $T$-matrix can be represented as a product of  delta-functions (point approximation)

$$\widehat {\bf{  T}}({\vec r_1},{\vec r_2},t,{\vec {r}_1^{\,\prime}},{\vec {r}_2^{\,\prime}},t')=
 \widehat{\bf {K}}\delta( {\vec r_2}-{\vec r_1})\delta({\vec {r}_1^{\,\prime}}-{\vec r_1}) \delta({\vec {r}_2^{\,\prime}}-{\vec r_1})
\delta( t'-t)+ O( \alpha)  $$

which corresponds to the 11-th order in critical coordinates representation:
 $$\widehat {\bf {T}}({\vec r_1},{\vec r_2},t,{\vec {r}_1^{\,\prime}},{\vec {r}_2^{\,\prime}}, t')=\alpha^{11}\widehat{\bf { K}} \delta(\vec {\underline r}_1-\underline {\vec  r}_2)\delta(\underline {\vec r}_1-\underline {\vec r}_1^{\,\prime})\,\delta(\underline {\vec r}_1-\underline {\vec r}_2^{\,\prime})\delta({\underline t}-{\underline t}')\left(1 + O( \alpha) \right)$$

\section*{Supplementary Note V. Calculation of the reaction rate matrix}

Let us define the reduced Green's function as
$$ \widehat {\bf G} (\vec r,\vec r^{\,\prime},\tau)=
   \int\
   \widehat {\bf G}(0,\vec r,\tau,\vec r_1^{\,\prime}-\vec r_1,\vec r_1^{\,\prime}-\vec r_1 +  \vec r^{\,\prime},0) \,d^3r_1'
$$
As follows from Eq.(\ref{GET}), the function $\widehat {\bf G} (\vec r,\vec r^{\,\prime},\tau)$ obeys equation
\begin{equation}
\label{GGET}
 \frac \partial{\partial \tau}
  \widehat {\bf G} (\vec r,\vec r^{\,\prime},\tau)
=( \widehat {\bf D}\Delta +\widehat {\bf V}( \vec r)+\widehat {\bf Q} )
 \widehat {\bf G} (\vec r,\vec r^{\,\prime},\tau)
\end{equation}
where
$$\qquad {D}_{ik,lm}=
  \delta_{il}\delta_{km} (D_i+D_k)$$
  and the generalized interaction operator $\widehat {\bf V}$ (Eq.(3.9) in the main article)
acts as follows
\begin{equation}
\label{eq:SphericalV}
\widehat {\bf V}(\vec r)  =  \widehat {\bf W}(\vec r) +   \widehat {\bf D}[\nabla\widehat{\bf U}(\vec r)]\nabla +  \widehat {\bf D}[\Delta \widehat{\bf U}(r)]
\end{equation}
with
$${ U}_{ik,lm}(\vec r)=\delta_{il}\delta_{km}U_{ik}(\vec r)$$
The  initial condition for Eq.(\ref{GGET}) is
$$ \frac \partial{\partial \tau}
  \widehat {\bf G} (\vec r,\vec r^{\,\prime},0)=\widehat  {\bf I} \delta(\vec r-\vec r^{\,\prime})$$

Now   Eq.(\ref{eq:K})  can be rewritten as

\begin{eqnarray}
\label{Kk}
 \widehat {\bf K} =\int  \widehat {\bf V}(\vec r) \,d^3r
+\int_0^\infty\,dt \int \int   \widehat {\bf V}(\vec r)
 \widehat {\bf G}(\vec r,\vec r^{\,\prime},t)
 \widehat {\bf V}(\vec r^{\,\prime})\,d^3r\,d^3r'
\end{eqnarray}
or, in another form

\begin{equation}
\label{eq:FinalK}
 \widehat {\bf K} =\int  \widehat {\bf W}(\vec r) \,d^3r + \int_0^\infty \,dt \int \int
  \widehat{\bf W}(\vec r) \widehat{\bf G}(\vec r,\vec r^{\,\prime},t)\widehat{\bf V}(\vec r^{\,\prime})
  \,d^3r\,d^3r'
\end{equation}

\section*{Supplementary Note VI. Modification of the collision integral series}
\label{supplementary:D}

To obtain a compact representation of the collision integral, we add a set of higher order diagrams in $\alpha$ to Eq.(6.13) in the main article

$$
 R_i(\vec r,t)=\quad \begin{picture}(2,1.5) \put(-0.2,0.6){\line(1,1){0.4}}
\put(-0.2,1.0){\line(1,-1){0.4}} \put(-0.2,-0.2){\line(1,1){0.4}} \put(-0.2,0.2){\line(1,-1){0.4}}
\put(0.4,0){\circle*{0.1}} \put(1.1,0){\circle*{0.1}} \multiput(0.4,0)(0,0.1){7}{\line(4,1){0.7}}
\put(0.4,0.8){\circle*{0.1}} \put(1.1,0.8){\circle*{0.1}} \put(0,0.8){\line(1,0){1.3}}
\put(0,0.){\line(1,0){1.3}} \put(0.4,0.8){\line(0,-1){0.8}} \put(1.1,0.8){\line(0,-1){0.8}}
\put(1.3,0.1){\makebox(0,0)[b]{$i$}} \put(1.1,-0.4){\makebox(0,0)[b]{$rt$}}
\end{picture}
\quad +\quad
\begin{picture}(2.9,1.5)
\put(-0.2,0.6){\line(1,1){0.4}} \put(-0.2,1.0){\line(1,-1){0.4}} \put(-0.2,-0.2){\line(1,1){0.4}}
\put(-0.2,0.2){\line(1,-1){0.4}} \put(0.4,0.8){\circle*{0.1}} \put(1.1,0.8){\circle*{0.1}}

\put(2,0){\circle*{0.1}} \put(2.7,0){\circle*{0.1}} \multiput(2,0)(0,0.1){7}{\line(4,1){0.7}}
\put(2,0.8){\circle*{0.1}} \put(2.7,0.8){\circle*{0.1}}

\put(0.4,0){\circle*{0.1}} \put(1.1,0){\circle*{0.1}} \multiput(0.4,0)(0,0.1){7}{\line(4,1){0.7}}

\put(0.4,0.8){\circle*{0.1}} \put(1.1,0.8){\circle*{0.1}} \put(0,0.8){\line(1,0){2.9}}
\put(0,0.){\line(1,0){2.9}} \put(2,0.8){\circle*{0.1}} \put(2,0){\circle*{0.1}}
\put(2,0.8){\line(0,-1){0.8}}\put(2.7,0.8){\line(0,-1){0.8}}
 \put(0.4,0.8){\line(0,-1){0.8}}
\put(1.1,0.8){\line(0,-1){0.8}}

\put(1.4,-0.15){\line(1,0){0.3}} \put(1.4,-0.15){\line(0,1){0.3}} \put(1.4,0.15){\line(1,0){0.3}}
\put(1.7,-0.15){\line(0,1){0.3}}

\put(2.9,0.1){\makebox(0,0)[b]{$i$}} \put(2.7,-0.4){\makebox(0,0)[b]{$rt$}}
\end{picture}\quad +\quad
\begin{picture}(2.9,1.5)
\put(-0.2,0.6){\line(1,1){0.4}} \put(-0.2,1.0){\line(1,-1){0.4}} \put(-0.2,-0.2){\line(1,1){0.4}}
\put(-0.2,0.2){\line(1,-1){0.4}} \put(0.4,0.8){\circle*{0.1}} \put(1.1,0.8){\circle*{0.1}}

\put(2,0){\circle*{0.1}} \put(2.7,0){\circle*{0.1}} \multiput(2,0)(0,0.1){7}{\line(4,1){0.7}}
\put(2,0.8){\circle*{0.1}} \put(2.7,0.8){\circle*{0.1}}

\put(0.4,0){\circle*{0.1}} \put(1.1,0){\circle*{0.1}} \multiput(0.4,0)(0,0.1){7}{\line(4,1){0.7}}

\put(0.4,0.8){\circle*{0.1}} \put(1.1,0.8){\circle*{0.1}} \put(0,0.8){\line(1,0){2.9}}
\put(0,0.){\line(1,0){2.9}} \put(2,0.8){\circle*{0.1}} \put(2,0){\circle*{0.1}}
\put(2,0.8){\line(0,-1){0.8}}\put(2.7,0.8){\line(0,-1){0.8}}
 \put(0.4,0.8){\line(0,-1){0.8}}
\put(1.1,0.8){\line(0,-1){0.8}}

\put(1.4,0.65){\line(1,0){0.3}} \put(1.4,0.65){\line(0,1){0.3}} \put(1.4,0.95){\line(1,0){0.3}}
\put(1.7,0.65){\line(0,1){0.3}}

\put(2.9,0.1){\makebox(0,0)[b]{$i$}} \put(2.7,-0.4){\makebox(0,0)[b]{$rt$}}
\end{picture}\quad +\quad
\begin{picture}(2.9,1.5)
\put(-0.2,0.6){\line(1,1){0.4}} \put(-0.2,1.0){\line(1,-1){0.4}} \put(-0.2,-0.2){\line(1,1){0.4}}
\put(-0.2,0.2){\line(1,-1){0.4}} \put(0.4,0.8){\circle*{0.1}} \put(1.1,0.8){\circle*{0.1}}

\put(2,0){\circle*{0.1}} \put(2.7,0){\circle*{0.1}} \multiput(2,0)(0,0.1){7}{\line(4,1){0.7}}
\put(2,0.8){\circle*{0.1}} \put(2.7,0.8){\circle*{0.1}}

\put(0.4,0){\circle*{0.1}} \put(1.1,0){\circle*{0.1}} \multiput(0.4,0)(0,0.1){7}{\line(4,1){0.7}}

\put(0.4,0.8){\circle*{0.1}} \put(1.1,0.8){\circle*{0.1}} \put(0,0.8){\line(1,0){2.9}}
\put(0,0.){\line(1,0){2.9}} \put(2,0.8){\circle*{0.1}} \put(2,0){\circle*{0.1}}
\put(2,0.8){\line(0,-1){0.8}}\put(2.7,0.8){\line(0,-1){0.8}}
 \put(0.4,0.8){\line(0,-1){0.8}}
\put(1.1,0.8){\line(0,-1){0.8}}

\put(1.4,-0.15){\line(1,0){0.3}} \put(1.4,-0.15){\line(0,1){0.3}} \put(1.4,0.15){\line(1,0){0.3}}
\put(1.7,-0.15){\line(0,1){0.3}} \put(1.4,0.65){\line(1,0){0.3}} \put(1.4,0.65){\line(0,1){0.3}}
\put(1.4,0.95){\line(1,0){0.3}} \put(1.7,0.65){\line(0,1){0.3}} \put(2.9,0.1){\makebox(0,0)[b]{$i$}}
\put(2.7,-0.4){\makebox(0,0)[b]{$rt$}}

\end{picture} \quad+\quad $$
$$
\begin{picture}(4.5,1.5)
\put(-0.2,0.6){\line(1,1){0.4}} \put(-0.2,1.0){\line(1,-1){0.4}} \put(-0.2,-0.2){\line(1,1){0.4}}
\put(-0.2,0.2){\line(1,-1){0.4}}

\put(2,0){\circle*{0.1}} \put(2.7,0){\circle*{0.1}} \multiput(2,0)(0,0.1){7}{\line(4,1){0.7}}
\put(2,0.8){\circle*{0.1}} \put(2.7,0.8){\circle*{0.1}}
\put(2,0.8){\line(0,-1){0.8}}\put(2.7,0.8){\line(0,-1){0.8}}

\put(0.4,0){\circle*{0.1}} \put(1.1,0){\circle*{0.1}} \multiput(0.4,0)(0,0.1){7}{\line(4,1){0.7}}
\put(0.4,0.8){\circle*{0.1}} \put(1.1,0.8){\circle*{0.1}}
 \put(0.4,0.8){\line(0,-1){0.8}}
\put(1.1,0.8){\line(0,-1){0.8}}

\put(3.6,0){\circle*{0.1}} \put(4.3,0){\circle*{0.1}} \multiput(3.6,0)(0,0.1){7}{\line(4,1){0.7}}
\put(3.6,0.8){\circle*{0.1}} \put(4.3,0.8){\circle*{0.1}}
\put(3.6,0.8){\line(0,-1){0.8}}\put(4.3,0.8){\line(0,-1){0.8}}

\put(0,0.8){\line(1,0){4.5}} \put(0,0.){\line(1,0){4.5}}

\put(1.4,-0.15){\line(1,0){0.3}} \put(1.4,-0.15){\line(0,1){0.3}} \put(1.4,0.15){\line(1,0){0.3}}
\put(1.7,-0.15){\line(0,1){0.3}}

\put(3,-0.15){\line(1,0){0.3}} \put(3,-0.15){\line(0,1){0.3}} \put(3,0.15){\line(1,0){0.3}}
\put(3.3,-0.15){\line(0,1){0.3}}

\put(4.5,0.1){\makebox(0,0)[b]{$i$}} \put(4.3,-0.4){\makebox(0,0)[b]{$rt$}}
\end{picture}\quad +\quad
\begin{picture}(4.5,1.5)
\put(-0.2,0.6){\line(1,1){0.4}} \put(-0.2,1.0){\line(1,-1){0.4}} \put(-0.2,-0.2){\line(1,1){0.4}}
\put(-0.2,0.2){\line(1,-1){0.4}}

\put(2,0){\circle*{0.1}} \put(2.7,0){\circle*{0.1}} \multiput(2,0)(0,0.1){7}{\line(4,1){0.7}}
\put(2,0.8){\circle*{0.1}} \put(2.7,0.8){\circle*{0.1}}
\put(2,0.8){\line(0,-1){0.8}}\put(2.7,0.8){\line(0,-1){0.8}}

\put(0.4,0){\circle*{0.1}} \put(1.1,0){\circle*{0.1}} \multiput(0.4,0)(0,0.1){7}{\line(4,1){0.7}}
\put(0.4,0.8){\circle*{0.1}} \put(1.1,0.8){\circle*{0.1}}
 \put(0.4,0.8){\line(0,-1){0.8}}
\put(1.1,0.8){\line(0,-1){0.8}}

\put(3.6,0){\circle*{0.1}} \put(4.3,0){\circle*{0.1}} \multiput(3.6,0)(0,0.1){7}{\line(4,1){0.7}}
\put(3.6,0.8){\circle*{0.1}} \put(4.3,0.8){\circle*{0.1}}
\put(3.6,0.8){\line(0,-1){0.8}}\put(4.3,0.8){\line(0,-1){0.8}}

\put(0,0.8){\line(1,0){4.5}} \put(0,0.){\line(1,0){4.5}}

\put(1.4,0.65){\line(1,0){0.3}} \put(1.4,0.65){\line(0,1){0.3}} \put(1.4,0.95){\line(1,0){0.3}}
\put(1.7,0.65){\line(0,1){0.3}}

\put(3,-0.15){\line(1,0){0.3}} \put(3,-0.15){\line(0,1){0.3}} \put(3,0.15){\line(1,0){0.3}}
\put(3.3,-0.15){\line(0,1){0.3}}

\put(4.5,0.1){\makebox(0,0)[b]{$i$}} \put(4.3,-0.4){\makebox(0,0)[b]{$rt$}}
\end{picture}\quad +\quad
\begin{picture}(4.5,1.5)
\put(-0.2,0.6){\line(1,1){0.4}} \put(-0.2,1.0){\line(1,-1){0.4}} \put(-0.2,-0.2){\line(1,1){0.4}}
\put(-0.2,0.2){\line(1,-1){0.4}}

\put(2,0){\circle*{0.1}} \put(2.7,0){\circle*{0.1}} \multiput(2,0)(0,0.1){7}{\line(4,1){0.7}}
\put(2,0.8){\circle*{0.1}} \put(2.7,0.8){\circle*{0.1}}
\put(2,0.8){\line(0,-1){0.8}}\put(2.7,0.8){\line(0,-1){0.8}}

\put(0.4,0){\circle*{0.1}} \put(1.1,0){\circle*{0.1}} \multiput(0.4,0)(0,0.1){7}{\line(4,1){0.7}}
\put(0.4,0.8){\circle*{0.1}} \put(1.1,0.8){\circle*{0.1}}
 \put(0.4,0.8){\line(0,-1){0.8}}
\put(1.1,0.8){\line(0,-1){0.8}}

\put(3.6,0){\circle*{0.1}} \put(4.3,0){\circle*{0.1}} \multiput(3.6,0)(0,0.1){7}{\line(4,1){0.7}}
\put(3.6,0.8){\circle*{0.1}} \put(4.3,0.8){\circle*{0.1}}
\put(3.6,0.8){\line(0,-1){0.8}}\put(4.3,0.8){\line(0,-1){0.8}}

\put(0,0.8){\line(1,0){4.5}} \put(0,0.){\line(1,0){4.5}}

\put(1.4,-0.15){\line(1,0){0.3}} \put(1.4,-0.15){\line(0,1){0.3}} \put(1.4,0.15){\line(1,0){0.3}}
\put(1.7,-0.15){\line(0,1){0.3}}

\put(1.4,0.65){\line(1,0){0.3}} \put(1.4,0.65){\line(0,1){0.3}} \put(1.4,0.95){\line(1,0){0.3}}
\put(1.7,0.65){\line(0,1){0.3}}

\put(3,-0.15){\line(1,0){0.3}} \put(3,-0.15){\line(0,1){0.3}} \put(3,0.15){\line(1,0){0.3}}
\put(3.3,-0.15){\line(0,1){0.3}}

\put(4.5,0.1){\makebox(0,0)[b]{$i$}} \put(4.3,-0.4){\makebox(0,0)[b]{$rt$}}
\end{picture}\quad +\quad
$$
$$
\begin{picture}(4.5,1.5)
\put(-0.2,0.6){\line(1,1){0.4}} \put(-0.2,1.0){\line(1,-1){0.4}} \put(-0.2,-0.2){\line(1,1){0.4}}
\put(-0.2,0.2){\line(1,-1){0.4}}

\put(2,0){\circle*{0.1}} \put(2.7,0){\circle*{0.1}} \multiput(2,0)(0,0.1){7}{\line(4,1){0.7}}
\put(2,0.8){\circle*{0.1}} \put(2.7,0.8){\circle*{0.1}}
\put(2,0.8){\line(0,-1){0.8}}\put(2.7,0.8){\line(0,-1){0.8}}

\put(0.4,0){\circle*{0.1}} \put(1.1,0){\circle*{0.1}} \multiput(0.4,0)(0,0.1){7}{\line(4,1){0.7}}
\put(0.4,0.8){\circle*{0.1}} \put(1.1,0.8){\circle*{0.1}}
 \put(0.4,0.8){\line(0,-1){0.8}}
\put(1.1,0.8){\line(0,-1){0.8}}

\put(3.6,0){\circle*{0.1}} \put(4.3,0){\circle*{0.1}} \multiput(3.6,0)(0,0.1){7}{\line(4,1){0.7}}
\put(3.6,0.8){\circle*{0.1}} \put(4.3,0.8){\circle*{0.1}}
\put(3.6,0.8){\line(0,-1){0.8}}\put(4.3,0.8){\line(0,-1){0.8}}

\put(0,0.8){\line(1,0){4.5}} \put(0,0.){\line(1,0){4.5}}

\put(1.4,-0.15){\line(1,0){0.3}} \put(1.4,-0.15){\line(0,1){0.3}} \put(1.4,0.15){\line(1,0){0.3}}
\put(1.7,-0.15){\line(0,1){0.3}}

 \put(3,0.65){\line(1,0){0.3}} \put(3,0.65){\line(0,1){0.3}} \put(3,0.95){\line(1,0){0.3}}
\put(3.3,0.65){\line(0,1){0.3}}

\put(4.5,0.1){\makebox(0,0)[b]{$i$}} \put(4.3,-0.4){\makebox(0,0)[b]{$rt$}}
\end{picture}\quad +\quad
\begin{picture}(4.5,1.5)
\put(-0.2,0.6){\line(1,1){0.4}} \put(-0.2,1.0){\line(1,-1){0.4}} \put(-0.2,-0.2){\line(1,1){0.4}}
\put(-0.2,0.2){\line(1,-1){0.4}}

\put(2,0){\circle*{0.1}} \put(2.7,0){\circle*{0.1}} \multiput(2,0)(0,0.1){7}{\line(4,1){0.7}}
\put(2,0.8){\circle*{0.1}} \put(2.7,0.8){\circle*{0.1}}
\put(2,0.8){\line(0,-1){0.8}}\put(2.7,0.8){\line(0,-1){0.8}}

\put(0.4,0){\circle*{0.1}} \put(1.1,0){\circle*{0.1}} \multiput(0.4,0)(0,0.1){7}{\line(4,1){0.7}}
\put(0.4,0.8){\circle*{0.1}} \put(1.1,0.8){\circle*{0.1}}
 \put(0.4,0.8){\line(0,-1){0.8}}
\put(1.1,0.8){\line(0,-1){0.8}}

\put(3.6,0){\circle*{0.1}} \put(4.3,0){\circle*{0.1}} \multiput(3.6,0)(0,0.1){7}{\line(4,1){0.7}}
\put(3.6,0.8){\circle*{0.1}} \put(4.3,0.8){\circle*{0.1}}
\put(3.6,0.8){\line(0,-1){0.8}}\put(4.3,0.8){\line(0,-1){0.8}}

\put(0,0.8){\line(1,0){4.5}} \put(0,0.){\line(1,0){4.5}}

\put(1.4,0.65){\line(1,0){0.3}} \put(1.4,0.65){\line(0,1){0.3}} \put(1.4,0.95){\line(1,0){0.3}}
\put(1.7,0.65){\line(0,1){0.3}}

\put(3,0.65){\line(1,0){0.3}} \put(3,0.65){\line(0,1){0.3}} \put(3,0.95){\line(1,0){0.3}}
\put(3.3,0.65){\line(0,1){0.3}}

\put(4.5,0.1){\makebox(0,0)[b]{$i$}} \put(4.3,-0.4){\makebox(0,0)[b]{$rt$}}
\end{picture}\quad +\quad
\begin{picture}(4.5,1.5)
\put(-0.2,0.6){\line(1,1){0.4}} \put(-0.2,1.0){\line(1,-1){0.4}} \put(-0.2,-0.2){\line(1,1){0.4}}
\put(-0.2,0.2){\line(1,-1){0.4}}

\put(2,0){\circle*{0.1}} \put(2.7,0){\circle*{0.1}} \multiput(2,0)(0,0.1){7}{\line(4,1){0.7}}
\put(2,0.8){\circle*{0.1}} \put(2.7,0.8){\circle*{0.1}}
\put(2,0.8){\line(0,-1){0.8}}\put(2.7,0.8){\line(0,-1){0.8}}

\put(0.4,0){\circle*{0.1}} \put(1.1,0){\circle*{0.1}} \multiput(0.4,0)(0,0.1){7}{\line(4,1){0.7}}
\put(0.4,0.8){\circle*{0.1}} \put(1.1,0.8){\circle*{0.1}}
 \put(0.4,0.8){\line(0,-1){0.8}}
\put(1.1,0.8){\line(0,-1){0.8}}

\put(3.6,0){\circle*{0.1}} \put(4.3,0){\circle*{0.1}} \multiput(3.6,0)(0,0.1){7}{\line(4,1){0.7}}
\put(3.6,0.8){\circle*{0.1}} \put(4.3,0.8){\circle*{0.1}}
\put(3.6,0.8){\line(0,-1){0.8}}\put(4.3,0.8){\line(0,-1){0.8}}

\put(0,0.8){\line(1,0){4.5}} \put(0,0.){\line(1,0){4.5}}

\put(1.4,-0.15){\line(1,0){0.3}} \put(1.4,-0.15){\line(0,1){0.3}} \put(1.4,0.15){\line(1,0){0.3}}
\put(1.7,-0.15){\line(0,1){0.3}}

\put(1.4,0.65){\line(1,0){0.3}} \put(1.4,0.65){\line(0,1){0.3}} \put(1.4,0.95){\line(1,0){0.3}}
\put(1.7,0.65){\line(0,1){0.3}}

\put(3,0.65){\line(1,0){0.3}} \put(3,0.65){\line(0,1){0.3}} \put(3,0.95){\line(1,0){0.3}}
\put(3.3,0.65){\line(0,1){0.3}}

\put(4.5,0.1){\makebox(0,0)[b]{$i$}} \put(4.3,-0.4){\makebox(0,0)[b]{$rt$}}
\end{picture}\quad +\quad
$$

$$
\begin{picture}(4.5,1.5)
\put(-0.2,0.6){\line(1,1){0.4}} \put(-0.2,1.0){\line(1,-1){0.4}} \put(-0.2,-0.2){\line(1,1){0.4}}
\put(-0.2,0.2){\line(1,-1){0.4}}

\put(2,0){\circle*{0.1}} \put(2.7,0){\circle*{0.1}} \multiput(2,0)(0,0.1){7}{\line(4,1){0.7}}
\put(2,0.8){\circle*{0.1}} \put(2.7,0.8){\circle*{0.1}}
\put(2,0.8){\line(0,-1){0.8}}\put(2.7,0.8){\line(0,-1){0.8}}

\put(0.4,0){\circle*{0.1}} \put(1.1,0){\circle*{0.1}} \multiput(0.4,0)(0,0.1){7}{\line(4,1){0.7}}
\put(0.4,0.8){\circle*{0.1}} \put(1.1,0.8){\circle*{0.1}}
 \put(0.4,0.8){\line(0,-1){0.8}}
\put(1.1,0.8){\line(0,-1){0.8}}

\put(3.6,0){\circle*{0.1}} \put(4.3,0){\circle*{0.1}} \multiput(3.6,0)(0,0.1){7}{\line(4,1){0.7}}
\put(3.6,0.8){\circle*{0.1}} \put(4.3,0.8){\circle*{0.1}}
\put(3.6,0.8){\line(0,-1){0.8}}\put(4.3,0.8){\line(0,-1){0.8}}

\put(0,0.8){\line(1,0){4.5}} \put(0,0.){\line(1,0){4.5}}

\put(1.4,-0.15){\line(1,0){0.3}} \put(1.4,-0.15){\line(0,1){0.3}} \put(1.4,0.15){\line(1,0){0.3}}
\put(1.7,-0.15){\line(0,1){0.3}}

\put(3,-0.15){\line(1,0){0.3}} \put(3,-0.15){\line(0,1){0.3}} \put(3,0.15){\line(1,0){0.3}}
\put(3.3,-0.15){\line(0,1){0.3}}

\put(3,0.65){\line(1,0){0.3}} \put(3,0.65){\line(0,1){0.3}} \put(3,0.95){\line(1,0){0.3}}
\put(3.3,0.65){\line(0,1){0.3}}

\put(4.5,0.1){\makebox(0,0)[b]{$i$}} \put(4.3,-0.4){\makebox(0,0)[b]{$rt$}}
\end{picture}\quad +\quad
\begin{picture}(4.5,1.5)
\put(-0.2,0.6){\line(1,1){0.4}} \put(-0.2,1.0){\line(1,-1){0.4}} \put(-0.2,-0.2){\line(1,1){0.4}}
\put(-0.2,0.2){\line(1,-1){0.4}}

\put(2,0){\circle*{0.1}} \put(2.7,0){\circle*{0.1}} \multiput(2,0)(0,0.1){7}{\line(4,1){0.7}}
\put(2,0.8){\circle*{0.1}} \put(2.7,0.8){\circle*{0.1}}
\put(2,0.8){\line(0,-1){0.8}}\put(2.7,0.8){\line(0,-1){0.8}}

\put(0.4,0){\circle*{0.1}} \put(1.1,0){\circle*{0.1}} \multiput(0.4,0)(0,0.1){7}{\line(4,1){0.7}}
\put(0.4,0.8){\circle*{0.1}} \put(1.1,0.8){\circle*{0.1}}
 \put(0.4,0.8){\line(0,-1){0.8}}
\put(1.1,0.8){\line(0,-1){0.8}}

\put(3.6,0){\circle*{0.1}} \put(4.3,0){\circle*{0.1}} \multiput(3.6,0)(0,0.1){7}{\line(4,1){0.7}}
\put(3.6,0.8){\circle*{0.1}} \put(4.3,0.8){\circle*{0.1}}
\put(3.6,0.8){\line(0,-1){0.8}}\put(4.3,0.8){\line(0,-1){0.8}}

\put(0,0.8){\line(1,0){4.5}} \put(0,0.){\line(1,0){4.5}}

\put(1.4,0.65){\line(1,0){0.3}} \put(1.4,0.65){\line(0,1){0.3}} \put(1.4,0.95){\line(1,0){0.3}}
\put(1.7,0.65){\line(0,1){0.3}}

\put(3,0.65){\line(1,0){0.3}} \put(3,0.65){\line(0,1){0.3}} \put(3,0.95){\line(1,0){0.3}}
\put(3.3,0.65){\line(0,1){0.3}}

\put(3,-0.15){\line(1,0){0.3}} \put(3,-0.15){\line(0,1){0.3}} \put(3,0.15){\line(1,0){0.3}}
\put(3.3,-0.15){\line(0,1){0.3}}

\put(4.5,0.1){\makebox(0,0)[b]{$i$}} \put(4.3,-0.4){\makebox(0,0)[b]{$rt$}}
\end{picture}\quad +\quad
\begin{picture}(4.5,1.5)
\put(-0.2,0.6){\line(1,1){0.4}} \put(-0.2,1.0){\line(1,-1){0.4}} \put(-0.2,-0.2){\line(1,1){0.4}}
\put(-0.2,0.2){\line(1,-1){0.4}}

\put(2,0){\circle*{0.1}} \put(2.7,0){\circle*{0.1}} \multiput(2,0)(0,0.1){7}{\line(4,1){0.7}}
\put(2,0.8){\circle*{0.1}} \put(2.7,0.8){\circle*{0.1}}
\put(2,0.8){\line(0,-1){0.8}}\put(2.7,0.8){\line(0,-1){0.8}}

\put(0.4,0){\circle*{0.1}} \put(1.1,0){\circle*{0.1}} \multiput(0.4,0)(0,0.1){7}{\line(4,1){0.7}}
\put(0.4,0.8){\circle*{0.1}} \put(1.1,0.8){\circle*{0.1}}
 \put(0.4,0.8){\line(0,-1){0.8}}
\put(1.1,0.8){\line(0,-1){0.8}}

\put(3.6,0){\circle*{0.1}} \put(4.3,0){\circle*{0.1}} \multiput(3.6,0)(0,0.1){7}{\line(4,1){0.7}}
\put(3.6,0.8){\circle*{0.1}} \put(4.3,0.8){\circle*{0.1}}
\put(3.6,0.8){\line(0,-1){0.8}}\put(4.3,0.8){\line(0,-1){0.8}}

\put(0,0.8){\line(1,0){4.5}} \put(0,0.){\line(1,0){4.5}}

\put(1.4,-0.15){\line(1,0){0.3}} \put(1.4,-0.15){\line(0,1){0.3}} \put(1.4,0.15){\line(1,0){0.3}}
\put(1.7,-0.15){\line(0,1){0.3}}

\put(1.4,0.65){\line(1,0){0.3}} \put(1.4,0.65){\line(0,1){0.3}} \put(1.4,0.95){\line(1,0){0.3}}
\put(1.7,0.65){\line(0,1){0.3}}

\put(3,0.65){\line(1,0){0.3}} \put(3,0.65){\line(0,1){0.3}} \put(3,0.95){\line(1,0){0.3}}
\put(3.3,0.65){\line(0,1){0.3}}

\put(3,-0.15){\line(1,0){0.3}} \put(3,-0.15){\line(0,1){0.3}} \put(3,0.15){\line(1,0){0.3}}
\put(3.3,-0.15){\line(0,1){0.3}}

\put(4.5,0.1){\makebox(0,0)[b]{$i$}} \put(4.3,-0.4){\makebox(0,0)[b]{$rt$}}
\end{picture}\quad +\quad
$$
\begin{equation}\label{DiagInBlc}
\begin{picture}(6.1,1.5)
\put(-0.2,0.6){\line(1,1){0.4}} \put(-0.2,1.0){\line(1,-1){0.4}} \put(-0.2,-0.2){\line(1,1){0.4}}
\put(-0.2,0.2){\line(1,-1){0.4}}

\put(2,0){\circle*{0.1}} \put(2.7,0){\circle*{0.1}} \multiput(2,0)(0,0.1){7}{\line(4,1){0.7}}
\put(2,0.8){\circle*{0.1}} \put(2.7,0.8){\circle*{0.1}}
\put(2,0.8){\line(0,-1){0.8}}\put(2.7,0.8){\line(0,-1){0.8}}

\put(0.4,0){\circle*{0.1}} \put(1.1,0){\circle*{0.1}} \multiput(0.4,0)(0,0.1){7}{\line(4,1){0.7}}
\put(0.4,0.8){\circle*{0.1}} \put(1.1,0.8){\circle*{0.1}}
 \put(0.4,0.8){\line(0,-1){0.8}}
\put(1.1,0.8){\line(0,-1){0.8}}

\put(3.6,0){\circle*{0.1}} \put(4.3,0){\circle*{0.1}} \multiput(3.6,0)(0,0.1){7}{\line(4,1){0.7}}
\put(3.6,0.8){\circle*{0.1}} \put(4.3,0.8){\circle*{0.1}}
\put(3.6,0.8){\line(0,-1){0.8}}\put(4.3,0.8){\line(0,-1){0.8}}

\put(5.2,0){\circle*{0.1}} \put(5.9,0){\circle*{0.1}} \multiput(5.2,0)(0,0.1){7}{\line(4,1){0.7}}
\put(5.2,0.8){\circle*{0.1}} \put(5.9,0.8){\circle*{0.1}}
\put(5.2,0.8){\line(0,-1){0.8}}\put(5.9,0.8){\line(0,-1){0.8}}

\put(0,0.8){\line(1,0){6.1}} \put(0,0.){\line(1,0){6.1}}

\put(1.4,-0.15){\line(1,0){0.3}} \put(1.4,-0.15){\line(0,1){0.3}} \put(1.4,0.15){\line(1,0){0.3}}
\put(1.7,-0.15){\line(0,1){0.3}}

\put(3,-0.15){\line(1,0){0.3}} \put(3,-0.15){\line(0,1){0.3}} \put(3,0.15){\line(1,0){0.3}}
\put(3.3,-0.15){\line(0,1){0.3}}

\put(4.6,-0.15){\line(1,0){0.3}} \put(4.6,-0.15){\line(0,1){0.3}} \put(4.6,0.15){\line(1,0){0.3}}
\put(4.9,-0.15){\line(0,1){0.3}}

 \put(6.1,0.1){\makebox(0,0)[b]{$i$}} \put(5.9,-0.4){\makebox(0,0)[b]{$rt$}}
\end{picture}\quad + \quad \cdots
\end{equation}

In addition to the diagrams given in  Eq.(6.13) in the main article, we added ladder-type diagrams in which the $T$-matrices are connected
with one line corresponding to either the Green's function  or the "Green’s function with interactions".
This expression can be simplified if we use Eq.(5.3) in the main article to replace the $T$-matrices. Reordering the terms, we can represent the result as an infinite series of diagrams:
$$ R_i(\vec r,t)=\quad
\begin{picture}(1.,1)
\put(0.0,0.8){\line(1,1){0.4}} \put(0.0,1.2){\line(1,-1){0.4}} \put(0.0,-0.2){\line(1,1){0.4}}
\put(0.0,0.2){\line(1,-1){0.4}} \put(0.2,0){\line(1,0){0.8}} \put(0.2,1){\line(1,0){0.8}}
\put(0.7,1){\circle*{0.1}} \put(0.7,0){\circle*{0.1}} \multiput(0.7,1)(0,-0.125){8}{\line(0,-1){0.1}}
\put(.9,0.2){\makebox(0,0)[b]{$i$}} \put(0.7,-0.4){\makebox(0,0)[b]{$rt$}}
\end{picture}\quad +\quad
\begin{picture}(1.6,1.5)
\put(0,0.8){\line(1,1){0.4}} \put(0.,1.2){\line(1,-1){0.4}} \put(0.,-0.2){\line(1,1){0.4}}
\put(0,0.2){\line(1,-1){0.4}}

 \put(0.2,0){\line(1,0){1.4}} \put(0.2,1){\line(1,0){1.4}}

\put(0.5,0){\circle*{0.1}} \put(0.5,1.){\circle*{0.1}} \multiput(0.5,1)(0,-0.125){8}{\line(0,-1){0.09}}

 \put(1.4,1){\circle*{0.1}}\put(1.4,0.){\circle*{0.1}}
\multiput(1.4,1)(0,-0.125){8}{\line(0,-1){0.09}}

 \put(1.6,0.2){\makebox(0,0)[b]{$i$}}
\put(1.4,-0.4){\makebox(0,0)[b]{$rt$}}
\end{picture}\quad + \quad
\begin{picture}(1.6,1.5)
\put(0,0.8){\line(1,1){0.4}} \put(0.,1.2){\line(1,-1){0.4}} \put(0.,-0.2){\line(1,1){0.4}}
\put(0,0.2){\line(1,-1){0.4}}

 \put(0.2,0){\line(1,0){1.4}} \put(0.2,1){\line(1,0){1.4}}

\put(0.5,0){\circle*{0.1}} \put(0.5,1.){\circle*{0.1}} \multiput(0.5,1)(0,-0.125){8}{\line(0,-1){0.09}}

 \put(1.4,1){\circle*{0.1}}\put(1.4,0.){\circle*{0.1}}
\multiput(1.4,1)(0,-0.125){8}{\line(0,-1){0.09}}

\put(0.8,-0.15){\line(1,0){0.3}} \put(0.8,-0.15){\line(0,1){0.3}} \put(0.8,0.15){\line(1,0){0.3}}
\put(1.1,-0.15){\line(0,1){0.3}}

 \put(1.6,0.2){\makebox(0,0)[b]{$i$}}
\put(1.4,-0.4){\makebox(0,0)[b]{$rt$}}
\end{picture}\quad + \quad
\begin{picture}(1.6,1.5)
\put(0,0.8){\line(1,1){0.4}} \put(0.,1.2){\line(1,-1){0.4}} \put(0.,-0.2){\line(1,1){0.4}}
\put(0,0.2){\line(1,-1){0.4}}

 \put(0.2,0){\line(1,0){1.4}} \put(0.2,1){\line(1,0){1.4}}

\put(0.5,0){\circle*{0.1}} \put(0.5,1.){\circle*{0.1}} \multiput(0.5,1)(0,-0.125){8}{\line(0,-1){0.09}}

 \put(1.4,1){\circle*{0.1}}\put(1.4,0.){\circle*{0.1}}
\multiput(1.4,1)(0,-0.125){8}{\line(0,-1){0.09}}

\put(0.8,0.85){\line(1,0){0.3}} \put(0.8,0.85){\line(0,1){0.3}} \put(0.8,1.15){\line(1,0){0.3}}
\put(1.1,0.85){\line(0,1){0.3}}

 \put(1.6,0.2){\makebox(0,0)[b]{$i$}}
\put(1.4,-0.4){\makebox(0,0)[b]{$rt$}}
\end{picture}\quad + \quad
\begin{picture}(1.6,1.5)
\put(0,0.8){\line(1,1){0.4}} \put(0.,1.2){\line(1,-1){0.4}} \put(0.,-0.2){\line(1,1){0.4}}
\put(0,0.2){\line(1,-1){0.4}}

 \put(0.2,0){\line(1,0){1.4}} \put(0.2,1){\line(1,0){1.4}}

\put(0.5,0){\circle*{0.1}} \put(0.5,1.){\circle*{0.1}} \multiput(0.5,1)(0,-0.125){8}{\line(0,-1){0.09}}

 \put(1.4,1){\circle*{0.1}}\put(1.4,0.){\circle*{0.1}}
\multiput(1.4,1)(0,-0.125){8}{\line(0,-1){0.09}}

\put(0.8,-0.15){\line(1,0){0.3}} \put(0.8,-0.15){\line(0,1){0.3}} \put(0.8,0.15){\line(1,0){0.3}}
\put(1.1,-0.15){\line(0,1){0.3}}

\put(0.8,0.85){\line(1,0){0.3}} \put(0.8,0.85){\line(0,1){0.3}} \put(0.8,1.15){\line(1,0){0.3}}
\put(1.1,0.85){\line(0,1){0.3}}

 \put(1.6,0.2){\makebox(0,0)[b]{$i$}}
\put(1.4,-0.4){\makebox(0,0)[b]{$rt$}}
\end{picture}\quad + \quad
$$

$$
\begin{picture}(2.5,1.5)
\put(0,0.8){\line(1,1){0.4}} \put(0.,1.2){\line(1,-1){0.4}} \put(0.,-0.2){\line(1,1){0.4}}
\put(0,0.2){\line(1,-1){0.4}}

 \put(0.2,0){\line(1,0){2.3}} \put(0.2,1){\line(1,0){2.3}}

\put(0.5,0){\circle*{0.1}} \put(0.5,1.){\circle*{0.1}} \multiput(0.5,1)(0,-0.125){8}{\line(0,-1){0.09}}

 \put(1.4,1){\circle*{0.1}}\put(1.4,0.){\circle*{0.1}}
\multiput(1.4,1)(0,-0.125){8}{\line(0,-1){0.09}}

\put(2.3,0){\circle*{0.1}} \put(2.3,1.){\circle*{0.1}} \multiput(2.3,1)(0,-0.125){8}{\line(0,-1){0.09}}

 \put(2.5,0.2){\makebox(0,0)[b]{$i$}}
\put(2.3,-0.4){\makebox(0,0)[b]{$rt$}}
\end{picture}\quad + \quad
\begin{picture}(2.5,1.5)
\put(0,0.8){\line(1,1){0.4}} \put(0.,1.2){\line(1,-1){0.4}} \put(0.,-0.2){\line(1,1){0.4}}
\put(0,0.2){\line(1,-1){0.4}}

 \put(0.2,0){\line(1,0){2.3}} \put(0.2,1){\line(1,0){2.3}}

\put(0.5,0){\circle*{0.1}} \put(0.5,1.){\circle*{0.1}} \multiput(0.5,1)(0,-0.125){8}{\line(0,-1){0.09}}

 \put(1.4,1){\circle*{0.1}}\put(1.4,0.){\circle*{0.1}}
\multiput(1.4,1)(0,-0.125){8}{\line(0,-1){0.09}}

\put(2.3,0){\circle*{0.1}} \put(2.3,1.){\circle*{0.1}} \multiput(2.3,1)(0,-0.125){8}{\line(0,-1){0.09}}

\put(0.8,-0.15){\line(1,0){0.3}} \put(0.8,-0.15){\line(0,1){0.3}} \put(0.8,0.15){\line(1,0){0.3}}
\put(1.1,-0.15){\line(0,1){0.3}}

 \put(2.5,0.2){\makebox(0,0)[b]{$i$}}
\put(2.3,-0.4){\makebox(0,0)[b]{$rt$}}
\end{picture}\quad + \quad
\begin{picture}(2.5,1.5)
\put(0,0.8){\line(1,1){0.4}} \put(0.,1.2){\line(1,-1){0.4}} \put(0.,-0.2){\line(1,1){0.4}}
\put(0,0.2){\line(1,-1){0.4}}

 \put(0.2,0){\line(1,0){2.3}} \put(0.2,1){\line(1,0){2.3}}

\put(0.5,0){\circle*{0.1}} \put(0.5,1.){\circle*{0.1}} \multiput(0.5,1)(0,-0.125){8}{\line(0,-1){0.09}}

 \put(1.4,1){\circle*{0.1}}\put(1.4,0.){\circle*{0.1}}
\multiput(1.4,1)(0,-0.125){8}{\line(0,-1){0.09}}

\put(2.3,0){\circle*{0.1}} \put(2.3,1.){\circle*{0.1}} \multiput(2.3,1)(0,-0.125){8}{\line(0,-1){0.09}}

\put(0.8,0.85){\line(1,0){0.3}} \put(0.8,0.85){\line(0,1){0.3}} \put(0.8,1.15){\line(1,0){0.3}}
\put(1.1,0.85){\line(0,1){0.3}}

 \put(2.5,0.2){\makebox(0,0)[b]{$i$}}
\put(2.3,-0.4){\makebox(0,0)[b]{$rt$}}
\end{picture}\quad + \quad
\begin{picture}(2.5,1.5)
\put(0,0.8){\line(1,1){0.4}} \put(0.,1.2){\line(1,-1){0.4}} \put(0.,-0.2){\line(1,1){0.4}}
\put(0,0.2){\line(1,-1){0.4}}

 \put(0.2,0){\line(1,0){2.3}} \put(0.2,1){\line(1,0){2.3}}

\put(0.5,0){\circle*{0.1}} \put(0.5,1.){\circle*{0.1}} \multiput(0.5,1)(0,-0.125){8}{\line(0,-1){0.09}}

 \put(1.4,1){\circle*{0.1}}\put(1.4,0.){\circle*{0.1}}
\multiput(1.4,1)(0,-0.125){8}{\line(0,-1){0.09}}

\put(2.3,0){\circle*{0.1}} \put(2.3,1.){\circle*{0.1}} \multiput(2.3,1)(0,-0.125){8}{\line(0,-1){0.09}}

\put(0.8,-0.15){\line(1,0){0.3}} \put(0.8,-0.15){\line(0,1){0.3}} \put(0.8,0.15){\line(1,0){0.3}}
\put(1.1,-0.15){\line(0,1){0.3}}

\put(0.8,0.85){\line(1,0){0.3}} \put(0.8,0.85){\line(0,1){0.3}} \put(0.8,1.15){\line(1,0){0.3}}
\put(1.1,0.85){\line(0,1){0.3}}

 \put(2.5,0.2){\makebox(0,0)[b]{$i$}}
\put(2.3,-0.4){\makebox(0,0)[b]{$rt$}}
\end{picture}\quad + \quad
$$

$$
\begin{picture}(2.5,1.5)
\put(0,0.8){\line(1,1){0.4}} \put(0.,1.2){\line(1,-1){0.4}} \put(0.,-0.2){\line(1,1){0.4}}
\put(0,0.2){\line(1,-1){0.4}}

 \put(0.2,0){\line(1,0){2.3}} \put(0.2,1){\line(1,0){2.3}}

\put(0.5,0){\circle*{0.1}} \put(0.5,1.){\circle*{0.1}} \multiput(0.5,1)(0,-0.125){8}{\line(0,-1){0.09}}

 \put(1.4,1){\circle*{0.1}}\put(1.4,0.){\circle*{0.1}}
\multiput(1.4,1)(0,-0.125){8}{\line(0,-1){0.09}}

\put(2.3,0){\circle*{0.1}} \put(2.3,1.){\circle*{0.1}} \multiput(2.3,1)(0,-0.125){8}{\line(0,-1){0.09}}

\put(1.7,-0.15){\line(1,0){0.3}} \put(1.7,-0.15){\line(0,1){0.3}} \put(1.7,0.15){\line(1,0){0.3}}
\put(2,-0.15){\line(0,1){0.3}}

 \put(2.5,0.2){\makebox(0,0)[b]{$i$}}
\put(2.3,-0.4){\makebox(0,0)[b]{$rt$}}
\end{picture}\quad + \quad
\begin{picture}(2.5,1.5)
\put(0,0.8){\line(1,1){0.4}} \put(0.,1.2){\line(1,-1){0.4}} \put(0.,-0.2){\line(1,1){0.4}}
\put(0,0.2){\line(1,-1){0.4}}

 \put(0.2,0){\line(1,0){2.3}} \put(0.2,1){\line(1,0){2.3}}

\put(0.5,0){\circle*{0.1}} \put(0.5,1.){\circle*{0.1}} \multiput(0.5,1)(0,-0.125){8}{\line(0,-1){0.09}}

 \put(1.4,1){\circle*{0.1}}\put(1.4,0.){\circle*{0.1}}
\multiput(1.4,1)(0,-0.125){8}{\line(0,-1){0.09}}

\put(2.3,0){\circle*{0.1}} \put(2.3,1.){\circle*{0.1}} \multiput(2.3,1)(0,-0.125){8}{\line(0,-1){0.09}}

\put(0.8,-0.15){\line(1,0){0.3}} \put(0.8,-0.15){\line(0,1){0.3}} \put(0.8,0.15){\line(1,0){0.3}}
\put(1.1,-0.15){\line(0,1){0.3}}

\put(1.7,-0.15){\line(1,0){0.3}} \put(1.7,-0.15){\line(0,1){0.3}} \put(1.7,0.15){\line(1,0){0.3}}
\put(2,-0.15){\line(0,1){0.3}}

 \put(2.5,0.2){\makebox(0,0)[b]{$i$}}
\put(2.3,-0.4){\makebox(0,0)[b]{$rt$}}
\end{picture}\quad + \quad
\begin{picture}(2.5,1.5)
\put(0,0.8){\line(1,1){0.4}} \put(0.,1.2){\line(1,-1){0.4}} \put(0.,-0.2){\line(1,1){0.4}}
\put(0,0.2){\line(1,-1){0.4}}

 \put(0.2,0){\line(1,0){2.3}} \put(0.2,1){\line(1,0){2.3}}

\put(0.5,0){\circle*{0.1}} \put(0.5,1.){\circle*{0.1}} \multiput(0.5,1)(0,-0.125){8}{\line(0,-1){0.09}}

 \put(1.4,1){\circle*{0.1}}\put(1.4,0.){\circle*{0.1}}
\multiput(1.4,1)(0,-0.125){8}{\line(0,-1){0.09}}

\put(2.3,0){\circle*{0.1}} \put(2.3,1.){\circle*{0.1}} \multiput(2.3,1)(0,-0.125){8}{\line(0,-1){0.09}}

\put(0.8,0.85){\line(1,0){0.3}} \put(0.8,0.85){\line(0,1){0.3}} \put(0.8,1.15){\line(1,0){0.3}}
\put(1.1,0.85){\line(0,1){0.3}}

\put(1.7,-0.15){\line(1,0){0.3}} \put(1.7,-0.15){\line(0,1){0.3}} \put(1.7,0.15){\line(1,0){0.3}}
\put(2,-0.15){\line(0,1){0.3}}

 \put(2.5,0.2){\makebox(0,0)[b]{$i$}}
\put(2.3,-0.4){\makebox(0,0)[b]{$rt$}}
\end{picture}\quad + \quad
\begin{picture}(2.5,1.5)
\put(0,0.8){\line(1,1){0.4}} \put(0.,1.2){\line(1,-1){0.4}} \put(0.,-0.2){\line(1,1){0.4}}
\put(0,0.2){\line(1,-1){0.4}}

 \put(0.2,0){\line(1,0){2.3}} \put(0.2,1){\line(1,0){2.3}}

\put(0.5,0){\circle*{0.1}} \put(0.5,1.){\circle*{0.1}} \multiput(0.5,1)(0,-0.125){8}{\line(0,-1){0.09}}

 \put(1.4,1){\circle*{0.1}}\put(1.4,0.){\circle*{0.1}}
\multiput(1.4,1)(0,-0.125){8}{\line(0,-1){0.09}}

\put(2.3,0){\circle*{0.1}} \put(2.3,1.){\circle*{0.1}} \multiput(2.3,1)(0,-0.125){8}{\line(0,-1){0.09}}

\put(0.8,-0.15){\line(1,0){0.3}} \put(0.8,-0.15){\line(0,1){0.3}} \put(0.8,0.15){\line(1,0){0.3}}
\put(1.1,-0.15){\line(0,1){0.3}}

\put(0.8,0.85){\line(1,0){0.3}} \put(0.8,0.85){\line(0,1){0.3}} \put(0.8,1.15){\line(1,0){0.3}}
\put(1.1,0.85){\line(0,1){0.3}}

\put(1.7,-0.15){\line(1,0){0.3}} \put(1.7,-0.15){\line(0,1){0.3}} \put(1.7,0.15){\line(1,0){0.3}}
\put(2,-0.15){\line(0,1){0.3}}

 \put(2.5,0.2){\makebox(0,0)[b]{$i$}}
\put(2.3,-0.4){\makebox(0,0)[b]{$rt$}}
\end{picture}\quad + \quad
$$

$$
\begin{picture}(2.5,1.5)
\put(0,0.8){\line(1,1){0.4}} \put(0.,1.2){\line(1,-1){0.4}} \put(0.,-0.2){\line(1,1){0.4}}
\put(0,0.2){\line(1,-1){0.4}}

 \put(0.2,0){\line(1,0){2.3}} \put(0.2,1){\line(1,0){2.3}}

\put(0.5,0){\circle*{0.1}} \put(0.5,1.){\circle*{0.1}} \multiput(0.5,1)(0,-0.125){8}{\line(0,-1){0.09}}

 \put(1.4,1){\circle*{0.1}}\put(1.4,0.){\circle*{0.1}}
\multiput(1.4,1)(0,-0.125){8}{\line(0,-1){0.09}}

\put(2.3,0){\circle*{0.1}} \put(2.3,1.){\circle*{0.1}} \multiput(2.3,1)(0,-0.125){8}{\line(0,-1){0.09}}

\put(1.7,0.85){\line(1,0){0.3}} \put(1.7,0.85){\line(0,1){0.3}} \put(1.7,1.15){\line(1,0){0.3}}
\put(2,0.85){\line(0,1){0.3}}

 \put(2.5,0.2){\makebox(0,0)[b]{$i$}}
\put(2.3,-0.4){\makebox(0,0)[b]{$rt$}}
\end{picture}\quad + \quad
\begin{picture}(2.5,1.5)
\put(0,0.8){\line(1,1){0.4}} \put(0.,1.2){\line(1,-1){0.4}} \put(0.,-0.2){\line(1,1){0.4}}
\put(0,0.2){\line(1,-1){0.4}}

 \put(0.2,0){\line(1,0){2.3}} \put(0.2,1){\line(1,0){2.3}}

\put(0.5,0){\circle*{0.1}} \put(0.5,1.){\circle*{0.1}} \multiput(0.5,1)(0,-0.125){8}{\line(0,-1){0.09}}

 \put(1.4,1){\circle*{0.1}}\put(1.4,0.){\circle*{0.1}}
\multiput(1.4,1)(0,-0.125){8}{\line(0,-1){0.09}}

\put(2.3,0){\circle*{0.1}} \put(2.3,1.){\circle*{0.1}} \multiput(2.3,1)(0,-0.125){8}{\line(0,-1){0.09}}

\put(0.8,-0.15){\line(1,0){0.3}} \put(0.8,-0.15){\line(0,1){0.3}} \put(0.8,0.15){\line(1,0){0.3}}
\put(1.1,-0.15){\line(0,1){0.3}}

\put(1.7,0.85){\line(1,0){0.3}} \put(1.7,0.85){\line(0,1){0.3}} \put(1.7,1.15){\line(1,0){0.3}}
\put(2,0.85){\line(0,1){0.3}}

 \put(2.5,0.2){\makebox(0,0)[b]{$i$}}
\put(2.3,-0.4){\makebox(0,0)[b]{$rt$}}
\end{picture}\quad + \quad
\begin{picture}(2.5,1.5)
\put(0,0.8){\line(1,1){0.4}} \put(0.,1.2){\line(1,-1){0.4}} \put(0.,-0.2){\line(1,1){0.4}}
\put(0,0.2){\line(1,-1){0.4}}

 \put(0.2,0){\line(1,0){2.3}} \put(0.2,1){\line(1,0){2.3}}

\put(0.5,0){\circle*{0.1}} \put(0.5,1.){\circle*{0.1}} \multiput(0.5,1)(0,-0.125){8}{\line(0,-1){0.09}}

 \put(1.4,1){\circle*{0.1}}\put(1.4,0.){\circle*{0.1}}
\multiput(1.4,1)(0,-0.125){8}{\line(0,-1){0.09}}

\put(2.3,0){\circle*{0.1}} \put(2.3,1.){\circle*{0.1}} \multiput(2.3,1)(0,-0.125){8}{\line(0,-1){0.09}}

\put(0.8,0.85){\line(1,0){0.3}} \put(0.8,0.85){\line(0,1){0.3}} \put(0.8,1.15){\line(1,0){0.3}}
\put(1.1,0.85){\line(0,1){0.3}}

\put(1.7,0.85){\line(1,0){0.3}} \put(1.7,0.85){\line(0,1){0.3}} \put(1.7,1.15){\line(1,0){0.3}}
\put(2,0.85){\line(0,1){0.3}}

 \put(2.5,0.2){\makebox(0,0)[b]{$i$}}
\put(2.3,-0.4){\makebox(0,0)[b]{$rt$}}
\end{picture}\quad + \quad
\begin{picture}(2.5,1.5)
\put(0,0.8){\line(1,1){0.4}} \put(0.,1.2){\line(1,-1){0.4}} \put(0.,-0.2){\line(1,1){0.4}}
\put(0,0.2){\line(1,-1){0.4}}

 \put(0.2,0){\line(1,0){2.3}} \put(0.2,1){\line(1,0){2.3}}

\put(0.5,0){\circle*{0.1}} \put(0.5,1.){\circle*{0.1}} \multiput(0.5,1)(0,-0.125){8}{\line(0,-1){0.09}}

 \put(1.4,1){\circle*{0.1}}\put(1.4,0.){\circle*{0.1}}
\multiput(1.4,1)(0,-0.125){8}{\line(0,-1){0.09}}

\put(2.3,0){\circle*{0.1}} \put(2.3,1.){\circle*{0.1}} \multiput(2.3,1)(0,-0.125){8}{\line(0,-1){0.09}}

\put(0.8,-0.15){\line(1,0){0.3}} \put(0.8,-0.15){\line(0,1){0.3}} \put(0.8,0.15){\line(1,0){0.3}}
\put(1.1,-0.15){\line(0,1){0.3}}

\put(0.8,0.85){\line(1,0){0.3}} \put(0.8,0.85){\line(0,1){0.3}} \put(0.8,1.15){\line(1,0){0.3}}
\put(1.1,0.85){\line(0,1){0.3}}

\put(1.7,0.85){\line(1,0){0.3}} \put(1.7,0.85){\line(0,1){0.3}} \put(1.7,1.15){\line(1,0){0.3}}
\put(2,0.85){\line(0,1){0.3}}

 \put(2.5,0.2){\makebox(0,0)[b]{$i$}}
\put(2.3,-0.4){\makebox(0,0)[b]{$rt$}}
\end{picture}\quad + \quad
$$
$$
\begin{picture}(2.5,1.5)
\put(0,0.8){\line(1,1){0.4}} \put(0.,1.2){\line(1,-1){0.4}} \put(0.,-0.2){\line(1,1){0.4}}
\put(0,0.2){\line(1,-1){0.4}}

 \put(0.2,0){\line(1,0){2.3}} \put(0.2,1){\line(1,0){2.3}}

\put(0.5,0){\circle*{0.1}} \put(0.5,1.){\circle*{0.1}} \multiput(0.5,1)(0,-0.125){8}{\line(0,-1){0.09}}

 \put(1.4,1){\circle*{0.1}}\put(1.4,0.){\circle*{0.1}}
\multiput(1.4,1)(0,-0.125){8}{\line(0,-1){0.09}}

\put(2.3,0){\circle*{0.1}} \put(2.3,1.){\circle*{0.1}} \multiput(2.3,1)(0,-0.125){8}{\line(0,-1){0.09}}

\put(1.7,0.85){\line(1,0){0.3}} \put(1.7,0.85){\line(0,1){0.3}} \put(1.7,1.15){\line(1,0){0.3}}
\put(2,0.85){\line(0,1){0.3}}

\put(1.7,-0.15){\line(1,0){0.3}} \put(1.7,-0.15){\line(0,1){0.3}} \put(1.7,0.15){\line(1,0){0.3}}
\put(2,-0.15){\line(0,1){0.3}}

 \put(2.5,0.2){\makebox(0,0)[b]{$i$}}
\put(2.3,-0.4){\makebox(0,0)[b]{$rt$}}
\end{picture}\quad + \quad
\begin{picture}(2.5,1.5)
\put(0,0.8){\line(1,1){0.4}} \put(0.,1.2){\line(1,-1){0.4}} \put(0.,-0.2){\line(1,1){0.4}}
\put(0,0.2){\line(1,-1){0.4}}

 \put(0.2,0){\line(1,0){2.3}} \put(0.2,1){\line(1,0){2.3}}

\put(0.5,0){\circle*{0.1}} \put(0.5,1.){\circle*{0.1}} \multiput(0.5,1)(0,-0.125){8}{\line(0,-1){0.09}}

 \put(1.4,1){\circle*{0.1}}\put(1.4,0.){\circle*{0.1}}
\multiput(1.4,1)(0,-0.125){8}{\line(0,-1){0.09}}

\put(2.3,0){\circle*{0.1}} \put(2.3,1.){\circle*{0.1}} \multiput(2.3,1)(0,-0.125){8}{\line(0,-1){0.09}}

\put(0.8,-0.15){\line(1,0){0.3}} \put(0.8,-0.15){\line(0,1){0.3}} \put(0.8,0.15){\line(1,0){0.3}}
\put(1.1,-0.15){\line(0,1){0.3}}

\put(1.7,0.85){\line(1,0){0.3}} \put(1.7,0.85){\line(0,1){0.3}} \put(1.7,1.15){\line(1,0){0.3}}
\put(2,0.85){\line(0,1){0.3}}

\put(1.7,-0.15){\line(1,0){0.3}} \put(1.7,-0.15){\line(0,1){0.3}} \put(1.7,0.15){\line(1,0){0.3}}
\put(2,-0.15){\line(0,1){0.3}}

 \put(2.5,0.2){\makebox(0,0)[b]{$i$}}
\put(2.3,-0.4){\makebox(0,0)[b]{$rt$}}
\end{picture}\quad + \quad
\begin{picture}(2.5,1.5)
\put(0,0.8){\line(1,1){0.4}} \put(0.,1.2){\line(1,-1){0.4}} \put(0.,-0.2){\line(1,1){0.4}}
\put(0,0.2){\line(1,-1){0.4}}

 \put(0.2,0){\line(1,0){2.3}} \put(0.2,1){\line(1,0){2.3}}

\put(0.5,0){\circle*{0.1}} \put(0.5,1.){\circle*{0.1}} \multiput(0.5,1)(0,-0.125){8}{\line(0,-1){0.09}}

 \put(1.4,1){\circle*{0.1}}\put(1.4,0.){\circle*{0.1}}
\multiput(1.4,1)(0,-0.125){8}{\line(0,-1){0.09}}

\put(2.3,0){\circle*{0.1}} \put(2.3,1.){\circle*{0.1}} \multiput(2.3,1)(0,-0.125){8}{\line(0,-1){0.09}}

\put(0.8,0.85){\line(1,0){0.3}} \put(0.8,0.85){\line(0,1){0.3}} \put(0.8,1.15){\line(1,0){0.3}}
\put(1.1,0.85){\line(0,1){0.3}}

\put(1.7,0.85){\line(1,0){0.3}} \put(1.7,0.85){\line(0,1){0.3}} \put(1.7,1.15){\line(1,0){0.3}}
\put(2,0.85){\line(0,1){0.3}}

\put(1.7,-0.15){\line(1,0){0.3}} \put(1.7,-0.15){\line(0,1){0.3}} \put(1.7,0.15){\line(1,0){0.3}}
\put(2,-0.15){\line(0,1){0.3}}

 \put(2.5,0.2){\makebox(0,0)[b]{$i$}}
\put(2.3,-0.4){\makebox(0,0)[b]{$rt$}}
\end{picture}\quad + \quad
\begin{picture}(2.5,1.5)
\put(0,0.8){\line(1,1){0.4}} \put(0.,1.2){\line(1,-1){0.4}} \put(0.,-0.2){\line(1,1){0.4}}
\put(0,0.2){\line(1,-1){0.4}}

 \put(0.2,0){\line(1,0){2.3}} \put(0.2,1){\line(1,0){2.3}}

\put(0.5,0){\circle*{0.1}} \put(0.5,1.){\circle*{0.1}} \multiput(0.5,1)(0,-0.125){8}{\line(0,-1){0.09}}

 \put(1.4,1){\circle*{0.1}}\put(1.4,0.){\circle*{0.1}}
\multiput(1.4,1)(0,-0.125){8}{\line(0,-1){0.09}}

\put(2.3,0){\circle*{0.1}} \put(2.3,1.){\circle*{0.1}} \multiput(2.3,1)(0,-0.125){8}{\line(0,-1){0.09}}

\put(0.8,-0.15){\line(1,0){0.3}} \put(0.8,-0.15){\line(0,1){0.3}} \put(0.8,0.15){\line(1,0){0.3}}
\put(1.1,-0.15){\line(0,1){0.3}}

\put(0.8,0.85){\line(1,0){0.3}} \put(0.8,0.85){\line(0,1){0.3}} \put(0.8,1.15){\line(1,0){0.3}}
\put(1.1,0.85){\line(0,1){0.3}}

\put(1.7,0.85){\line(1,0){0.3}} \put(1.7,0.85){\line(0,1){0.3}} \put(1.7,1.15){\line(1,0){0.3}}
\put(2,0.85){\line(0,1){0.3}}

\put(1.7,-0.15){\line(1,0){0.3}} \put(1.7,-0.15){\line(0,1){0.3}} \put(1.7,0.15){\line(1,0){0.3}}
\put(2,-0.15){\line(0,1){0.3}}

 \put(2.5,0.2){\makebox(0,0)[b]{$i$}}
\put(2.3,-0.4){\makebox(0,0)[b]{$rt$}}
\end{picture}\quad + \quad
$$
\begin{equation} \label{BlcLadder}
\begin{picture}(3.4,1.5)
\put(0,0.8){\line(1,1){0.4}} \put(0.,1.2){\line(1,-1){0.4}} \put(0.,-0.2){\line(1,1){0.4}}
\put(0,0.2){\line(1,-1){0.4}}

 \put(0.2,0){\line(1,0){3.2}} \put(0.2,1){\line(1,0){3.2}}

\put(0.5,0){\circle*{0.1}} \put(0.5,1.){\circle*{0.1}} \multiput(0.5,1)(0,-0.125){8}{\line(0,-1){0.09}}

 \put(1.4,1){\circle*{0.1}}\put(1.4,0.){\circle*{0.1}}
\multiput(1.4,1)(0,-0.125){8}{\line(0,-1){0.09}}

\put(2.3,0){\circle*{0.1}} \put(2.3,1.){\circle*{0.1}} \multiput(2.3,1)(0,-0.125){8}{\line(0,-1){0.09}}

\put(3.2,0){\circle*{0.1}} \put(3.2,1.){\circle*{0.1}} \multiput(3.2,1)(0,-0.125){8}{\line(0,-1){0.09}}

 \put(3.4,0.2){\makebox(0,0)[b]{$i$}}
\put(3.2,-0.4){\makebox(0,0)[b]{$rt$}}
\end{picture}\quad + \quad
\begin{picture}(3.4,1.5)
\put(0,0.8){\line(1,1){0.4}} \put(0.,1.2){\line(1,-1){0.4}} \put(0.,-0.2){\line(1,1){0.4}}
\put(0,0.2){\line(1,-1){0.4}}

 \put(0.2,0){\line(1,0){3.2}} \put(0.2,1){\line(1,0){3.2}}

\put(0.5,0){\circle*{0.1}} \put(0.5,1.){\circle*{0.1}} \multiput(0.5,1)(0,-0.125){8}{\line(0,-1){0.09}}

 \put(1.4,1){\circle*{0.1}}\put(1.4,0.){\circle*{0.1}}
\multiput(1.4,1)(0,-0.125){8}{\line(0,-1){0.09}}

\put(2.3,0){\circle*{0.1}} \put(2.3,1.){\circle*{0.1}} \multiput(2.3,1)(0,-0.125){8}{\line(0,-1){0.09}}

\put(3.2,0){\circle*{0.1}} \put(3.2,1.){\circle*{0.1}} \multiput(3.2,1)(0,-0.125){8}{\line(0,-1){0.09}}

\put(0.8,-0.15){\line(1,0){0.3}} \put(0.8,-0.15){\line(0,1){0.3}} \put(0.8,0.15){\line(1,0){0.3}}
\put(1.1,-0.15){\line(0,1){0.3}}

 \put(3.4,0.2){\makebox(0,0)[b]{$i$}}
\put(3.2,-0.4){\makebox(0,0)[b]{$rt$}}
\end{picture}\quad + \quad \cdots
\end{equation}

By introducing the effective Green's function (Eq.(6.16) in the main paper), we can rewrite the series Eq.(\ref{BlcLadder}) in the simpler form given by Eq.(6.15) in the main paper.

\section*{Supplementary Note VII.   Self-consistent and steady-state rates for multistage reactive system}

Let's consider the case of the electron transfer from an excited particle $D^*$ to the acceptor $A$ in the bulk, generating a geminate pair $D^+$ and $A^-$.
\begin{equation}
 \begin{array}{ccc}
D^*  + A\rightarrow& [D^+ \dots A^-]& \rightarrow D + A\\
\\
&\downarrow& \\
\\
&D^+ + A^-& \rightarrow D + A
\end{array}
\label{Dorf}
\end{equation}

The $D^+$ and $A^-$ particles can then recombine via reverse electron transfer during geminate kinetics or can separate and subsequently recombine in a bulk reaction.
The kinetics of this reactive system is simulated within CMET in subsection X.D of the main article.
According to the procedure described in Supplementary Note V, there are three non-zero non-diagonal elements of the matrix   $\widehat {\bf K}$ , corresponding to the following reactions:

\begin{eqnarray}
 &K_{D^+A^-,D^*A} &\quad \mbox{to} \quad D^*  + A \rightarrow D^+ + A^- \nonumber\\
&K_{DA,D^*A} & \quad \mbox{to}\quad D^*  + A \rightarrow D + A \nonumber\\
& K_{DA,D^+A^-} & \quad \mbox{to}\quad D^+ + A^-  \rightarrow D + A \nonumber\\
\end{eqnarray}

In order to write down SCRTA equations \cite{GopichSzaboJCP2002}  for this system one has to find these reaction rates. It is suggested in Ref. \onlinecite{GopichSzaboJCP2002} that steady-state or self-consistent reaction rates can be used for this purpose.
As can be concluded from Fig. 3 in the main article the kinetics of the system (\ref{Dorf}) can not be described by LMA equations.
Thus, it is not clear how  to define the steady-state reaction rates in this case.

The self-consistent SCRTA rate constants for the reaction $  A+B \leftrightarrow C + D$ are defined in Ref. \cite{GopichSzaboJCP2002} via the integral of the relaxation function:
\begin{equation}
    R(t) = \frac{C_{A}(t) - C_{A}(\infty)}{C_{A}(0) - C_{A}(\infty)}=\frac{C_B(t) - C_{B}(\infty)}{C_{B}(0) - C_{B}(\infty)}=\frac{C_{C}(t) - C_{C}(\infty)}{C_{C}(0) - C_{C}(\infty)}
    =\frac{C_{D}(t) - C_{D}(\infty)}{C_{D}(0) - C_{D}(\infty)}
\end{equation}

In case of  reactive system (\ref{Dorf}) one has to deal with three different relaxation functions instead of one:
\begin{equation}
    R_{D*}(t) = \frac{C_{D^*}(t) - C_{D^*}(\infty)}{C_{D^*}(0) - C_{D^*}(\infty)}
\end{equation}
\begin{equation}
    R_{A}(t) = \frac{C_{A}(t) - C_{A}(\infty)}{C_{A}(0) - C_{A}(\infty)}
\end{equation}
\begin{equation}
    R_{D+}(t) = \frac{C_{D^+}(t) - C_{D^+}(\infty)}{C_{D^+}(0) - C_{D^+}(\infty)}
\end{equation}

Unfortunately,   $R_{D+}(t)$  and $R_{A}(t)$ are defined incorrectly in this way,  since $C_{D^+}(0) = C_{D^+}(\infty) = 0$ and $C_{A}(0) = C_{A}(\infty)$.
Thus, the self-consistent rates in this case can not be correctly defined.

\newpage
\section*{Supporting Figures}

\begin{figure}[h!]
    \centering
    \includegraphics[width=0.8\linewidth]{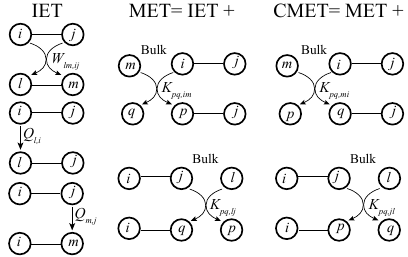}
    \caption{Schematic representation of processes contributing to pair correlation dynamics/interconversion in IET, MET, and CMET. Correlations are represented by solid lines connecting two circles with particle indices. Chemical reactions (labeled by arrows) result in correlation interconversions.  \textbf{Left Column -- Correlation interconversions in IET (top to bottom):} Bimolecular reaction $A_i+A_j\to A_l +A_m$ converts correlation in a pair $(i,j)$ to pair $(l,m)$; Monomolecular reaction $A_i\to A_l$ converts correlation in a pair $(i,j)$ to pair $(l,j)$; Monomolecular reaction $A_j\to A_m$ converts correlation in a pair $(i,j)$ to pair $(i,m)$. \textbf{Middle Column -- Additional correlation interconversions in MET (top to bottom):} Bulk bimolecular reaction $A_i+A_m\to A_p +A_q$ converts correlation in a pair $(i,j)$ to a pair $(p,j)$; Bulk bimolecular reaction $A_j+A_l\to A_q +A_p$ converts correlation in a pair $(i,j)$ to a pair $(i,q)$. \textbf{Right Column -- Additional correlation interconversions in CMET (top to bottom):} Bulk bimolecular reaction $A_i+A_m\to A_p +A_q$ also converts correlation in a pair $(i,j)$ to a pair $(q ,j)$; Bulk bimolecular reaction $A_j+A_l\to A_q +A_p$ also converts correlation in a pair $(i,j)$ to a pair $(i,p)$
    }
    \label{fig:S1}
\end{figure}

\bibliographystyle{unsrt}
\bibliography{bib}